\documentclass[superscriptaddress,twocolumn,showpacs,a4paper,
amssymb,amsmath,nobibnotes,aps,prd,
showkeys,
nofootinbib,notitlepage]{revtex4-1}
\usepackage{verbatim}
\usepackage[T1]{fontenc}
\usepackage[utf8]{inputenc}
\usepackage[american]{babel}
\usepackage{epsfig}
\usepackage{booktabs}
\usepackage{multirow}
\usepackage{dcolumn}
\usepackage{amsmath}
\usepackage{mathtools}
\usepackage{amsfonts}
\usepackage{amssymb}
\usepackage{ulem}
\usepackage{epstopdf}
\usepackage{bm}
\usepackage{siunitx}
\usepackage{braket}
\usepackage{enumitem}
\usepackage{soul}
\usepackage[table]{xcolor}
\usepackage{color}
\usepackage{transparent}
\usepackage{pifont}
\usepackage{enumitem}

\definecolor{navyblue}{rgb}{0.0, 0.0, 0.5}
\definecolor{royalblue}{rgb}{0.25, 0.41, 0.88}
\definecolor{cadmiumgreen}{rgb}{0.0, 0.42, 0.24}
\definecolor{blue-violet}{rgb}{0.54, 0.17, 0.89}
\definecolor{darkviolet}{rgb}{0.58, 0.0, 0.83}
\definecolor{orange(colorwheel)}{rgb}{1.0, 0.5, 0.0}

\usepackage{hyperref}
\hypersetup{
    colorlinks=true, 
    linkcolor=royalblue, 
    citecolor=magenta}

\usepackage{booktabs}
\usepackage{multirow}
\usepackage{dcolumn}
\usepackage{colortbl}

\begin{document}

\title{Inflation, the Hubble Tension and Early Dark Energy: an alternative overview}

\author{William Giar\`{e}}
\email{w.giare@sheffield.ac.uk}
\affiliation{School of Mathematics and Statistics, University of Sheffield, Hounsfield Road, Sheffield S3 7RH, United Kingdom \looseness=-1}

\date{\today}

\begin{abstract}
\noindent 
I review and discuss the possible implications for inflation resulting from considering new physics in light of the Hubble tension. My study is motivated by a simple argument that the constraints on inflationary parameters, most typically the spectral index $n_s$, depend to some extent on the cosmological framework. To avoid broadening the uncertainties resulting from marginalizing over additional parameters (typical in many alternative models), I first adopt the same alternative viewpoint of previous studies and analyze what happens if a physical theory can \textit{fix} extra parameters to non-standard values. Focusing on the dark energy equation of state $w$ and the effective number of relativistic species $N_{\rm{eff}}$, I confirm that physical theories able to fix $w \approx -1.2$ or $N_{\rm{eff}} \approx 3.9$ produce values of $H_0$ from CMB and BAO in line with the local distance ladder estimate. While in the former case I do not find any relevant implications for inflation, in the latter scenarios, I observe a shift towards $n_s \approx 1$. From a model-selection perspective, both cases are strongly disfavored compared to $\Lambda$CDM. However, models with $N_{\rm{eff}} \approx 3.3 - 3.4$ could bring the $H_0$ tension down to $\sim 3\sigma$ while being moderately disfavored. Yet, this is enough to change the constraints on inflation so that the most accredited models (e.g., Starobinsky inflation) would no longer be favored by data. I then focus on Early Dark Energy (EDE), arguing that an EDE fraction $f_{\rm{EDE}}\sim 0.04 - 0.06$ (only able to mildly reduce the $H_0$-tension down to $\sim 3\sigma$) could already require a similar change in perspective on inflation. In fact, performing a full joint analysis of EDE and Starobinsky inflation, I find that the two models can hardly coexist for $f_{\rm{EDE}}\gtrsim 0.06$.
\end{abstract}

\maketitle


\section{Introduction}
\label{sec:introduction}
Cosmological inflation, a phase of rapid accelerated expansion of space-time in a nearly de Sitter background geometry, is widely recognized as the leading theory to set the initial conditions in the very early Universe.  While it was originally proposed to account for various observational issues, including spatial flatness, the horizon and entropy problems, and the apparent lack of topological defects~\cite{Guth:1980zm, Linde:1981mu, Albrecht:1982wi,Vilenkin:1983xq}, inflation offers also an elegant framework to explain the physical origin of the first fluctuations in the Universe which eventually gave rise to observed structures such as galaxies and clusters of galaxies~\cite{Mukhanov:1981xt, Bardeen:1983qw, Hawking:1982cz, Guth:1982ec}. 

The significant advancements in observational astrophysics and cosmology, culminating in a vast array of experiments probing different cosmic epochs and scales~\cite{WMAP:2012fli,WMAP:2012nax,Planck:2018nkj,Planck:2018vyg,SPT:2004qip,SPT-3G:2021eoc,ACT:2020frw,ACT:2020gnv,ACT:2023kun,BOSS:2012dmf,BOSS:2013rlg,BOSS:2014hwf,BOSS:2016wmc,BOSS:2013uda,eBOSS:2020yzd,SDSS:2003eyi,SDSS:2004kqt,SDSS:2006lmn,SDSS:2014iwm,DES:2016jjg,DES:2017qwj,DES:2021wwk,DES:2022ccp,KiDS:2020suj,deJong:2012zb,KiDS:2020ghu,Kilo-DegreeSurvey:2023gfr,Pan-STARRS1:2017jku,Brout:2022vxf,LIGO_SGWB-2017,LIGO_SGWB-2019,TheLIGOScientific:2016dpb}, have allowed for precise constraints on the inflationary Universe and its related observables. Over the last few decades, several inflationary models and theories have been tested against a wide range of available data, including Cosmic Microwave Background (CMB), Big Bang Nucleosynthesis (BBN), and Gravitational Wave (GW) measurements\footnote{With no claim to be exhaustive, for works in this direction see, e.g., Refs.~\cite{Leach:2002ar,Boubekeur:2005zm,Martin:2006rs,Moss:2007qd,Bezrukov:2010jz,Zhao:2011zb,Martin:2013nzq,Martin:2014rqa,Martin:2014lra,Carrillo-Gonzalez:2014tia,Creminelli:2014oaa,DiValentino:2016nni,DiValentino:2016ziq,Campista:2017ovq,Giare:2019snj,Forconi:2021que,Dai:2019ejv,Baumann:2015xxa,Odintsov:2020ilr,Giare:2020plo,Oikonomou:2021kql,Odintsov:2022cbm,Namba:2015gja,Peloso:2016gqs,Pi:2019ihn,Ozsoy:2020ccy,Stewart:2007fu, Mukohyama:2014gba,Giovannini:2015kfa,Giovannini:2018dob,Giovannini:2018nkt,Giovannini:2018zbf,Giare:2020vhn,Giare:2020vss,Giare:2020plo,Giare:2022wxq,Baumgart:2021ptt,Franciolini:2018ebs,DEramo:2019tit,Giare:2019snj,Caldwell:2018giq,Clarke:2020bil,Caprini:2018mtu,Allen:1997ad,Smith:2006nka,Boyle:2007zx,Kuroyanagi:2014nba,Ben-Dayan:2019gll,Aich:2019obd,Cabass:2015jwe,Vagnozzi:2020gtf,Benetti:2021uea,Calcagni:2020tvw,Oikonomou:2022ijs,Barrow:1993ad,Peng:2021zon,Ota:2022hvh,Odintsov:2022sdk,Baumgart:2021ptt,Capurri:2020qgz,Canas-Herrera:2021sjs,Odintsov:2023aaw,Oikonomou:2023bah,Fronimos:2023tim,Fronimos:2023tim,Cai:2022lec,Oikonomou:2022irx,Gangopadhyay:2022vgh,Odintsov:2022hxu,Odintsov:2022cbm,Odintsov:2020mkz,Galloni:2022mok,Braglia:2022phb,Giare:2023kiv,Antoniadis:2023zhi,NANOGrav:2023gor,NANOGrav:2023hvm,Vagnozzi:2023lwo,Oikonomou:2023qfz,Iacconi:2020yxn,Iacconi:2021ltm,Iacconi:2023mnw,Santos:2023bnu,Pozo:2023dto,Giare:2024sdl,Cecchini:2024xoq,Wang:2024vfv} and references therein. See also the recent Ref.~\cite{Martin:2024qnn} for a detailed Bayesian model comparison involving nearly three hundred models of single-field slow-roll inflation. }. Despite embedding inflation within a more fundamental theory remains an open problem~\cite{Lyth:1998xn,Linde:2007fr,Martin:2013nzq,Baumann:2014nda} and a large plethora of proposed models and mechanisms can be deemed equally viable in describing current observations~\cite{Martin:2013nzq,Planck:2018jri}, a few general guidelines can be drawn from existing data. 

For instance, while multifield models remain theoretically appealing\footnote{Low-energy effective field theories inspired by theories of particle physics beyond the Standard Model or quantum gravity, often incorporate multiple scalar degrees of freedom and suggest that inflation could be driven by multiple fields, potentially featuring non-minimal couplings, see e.g., Refs.~\cite{Starobinsky:2001xq,Tsujikawa:2002qx,DiMarco:2002eb,Kaiser:2010ps,Achucarro:2010da,vandeBruck:2010yw,Kaiser:2013sna,vandeBruck:2015tna,vandeBruck:2015xpa,vandeBruck:2016vlw,Carrilho:2018ffi,Achucarro:2019pux,Pinol:2020kvw,Achucarro:2018vey,vandeBruck:2021xkm,DeAngelis:2023fdu,Tsujikawa:2000wc,Weinberg:2004kf,Kaiser:2010yu,Frazer:2013zoa,Achucarro:2012fd,vandeBruck:2014ata,Dias:2015rca,Dias:2016rjq,Braglia:2020fms,Braglia:2021ckn,Cabass:2022ymb,Geller:2022nkr,Wang:2022eop,Iacconi:2023slv,DeAngelis:2023fdu,Qin:2023lgo,Freytsis:2022aho,Geller:2022nkr,Cicoli:2021yhb,Pinol:2020kvw,Guerrero:2020lng,Garcia-Saenz:2019njm,Nguyen:2019kbm,Li:2019zbk,Bernardeau:2002jy,Kaiser:2012ak,McAllister:2012am,Peterson:2011yt,Dias:2012nf,Kehagias:2012td,Leung:2012ve,Meyers:2010rg,Price:2014ufa,Battefeld:2011yj,Kaiser:2015usz,Ashcroft:2002ap,Paliathanasis:2021fxi,Paliathanasis:2020abu,Paliathanasis:2020wjl,Christodoulidis:2021vye,Piao:2006nm,Rinaldi:2023mdf,Ijaz:2023cvc,Giare:2023kiv}.}, the absence of evidence for isocurvature modes in the CMB and the high level of Gaussianity observed in primordial perturbations~\cite{Planck:2018jri,Planck:2019kim} suggest that inflation might be well described in terms of a single scalar field undergoing slow-roll evolution on its potential. Within this framework, the physical properties of primordial (scalar and tensor) perturbations are captured by the two-point correlation functions or primordial power spectra\footnote{For a generic Gaussian random field $\psi_k$, the power spectrum is defined in terms of its two-point correlation function as $\langle \psi_k\,\psi_{k\prime}\rangle\doteq (2\pi)^3\delta^{3}_{k+k\prime} \,P_{\psi}(k)$. A \textit{dimensionless} version of the power spectrum can be defined as $\mathcal P_{\psi}(k)\doteq (k^3/2\pi^2)\,P_{\psi}(k)$.}. 

Since the spectrum of the quantum fluctuations of a (massless) scalar field in a de Sitter background is flat, a broadly expected outcome of single-field models is that the power spectra of scalar and tensor perturbations in a quasi de Sitter geometry should be nearly, but not precisely, flat. The residual scale-dependence can be quantified by the so-called scalar and tensor tilts, $n_s$ and $n_{\rm t}$, respectively.  For scalar perturbations, the most recent observations of the CMB temperature and polarization anisotropies — echoes of the Big Bang — provided by the Planck satellite~\cite{Planck:2018vyg,Planck:2018nkj}, measure $n_s=0.9649\pm0.0044$, ruling out a scale-invariant Harrison-Zel'dovich spectrum~\cite{Harrison:1969fb,Zeldovich:1972zz,Peebles:1970ag} (corresponding to $n_{\rm s}= 1$) at approximately 8 standard deviations. This is universally recognized as one of the (if not the) most important indirect observational evidence supporting inflation as the favored early Universe theory.  

Concerning tensor modes, within single-field slow-roll inflation minimally coupled to gravity, the tensor amplitude and tilt can be linked by the well-known slow-roll consistency relation~\cite{Martin:2013tda,Caprini:2018mtu} $r=-8\,n_{\rm t}$ (where $r=A_{\rm t} / A_s$ is the tensor-to-scalar-ratio). While other physical mechanisms beyond inflation have been proposed, predicting similar scale-dependence for the scalar spectrum, this relation remains a unique testable prediction of single-field models. For this reason, the detection of primordial gravitational waves is widely regarded as another necessary step to provide direct smoking-gun evidence for inflation and eliminate any competing models. However, despite the best efforts, a detection of primordial tensor modes remains elusive, and the combined analysis of current Planck and BICEP/Keck array (BK18) data only sets an upper bound $r < 0.037$ at a $95\%$ confidence level on the amplitude of primordial gravitational waves~\cite{BICEP:2021xfz}\footnote{Upcoming CMB experiments~\cite{BICEP3,CLASS,SPT-3G,ACTPol,LBIRD,CMB-S4} are expected to reach a sensitivity $r \sim 0.01 - 0.001$, potentially leading to the first detection of tensor modes for sufficiently high-scale inflation.}. 

The absence of direct detection of B-mode polarization remains a clear and important limitation for inflation. However, at the model selection level, current constraints on the amplitude of primordial gravitational waves and the spectral index of scalar modes derived from the analysis of Planck satellite data and B-mode polarization measurements released by the Bicep/Keck collaboration impose significant limitations on the category of models able to explain CMB observations at large angular scales.  Notably, the model proposed by Starobinsky~\cite{Starobinsky:1980te} (where inflation is obtained by including an additional $R^2$ term in the Einstein-Hilbert action) or its $\alpha$-attractor extension~\cite{Kallosh:2013yoa,Kehagias:2013mya} represent benchmark scenarios, being well-motivated from a theoretical standpoint and among the best-fit options for current large-scale CMB observations~\cite{Planck:2018jri,Giare:2023wzl}.

Given the remarkable success and overall good health of the inflationary paradigm, there is a widespread sense in the cosmological community that a solid understanding of the inflationary Universe and its related properties has been attained. While this is of course true, I wish to emphasize that, as often happens in cosmology, much of our knowledge is based on indirect evidence. We rely on results obtained from the analysis of specific datasets under certain precise theoretical frameworks. For instance, the prevailing narrative thus far relies solely on the analysis of Planck (and BK18) data, assuming a standard $\Lambda$CDM cosmology. On one side, while this model offers a robust description of our observable Universe supported by a wide range of cosmological and astrophysical experiments, several discrepancies between various observations have emerged in recent decades, questioning its validity across all possible cosmological epochs and scales~\cite{Perivolaropoulos:2021jda,DiValentino:2022fjm,Abdalla:2022yfr}. On the other side, when looking at CMB experiments other than Planck, the situation becomes somewhat confusing, with several emerging discrepancies among different experiments being reported by several independent groups~\cite{Lin:2019zdn,Forconi:2021que,Handley:2020hdp,LaPosta:2022llv,DiValentino:2022rdg,DiValentino:2022oon,Giare:2022rvg,Calderon:2023obf,Giare:2023xoc}.

In this context, with no claim of disputing our understanding of the inflationary Universe or diminishing the efforts that have enabled such knowledge, I would like to draw attention to some aspects that, when considered together, might inspire reflections on the present state of inflationary cosmology.

Firstly, the aforementioned discrepancies among different CMB experiments primarily involve inflationary parameters~\cite{Forconi:2021que, Giare:2022rvg, DiValentino:2022oon,Gariazzo:2024sil}. Unexpectedly, small-scale CMB measurements provided by the Data Release 4 of the Atacama Cosmology Telescope (ACT) show an agreement with a scale invariant Harrison-Zel'dovich spectrum ($n_s=1.009 \pm 0.015$), introducing a tension with a significance of $99.3\%$ confidence level with the results from the Planck satellite~\cite{Giare:2022rvg}. If we set aside observational systematics and consider the differences between ACT and Planck as genuine, the analysis of small-scale CMB data lead to different predictions for inflationary models~\cite{Giare:2023wzl}.

Secondly, and most importantly, as suggested by several scattered studies~\cite{DiValentino:2018zjj,Ye:2022efx,Jiang:2022uyg, Jiang:2022qlj,Takahashi:2021bti,Lin:2022gbl,Hazra:2022rdl,Braglia:2021sun,Keeley:2020rmo,Jiang:2023bsz}, the well-known Hubble tension~\cite{Verde:2019ivm,DiValentino:2021izs,Kamionkowski:2022pkx,Khalife:2023qbu} -- i.e., a $\sim 5 \sigma$ discrepancy between the values of the Hubble constant measured by the SH0ES collaboration using the luminosity distances of Type Ia supernovae~\cite{Riess:2021jrx} ($H_0=73\pm1$ km/s/Mpc) and the value derived from the Planck satellite~\cite{Planck:2018vyg} ($H_0=67.4\pm0.5$ km/s/Mpc) -- can represent another layer of uncertainty in our understanding of inflation. 

As many have argued, resolving the Hubble tension might require a paradigm shift in cosmology, introducing new physics into the standard cosmological model.  Although it seems unlikely that the new physics needed to resolve the Hubble tension would come entirely from the inflationary sector of the theory, in general, considering new physics within the cosmological model could lead to changes in the values of cosmological parameters inferred from data. In this regard, I would like to highlight one more time that, just like with $H_0$, the results concerning inflationary parameters discussed so far (most typically $n_s$) are derived from the analysis of CMB data within the standard cosmological model. Considering different models may affect -- although indirectly -- our understanding of the inflationary Universe. A striking example supporting this claim comes from Early Dark Energy (EDE), a highly studied theoretical extension representing a possible solution to the Hubble tension~\cite{Wetterich:2004pv,Doran:2006kp,Hollenstein:2009ph,Calabrese:2010uf,Calabrese:2011hg,Calabrese:2011hg,Pettorino:2013ia,Archidiacono:2014msa,Poulin:2023lkg,Poulin:2018dzj,Poulin:2018zxs,Smith:2019ihp,Niedermann:2019olb,Niedermann:2020dwg, Murgia:2020ryi,Ye:2020btb,Klypin:2020tud,Hill:2020osr,Herold:2021ksg,Herold:2022iib,Reeves:2022aoi,Jiang:2022uyg,Simon:2022adh,Smith:2022hwi,Nakagawa:2022knn,Kamionkowski:2022pkx,Niedermann:2023ssr,Cruz:2023lmn,Eskilt:2023nxm,Smith:2023oop,Sharma:2023kzr,Efstathiou:2023fbn,Gsponer:2023wpm,Goldstein:2023gnw,Forconi:2023hsj,Fu:2023tfo}. No matter whether analyzed in light of Planck or ACT data, this model consistently yields a higher fitting value for the scalar spectral index~\cite{DiValentino:2018zjj,Ye:2021nej,Ye:2022efx,Jiang:2022uyg,Jiang:2022qlj,Takahashi:2021bti,Jiang:2023bsz,Peng:2023bik,Forconi:2023hsj,Fu:2023tfo}. This suggests that, in EDE cosmology, resolving the Hubble tension requires moving towards the same direction favored by small-scale CMB experiments (i.e., towards larger values of $n_s$)\footnote{Notably, evidence at a 3 standard deviation level in favor of EDE is obtained precisely from such experiments, reflecting, among other things, this preference for a higher $n_s$~\cite{Hill:2020osr}.}.

Although some aspects have already been discussed in the literature regarding the preference for higher values of $n_s$ in theoretical models aimed at resolving the Hubble tension, I feel that certain points remain partially unanswered or not entirely clear. For example, I should note that in many of these theoretical solutions involving new physics beyond the standard cosmological model (including EDE), additional parameters are also introduced compared to $\Lambda$CDM. On the one hand, marginalizing over additional parameters has the effect of broadening the uncertainties within which we can extrapolate their values. On the other hand, we introduce additional degeneracy lines among the parameters themselves. That is, very often, simultaneously varying different parameters can lead to very similar effects on cosmological observables (such as the CMB angular power spectra), thus resulting in correlations along which the total effects cancel out. This raises the question of whether the shift observed in the inflationary parameter is a genuine consequence of introducing new physics in the cosmological model or more an artifact of the enlarged parameter space producing additional correlations. Additionally, very often, when taking the results on the inflationary parameters inferred in alternative cosmological models at face value, we remain in agreement with the theoretical predictions of benchmark scenarios of inflation just because of the larger uncertainties, despite observing significant shifts in their central values. This makes it somewhat unclear to what extent the Hubble tension represents an additional layer of uncertainty for inflationary cosmology.

To the best of my knowledge, in Ref.~\cite{Vagnozzi:2019ezj} similar concerns were raised against extended cosmological models with many parameters proposed as solutions of the $H_0$-tension. In that work, this problem was studied by adopting a somewhat different perspective, namely considering theoretical scenarios able to introduce new physics by \textit{fixing} cosmological parameters to non-standard values.  In practice, it was analyzed what happens to the Hubble tension if a physical theory is able to fix (or approximately fix) the dark energy equation of state ($w$) or the effective number of relativistic species ($N_{\text{eff}}$) to a specific set of non-standard values, embodying phantom dark energy ($w < -1$) or extra relativistic particles before recombination ($N_{\text{eff}} > 3.046$). It was argued that models able to predict $N_{\text{eff}} \simeq 3.45$ are able to reduce the tension down to $2\sigma$ while only being weakly disfavored with respect to $\Lambda$CDM from a model comparison point of view. However, nothing was said about the possible implications for other cosmological parameters, nor were the implications for inflation discussed altogether. 

More recently, a similar approach has been proposed in Ref.~\cite{Pedreira:2023qqt} as an empirical method to visualize and investigate the relations between the $H_0$ tension and the more controversial tensions surrounding matter clustering parameters~\cite{DiValentino:2020vvd}, most notably $\sigma_8$. The method consists of comparing the values extrapolated for $H_0$ and $\sigma_8$ from parameter inference analyses concerning different theoretical models beyond $\Lambda$CDM where one or more of the new-physics parameters (that represent the $\Lambda$CDM extension) are fixed to reference values (see, for example, Figures 1-3 in Ref.~\cite{Pedreira:2023qqt}).

Following the very same alternative way of thinking proposed in these two works, in this paper I first extend the analysis performed in Ref.~\cite{Vagnozzi:2019ezj} to closely focus on the implications for inflation. The first part of my analysis is aimed at clarifying if the shift towards larger values of the spectral index is found by also keeping a constraining power and number of parameters identical to the standard case. I will quantify the implications for single-field slow-roll inflation by examining constraints on the slow-roll parameters in these alternative scenarios and investigate the implications for Starobinsky inflation. My results indicate that the mild deviation proposed in Ref.~\cite{Vagnozzi:2019ezj} (i.e., $N_{\text{eff}} = 3.45$) would fundamentally alter our understanding of the inflationary Universe. When this value is assumed in the cosmological model, current data no longer support the most accredited models of inflation, such as Starobinsky inflation. Conversely, considering a phantom $w$ barely alters the results on inflationary parameters. This underscores how the presence of the Hubble tension represents a source of uncertainty, particularly in the context of theoretical attempts to introduce new physics at early times. Motivated by these findings, I shift my focus to specific realizations of early-time new physics. Focusing on EDE, I demonstrate that even a small fraction of the total energy-density of the Universe in EDE -- ranging between 4\% and 6\% (which could only mildly reduce the $H_0$-tension down to $\sim 3\sigma$) -- , could already necessitate a similar shift in perspective regarding favored inflationary models based on current data. Additionally, through a traditional full joint analysis of EDE and Starobinsky inflation (i.e., by performing a canonical Markov Chain Monte Carlo analysis where \textit{all} model parameters are left free to vary), I conclusively prove that the two models can hardly coexist when the EDE fraction exceeds $\sim 6\%$.

Before proceeding further, I would like to outline some important \textit{do's} and \textit{don'ts} of the present work, aimed at preventing any potential misunderstanding of my intentions in this article:

\begin{itemize}[leftmargin=*]

\item I \textit{do not} attempt to avoid additional degrees of freedom. I \textit{do not} propose solving the Hubble tension by fixing alternative parameters to non-standard values. I \textit{do not} suggest specific models that can fix parameters to non-standard values (although they \textit{do} exist, as discussed in Section IV.D of Ref.~\cite{Vagnozzi:2019ezj} and Ref.~\cite{Pedreira:2023qqt}, as well). I \textit{do not} delve deeper into how well these non-standard scenarios may be physically motivated (although many of them \textit{do} have solid theoretical backgrounds~\cite{Vagnozzi:2019ezj,Pedreira:2023qqt}).

\item In the first part of the analysis, I \textit{do} use this approach to parameterize the effects left by early-time and late-time new physics without additional parameters. I \textit{do} use this approach to extend results already documented in the literature, thereby exploring the implications for inflation in the presence of new physics that can potentially adjust $H_0$, while maintaining a level of constraining power similar to the standard model. I \textit{do} use this approach to gain insight into how sensitive our constraints on inflation are to the cosmological model. I \textit{do} extend my major findings obtained within this (maybe non-canonical) alternative thinking to more acceptable scenarios such as EDE. For EDE I \textit{do} follow also a canonical approach leaving all parameters free to vary. I \textit{do} ensure that my main results \textit{do not} depend on the alternative perspective adopted.

\end{itemize}

The manuscript is organized as follows. In \autoref{sec:general} I review the theoretical framework I adopt to parameterize the effects of new physics in the cosmological model. In \autoref{sec:Methods} I outline the methodology and the data underlying my study. In \autoref{sec:results:early} and \autoref{sec:results:late} I discuss the implications of early-time and late-time new physics for inflation, respectively. In \autoref{sec:EDE} I focus on EDE and discuss inflation in EDE cosmology. Finally, in \autoref{sec:Conclusion}, I summarize my main conclusions.

\section{Theory}
\label{sec:general}
Inferring the value of the present-day expansion rate of the Universe from the measured spectra of temperature and polarization anisotropies in the cosmic photon background requires assuming a theoretical model to describe the evolution of the baryon-photon fluid at early times (i.e., prior to recombination) as well as the evolution of the Universe at later times (i.e., post-recombination). Therefore, the value of $H_0$ estimated via CMB observations relies on the standard $\Lambda$CDM model, and several potential theoretical alternatives have been proposed to obtain a higher value of $H_0$ consistent with local distance ladder measurements~\cite{Abdalla:2022yfr,Knox:2019rjx}. Without claiming to be exhaustive, we can categorize these efforts into two main categories: \textit{early-time solutions} and \textit{late-time solutions}. In the first case, the general idea is to introduce new physical components that act prior to recombination. Solutions of this kind often aim to reduce the value of the sound horizon by increasing the energy-density of the primordial Universe. Conversely, late-time solutions seek to modify physics after recombination, thereby affecting the value of $H_0$ derived from the angular distance from the CMB. Both of these approaches have their advantages and disadvantages. While late-time solutions typically find good room in CMB measurements (also due to a strong geometrical degeneracy of parameters), precise measurements of Baryon Acoustic Oscillations (BAO)~\cite{eBOSS:2020yzd} and Type Ia supernovae~\cite{Brout:2022vxf} impose significant constraints on the possibility of introducing new physics at late times. Consequently, it appears unlikely that a definitive solution can be solely based on late-time modifications~\cite{Krishnan:2021dyb,Keeley:2022ojz}\footnote{Recently, in Ref.~\cite{Vagnozzi:2023nrq}, seven hints were proposed, suggesting that a compelling definitive solution might involve combining early and late-time solutions together.}. On the other hand, early-time solutions typically need to operate during the matter-dominated era, specifically near recombination when photons begin to decouple from the baryon-photon fluid. This requires a moderate level of fine-tuning and introduces significant implications for the anisotropies in the CMB and the growth of primordial density perturbations. 

Here, I set aside the issues that have been fairly discussed in the literature and briefly review the two possible ways of addressing the $H_0$ tension that I adopt in this work. In particular I discuss how and why the presence of extra relativistic species in the early Universe (\autoref{sec:ealry}) or a phantom dark energy component at late times (\autoref{sec:late}) moves us in the right direction towards resolving the $H_0$ tension.

\subsection{Early-time new physics}
\label{sec:ealry}

I start from early-time (new) physics, namely physical theories that can act before recombination (i.e., before photons decoupled from the primordial plasma). During this period, physics is largely governed by the gravitational forces experienced by the baryon-photon fluid. The interplay between gravity and radiation pressure gives raise to sound waves, commonly known as Baryon Acoustic Oscillations. These sound waves, propagating within the baryon-photon plasma, left characteristic imprints in the Cosmic Microwave Background radiation, predicting a series of (damped) acoustic peaks in the spectrum of temperature anisotropies, which have been accurately measured by several CMB experiments and most precisely by the Planck satellite.

Particularly relevant physical information is captured by the angular scale associated to the position of the first acoustic peak in the spectrum of temperature anisotropies. This peak corresponds to the oscillation mode that had just enough time to complete one full compression cycle before the photons decoupled shortly after recombination. When photons decoupled, they essentially "froze" in place, preserving the pattern of sound waves that were traveling through the early Universe. In practice, by measuring the multipole corresponding to the position of the first acoustic peak ($\ell_{\rm{peak}}$), one can determine the angular size of the sound horizon $\theta_{*}\simeq \pi/\ell_{\rm{peak}}$ which is related to the comoving sound horizon $r_s\left(z_{*}\right)$
\begin{equation}
r_s\left(z_{*}\right)=\int_{z_{*}}^{\infty} d z \frac{c_s(z)}{H(z)},
\label{eq:rs}
\end{equation}
and the angular diameter distance $D_A(z_{*})$ 
\begin{equation}
D_A\left(z_{*}\right)=\frac{1}{1+z_{*}} \int_0^{z_{*}} d z \frac{1}{H(z)} .
\label{eq:DA}
\end{equation}
(both evaluated at redshift corresponding to the last scattering surface $z_{*}$) by very simple trigonometric considerations:
\begin{equation}
\theta_{*} =  \frac{r_s(z_{*})}{D_A(z_{*})},
\label{eq:theta_star}
\end{equation}

Given its strong geometric interpretation, the value of $\theta_{*}$ shows weak dependence on the specific cosmological model and stands as the most precisely measured parameter from Planck. Consequently, any attempt to introduce new physics into the cosmological model typically should conserve this quantity. To preserve $\theta_{*}$ while simultaneously increasing the value of $H_0$, our focus remains on the two model-dependent quantities $r_s(z_*)$ and $D_A(z_*)$. Looking at Eq.\eqref{eq:DA}, it is easy to see that $D_A(z_*)$ contains information about the expansion history of the Universe from $z_{*}$ (i.e., from recombination) to the present day ($z=0$). Therefore, solutions aiming to modify the angular diameter distance are commonly referred to as late-time solutions and will be discussed in the next subsections. Instead, here I focus on the value of the sound horizon. 

$r_s(z_*)$ is determined by the expansion history before recombination, from the Big Bang singularity ($z\to\infty$) all the way up to $z_{*}$. For this reason solutions aiming to increase $H_0$ by reducing the value of $r_{s}(z_*)$ are referred to as early-time solutions. My goal is to parameterize such early-time solutions without introducing specific model dependencies. Looking at Eq.~\eqref{eq:rs}, I am left with only two possibilities: working on the sound speed of the photon-baryon fluid $c_s(z)$, or changing the rate of expansion $H(z)$. Before undergoing a rapid drop when matter begins to dominate, for most of the expansion history before recombination $c_s^2(z) \simeq 1/3$. Therefore, to follow the first possibility, one would need to develop a model which is not my goal here. On the other hand, the expansion rate before recombination can be expressed as:
\begin{equation}
\left(\frac{H(z)}{H_0}\right)^2 = \left(\Omega_c+\Omega_b\right)(1+z)^3+\Omega_\gamma\left(1+0.23 N_{\mathrm{eff}}\right)(1+z)^4
\end{equation}
where $N_{\rm eff}$ is the effective number of relativistic particles, while $\Omega_c$, $\Omega_b$, and $\Omega_\gamma$ are the cold dark matter, baryons, and photons energy densities, respectively. The energy-density of various species can be accurately measured within a narrow margin of error, both through the effects in the CMB and through independent observations such as the BBN (for $\Omega_b$) or BAO measurements (for $\Omega_{c}$). Therefore, as also highlighted in Ref.~\cite{Vagnozzi:2019ezj}, a simple (yet physically motivated) method to parameterize new physics able to increase the rate of expansion of the Universe before recombination consists of increasing the value of the effective number of relativistic particles, $N_{\text{eff}}$. Within the standard cosmological model, this parameter takes the reference value of $N_{\rm eff}=3.044$~\cite{Mangano:2005cc,Akita:2020szl,Froustey:2020mcq,Bennett:2020zkv}, accounting for three different families of relativistic neutrinos along with an additional contribution arising from non-instantaneous neutrino decoupling. However, given its flexibility in quantifying each relativistic component in the early Universe, this value can be changed by a wide range of phenomenological extensions spanning from modifications to the standard model of particle physics that involve new relativistic degrees of freedom active in the early Universe, to models that incorporate additional scalar degrees of freedom or significant graviton cosmic background, see e.g., Refs~\cite{Giare:2020vzo,Giare:2021cqr,DEramo:2022nvb,Giare:2022wxq,Papanikolaou:2023oxq,Garny:2024ums}.

\subsection{Late-time new physics}
\label{sec:late}

The second option is considering late-time solutions, namely physical theories that can change the expansion history of the Universe after recombination. Notice that such mechanisms, leaving the early-time expansion rate unchanged, typically do not change the value of the sound horizon. In addition, as I pointed out in the previous subsection, $\theta_{*}$ should also remain fixed given its strong geometrical interpretation and weak model dependence. If we take a look at Eq.~\eqref{eq:theta_star}, we see that this implies that $D_A(z_{*})$ must also remain unchanged. Therefore, in practice, a successful late-time solution should increase $H_0$ while maintaining $D_A(z_*)$ unchanged to preserve $\theta_{*}$.

To take a step forward, I write down the equation that describes the expansion rate of the Universe at late times for a generic model of cosmology described by a non-flat FRW metric 
\begin{equation}
\left(\frac{H(z)}{H_0}\right)^2 \simeq \Omega_m (1+z)^3 + \Omega_k (1+z)^2 + \Omega_{\rm de} (1+z)^{3(1+w)}
\label{eq:H_late_general}
\end{equation}
where $\Omega_m=\Omega_b+\Omega_c+\Omega_{\nu}$ represents the matter energy-density (encompassing baryons, cold dark matter, and neutrinos), $\Omega_k$ denotes the curvature density parameter, and $\Omega_{\rm{DE}}$ describes the energy-density component associated with Dark Energy (DE). Notice that I assume the DE equation of state $w \equiv P_{\rm{DE}}/\rho_{\rm{DE}}$ (i.e., the ratio between the pressure $P_{\text{DE}}$ and energy-density $\rho_{\rm{DE}}$ of DE) to be constant. Within the standard $\Lambda$CDM model of cosmology, this parameter is fixed to $w=-1$, corresponding to a cosmological constant case $\Omega_{\rm{DE}}=\Omega_{\Lambda}$. Within $\Lambda$CDM the spatial background geometry is assumed to be flat, $\Omega_k=0$. Therefore, if we aim to use the simple Eq.~\eqref{eq:H_late_general} to parameterize the effects of new physics beyond the standard cosmological model, we are left with only two possibilities: working on the background geometry or on the DE equation of state\footnote{Notice that many other attempts to solve the Hubble tension in the context of late-time new physics have been discussed in the literature, introducing modifications to the background dynamics parameterized by Eq.~\eqref{eq:H_late_general}. Just to mention a few concrete examples, a non-exhaustive list of possibilities involves considering modified gravity theories and or exotic/phenomenological theoretical frameworks such as Interacting Dark Energy or sign-switching cosmology. With no claim of covering all possibilities, for works in this direction, the interested reader can refer to, e.g., Refs.~\cite{Gavela:2009cy,DiValentino:2017iww, Kumar:2017dnp, Wang:2016lxa, DiValentino:2019ffd, Pan:2019jqh, Yang:2018euj, Murgia:2016ccp, Yang:2019uzo, Pan:2019gop, DiValentino:2019jae, Lucca:2020zjb, Gomez-Valent:2020mqn, DiValentino:2020vnx, Yang:2021hxg, Gariazzo:2021qtg,Bernui:2023byc,Zhai:2023yny,Montani:2023xpd,Schiavone:2022wvq,Montani:2023ywn,Akarsu:2021fol,Akarsu:2022typ,Akarsu:2023mfb,Gomez-Valent:2023uof,Giare:2024ytc,Giare:2024smz} and references therein.}. Both of these parameters can be well constrained by a multitude of observational data and have been the subject of intense attention and debate for a variety of reasons\footnote{See, e.g., Refs.~\cite{Park:2017xbl,Handley:2019tkm,DiValentino:2019qzk,Efstathiou:2020wem,DiValentino:2020hov,Benisty:2020otr,Vagnozzi:2020rcz,Vagnozzi:2020dfn,DiValentino:2020kpf,Yang:2021hxg,Cao:2021ldv,Dhawan:2021mel,Dinda:2021ffa,Gonzalez:2021ojp,Akarsu:2021max,Cao:2022ugh,Glanville:2022xes,Bel:2022iuf,Yang:2022kho,Stevens:2022evv,Favale:2023lnp} for discussions surrounding the curvature parameter and Refs.~\cite{Escamilla:2023oce,Planck:2018vyg,eBOSS:2020yzd,Semenaite:2022unt,Carrilho:2022mon,Chudaykin:2020ghx,DAmico:2020kxu,Brieden:2022lsd,Pan-STARRS1:2017jku,Brout:2022vxf,DES:2018ufa,KiDS:2020ghu,DES:2022ccp,Moresco:2016nqq,Vagnozzi:2020dfn,Yang:2021flj,DiValentino:2020vnx,Moresco:2022phi,Vagnozzi:2021tjv,Bargiacchi:2021hdp,Moresco:2022phi,Grillo:2020yvj,Cao:2021cix,Hogg:2023khs,DESI:2024mwx} for recent constraints on the DE equation of state from a multitude of astrophysical and cosmological probes.}. Again, I will not take part in the debate as my goal is to parameterize new physics able to alleviate the Hubble tension and study its implications for inflation. Since one of the most typical predictions of inflation is that the observable Universe should be locally flat, I fix the background geometry to $\Omega_k=0$, focusing on the potential realizations of new theories capable of predicting non-standard values for $w$. 

\section{Methodology}
\label{sec:Methods}

\subsection{Early-time solutions}
I study a class of models described by 7 free parameters (one more than in $\Lambda$CDM) namely: the baryon $\omega_{\rm b}\equiv \Omega_{\rm b}h^2$ and cold dark matter $\omega_{\rm c}\equiv\Omega_{\rm c}h^2$ energy densities, the angular size of the horizon at the last scattering surface $\theta_{*}$, the optical depth $\tau$, the amplitude of primordial scalar perturbation $\log(10^{10}A_{\rm s})$, the scalar spectral index $n_{\rm s}$ and the tensor amplitude $r$. Notice that I am considering the tensor amplitude $r$ as a free parameter because my goal here is to study the implications for inflationary models that can be well described in terms of their joint predictions for $n_s$ and $r$. Following Ref.~\cite{Vagnozzi:2019ezj}, I assume that a physical theory can set $N_{\rm eff}$ to non-standard values, such that $N_{\rm eff} > 3.044$. Specifically, I consider 11 different cases where $\Delta N_{\rm eff} = N_{\rm eff} - 3.044$ ranges from 0 (corresponding to $\Lambda$CDM) to 1 (corresponding to $N_{\rm eff} \approx 4$) in steps of $\Delta N_{\rm eff} = 0.1$. For each of these steps, I perform a full MCMC analysis using the publicly available code \texttt{cobaya}~\cite{Torrado:2020dgo} and computing the cosmological model using the publicly available Boltzmann integrator code \texttt{CAMB}~\cite{Lewis:1999bs,Howlett:2012mh}. Within each MCMC run, I keep $N_{\rm eff}$ fixed while varying all the other 7 cosmological parameters. 

\subsection{Late-time solutions}

I study a class of models with 7 free parameters (the same listed above), assuming now that a physical theory can fix $w$ to non-standard values. Specifically, just like Ref.~\cite{Vagnozzi:2019ezj}, to raise $H_0$, I consider 5 different cases where $w$ ranges from $-1$ (i.e., the cosmological constant value) up to $-1.2$ in steps of $\Delta w = -0.05$. Therefore $w$ remains confined in the phantom regime where $w < -1$.  Following the same methodology outlined when I studied early-time solutions, for each of these steps, I perform a full MCMC analysis. Within each run, I keep $w$ fixed while varying all the other 7 cosmological parameters.

\subsection{Datasets}
The datasets involved in my analysis are\footnote{Concerning the choice of likelihoods employed in my study, I conservatively used the very same combinations of observables analyzed in the Planck 2018 paper on inflation~\cite{Planck:2018jri}. Namely the same Planck 2018 CMB likelihoods (including the lensing potential), B-mode polarization measurements, and baryon acoustic oscillations.}:
\begin{itemize}
\item Measurements of the Planck CMB temperature anisotropy and polarization power spectra, their cross-spectra, and the lensing power spectrum. All the Planck CMB likelihoods employed in this work are listed below:
\begin{itemize}
\item[(i)] Measurements of the power spectra of temperature and polarization anisotropies, $C_{\ell}^{TT}$, $C_{\ell}^{TE}$, and $C_{\ell}^{EE}$, at small scales ($\ell>30$), obtained by the Planck \texttt{plik} likelihood~\cite{Planck:2018vyg,Planck:2019nip};
\item[(ii)] Measurements of the spectrum of temperature anisotropies, $C_{\ell}^{TT}$, at large scales ($2 \leq \ell \leq 30$), obtained by the Planck \texttt{Commander} likelihood~\cite{Planck:2018vyg,Planck:2019nip};
\item[(iii)] Measurements of the spectrum of E-mode polarization, $C_{\ell}^{EE}$, at large scales ($2 \leq \ell \leq 30$), obtained by the Planck \texttt{SimAll} likelihood~\cite{Planck:2018vyg,Planck:2019nip};
\item[(iv)] Reconstruction of the spectrum of the lensing potential, obtained by the Planck \texttt{plik} likelihood~\cite{Planck:2018lbu}.
\end{itemize}
\item B-modes CMB polarization data released by the BICEP/Keck Collaboration~\cite{BICEP:2021xfz}.
\item Baryon Acoustic Oscillation (BAO) and Redshift-Space Distortions (RSD) measurements from the completed SDSS-IV eBOSS survey. These include isotropic and anisotropic distance and expansion rate measurements and are summarized in Table~3 of Ref.~\cite{eBOSS:2020yzd}.
\end{itemize}
I refer to the full combinations of these datasets as \textit{Planck+BK18+BAO}.

\begin{figure*}[htp!]
    \centering
    \includegraphics[width=\textwidth]{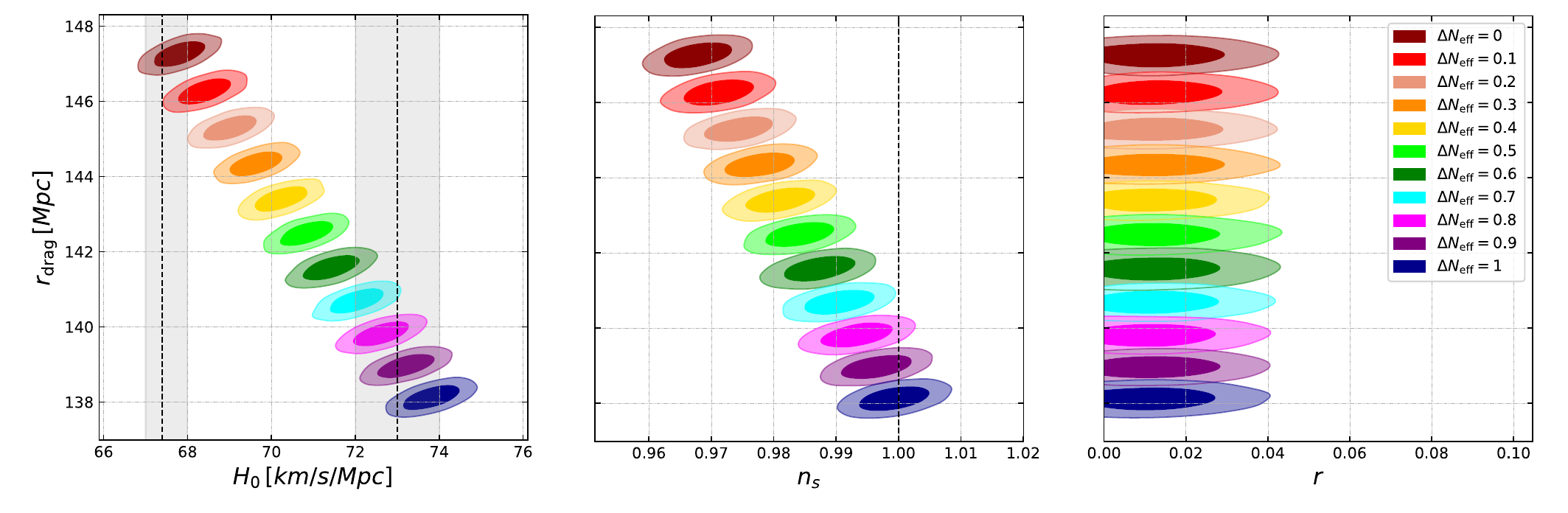}
    \caption{Two-dimensional correlations between the sound horizon at the drag epoch $r_{\rm{drag}}$, the Hubble constant $H_0$ (left panel), $n_s$ (middle panel), and $r$ (right panel). The different contours are obtained by considering various realizations of early-time (new) physics altering the expansion history of the Universe before recombination, whose effects are parameterized in terms of corrections to the effective number of relativistic degrees of freedom.}
    \label{fig:1}
\end{figure*}

\section{Early-time new physics and Inflation}
\label{sec:results:early}

In this section, I study the implications for inflation resulting from early-time new physics parametrized by imposing $N_{\rm eff}>3.044$. I divide the section into two subsections: in \autoref{sec:ealry.single}, I discuss the implications for generic single-field models of inflation where the inflationary potential is parameterized in terms of the slow-roll parameters. Conversely, in \autoref{sec:ealry.Starobinsky}, I will focus on the implications of one particular benchmark scenario: the model proposed by A. Starobinsky~\cite{Starobinsky:1980te}.

\subsection{Implications for single field inflation}
\label{sec:ealry.single}

I start studying the implications resulting from introducing early-time new physics, considering generic single-field models of inflation. Specifically, I assume inflation to be driven by a scalar field $\phi$ minimally coupled to gravity. The action reads: 
\begin{equation}
S=\int d^4x\,\sqrt{-g}\left[\frac{M _ { \rm { pl } } ^ { 2 } }{2}\,R+\frac{1}{2}\,g^{\mu\nu}\partial_{\mu}\phi\partial_{\nu}\phi-V(\phi)\right],
\label{minimal coupled action}
\end{equation}
with $M _ { \rm { pl } }=1/\sqrt{8\pi G}=2.4\times 10^{18}\,\rm{GeV}$ the reduced Planck mass in the natural units ($c=\hslash=1$) and $R$ the Ricci scalar. For the moment, I do not assume any specific inflationary potentials and parameterize the spectra of primordial scalar and tensor perturbations by adopting the usual power-law form 
\begin{equation}
\log \mathcal{P}_{\rm s}(k)=\log A_{\mathrm{s}}+\left(n_{\mathrm{s}}-1\right) \log \left(k / k_{*}\right) 
\label{PLS}
\end{equation}
\begin{equation}
\log \mathcal{P}_{\rm T}(k)=\log \left(r\,A_{\mathrm{s}}\right)+\left(n_{\mathrm{T}}\right) \log \left(k / k_{*}\right) 
\label{PLT}
\end{equation}
where $k_*$ denotes an arbitrary \textit{pivot scale} that I fix to $k_{*}=0.05\, \rm{Mpc}^{-1}$. However, I express the scalar and tensor tilts in terms of slow-roll parameters 
\begin{equation}
\epsilon \doteq  \frac{M _ { \rm { pl } } ^ { 2 }}{2} \left(\frac { V _ { \phi } ^ { 2 } } { V ^ { 2 } }\right), \quad \eta  \doteq M _ { \rm { pl } } ^ { 2 } \left(\frac { V _ { \phi \phi } } { V }\right)
\label{eq:slow-roll}
\end{equation}
that quantifies the derivatives of the inflationary potentials with respect to the filed ($V_{\phi\dots\phi}$). In particular, I assume the well-known (consistency) relations~\cite{Martin:2013tda}:
\begin{equation}
n_{s}-1=2\eta-6\epsilon, \quad n_{\rm t}=-2\epsilon=-r/8,
\label{Spectral}
\end{equation}
where, as usual, the slow-roll parameters $\epsilon$ and $\eta$ are calculated for specific field values $\phi_*$ corresponding to when perturbations cross the horizon during inflation.

\begin{figure*}[htp!]
    \centering
    \includegraphics[width=0.7\textwidth]{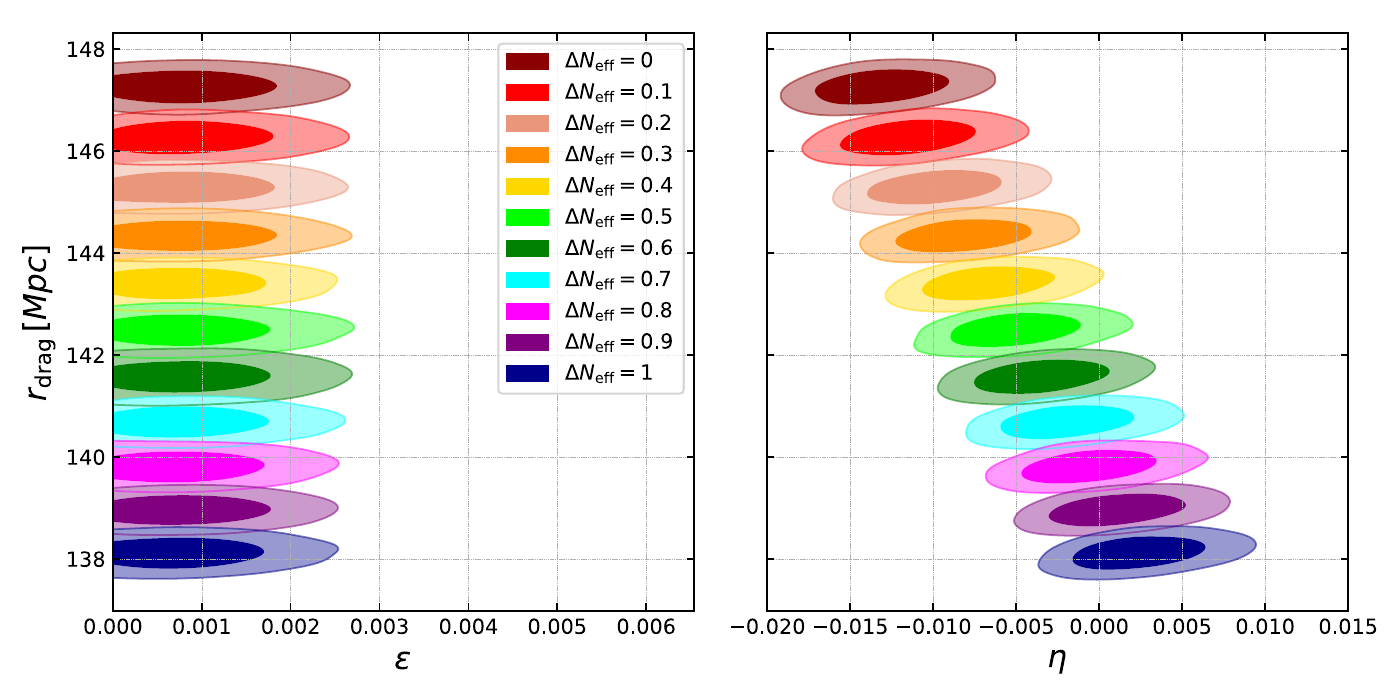}
    \caption{Two-dimensional correlations between the sound horizon at the drag epoch $r_{\rm{drag}}$ and the slow-roll parameters $\epsilon$ (left panel) and $\eta$ (right panel). The different contours are obtained by considering various realizations of early-time (new) physics altering the expansion history of the Universe before recombination, whose effects are parameterized in terms of corrections to the effective number of relativistic degrees of freedom.}
    \label{fig:2}
\end{figure*}

The most significant result of this analysis is visually represented in~\autoref{fig:1}. The figure shows the two-dimensional correlations among different parameters relevant to the discussion. In particular, the first panel on the left depicts the correlation between the value of the sound horizon at the drag recombination epoch ($r_{\rm{drag}}$)~\footnote{It is worth noting that BAO measurements are sensitive to the sound horizon evaluated at the baryon drag epoch, typically denoted by $r_{\text{drag}}$~\cite{Aubourg:2014yra}, while, as explained in \autoref{sec:general}, the scale relevant for the acoustic peaks in the CMB is the sound horizon evaluated at recombination, typically denoted by $r_{*}$~\cite{Hu:2001bc}. The two epochs are separated in redshift by $\Delta z = z_{\text{drag}} - z_{*} \sim 30$, so the difference between the two sound horizons is important. Since the results in the literature are commonly given in terms of $r_{\text{drag}}$, in this work I will adhere to the most commonly used practice for direct comparison.} and the present-day rate of expansion of the Universe for all 11 different scenarios considered. The results distinctly indicate that incorporating new physics able to increase the energy budget in the early Universe concerning the standard cosmological model, diminishes the value of the sound horizon, thereby leading to a higher expansion rate of the Universe in the current epoch. This is an acknowledged mechanism that has spurred the development of several models, eventually proposed as potential resolutions for the Hubble tension. More precisely, to achieve values of $H_0$ consistent with the SH0ES result, one needs to reduce the value of the sound horizon by approximately 5\%, transitioning from $r_{\rm{drag}}\sim147$ Mpc (value obtained within the $\Lambda$CDM model) to $r_{\rm{drag}}\sim140 - 138$ Mpc. Within the parameterization employed in this section, achieving such a shift requires considering corrections $\Delta N_{\rm eff}\gtrsim0.7$ as it has been already documented in the literature (see, for instance, the discussion in Ref.~\cite{Vagnozzi:2019ezj}).

In the second and third panels of~\autoref{fig:1}, I focus on the correlation between $r_{\rm{drag}}$ and $n_s$ (second panel) and between $r_{\rm{drag}}$ and $r$ (third panel). Concerning the spectral index, we see that increasing $\Delta N_{\rm eff}$ results in a significant shift towards $n_s \sim 1$. Reducing the sound horizon value by about 5\% to resolve the Hubble tension imposes constraints on the spectral index $n_s \gtrsim 0.99$. On the other hand, the constraints on the tensor amplitude remain substantially unchanged.

In the context of single-field slow-roll inflation, the values of the spectral index and tensor amplitude can be directly correlated to the values of derivatives of the inflationary potential through the relationships indicated in Eq.~\eqref{Spectral}. Therefore, another point I would like to briefly discuss concerns the possible implications of the Hubble tension for inflationary models. Keeping the discussion as general as possible, in \autoref{fig:2}, I depict the constraints on the two slow-roll parameters, $\epsilon$ and $\eta$, as well as their correlation with the value of $r_{\rm{drag}}$. As highlighted in the figure, the parameter $\epsilon$ is not sensitive to the value of the sound horizon. This parameter is directly related to the amplitude of tensor perturbations by the slow roll relation $r = 16\epsilon$. B-mode polarization measurements from BICEP/Keck likelihood set very stringent constraints on the tensor amplitude $r$, and these constraints are essentially unchanged with the cosmological model\footnote{The limit on the tensor amplitude is primarily due to measurements of the BB angular power spectrum at large angular scales ($\ell \lesssim 100$), where the lensing contribution in the BB spectrum is subdominant compared to the primordial tensor counterpart. In this range of the spectrum, the only significant parameter to gauge the amplitude of the signal is the tensor-to-scalar ratio $r$, significantly reducing the dependence of the constraints on the background cosmology.}. As a result, for all cases considered, I obtain a very tight limit of $\epsilon \lesssim 0.002$, making its contribution in Eq.~\eqref{Spectral} subdominant compared to $\eta$\footnote{I note that in the minimal theoretical setup described by single field slow-roll inflation, it must be $\epsilon > 0$ to preserve the Null Energy Condition. Therefore, this parameter is bounded in the range $\epsilon \in [0, 0.002]$.}. Therefore, the significant shift in the value of $n_s$ is recast into a variation in the inferred value for the second slow-roll parameter, $\eta$.
Within the standard cosmological model, $\eta$ is measured to be $\eta=-0.0130^{+0.0024}_{-0.0029}$, namely $\eta<0$ and such that $1 \gg |\eta| \gg \epsilon$. Being linked to derivatives of the inflationary potential, such a hierarchy poses a minor fine-tuning problem since there exists no fundamental reason or symmetry in nature that can justify it. In fact, it simply stands as a prediction of current data imposing strong constraints on the inflationary models that can be deemed suitable to explain current observations. As a matter of fact, a few well-motivated inflation models that are able to predict such a hierarchy have become along the years benchmark scenarios to gauge future CMB experiments. This is, for instance, the case of the model proposed by Alexei Starobinsky in Ref.~\cite{Starobinsky:1980te}, which is able to naturally explain such a hierarchy between the inflationary slow-roll parameters. That said, observing \autoref{fig:2} again, we can notice how this hierarchy is called into question when considering the possibility of introducing new physical mechanisms that could resolve the Hubble tension. As soon as the value of the sound horizon gets smaller (and $H_0$ higher), the parameter $\eta$ shifts towards $\eta \to 0$, allowing in some cases values close to 0 or even crossing the positive region\footnote{To provide concrete numbers quantifying this shift, I observe that within the $\Lambda$CDM model, $\eta=0$ is ruled out at $5.4\sigma$, whereas for $\Delta N_{\rm eff} \gtrsim 0.6$, it is well allowed within the 68\% confidence level contours.}. While this could reduce the fine-tuning between the two slow-roll parameters, it has important implications in terms of model selection. Referring back to Eq.~\eqref{eq:slow-roll} we see that determining whether $\eta$ is positive or negative is equivalent to determining the sign of the second derivative of the inflationary potential $V''(\phi)$, and thus whether the inflationary potential is concave or convex. Maintaining a standard cosmological model, the various datasets analyzed in this study produce probability contours in the concave region ($\eta < 0$), while altering the cosmological model the results gradually shift towards the region of convex potentials ($\eta > 0$), thereby challenging some well-established benchmark models, as discussed in the next section.

\subsection{Implications for Starobinsky inflation}

\begin{figure}[tbp!]
    \centering
    \includegraphics[width=0.9\columnwidth]{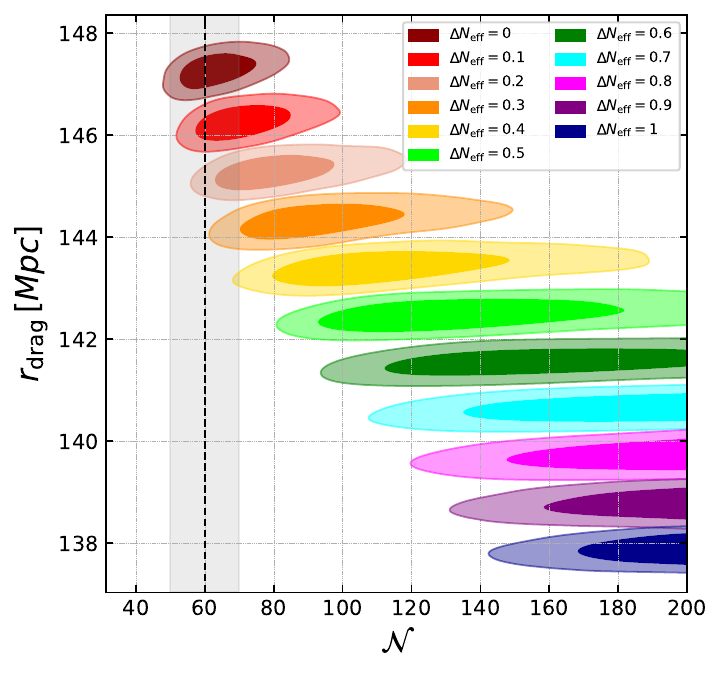}
    \caption{Two-dimensional correlations between the sound horizon at the drag epoch $r_{\rm{drag}}$ and the number of e-folds $\mathcal N$ in Starobinsky Inflation. The different contours are obtained by considering various realizations of early-time (new) physics altering the expansion history of the Universe before recombination, whose effects are parameterized in terms of corrections to the effective number of relativistic degrees of freedom.}
    \label{fig:3}
\end{figure}

\label{sec:ealry.Starobinsky}
To further clarify the implications resulting from early-time new physics in the cosmological model, now I explicitly consider the benchmark scenario proposed by Alexei Starobinsky~\cite{Starobinsky:1980te} where the inflationary sector of the theory is described by the following action:
\begin{equation}
S=\frac{M _ { \rm { pl } } ^ { 2 }}{2} \int d^4x\,\sqrt{-g}\, \left(\,R+\frac{R^2}{6m^2}\right).
\label{staro_action}
\end{equation}
In this model, inflation is achieved by considering $f(R)$-gravity with $f(R) = R + R^2/6m^2$ ($m$ is a constant with the unit of mass). However, one can switch from the Jordan to the Einstein frame where Starobinsky inflation is equivalently described in terms of a scalar field ($\varphi$) whose potential reads
\begin{eqnarray}
    V (\varphi) = V_0 \left(1 - e^{-\sqrt{\frac{2}{3}}\varphi/ M _ { \rm { pl } } }\right)^2.
    \label{eq:pot_Starob}
\end{eqnarray}
Plugging Eq.~\eqref{eq:pot_Starob} into the relations for the slow-roll parameter $\epsilon$ and $\eta$ given by Eq.\eqref{eq:slow-roll}, one gets
\begin{equation}
\epsilon\cong \frac{4}{3}e^{-2\sqrt{\frac{2}{3}}\varphi_*/M _ { \rm { pl } }  }, \quad \eta\cong -\frac{4}{3}e^{-\sqrt{\frac{2}{3}}\varphi_*/M _ { \rm { pl } }  }  ,
\end{equation}
where the field is evaluated at a given time $t_{*}$, corresponding to when perturbations cross the horizon during inflation. For the following discussion, it is useful to work in terms of the number of e-folds $\mathcal N$ between the horizon crossing and the end of inflation, which can be expressed as
\begin{eqnarray}\label{N_0}
{\mathcal N}=\frac{1}{M _ { \rm { pl } } }\int_{\varphi_*}^{\varphi_{end}}\frac{1}{\sqrt{2\epsilon}}d\varphi
\cong\frac{3}{4}e^{\sqrt{\frac{2}{3}}\varphi_*/M _ { \rm { pl } } }.
\end{eqnarray}
Using this result, one can derive a set of well-known relations that link the number of $e$-folds $\mathcal{N}$ to the spectral index of the scalar perturbations $n_s$ and the tensor-to-scalar ratio $r$. Namely:
\begin{eqnarray}
  n_s\cong 1- \frac{2} {\mathcal N}, \qquad r\cong \frac{12}{{\mathcal N}^2}.
    \label{eq.Starob}
\end{eqnarray}
Notice that, in general, modes with frequency $f=k/2\pi$ will cross the horizon $\mathcal N$ e-folds before the end of inflation given by the following relation~\cite{Martin:2013tda}
\begin{align} 
\nonumber \mathcal N \simeq&-\ln \left(\frac{k}{a_{0} H_{0}}\right)+\ln \left(\frac{H_{*}}{H_{\rm {end}}}\right)-\frac{2}{3}\ln\left(\frac{T_{\rm RH}}{\Lambda}\right) \\ &\nonumber +\ln \left(\frac{T_{\rm{RH}}}{T_{\rm {\rm{eq}}}}\right)+\frac{1}{3} \ln \left(\frac{g_{* S}\left(T_{R H}\right)}{g_{* S}\left(T_{\rm{eq}}\right)}\right) \\ &+\ln \left(\frac{a_{\rm{eq}}H_{\rm{eq}}}{a_0 H_0}\right) .
\label{Nk}
\end{align}
where $a_0H_0 \simeq 2.248 \times 10^{-4} \rm{Mpc}^{-1}$ represents the inverse of the comoving horizon size in the current Universe, $H_*$ is the value of the Hubble parameter at the horizon exit, $H_{\rm end}$ is the Hubble parameter at the end of inflation, $\Lambda$ is the energy scale of inflation and the subscript "eq" denotes quantities evaluated at matter-radiation equality. Approximating $H_{\rm end}\simeq H_{*}$ and recalling that in single field models of inflation the energy-scale can be related to the amplitude of tensor perturbations by $\Lambda\simeq 3.3 \times r^{1/4}\times 10^{16}\rm{GeV}$, one can get a rough idea of the value of $\mathcal N$ at $k=k_{*}=0.05$ Mpc$^{-1}$. In particular, fixing a $\Lambda$CDM-like history after inflation, the typical range allowed for $\mathcal{N}$ within single-field models falls in the range $\mathcal{N} \in [55,65]$ ~\cite{Liddle:2003as}. Plugging these values into Eq.~\eqref{eq.Starob} makes it easy to see that the value obtained for the inflationary parameters perfectly aligns with those measured by the Planck collaboration (as well as with the upper bound on $r$ obtained by BK18 data) when a $\Lambda$CDM model of cosmology is assumed.

To study the possible effects of non-standard models as a proxy for new physics in the early Universe, I repeat the same exercise of considering 11 different cases where $\Delta N_{\rm eff}$ ranges from 0 to 1 in steps of $\Delta N_{\rm eff} = 0.1$. However in this case, the inflationary sector of the theory is fixed to the Starobinsky model that I incorporate from the onset into the cosmological framework and the data analysis process. Specifically, I still use conventional power-law parametrizations for the scalar and tensor spectrum given by Eq.~\eqref{PLS} and Eq.~\eqref{PLT}. However, I impose the theoretical relationships between the inflationary parameters $n_{s}$ and $r$ as defined in Eq.~\eqref{eq.Starob}. It is worth noting that these equations reduce the number of free degrees of freedom in the model from 7 to 6 (i.e., the same number of free parameters as the standard picture). Among other benefits, this facilitates a more precise analysis. For each of these steps, I perform a full MCMC analysis using the same methodology and datasets described in \autoref{sec:Methods}. 

The results I want to highlight are depicted in \autoref{fig:3} where one can see the correlation between $r_{\rm{drag}}$ and $\mathcal{N}$. From the figure, it is evident that assuming a standard cosmology (i.e., $\Delta N_{\rm eff}\simeq 0$), Starobinsky inflation remarkably describes current data, and that $\mathcal{N} = 63.8^{+5.5}_{-8.9}$ aligns well with the theoretically expected value for this parameter. However, when different pre-recombination expansion histories of the Universe are considered, the constraints on $\mathcal{N}$ undergo significant shifts. Simulating the phenomenology needed to alleviate the Hubble tension, the number of e-folds required to achieve a good fit with the current data becomes excessively high, unequivocally confirming that this model may no longer be well supported by observations in these alternatives scenarios. Interestingly, I note here that a contribution $\Delta N_{\rm eff} \sim 0.3 - 0.4$ would be already enough for this model to become somewhat unsupported by current CMB and BAO data, postponing the discussion of whether such values of $\Delta N_{\rm eff}$ are actually allowed to the next subsection.

\subsection{Discussions and Concluding Remarks}
To conclude this section, I want to address one last, but not least, aspect. In light of all the extensively discussed effects regarding the consequences of introducing new physics at early times, it is worth asking to what extent models that \textit{fix} $\Delta N_{\text{eff}}$ to non-standard values are able to provide a good explanation of the CMB and BAO data and simultaneously offer a good description of the SH0ES measurement of the current expansion rate of the Universe (and eventually quantify the implications for inflation further).

To begin with, we can certainly make a list of features that we believe need to be preserved along with the appropriate statistical metrics introduced and used to quantitatively test whether and to what extent such conditions are effectively satisfied.

\begin{enumerate}[leftmargin=*]

   \item Firstly, we would like to solve or at least reduce the Hubble tension. To quantify the level of agreement between the Hubble parameter value obtained from CMB data within models introducing new physics at early times ($H_0^{\text {CMB}}$) and the value of the same parameter measured by the SH0ES collaboration ($H_0^{\text {SH0ES}}$), I use the simplest and most intuitive measure of the degree of tension in terms of the number of standard deviations ($\# \sigma$) defined as:
   \begin{equation}
    \# \sigma = \frac{\left|H_0^{\text {CMB}}-H_0^{\text {SH0ES}}\right|}{\sqrt{\sigma_{\text {CMB}}^2+\sigma_{\text {SH0ES}}^2}}
    \label{eq:H0}
    \end{equation}

   \item The introduction of new physics in the early Universe produces a series of almost inevitable consequences in the spectra of CMB temperature and polarization anisotropies. Therefore, a second requirement we need to impose is that the model of new physics under consideration should not excessively degrade the fit to CMB data. In non-standard models, I compute the total $\Delta\chi^2$ resulting from the three likelihoods considered in this study against (\textit{Planck+BK18+BAO}) with respect to $\Lambda$CDM. I define $\Delta\chi^2$ in such a way that negative values indicate an improvement in the fit over $\Lambda$CDM, while positive values indicate a worsening of the fit compared to $\Lambda$CDM.
   
   \item As a third requirement I ensure that the model of new physics is not too much disfavored from a statistical perspective compared to the standard cosmological model. To quantify this, I calculate the Bayesian Evidence (with respect to $\Lambda$CDM). Particularly, when provided with a dataset $\mathcal{D}$ and two models, $\mathcal{M}_i$ and $\mathcal{M}_j$, characterized by the parameters $\boldsymbol{\theta}_i$ and $\boldsymbol{\theta}_j$ respectively, the logarithm of the Bayes factor for model $\mathcal{M}_i$ with respect to model $\mathcal{M}_j$ is calculated as follows:
   \begin{equation}
   \ln B_{i j} = \ln \left[\frac{\int d \boldsymbol{\theta}_i \pi\left(\boldsymbol{\theta}_i \mid \mathcal{M}_i\right) \mathcal{L}\left(\mathcal{D} \mid \boldsymbol{\theta}_i, \mathcal{M}_i\right)}{\int d \boldsymbol{\theta}_j \pi\left(\boldsymbol{\theta}_j \mid \mathcal{M}_j\right) \mathcal{L}\left(\mathcal{D} \mid \boldsymbol{\theta}_j, \mathcal{M}_j\right)} \right]
   \label{eq:BE}
   \end{equation}
   Where $\pi\left(\boldsymbol{\theta}_i \mid \mathcal{M}_i\right)$ is the prior for the parameters $\theta_i$ while $\mathcal{L}\left(\mathcal{D} \mid \boldsymbol{\theta}_i, \mathcal{M}_i\right)$ is the likelihood of the data $\mathcal D$ given the model parameters $\theta_i$. Notice also that I am assuming equal prior probabilities for the two models.  In practice, to evaluate  $\ln B_{i j}$, I employ the \texttt{MCEvidence} package, which is publicly available~\cite{Heavens:2017hkr,Heavens:2017afc}\footnote{The \texttt{MCEvidence} package can be accessed at the following link: \url{https://github.com/yabebalFantaye/MCEvidence}.}. I use the convention that a negative value of $\ln B_{i j}$ means a preference for the non-standard model(s) against $\Lambda$CDM while positive values of $\ln B_{i j}$ means a preference for $\Lambda$CDM against non-standard model(s). I refer to the revised Jeffrey's scale~\cite{Kass:1995loi,Trotta:2008qt}, to interpret the results and will say that the evidence is inconclusive if $0 \leq | \ln B_{ij}|  < 1$, weak if $1 \leq | \ln B_{ij}|  < 2.5$, moderate if $2.5 \leq | \ln B_{ij}|  < 5$, strong if $5 \leq | \ln B_{ij}|  < 10$, and very strong if $| \ln B_{ij} | \geq 10$.

   \item  Finally, even when limiting ourselves to mechanisms that can move towards the SH0ES value without significantly worsening the fit to the CMB, we need to quantify the changes that might be needed to consider to get an inflation model able to provide an explanation for the observed high values of $n_s$, while maintaining a tensor amplitude in agreement with the latest BK18 measurements of B-mode polarization. Since, as demonstrated in the last subsection, when considering early-time new physics, Starobinsky inflation somewhat struggles to provide a good description of the data, here I explore a purely phenomenological extension where $n_s=1-(2n/\mathcal{N})$. It is noteworthy that for $n=1$, Eq.~\eqref{eq.Starob} is recovered. However, introducing a new parameter $n$ can offer the benefit of providing us with more flexibility to maintain an acceptable number of last e-folds of inflation $\mathcal N$ when $n_s$ shifts towards larger values. This would entail considering values of $n$ less than 1 and quantify the extent to which one needs to deviate from Starobinsky, gauging inflationary models for early-time new physics.

\end{enumerate}

\begin{figure}[tbp!]
    \centering
    \includegraphics[width=\columnwidth]{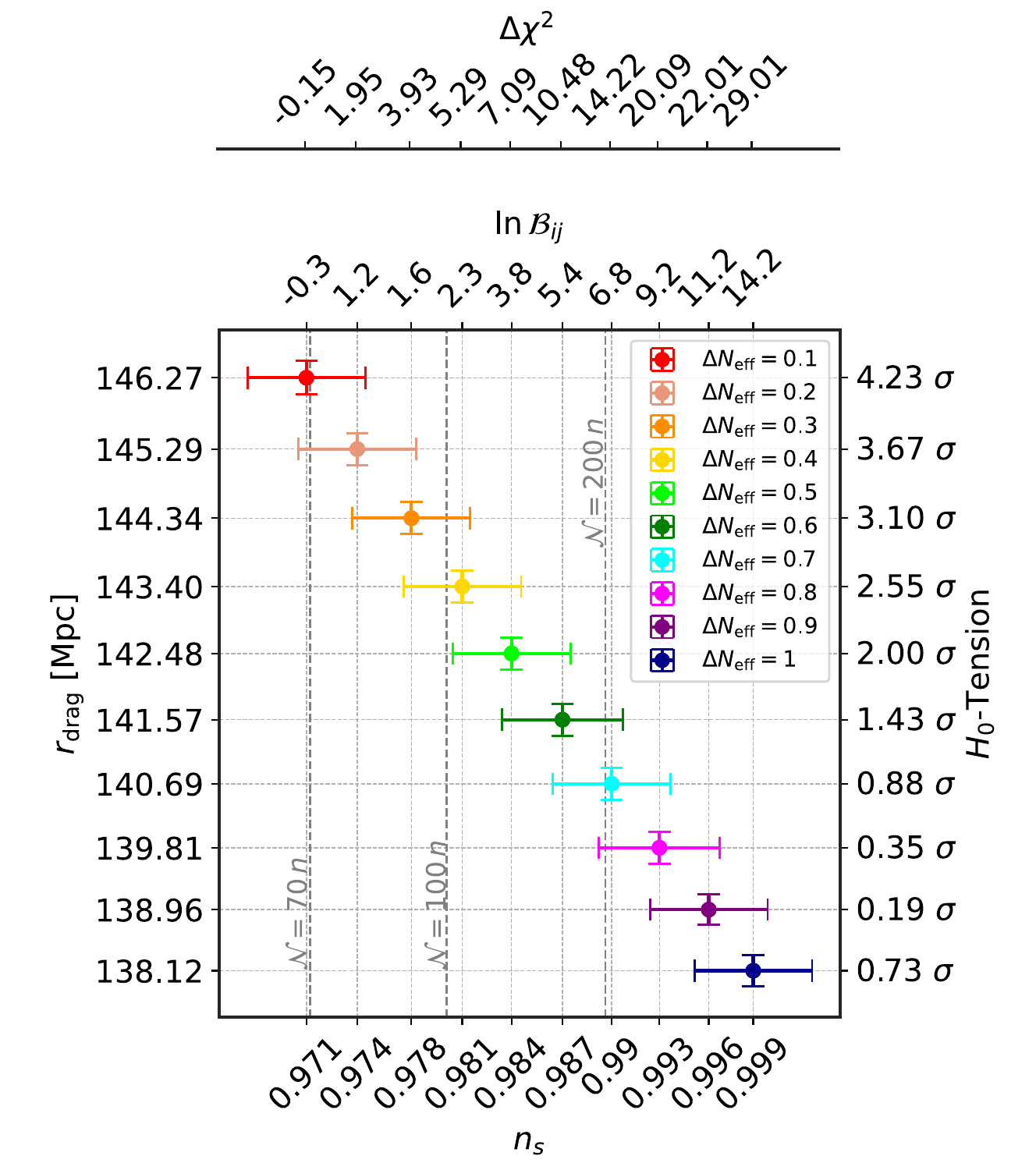}
    \caption{Each point of different color represents predictions in the plane ($n_s$ , $r_{\rm{drag}}$) along with corresponding error bars for the two parameters from models that parameterize the effects of new physics at early times in terms of corrections to the effective number of relativistic degrees of freedom, $\Delta N_{\rm eff}=\{0.1, \dots, 1\}$. On the y-axis, on the right side of the figure, for each point, the Hubble tension is quantified using Eq.~\eqref{eq:H0}. Conversely, on the x-axis at the top of the figure, the Bayes factors compared to the standard cosmological model, $\ln B_{ij}$ are obtained by Eq.~\eqref{eq:BE}. The $\Delta \chi^2$ is also shown at the top of the figure. The gray dashed vertical lines represent the number of e-folds $\mathcal N$ required for models where $n_s\simeq 1-(2n/\mathcal{N})$.}
    \label{fig:4}
\end{figure}

To assess the extent to which we can simultaneously satisfy these requirements, one can refer to the graphical representation that illustrates all these metrics simultaneously given in \autoref{fig:4}. In particular, each point of a different color represents predictions in the plane $n_s$-$r_{\rm{drag}}$ along with corresponding error bars for the two parameters for the 10 models considered in this section. These models parameterize the effects of new physics at primordial times in terms of corrections to the effective number of relativistic degrees of freedom, $\Delta N_{\rm eff}=\{0.1, \dots, 1\}$. On the right-side y-axis in the figure, for each point, I quantify the Hubble tension with respect to the value measured by SH0ES ($H_0^{\text {SH0ES}}=73\pm 1$ km/s/Mpc) using Eq.~\eqref{eq:H0}. Conversely, on the x-axis at the top of the figure, I show $\ln B_{ij}$ obtained by Eq.\eqref{eq:BE} where each model is compared to the standard cosmological model ($\Delta N_{\rm eff}=0$). As extensively discussed in the previous section, increasing the expansion rate before recombination shifts us in the direction of reducing the sound horizon and, consequently, decreasing the Hubble tension. However, moving in this direction leads to a significant deterioration in the fit to the CMB data. Models that are able to reduce the Hubble tension correspond to increasingly larger and positive values of $\ln B_{ij}$, indicating that these scenarios are significantly penalized compared to the standard cosmological model. To these two effects, we add the shift towards scale invariance ($n_s\to 1$), which, in terms of implications for inflationary models, is quantified by the number of e-folds $\mathcal N$ obtained for models where $n_s\simeq 1-(2n/\mathcal{N})$. These values are shown in the figure by gray dashed vertical lines.

From the results depicted in the figure, it may seem like we are ensnared in a typical "short blanket" scenario, where covering one aspect exposes another, underscoring a challenge of balance and compromise. For instance, as also highlighted in Ref.~\cite{Vagnozzi:2019ezj}, models with $\Delta N_{\rm eff} \sim 0.3 - 0.4$ can reduce the Hubble tension from approximately $5\sigma$ to $\lesssim 3\sigma$. These models are moderately disfavored compared to $\Lambda$CDM ($\ln B_{ij} \lesssim 2.5$) but nevertheless show a worsening in the fit to CMB and BAO data, leading to $\Delta \chi^2 \simeq 3 - 5$. Conversely, models with $\Delta N_{\rm eff} \geq 0.6$ are effective in alleviating the Hubble tension but are strongly or very strongly disfavored from the Bayesian evidence point of view, leading to a significant worsening in the $\chi^2$, as well (see also \autoref{fig:4}). Nevertheless, when considering the implications for inflationary models, scenarios with $\Delta N_{\rm eff} \sim 0.3$ would require a number of e-folds $\mathcal N \gtrsim 100n$. Setting $n=1$ (i.e., Starobinsky and $\alpha$-attractor inflation) would lead to a number of e-folds clearly too large. Therefore, one important conclusion, extending the results of~Ref.~\cite{Vagnozzi:2019ezj} to inflation, is that models predicting $\Delta N_{\rm eff} \simeq 0.3 - 0.4$ can indeed reduce the tension to $\sim 3\sigma$ while only being moderately  disfavored compared to $\Lambda$CDM. However, this is already enough to make most typical inflationary models, such as Starobinsky inflation, or more broadly any potential model predicting a relation $n_s \propto 1-(2n/\mathcal{N})$ with $n\geq 1$, inadequate for describing the data, implying reconsidering scenarios in which $n<1$, which are in turn significantly disfavored within the standard $\Lambda$CDM framework. This is direct evidence that even in very simple models with a number of degrees of freedom and a parameter constraining power similar to the standard case, mild deviations from an early-time $\Lambda$CDM cosmology (partially supported by current data) could lead to very significant differences in our understanding of the inflationary Universe, potentially to a greater degree than commonly realized.

\begin{figure*}[tbp!]
    \centering
    \includegraphics[width=0.75\textwidth]{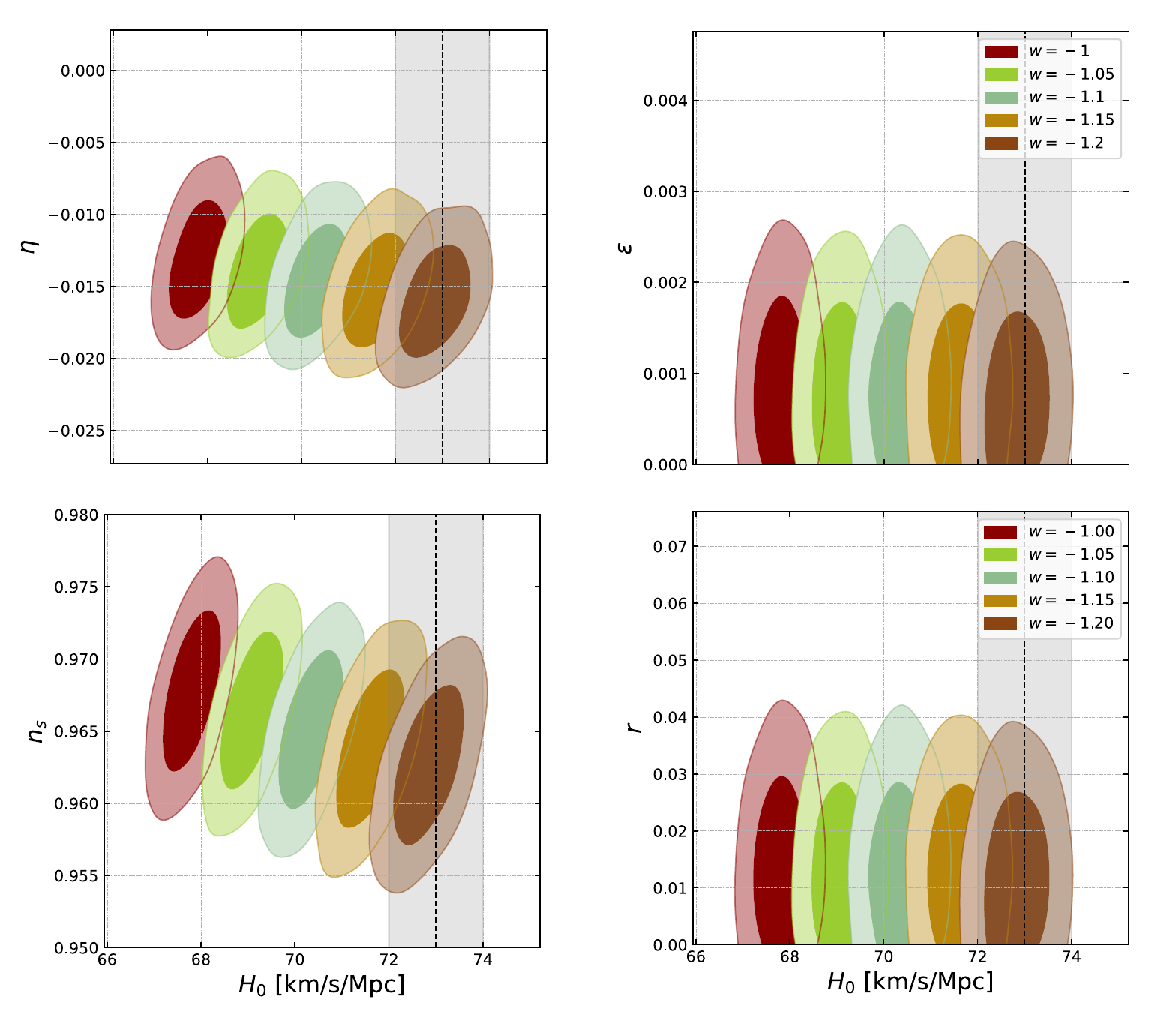}
    \caption{\textit{Bottom panels:} Two-dimensional correlations between $H_0$ and the inflationary parameters $n_s$ (left) and $r$ (right). \textit{Top panels:}  Two-dimensional correlations between the slow-roll parameters $\eta$ (left) and $\epsilon$ (right). The different contours are obtained by considering various realizations of late-time (new) physics altering the expansion history of the Universe after recombination, whose effects are parameterized in terms of a phantom Dark Energy component $w<-1$.}
    \label{fig:5}
\end{figure*}

\section{Late-time new physics and inflation}
\label{sec:results:late}
In this section, I study the implications for inflation resulting from models able to fix $w<-1$. As usual, I divide the section into two subsections: in \autoref{sec:Late.single}, I discuss the implications for generic single-field models while in \autoref{sec:Late.Starobinsky}, I focus on Starobinsky inflation.\\
\\

\subsection{Implications for Single field inflation}
\label{sec:Late.single}

In this case, the main results and correlations between various parameters of interest are displayed in~\autoref{fig:5}, which corroborate findings already extensively documented in the literature. Namely, a shift in the equation of state toward the phantom regime can alleviate the Hubble tension. Without delving into the debate of whether such physical realizations could occur in nature, I assume this parameterization as an effective representation of a late-time solution and explore its effects on inflation. Unlike when considering parameterizations of new physics at early times, in this case, the implications for inflationary parameters are much more modest. The bottom panels in \autoref{fig:5} show the correlation between $H_0$, the spectral index $n_s$, and the tensor tilt $r$. Moving towards higher values of $H_0$, the two-dimensional contours for $n_s$ remain almost unchanged, albeit with a slight shift. A remarkable thing is that, contrary to early-time solutions, here the shift here occurs towards lower values of $n_s$. On the other hand, focusing on the second panel (the bottom right one) of the same figure, we note that the constraints on the tensor amplitude remain unchanged in this scenario as well. When interpreting these results in relation to the slow-roll parameters $\epsilon$ and $\eta$ (depicted in the top panels of \autoref{fig:5}), it becomes evident that the upper limit on $\epsilon$ (which proportionally reflects the upper limit on the amplitude of tensor perturbations) is not sensitive to the introduction of new physics in the cosmological model. However, in the case of the parameter $\eta$, akin to the behavior observed in $n_s$, we note a modest shift resulting from the inclusion of new physics at later times. Notably, with $n_s$ shifting towards smaller values, $\eta$ now moves towards more negative values. Therefore, although the overall effect is very modest, I observe that the hierarchy between the slow-roll parameters $1 \gg |\eta| \gg \epsilon$ seems further favored based on these results\footnote{Although I do not discuss it in detail (not being relevant for inflation), it is worth noting that, just like Ref.~\cite{Vagnozzi:2019ezj}, I find that models with $w \simeq -1.2$ are strongly disfavored from a statistical point of view compared to $\Lambda$CDM.}.

\subsection{Implications for Starobinsky inflation}
\label{sec:Late.Starobinsky}

For the sake of completeness, I now turn to briefly study the implications for Starobinsky inflation. I consider 5 different cases where $w$ ranges from $w=-1$ up to $w=-1.2$ in steps of $\Delta w = -0.05$ adopting the theoretical relationships between $n_{s}$ and $r$ as defined in Eq.~\eqref{eq.Starob} from the beginning of the analysis. Consequently, the cosmological model retains 6 free parameters. I always examine the same combination of data, namely \textit{Planck+BK18+BAO}. 

\begin{figure}[tbp!]
    \centering
    \includegraphics[width=0.9\columnwidth]{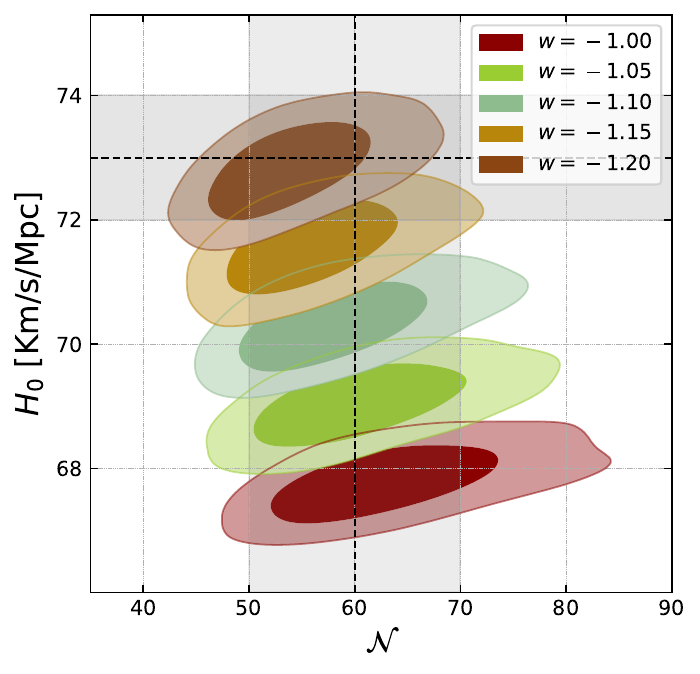}
    \caption{Two-dimensional correlations between $H_0$ and the number of e-folds $\mathcal N$ in Starobinsky Inflation. The different contours are obtained by considering various realizations of late-time (new) physics altering the expansion history of the Universe post-to recombination, whose effects are parameterized in terms of a phantom Dark Energy component.}
    \label{fig:6}
\end{figure}

The most interesting results I want to discuss are depicted in \autoref{fig:6}. The figure shows the correlation between $H_0$ and the number of e-folds $\mathcal{N}$ in Starobinsky Inflation across the aforementioned five cases. The shift in the value of $n_s$ produces a minor shift of $\mathcal{N}$ towards lower values compared to those obtained by fixing a standard background cosmology. However, contrary to what occurs within parameterizations of early-time new physics (where for sufficiently large corrections to $\Delta N_{\text{eff}}$, the model becomes inadequate in providing an optimal description of data), the model remains in good agreement with the observations. The value of $\mathcal{N}$ is consistent with the expected value based on theoretical arguments for this parameter.

\section{Early Dark Energy and Inflation}
\label{sec:EDE}

Up until now, I have been studying the implications of (early-time and late-time) new physics, whether for generic inflationary models or for Starobinsky inflation. The peculiarity of the approach I have taken lies in hypothesizing that a generic theory of new physics could set the values of parameters describing the expansion of the Universe (either at early or late times) to non-standard values beyond those set in $\Lambda$CDM. 

The advantage of this approach -- apart from remaining blind to the specific model of new physics -- is given by the limited number of free parameters involved in the analysis. In fact, in all cases discussed thus far, I considered models with either 7 or 6 free parameters: the same number or one more than in the standard cosmological model. This has allowed me to explore the implications of the Hubble tension for inflation following within a level of constraining power similar to $\Lambda$CDM. 

While I believe this analysis has potentially clarified and highlighted how the Hubble tension can represent an element of uncertainty surrounding our knowledge of cosmological inflation, the results rely in part on assumptions: primarily, the existence of a theory able to fix parameters to non-standard values without introducing additional degrees of freedom. An interesting difference compared to Ref.~\cite{Vagnozzi:2019ezj} (a work upon which the analysis conducted so far was largely inspired), is that these minimal scenarios seem to be somewhat "more disfavored" in light of updated data. Therefore, I believe it is imperative to extend further the analysis to more realistic potential solutions of the Hubble tension. 

Here I take an additional step forward and extend the discussion of inflation to axion EDE. After briefly describing the theoretical and numerical implementation of EDE (\autoref{sec:EDE_Implementation}), following an approach very similar to what I have been doing so far, I point out the effects resulting from \textit{assuming} the existence of a non-vanishing fraction of EDE (\autoref{sec:EDE-preliminary}). Since my analysis confirms that EDE might bring non-negligible implications for inflation, in \autoref{sec:general_EDE} I adopt a more traditional approach and discuss how the constraints on generic single-field models change in EDE cosmology allowing \textit{all} parameters to freely vary. Finally, another aspect that I would like to clarify is the following one: several scattered studies have suggested that Starobinsky inflation appears to be disfavored in the presence of EDE. However, none of the analyses in the literature have accounted for both paradigms simultaneously. In other words, an analysis that incorporates EDE in the early Universe while assuming Starobinsky inflation is lacking. This gap, to the best of my knowledge, remains unaddressed in the literature to date. Therefore, in \autoref{sec:Starobinsky_EDE}, I examine what current data can reveal about EDE and Starobinsky inflation, aiming to answer the fundamental question: can Starobinsky Inflation coexist within the framework of EDE cosmology?

\subsection{Axion Early Dark Energy}
\label{sec:EDE_Implementation}

I adopt the same EDE implementation detailed in Ref.~\cite{Hill:2020osr}. In this model, a light scalar field $\phi_{\text{EDE}}$ behaves as an effective cosmological constant that dynamically decays by oscillating in its potential $V(\phi_{\text{EDE}})$. To employ such a mechanism to reduce the value of the sound horizon and alleviate the Hubble tension, the field must begin to oscillate roughly around recombination, requiring some fine-tuning. For instance, around recombination, the Hubble parameter is $H \propto {T^2}/{M_{\text{pl}}}\sim 10^{-27}\text{ eV}$. Consequently, the scalar field must be extremely light to remain frozen for such a long time. From a particle physics standpoint, the only known example is the axion. For this reason, this model is usually referred to as axion-EDE. In order to avoid spoiling late-time cosmology, the vacuum energy associated with the field must redshift away faster than matter (i.e., faster than $a^{-3}$), and the field should behave as a subdominant component of the Universe. This requires significant restrictions on the potential. I adopt the same potential studied in Ref.~\cite{Hill:2020osr}
\begin{equation}
V(\phi_{\text{EDE}})\propto\left[1-\cos (\phi_{\text{EDE}} / f)\right]^{\gamma}
\end{equation}
(which can arise from instanton actions, although requiring a certain level of fine-tuning) by fixing $\gamma=3$.

From the numerical point of view, the EDE model is implemented within the Boltzmann solver code \texttt{CLASS}, making the necessary modifications so that the code solves for the evolution of the scalar field perturbations directly using the perturbed Klein–Gordon equation as described in Section III of Ref.~\cite{Hill:2020osr}. The set of parameters used to describe EDE consists of the fractional contribution to the total energy-density of the Universe, $f_{\text{EDE}}(z) \equiv \rho_{\text{EDE}}(z)/\rho_{\text{tot}}(z)$ evaluated at the critical redshift $z_c$ at which it reaches the maximum value, and $\theta_i$, which is the parameter that usually describes the initial field displacement. Along with the free EDE parameters, the other free parameters of the cosmological model always involve the baryon $\omega_{\text{b}}\equiv \Omega_{\text{b}}h^2$ and cold dark matter $\omega_{\text{c}}\equiv\Omega_{\text{c}}h^2$ energy densities, the angular size of the horizon at the last scattering surface $\theta_{*}$, the optical depth $\tau$, and the amplitude of primordial scalar perturbation $\log(10^{10}A_{\text{s}})$. Instead, as concerns other inflationary parameters such as the scalar spectral index $n_{\text{s}}$ and the tensor amplitude $r$, the parameterization adopted in this case depends on the different scenarios considered in the following subsections where different models of inflation are examined. The datasets involved in my analysis are the same as described in \autoref{sec:Methods} with the additional inclusion of a prior on $H_0=73\pm 1$ km/s/Mpc that for every case I include and exclude from the analysis.

\subsection{Preliminary observations}
\label{sec:EDE-preliminary}

\begin{figure}[htbp!]
    \centering
    \includegraphics[width=\columnwidth]{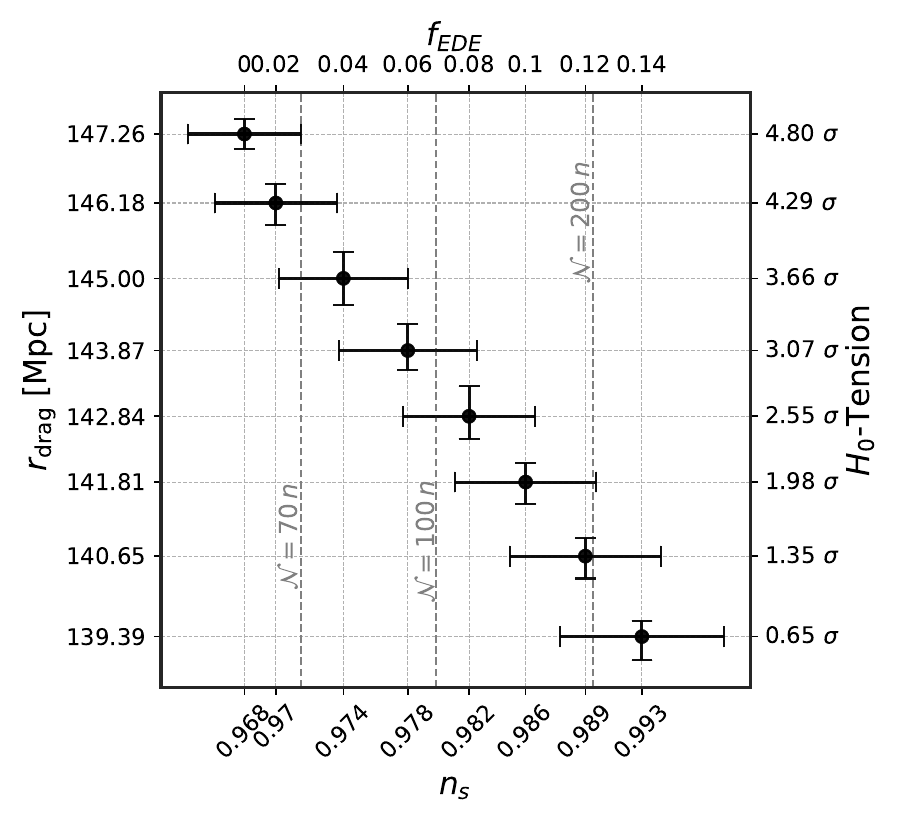}
    \caption{Each point of different color represents predictions in the plane ($n_s$ , $r_{\text{drag}}$) along with corresponding error bars for these two parameters from EDE models where $f_{\text{EDE}}$ is fixed to different values $f_{\text{EDE}}=\{0.02, \dots, 0.14\}$ displayed on the top x-axis. On the y-axis, on the right side of the figure, for each point, the Hubble tension is quantified using Eq.~\eqref{eq:H0}. The gray dashed vertical lines represent the number of e-folds $\mathcal{N}$ required for models where $n_s \simeq 1 - (2n/\mathcal{N})$.}
    \label{Fig:7}
\end{figure}

A non-negligible contribution from an EDE component in the early Universe implies a larger amount of energy-density just prior to recombination. As a results this mechanism leads to a reduction of the sound horizon and, consequently, to a larger value of the Hubble constant inferred by CMB observations. For this reason, EDE gained significant research attention due to its potential role in addressing the Hubble tension~\cite{Kamionkowski:2022pkx,Poulin:2023lkg,Abdalla:2022yfr}. 

This is very clear from \autoref{Fig:7}. To produce this figure, I followed an approach somewhat similar to what was done in the previous sections. Namely, I am fixing by hand the total fraction of EDE at values $f_{\text{EDE}} \in [0, 0.14]$ at steps of $\Delta f_{\text{EDE}}=0.02$. For each of these steps, I vary all the other 9 cosmological parameters listed in \autoref{sec:EDE_Implementation} and perform a full MCMC analysis (following the usual methodology). From the figure, we can clearly see how increasing the fraction of EDE (top x-axis) decreases the value of the sound horizon (left y-axis) and the Hubble tension (right y-axis). Nevertheless, when we consider increasingly higher $f_{\text{EDE}}$, we also move towards $n_s \to 1$ (bottom x-axis). As usual, the merit of this approach, on the one hand, is to obtain constraints with a precision comparable to the baseline cosmological model, thereby limiting the effects of degeneracy  among different parameters (most notably between $n_s$ and $f_{\rm EDE}$). On the other hand, it spots the effects of considering a non-vanishing fraction of EDE in the cosmological model and how it can (indirectly) change our constraints on inflationary parameters and possibly our understanding of cosmic inflation.

In this regard, I note that the analysis confirms the overall hints derived in the previous sections. New physics in the early Universe (potentially needed to solve the Hubble tension) might bring non-negligible implications for inflation, to a degree possiblly larger than expected. In \autoref{Fig:7} I quantify this effects in terms of the number of e-folds required for models where $n_s \simeq 1 - (2n/\mathcal{N})$. I find that an EDE fraction $f_{\text{EDE}} \sim 0.4 - 0.6$ (that could only mildly reduce the $H_0$-tension down to $\sim 3\sigma$) would already require a number of e-folds $\mathcal{N} \simeq 100n$. This value would be too large for Starobinsky-like models of inflation (where I recall $n \geq 1$), pointing instead towards models with $n < 1$ (largely disfavored within a minimal $\Lambda$CDM cosmology).

Motivated by these preliminary findings, I will now focus on EDE and inflation and, following a more traditional approach where all parameters are left free to vary, I will comprehensively review the constraints on inflationary parameters that we can obtain in the context of EDE cosmology. Following the same narrative thread adopted previously in this study, I will first consider the implications for generic single-field models, pointing out the implications for the inflationary parameters $\epsilon$ and $\eta$. Then I will focus on Starobinsky Inflation, answering the question of whether one of our best benchmark models of inflation is or is not in agreement with one of our best theoretical attempts to solve the Hubble tension.

\subsection{Early Dark Energy and Single Field Inflation}
\label{sec:general_EDE}

I start by considering generic single-field inflation described by Eq.\eqref{minimal coupled action}, allowing \textit{all} cosmological parameters to freely vary. As a result, the model I consider in this subsection has a total number of 10 parameters (listed in \autoref{sec:EDE_Implementation}). I perform a full MCMC analysis using the following two datasets:  \textit{Planck+BK18+BAO} and \textit{Planck+BK18+BAO+$H_0$}. The first one contains all the different likelihoods listed in \autoref{sec:Methods}. The second one includes also a Gaussian prior $H_0=73\pm 1$ km/s/Mpc coming from the SH0ES collaboration~\cite{Riess:2021jrx}.

The main point I want to address in this subsection is to discuss what happens to the slow-roll parameters $\epsilon$ and $\eta$ in EDE cosmology. The most interesting findings are shown in \autoref{fig:8}, where I display the two-dimensional contours and correlations between $\epsilon$ and $\eta$ for three different cases:

\begin{figure}[htbp!]
    \centering
    \includegraphics[width=\columnwidth]{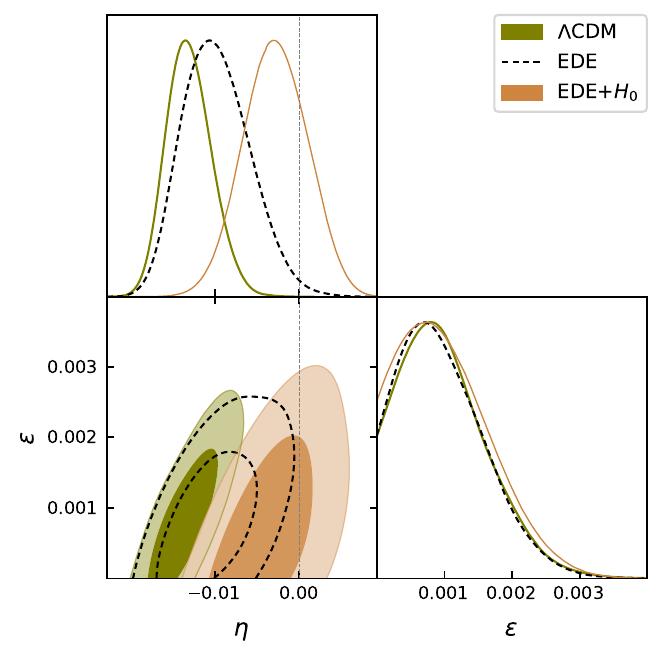}
    \caption{Two-dimensional correlations and one-dimensional posterior distributions for the slow-roll parameters $\epsilon$ and $\eta$ as obtained within $\Lambda$CDM and EDE (including and not including a SH0ES prior on $H_0$).}
    \label{fig:8}
\end{figure}

\begin{figure*}[htbp!]
    \centering
    \includegraphics[width=0.7\textwidth]{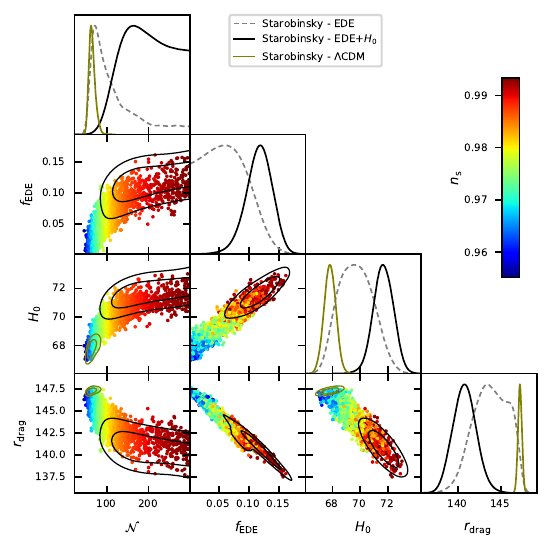}
    \caption{Two-dimensional marginalized contours for the number of efolds $\mathcal N$, the EDE fraction $f_{\rm EDE}$, the Hubble parameter $H_0$ and the sound horizon $r_{\rm{drag}}$ in a cosmological model where the inflationary sector of the theory is described by Starobinsky inflation. The color map in the figure represents the correlation between the value of $n_s$ and the other parameters as obtained by considering exclusively the combination of data Planck+BK18+BAO in EDE cosmology. The black contours in the figure show the results in the parameter space for the same model, assuming an additional prior on $H_0$ from local distance ladder measurements (referred to as Starobinsky-EDE+$H_0$ in the legend). To highlight the effects of EDE, I show in olive green the results obtained assuming Starobinsky inflation in a standard $\Lambda$CDM cosmology.}
    \label{fig:9}
\end{figure*}

\begin{itemize}[leftmargin=*]

\item \boldsymbol{$\Lambda$}\textbf{CDM}: In the context of the $\Lambda$CDM model, the results of the slow-roll parameters derived within a standard cosmological model for \textit{Planck+BK18+BAO} are represented by the green contours in the figure. These results were already presented in the earlier \autoref{sec:ealry.single} and are reiterated here. Despite having been extensively discussed before, I recall that within the standard cosmological model $\epsilon \lesssim 0.002$ while $\eta$ is measured to be $\eta=-0.0130^{+0.0024}_{-0.0029}$, namely negative and such that $1 \gg |\eta| \gg \epsilon$.

\item \textbf{EDE}: The results obtained from the analysis of the same dataset, \textit{Planck+BK18+BAO}, considering an evolutionary history of the primordial Universe that incorporates a fraction of Early Dark Energy (free to vary in the model), are represented by the dashed black contours in \autoref{fig:8}. As evident from the plot, there is a noticeable broadening of uncertainties for the parameter $\eta = -0.0098^{+0.0035}_{-0.0045}$. On the other hand, the results for the parameter $\epsilon$ remain mostly stable. This increase in the uncertainty of $\eta$ is partly attributable to the correlation between this parameter and those characteristics of EDE. The most significant correlation is undoubtedly between $\eta$ and $f_{\text{EDE}}$. Therefore, the constraints on the slow-roll parameters leave a wider margin to accommodate many inflationary models ruled out by the same data assuming a standard cosmological model. On a side note, I would like to stress that this broadening of uncertainties observed when all parameters are left to vary is precisely the reason for which I considered alternative approaches so far in the analysis.

\item \textbf{EDE+}\boldsymbol{$H_0$}: While EDE has received considerable attention for its potential to mitigate or resolve the Hubble tension, it is important to note that upon analysis with \textit{Planck+BK18+BAO} data, this model, although allowing for slightly higher values of $H_0$ compared to the standard cosmological model, does not inherently favor the value measured by the SH0ES collaboration. While this has raised doubts about the genuine effectiveness of this model in resolving the Hubble tension, an argument frequently advocated by the EDE community is that by incorporating a prior on $H_0$ in the data analysis, EDE notably enhances its fit over $\Lambda$CDM. Consequently, an alternative viewpoint is to acknowledge that when considering all the data collectively, EDE provides a better explanation than $\Lambda$CDM. While I am not delving deeper into this (somewhat tricky) debate (for more details, see e.g., the discussion on page 25 of Ref.~\cite{Poulin:2023lkg}), I would like to note that the introduction of a prior on $H_0$ leads to significant changes in the constraints on slow-roll parameters. The orange contours in \autoref{fig:8} depict the results obtained for \textit{Planck+BK18+BAO+$H_0$}. As one can see, the parameter $\epsilon$ remains somehow unchanged while the parameter $\eta = -0.0028 \pm 0.0037$ notably shifts. Values of $|\eta|\sim 0$ are now well within one standard deviation, indicating that hierarchies $1 \gg |\eta| \sim \epsilon$ (severely ruled out within $\Lambda$CDM) in this case are fully supported by data.
\end{itemize}

\subsection{Early Dark Energy and Starobinsky Inflation}
\label{sec:Starobinsky_EDE}

Based on previous efforts, I would like to examine a specific cosmological model that combines the Starobinsky action Eq.~\eqref{staro_action} to describe the inflationary Universe and simultaneously proposes EDE as a possible mechanism to mitigate the Hubble tension. My question is whether these two models can coexist based on current observations. In other words, is it feasible to resolve or at least alleviate the Hubble tension within the context of Early Dark Energy while maintaining agreement with Starobinsky inflation?

Before proceeding further, I would like to emphasize that this question has been already partially discussed in the literature, although with different levels of detail. In particular, in Refs.~\cite{Jiang:2022qlj,Ye:2022efx,Ye:2022afu,Fu:2023tfo}, it has been argued that reducing the Hubble constant discrepancy within EDE requires a significant change in $n_s$ that is in overall disagreement with the theoretical predictions of the aforementioned inflation model. That being said, it is important to note that these results were obtained without assuming any specific inflation model, but only comparing the results of inflationary parameters with those predicted by Eqs.~\eqref{eq.Starob} \textit{afterwards}. Here, I want to follow a different approach and consider both these models together from the beginning of the analysis. In other words, I assume the theoretical relations between $n_s$, $r$, and $\mathcal N$ predicted by the Starobinsky model from scratch, thereby reducing the number of free parameters. For EDE, I consider the same numerical implementation discussed in ~\autoref{sec:EDE_Implementation}. As a result, the cosmological model I am considering in this subsection has 9 free parameters: \{$\omega_{\rm b}\equiv \Omega_{\rm b}h^2$,  $\omega_{\rm c}\equiv\Omega_{\rm c}h^2$, $\theta_{*}$,  $\tau$, $\log(10^{10}A_{\rm s})$, $\mathcal N$, $f_{\rm EDE}(z_c)$, $z_c$, $\theta_i$\}. I analyze this model in light of observations from \textit{Planck+BK18+BAO} and \textit{Planck+BK18+BAO+$H_0$}.

The two-dimensional marginalized contours for the parameters of interest are illustrated in \autoref{fig:9}. The color map in the figure represents the correlation between the value of $n_s$ and the other parameters as obtained by considering exclusively the combination of data \textit{Planck+BK18+BAO}. The black contours in the figure show the results in the parameter space for the same model, assuming an additional prior on $H_0$ from local distance ladder measurements (referred to in the legend as Starobinsky-EDE+$H_0$). Furthermore, to highlight the effects of EDE, I show in olive green the results obtained assuming Starobinsky inflation in a standard $\Lambda$CDM cosmology.

\begin{figure}[htbp!]
    \centering
    \includegraphics[width=\columnwidth]{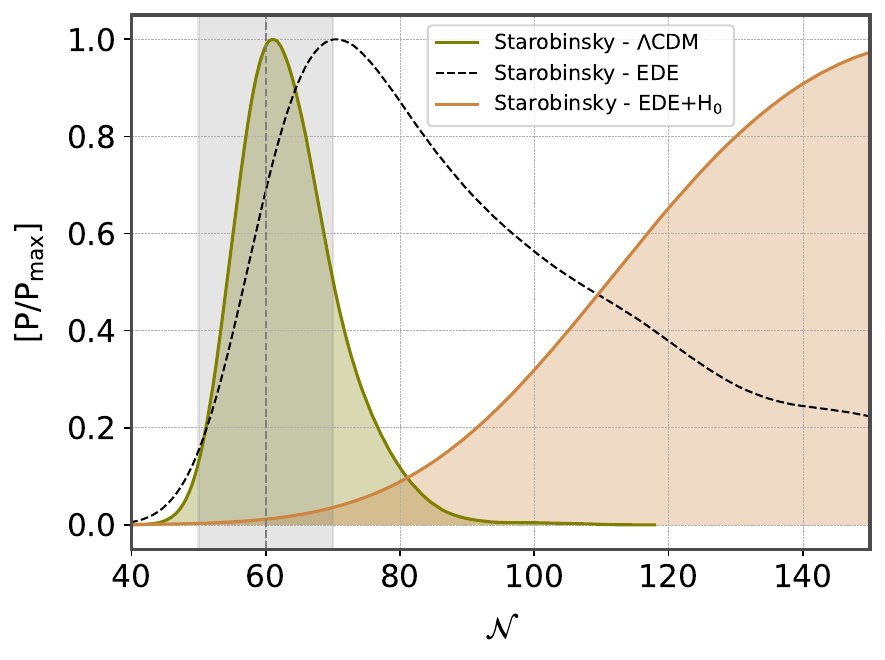}
    \caption{One-dimensional posterior distributions for the number of e-folds $\mathcal{N}$ in Starobinsky inflation as obtained within $\Lambda$CDM and EDE (including and not including a SH0ES prior on $H_0$).}
    \label{fig:10}
\end{figure}

Concerning the only free parameter in the inflationary sector of the theory -- the number of e-folds $\mathcal{N}$ -- I find that in EDE cosmology, I can get values that are compatible with those expected on the basis of theoretical arguments. So, Starobinsky inflation is not necessarily ruled out in EDE cosmology. This is shown clearly in \autoref{fig:10}, where I compare the 1D posterior of $\mathcal N$ for the three cases described earlier (i.e., $\Lambda$CDM, EDE, and EDE+$H_0$). I note that the substantial difference of assuming EDE or $\Lambda$CDM is that the 1D posterior distribution in the case of EDE has a much broader right tail compared to what is observed assuming the standard cosmological model. Returning to \autoref{fig:9}, it is easy to see that this probability tail is due to the correlation between $f_{\text{EDE}}$, $r_{\text{drag}}$, $H_0$, and $n_s$. In particular, a significant fraction of EDE leads to a reduction in the sound horizon value, thereby increasing $H_0$. As extensively discussed in the previous section, this results in higher values of $n_s$ and, consequently, larger values of $\mathcal{N}$. Therefore, from the same figure, it is clear that maintaining consistency with Starobinsky inflation would require having only a very tiny fraction of EDE $f_{\text{EDE}} \lesssim 0.06$, in line with what I discussed at the beginning of this section.

When considering a prior on $H_0$ from SH0ES, one expects the contribution of models with larger $f_{\text{EDE}}$ to become much more significant, as these models can increase $H_0$ and better explain data. This is what I observe in \autoref{fig:9}: assuming a prior on $H_0=73\pm1$ km/s/Mpc, the black contours shift significantly in the direction of a non-negligible fraction of Early Dark Energy and a lower value of the sound horizon. This implies higher values of $n_s$ and, within the Starobinsky model, higher values of $\mathcal N$. Indeed, as evident from \autoref{fig:10}, values of the number of e-folds in the range $\mathcal N\in[50,70]$ are now excluded at more than 3 standard deviations. In this regard, it is worth noting that the exact number of e-folds of inflation $\mathcal{N}$, generally depends on the model itself and the details of the subsequent post-inflationary evolution through the reheating temperature~\cite{Lidsey:1995np, Liddle:2003as}. Although under some circumstances it is theoretically possible to achieve models with a greater number of e-folds\footnote{In general, to achieve a greater number of $e$-folds of expansion, one needs to reduce the energy density during inflation or significantly lower the energy scale at the end of inflation. In these non-standard scenarios, it is possible to achieve an extreme upper bound of $\mathcal{N} \lesssim 100$; see Ref.~\cite{Liddle:2003as}.}, these realizations are typically not applicable to the minimal $R^2$-inflation considered in this work. In realistic realizations of reheating in Starobinsky inflation~\cite{Gorbunov:2010bn,Bezrukov:2011gp}, the inflaton coupling to all fields is Planck scale suppressed, and reheating mainly occurs via its decays into the Standard Model Higgs bosons (which immediately rescatter into Standard Model particles) since the inflaton couplings to all other fields are further suppressed due to conformal symmetry~\cite{Starobinsky:1980te,Vilenkin:1985md}. Consequently, the reheating temperature in Starobinsky inflation is usually lower than in other benchmark scenarios and can be estimated to be of the order of $T_{\text{rh}} \sim 10^9$ GeV. Therefore even in the most optimistic -- yet realistic -- setup, one gets a number of e-folds not greater than $\mathcal{N} \sim 55 - 60$, see e.g, Ref~\cite{Bezrukov:2011gp} (or Ref.~\cite{Planck:2018jri}). This unequivocally confirms that the fraction of Early Dark Energy allowed to maintain agreement with the theoretical predictions of the Starobinsky model cannot be as large as required to resolve the Hubble tension. Therefore, the conclusion I draw is that resolving the Hubble tension in the context of Early Dark Energy excludes Starobinsky Inflation at more than $3\sigma$ as the mechanism describing inflation.

\section{Conclusions}
\label{sec:Conclusion}
The Hubble constant tension could be pointing towards new physics beyond the standard $\Lambda$CDM model of cosmology. In this paper, I reviewed and discussed the possible implications for inflationary cosmology resulting from considering new physics in the cosmological framework. My study is motivated by a simple argument that the constraints on inflationary parameters (most typically the scalar spectral index $n_s$) depend to some extent on the cosmological model adopted. Despite a few scattered studies having argued that considering new physics in light of the Hubble tension could shift the constraints on inflation, many alternative theoretical models introduce extra degrees of freedom beyond the standard 6, typical of $\Lambda$CDM. Marginalizing over additional parameters has the effect of broadening the uncertainties. Therefore, taking the results on the inflationary parameters inferred in alternative cosmological models at face value, very often we remain in agreement with the theoretical predictions of benchmark scenarios just because of the larger uncertainties, while observing significant shifts in their central values. This makes it somewhat unclear to what extent the shift itself is resulting from correlations among (many) free parameters and more broadly to what extent the Hubble tension represents an actual additional layer of uncertainty for inflationary cosmology.

To clarify this matter, following Ref.~\cite{Vagnozzi:2019ezj}, I first adopted an alternative viewpoint answering the following question: what happens to our best understanding of inflation in the presence of a physical theory able to \textit{fix} extra parameters to a specific set of non-standard values? In this case, the degrees of freedom of the alternative models are reduced compared to the most typically proposed solutions, remaining comparable to the degrees of freedom of the baseline case. Similarly to Ref.~\cite{Vagnozzi:2019ezj}, I focused on the dark energy equation of state $w$ and the effective number of relativistic species $N_{\text{eff}}$. I found that physical theories able to fix $w \approx -1.2$ or $N_{\text{eff}} \approx 3.9$ would lead to an estimate of $H_0$ from CMB and BAO data in agreement with the local distance ladder measurements from the SH0ES collaboration. However, I showed that the latter also requires a shift towards a scale-invariant spectrum of primordial perturbations $n_s \to 1$, while the former does not have any particular implications for inflation. From a model-selection perspective, these two non-standard scenarios are strongly disfavored compared to $\Lambda$CDM. However, models that predict $N_{\text{eff}} \approx 3.3 - 3.4$ would be able to bring the tension down to 3-2$\sigma$ while "only" being moderately disfavored over $\Lambda$CDM. Nevertheless, my analysis suggests that considering scenarios with $N_{\text{eff}} \approx 3.3$ is already enough to change the constraints inferred for inflationary parameters in such a way that the most accredited models (most notably Starobinsky inflation) would no longer be supported by CMB and BAO data. 

More broadly, the first part of my analysis suggests that early-time new physics can have the larger impact on inflation, essentially calling into question the hierarchy between the slow-roll parameters $\epsilon$ and $\eta$. For all the cases considered, I obtain a limit $\epsilon \lesssim 0.002$. Conversely, the constraints on the slow-roll parameter $\eta$ are significantly more model dependent. Within the standard cosmological model, $\eta$ is measured to be $\eta=-0.0130^{+0.0024}_{-0.0029}$, namely negative and such that $1 \gg |\eta| \gg \epsilon$. This hierarchy might change when considering new physical mechanisms that could resolve the Hubble tension for instance increasing the expansion rate of the Universe prior to recombination. 

Motivated by these findings, I studied more likely proposed solutions to the $H_0$-tension and provide a comprehensive overview of the constraints on inflation in EDE cosmology. In particular, I initially adopted a similar perspective and explore how the constraints on inflation are influenced by \textit{assuming} a non-vanishing fraction of EDE in the early Universe. My analysis reveals that even a relatively modest EDE component -- of the order of $\sim 4\% - 6\%$ of the total energy-density of the Early Universe, $f_{\text{EDE}} \sim 0.04 - 0.06$ -- could already require a change in perspective in terms of which inflationary models are favored or disfavored based on current data while only moderately alleviating the $H_0$-tension to around $\sim 3.5\sigma$.  In \autoref{Fig:7}, I quantify these effects in terms of the number of e-folds required for models where $n_s \simeq 1 - (2n/\mathcal{N})$. For $f_{\text{EDE}} \sim 0.04 - 0.06$, I find a number of e-folds $\mathcal{N} \simeq 100n$. This value would be too large within Starobinsky-like models (where $n \geq 1$), pointing instead towards models with $n < 1$ (largely disfavored within a minimal $\Lambda$CDM cosmology). I confirm my findings by performing a full joint analysis of EDE and Starobinsky inflation where the two models are simultaneously assumed in the cosmological framework and \textit{all} model parameters are left free to vary. In this case, I conclusively demonstrate that the two paradigms can hardly coexist if $f_{\text{EDE}} \gtrsim 0.06$. Such values of $f_{\rm EDE}$ would in fact demand an excessively high number of e-folds before the end of inflation in Starobinsky Inflation. Therefore, the conclusion I draw is that addressing the Hubble tension in the context of EDE could exclude Starobinsky-like models at more than $3\sigma$ as the mechanism describing inflation.

Overall, based on the results obtained in this work, there is solid ground to conclude that the presence of the Hubble tension could indeed represent a non-negligible source of uncertainty for inflation. In this regard, I wish to emphasize that within scenarios in which we maintain a constraining power comparable to the standard cosmological model even \textit{mild} deviations from an early-time $\Lambda$CDM cosmology (potentially supported by current data) could have \textit{significant} implications for inflation. This highlights how constraints on inflation may be influenced by the cosmological model, potentially to a greater degree than commonly realized.

\begin{acknowledgments}
\noindent I thank Carsten van de Bruck, Mariaveronica De Angelis, Eleonora Di Valentino, Matteo Forconi, Elsa Teixeira and Sunny Vagnozzi for many interesting discussions and suggestions around the subject of this article. This article is based upon work from COST Action CA21136 Addressing observational tensions in cosmology with systematics and fundamental physics (CosmoVerse) supported by COST (European Cooperation in Science and Technology).I acknowledge IT Services at The University of Sheffield for the provision of services for High Performance Computing.
\end{acknowledgments}

\bibliographystyle{apsrev4-1}
\bibliography{main.bib}

\begin{thebibliography}{357}%
\makeatletter
\providecommand \@ifxundefined [1]{%
 \@ifx{#1\undefined}
}%
\providecommand \@ifnum [1]{%
 \ifnum #1\expandafter \@firstoftwo
 \else \expandafter \@secondoftwo
 \fi
}%
\providecommand \@ifx [1]{%
 \ifx #1\expandafter \@firstoftwo
 \else \expandafter \@secondoftwo
 \fi
}%
\providecommand \natexlab [1]{#1}%
\providecommand \enquote  [1]{``#1''}%
\providecommand \bibnamefont  [1]{#1}%
\providecommand \bibfnamefont [1]{#1}%
\providecommand \citenamefont [1]{#1}%
\providecommand \href@noop [0]{\@secondoftwo}%
\providecommand \href [0]{\begingroup \@sanitize@url \@href}%
\providecommand \@href[1]{\@@startlink{#1}\@@href}%
\providecommand \@@href[1]{\endgroup#1\@@endlink}%
\providecommand \@sanitize@url [0]{\catcode `\\12\catcode `\$12\catcode `\&12\catcode `\#12\catcode `\^12\catcode `\_12\catcode `\%12\relax}%
\providecommand \@@startlink[1]{}%
\providecommand \@@endlink[0]{}%
\providecommand \url  [0]{\begingroup\@sanitize@url \@url }%
\providecommand \@url [1]{\endgroup\@href {#1}{\urlprefix }}%
\providecommand \urlprefix  [0]{URL }%
\providecommand \Eprint [0]{\href }%
\providecommand \doibase [0]{http://dx.doi.org/}%
\providecommand \selectlanguage [0]{\@gobble}%
\providecommand \bibinfo  [0]{\@secondoftwo}%
\providecommand \bibfield  [0]{\@secondoftwo}%
\providecommand \translation [1]{[#1]}%
\providecommand \BibitemOpen [0]{}%
\providecommand \bibitemStop [0]{}%
\providecommand \bibitemNoStop [0]{.\EOS\space}%
\providecommand \EOS [0]{\spacefactor3000\relax}%
\providecommand \BibitemShut  [1]{\csname bibitem#1\endcsname}%
\let\auto@bib@innerbib\@empty
\bibitem [{\citenamefont {Guth}(1981)}]{Guth:1980zm}%
  \BibitemOpen
  \bibfield  {author} {\bibinfo {author} {\bibfnamefont {A.~H.}\ \bibnamefont {Guth}},\ }\href {\doibase 10.1103/PhysRevD.23.347} {\bibfield  {journal} {\bibinfo  {journal} {Phys. Rev. D}\ }\textbf {\bibinfo {volume} {23}},\ \bibinfo {pages} {347} (\bibinfo {year} {1981})}\BibitemShut {NoStop}%
\bibitem [{\citenamefont {Linde}(1982)}]{Linde:1981mu}%
  \BibitemOpen
  \bibfield  {author} {\bibinfo {author} {\bibfnamefont {A.~D.}\ \bibnamefont {Linde}},\ }\href {\doibase 10.1016/0370-2693(82)91219-9} {\bibfield  {journal} {\bibinfo  {journal} {Phys. Lett. B}\ }\textbf {\bibinfo {volume} {108}},\ \bibinfo {pages} {389} (\bibinfo {year} {1982})}\BibitemShut {NoStop}%
\bibitem [{\citenamefont {Albrecht}\ and\ \citenamefont {Steinhardt}(1982)}]{Albrecht:1982wi}%
  \BibitemOpen
  \bibfield  {author} {\bibinfo {author} {\bibfnamefont {A.}~\bibnamefont {Albrecht}}\ and\ \bibinfo {author} {\bibfnamefont {P.~J.}\ \bibnamefont {Steinhardt}},\ }\href {\doibase 10.1103/PhysRevLett.48.1220} {\bibfield  {journal} {\bibinfo  {journal} {Phys. Rev. Lett.}\ }\textbf {\bibinfo {volume} {48}},\ \bibinfo {pages} {1220} (\bibinfo {year} {1982})}\BibitemShut {NoStop}%
\bibitem [{\citenamefont {Vilenkin}(1983)}]{Vilenkin:1983xq}%
  \BibitemOpen
  \bibfield  {author} {\bibinfo {author} {\bibfnamefont {A.}~\bibnamefont {Vilenkin}},\ }\href {\doibase 10.1103/PhysRevD.27.2848} {\bibfield  {journal} {\bibinfo  {journal} {Phys. Rev. D}\ }\textbf {\bibinfo {volume} {27}},\ \bibinfo {pages} {2848} (\bibinfo {year} {1983})}\BibitemShut {NoStop}%
\bibitem [{\citenamefont {Mukhanov}\ and\ \citenamefont {Chibisov}(1981)}]{Mukhanov:1981xt}%
  \BibitemOpen
  \bibfield  {author} {\bibinfo {author} {\bibfnamefont {V.~F.}\ \bibnamefont {Mukhanov}}\ and\ \bibinfo {author} {\bibfnamefont {G.~V.}\ \bibnamefont {Chibisov}},\ }\href@noop {} {\bibfield  {journal} {\bibinfo  {journal} {JETP Lett.}\ }\textbf {\bibinfo {volume} {33}},\ \bibinfo {pages} {532} (\bibinfo {year} {1981})}\BibitemShut {NoStop}%
\bibitem [{\citenamefont {Bardeen}\ \emph {et~al.}(1983)\citenamefont {Bardeen}, \citenamefont {Steinhardt},\ and\ \citenamefont {Turner}}]{Bardeen:1983qw}%
  \BibitemOpen
  \bibfield  {author} {\bibinfo {author} {\bibfnamefont {J.~M.}\ \bibnamefont {Bardeen}}, \bibinfo {author} {\bibfnamefont {P.~J.}\ \bibnamefont {Steinhardt}}, \ and\ \bibinfo {author} {\bibfnamefont {M.~S.}\ \bibnamefont {Turner}},\ }\href {\doibase 10.1103/PhysRevD.28.679} {\bibfield  {journal} {\bibinfo  {journal} {Phys. Rev. D}\ }\textbf {\bibinfo {volume} {28}},\ \bibinfo {pages} {679} (\bibinfo {year} {1983})}\BibitemShut {NoStop}%
\bibitem [{\citenamefont {Hawking}(1982)}]{Hawking:1982cz}%
  \BibitemOpen
  \bibfield  {author} {\bibinfo {author} {\bibfnamefont {S.~W.}\ \bibnamefont {Hawking}},\ }\href {\doibase 10.1016/0370-2693(82)90373-2} {\bibfield  {journal} {\bibinfo  {journal} {Phys. Lett. B}\ }\textbf {\bibinfo {volume} {115}},\ \bibinfo {pages} {295} (\bibinfo {year} {1982})}\BibitemShut {NoStop}%
\bibitem [{\citenamefont {Guth}\ and\ \citenamefont {Pi}(1982)}]{Guth:1982ec}%
  \BibitemOpen
  \bibfield  {author} {\bibinfo {author} {\bibfnamefont {A.~H.}\ \bibnamefont {Guth}}\ and\ \bibinfo {author} {\bibfnamefont {S.~Y.}\ \bibnamefont {Pi}},\ }\href {\doibase 10.1103/PhysRevLett.49.1110} {\bibfield  {journal} {\bibinfo  {journal} {Phys. Rev. Lett.}\ }\textbf {\bibinfo {volume} {49}},\ \bibinfo {pages} {1110} (\bibinfo {year} {1982})}\BibitemShut {NoStop}%
\bibitem [{\citenamefont {Bennett}\ \emph {et~al.}(2013)\citenamefont {Bennett} \emph {et~al.}}]{WMAP:2012fli}%
  \BibitemOpen
  \bibfield  {author} {\bibinfo {author} {\bibfnamefont {C.~L.}\ \bibnamefont {Bennett}} \emph {et~al.} (\bibinfo {collaboration} {WMAP}),\ }\href {\doibase 10.1088/0067-0049/208/2/20} {\bibfield  {journal} {\bibinfo  {journal} {Astrophys. J. Suppl.}\ }\textbf {\bibinfo {volume} {208}},\ \bibinfo {pages} {20} (\bibinfo {year} {2013})},\ \Eprint {http://arxiv.org/abs/1212.5225} {arXiv:1212.5225 [astro-ph.CO]} \BibitemShut {NoStop}%
\bibitem [{\citenamefont {Hinshaw}\ \emph {et~al.}(2013)\citenamefont {Hinshaw} \emph {et~al.}}]{WMAP:2012nax}%
  \BibitemOpen
  \bibfield  {author} {\bibinfo {author} {\bibfnamefont {G.}~\bibnamefont {Hinshaw}} \emph {et~al.} (\bibinfo {collaboration} {WMAP}),\ }\href {\doibase 10.1088/0067-0049/208/2/19} {\bibfield  {journal} {\bibinfo  {journal} {Astrophys. J. Suppl.}\ }\textbf {\bibinfo {volume} {208}},\ \bibinfo {pages} {19} (\bibinfo {year} {2013})},\ \Eprint {http://arxiv.org/abs/1212.5226} {arXiv:1212.5226 [astro-ph.CO]} \BibitemShut {NoStop}%
\bibitem [{\citenamefont {Aghanim}\ \emph {et~al.}(2020{\natexlab{a}})\citenamefont {Aghanim} \emph {et~al.}}]{Planck:2018nkj}%
  \BibitemOpen
  \bibfield  {author} {\bibinfo {author} {\bibfnamefont {N.}~\bibnamefont {Aghanim}} \emph {et~al.} (\bibinfo {collaboration} {Planck}),\ }\href {\doibase 10.1051/0004-6361/201833880} {\bibfield  {journal} {\bibinfo  {journal} {Astron. Astrophys.}\ }\textbf {\bibinfo {volume} {641}},\ \bibinfo {pages} {A1} (\bibinfo {year} {2020}{\natexlab{a}})},\ \Eprint {http://arxiv.org/abs/1807.06205} {arXiv:1807.06205 [astro-ph.CO]} \BibitemShut {NoStop}%
\bibitem [{\citenamefont {Aghanim}\ \emph {et~al.}(2020{\natexlab{b}})\citenamefont {Aghanim} \emph {et~al.}}]{Planck:2018vyg}%
  \BibitemOpen
  \bibfield  {author} {\bibinfo {author} {\bibfnamefont {N.}~\bibnamefont {Aghanim}} \emph {et~al.} (\bibinfo {collaboration} {Planck}),\ }\href {\doibase 10.1051/0004-6361/201833910} {\bibfield  {journal} {\bibinfo  {journal} {Astron. Astrophys.}\ }\textbf {\bibinfo {volume} {641}},\ \bibinfo {pages} {A6} (\bibinfo {year} {2020}{\natexlab{b}})},\ \bibinfo {note} {[Erratum: Astron.Astrophys. 652, C4 (2021)]},\ \Eprint {http://arxiv.org/abs/1807.06209} {arXiv:1807.06209 [astro-ph.CO]} \BibitemShut {NoStop}%
\bibitem [{\citenamefont {Ruhl}\ \emph {et~al.}(2004)\citenamefont {Ruhl} \emph {et~al.}}]{SPT:2004qip}%
  \BibitemOpen
  \bibfield  {author} {\bibinfo {author} {\bibfnamefont {J.~E.}\ \bibnamefont {Ruhl}} \emph {et~al.} (\bibinfo {collaboration} {SPT}),\ }\href {\doibase 10.1117/12.552473} {\bibfield  {journal} {\bibinfo  {journal} {Proc. SPIE Int. Soc. Opt. Eng.}\ }\textbf {\bibinfo {volume} {5498}},\ \bibinfo {pages} {11} (\bibinfo {year} {2004})},\ \Eprint {http://arxiv.org/abs/astro-ph/0411122} {arXiv:astro-ph/0411122} \BibitemShut {NoStop}%
\bibitem [{\citenamefont {Dutcher}\ \emph {et~al.}(2021)\citenamefont {Dutcher} \emph {et~al.}}]{SPT-3G:2021eoc}%
  \BibitemOpen
  \bibfield  {author} {\bibinfo {author} {\bibfnamefont {D.}~\bibnamefont {Dutcher}} \emph {et~al.} (\bibinfo {collaboration} {SPT-3G}),\ }\href {\doibase 10.1103/PhysRevD.104.022003} {\bibfield  {journal} {\bibinfo  {journal} {Phys. Rev. D}\ }\textbf {\bibinfo {volume} {104}},\ \bibinfo {pages} {022003} (\bibinfo {year} {2021})},\ \Eprint {http://arxiv.org/abs/2101.01684} {arXiv:2101.01684 [astro-ph.CO]} \BibitemShut {NoStop}%
\bibitem [{\citenamefont {Choi}\ \emph {et~al.}(2020)\citenamefont {Choi} \emph {et~al.}}]{ACT:2020frw}%
  \BibitemOpen
  \bibfield  {author} {\bibinfo {author} {\bibfnamefont {S.~K.}\ \bibnamefont {Choi}} \emph {et~al.} (\bibinfo {collaboration} {ACT}),\ }\href {\doibase 10.1088/1475-7516/2020/12/045} {\bibfield  {journal} {\bibinfo  {journal} {JCAP}\ }\textbf {\bibinfo {volume} {12}},\ \bibinfo {pages} {045} (\bibinfo {year} {2020})},\ \Eprint {http://arxiv.org/abs/2007.07289} {arXiv:2007.07289 [astro-ph.CO]} \BibitemShut {NoStop}%
\bibitem [{\citenamefont {Aiola}\ \emph {et~al.}(2020)\citenamefont {Aiola} \emph {et~al.}}]{ACT:2020gnv}%
  \BibitemOpen
  \bibfield  {author} {\bibinfo {author} {\bibfnamefont {S.}~\bibnamefont {Aiola}} \emph {et~al.} (\bibinfo {collaboration} {ACT}),\ }\href {\doibase 10.1088/1475-7516/2020/12/047} {\bibfield  {journal} {\bibinfo  {journal} {JCAP}\ }\textbf {\bibinfo {volume} {12}},\ \bibinfo {pages} {047} (\bibinfo {year} {2020})},\ \Eprint {http://arxiv.org/abs/2007.07288} {arXiv:2007.07288 [astro-ph.CO]} \BibitemShut {NoStop}%
\bibitem [{\citenamefont {Madhavacheril}\ \emph {et~al.}(2023)\citenamefont {Madhavacheril} \emph {et~al.}}]{ACT:2023kun}%
  \BibitemOpen
  \bibfield  {author} {\bibinfo {author} {\bibfnamefont {M.~S.}\ \bibnamefont {Madhavacheril}} \emph {et~al.} (\bibinfo {collaboration} {ACT}),\ }\href@noop {} {\  (\bibinfo {year} {2023})},\ \Eprint {http://arxiv.org/abs/2304.05203} {arXiv:2304.05203 [astro-ph.CO]} \BibitemShut {NoStop}%
\bibitem [{\citenamefont {Dawson}\ \emph {et~al.}(2013)\citenamefont {Dawson} \emph {et~al.}}]{BOSS:2012dmf}%
  \BibitemOpen
  \bibfield  {author} {\bibinfo {author} {\bibfnamefont {K.~S.}\ \bibnamefont {Dawson}} \emph {et~al.} (\bibinfo {collaboration} {BOSS}),\ }\href {\doibase 10.1088/0004-6256/145/1/10} {\bibfield  {journal} {\bibinfo  {journal} {Astron. J.}\ }\textbf {\bibinfo {volume} {145}},\ \bibinfo {pages} {10} (\bibinfo {year} {2013})},\ \Eprint {http://arxiv.org/abs/1208.0022} {arXiv:1208.0022 [astro-ph.CO]} \BibitemShut {NoStop}%
\bibitem [{\citenamefont {Anderson}\ \emph {et~al.}(2014)\citenamefont {Anderson} \emph {et~al.}}]{BOSS:2013rlg}%
  \BibitemOpen
  \bibfield  {author} {\bibinfo {author} {\bibfnamefont {L.}~\bibnamefont {Anderson}} \emph {et~al.} (\bibinfo {collaboration} {BOSS}),\ }\href {\doibase 10.1093/mnras/stu523} {\bibfield  {journal} {\bibinfo  {journal} {Mon. Not. Roy. Astron. Soc.}\ }\textbf {\bibinfo {volume} {441}},\ \bibinfo {pages} {24} (\bibinfo {year} {2014})},\ \Eprint {http://arxiv.org/abs/1312.4877} {arXiv:1312.4877 [astro-ph.CO]} \BibitemShut {NoStop}%
\bibitem [{\citenamefont {Delubac}\ \emph {et~al.}(2015)\citenamefont {Delubac} \emph {et~al.}}]{BOSS:2014hwf}%
  \BibitemOpen
  \bibfield  {author} {\bibinfo {author} {\bibfnamefont {T.}~\bibnamefont {Delubac}} \emph {et~al.} (\bibinfo {collaboration} {BOSS}),\ }\href {\doibase 10.1051/0004-6361/201423969} {\bibfield  {journal} {\bibinfo  {journal} {Astron. Astrophys.}\ }\textbf {\bibinfo {volume} {574}},\ \bibinfo {pages} {A59} (\bibinfo {year} {2015})},\ \Eprint {http://arxiv.org/abs/1404.1801} {arXiv:1404.1801 [astro-ph.CO]} \BibitemShut {NoStop}%
\bibitem [{\citenamefont {Alam}\ \emph {et~al.}(2017)\citenamefont {Alam} \emph {et~al.}}]{BOSS:2016wmc}%
  \BibitemOpen
  \bibfield  {author} {\bibinfo {author} {\bibfnamefont {S.}~\bibnamefont {Alam}} \emph {et~al.} (\bibinfo {collaboration} {BOSS}),\ }\href {\doibase 10.1093/mnras/stx721} {\bibfield  {journal} {\bibinfo  {journal} {Mon. Not. Roy. Astron. Soc.}\ }\textbf {\bibinfo {volume} {470}},\ \bibinfo {pages} {2617} (\bibinfo {year} {2017})},\ \Eprint {http://arxiv.org/abs/1607.03155} {arXiv:1607.03155 [astro-ph.CO]} \BibitemShut {NoStop}%
\bibitem [{\citenamefont {Beutler}\ \emph {et~al.}(2014)\citenamefont {Beutler} \emph {et~al.}}]{BOSS:2013uda}%
  \BibitemOpen
  \bibfield  {author} {\bibinfo {author} {\bibfnamefont {F.}~\bibnamefont {Beutler}} \emph {et~al.} (\bibinfo {collaboration} {BOSS}),\ }\href {\doibase 10.1093/mnras/stu1051} {\bibfield  {journal} {\bibinfo  {journal} {Mon. Not. Roy. Astron. Soc.}\ }\textbf {\bibinfo {volume} {443}},\ \bibinfo {pages} {1065} (\bibinfo {year} {2014})},\ \Eprint {http://arxiv.org/abs/1312.4611} {arXiv:1312.4611 [astro-ph.CO]} \BibitemShut {NoStop}%
\bibitem [{\citenamefont {Alam}\ \emph {et~al.}(2021)\citenamefont {Alam} \emph {et~al.}}]{eBOSS:2020yzd}%
  \BibitemOpen
  \bibfield  {author} {\bibinfo {author} {\bibfnamefont {S.}~\bibnamefont {Alam}} \emph {et~al.} (\bibinfo {collaboration} {eBOSS}),\ }\href {\doibase 10.1103/PhysRevD.103.083533} {\bibfield  {journal} {\bibinfo  {journal} {Phys. Rev. D}\ }\textbf {\bibinfo {volume} {103}},\ \bibinfo {pages} {083533} (\bibinfo {year} {2021})},\ \Eprint {http://arxiv.org/abs/2007.08991} {arXiv:2007.08991 [astro-ph.CO]} \BibitemShut {NoStop}%
\bibitem [{\citenamefont {Tegmark}\ \emph {et~al.}(2004)\citenamefont {Tegmark} \emph {et~al.}}]{SDSS:2003eyi}%
  \BibitemOpen
  \bibfield  {author} {\bibinfo {author} {\bibfnamefont {M.}~\bibnamefont {Tegmark}} \emph {et~al.} (\bibinfo {collaboration} {SDSS}),\ }\href {\doibase 10.1103/PhysRevD.69.103501} {\bibfield  {journal} {\bibinfo  {journal} {Phys. Rev. D}\ }\textbf {\bibinfo {volume} {69}},\ \bibinfo {pages} {103501} (\bibinfo {year} {2004})},\ \Eprint {http://arxiv.org/abs/astro-ph/0310723} {arXiv:astro-ph/0310723} \BibitemShut {NoStop}%
\bibitem [{\citenamefont {Seljak}\ \emph {et~al.}(2005)\citenamefont {Seljak} \emph {et~al.}}]{SDSS:2004kqt}%
  \BibitemOpen
  \bibfield  {author} {\bibinfo {author} {\bibfnamefont {U.}~\bibnamefont {Seljak}} \emph {et~al.} (\bibinfo {collaboration} {SDSS}),\ }\href {\doibase 10.1103/PhysRevD.71.103515} {\bibfield  {journal} {\bibinfo  {journal} {Phys. Rev. D}\ }\textbf {\bibinfo {volume} {71}},\ \bibinfo {pages} {103515} (\bibinfo {year} {2005})},\ \Eprint {http://arxiv.org/abs/astro-ph/0407372} {arXiv:astro-ph/0407372} \BibitemShut {NoStop}%
\bibitem [{\citenamefont {Tegmark}\ \emph {et~al.}(2006)\citenamefont {Tegmark} \emph {et~al.}}]{SDSS:2006lmn}%
  \BibitemOpen
  \bibfield  {author} {\bibinfo {author} {\bibfnamefont {M.}~\bibnamefont {Tegmark}} \emph {et~al.} (\bibinfo {collaboration} {SDSS}),\ }\href {\doibase 10.1103/PhysRevD.74.123507} {\bibfield  {journal} {\bibinfo  {journal} {Phys. Rev. D}\ }\textbf {\bibinfo {volume} {74}},\ \bibinfo {pages} {123507} (\bibinfo {year} {2006})},\ \Eprint {http://arxiv.org/abs/astro-ph/0608632} {arXiv:astro-ph/0608632} \BibitemShut {NoStop}%
\bibitem [{\citenamefont {Betoule}\ \emph {et~al.}(2014)\citenamefont {Betoule} \emph {et~al.}}]{SDSS:2014iwm}%
  \BibitemOpen
  \bibfield  {author} {\bibinfo {author} {\bibfnamefont {M.}~\bibnamefont {Betoule}} \emph {et~al.} (\bibinfo {collaboration} {SDSS}),\ }\href {\doibase 10.1051/0004-6361/201423413} {\bibfield  {journal} {\bibinfo  {journal} {Astron. Astrophys.}\ }\textbf {\bibinfo {volume} {568}},\ \bibinfo {pages} {A22} (\bibinfo {year} {2014})},\ \Eprint {http://arxiv.org/abs/1401.4064} {arXiv:1401.4064 [astro-ph.CO]} \BibitemShut {NoStop}%
\bibitem [{\citenamefont {Abbott}\ \emph {et~al.}(2016)\citenamefont {Abbott} \emph {et~al.}}]{DES:2016jjg}%
  \BibitemOpen
  \bibfield  {author} {\bibinfo {author} {\bibfnamefont {T.}~\bibnamefont {Abbott}} \emph {et~al.} (\bibinfo {collaboration} {DES}),\ }\href {\doibase 10.1093/mnras/stw641} {\bibfield  {journal} {\bibinfo  {journal} {Mon. Not. Roy. Astron. Soc.}\ }\textbf {\bibinfo {volume} {460}},\ \bibinfo {pages} {1270} (\bibinfo {year} {2016})},\ \Eprint {http://arxiv.org/abs/1601.00329} {arXiv:1601.00329 [astro-ph.CO]} \BibitemShut {NoStop}%
\bibitem [{\citenamefont {Troxel}\ \emph {et~al.}(2018)\citenamefont {Troxel} \emph {et~al.}}]{DES:2017qwj}%
  \BibitemOpen
  \bibfield  {author} {\bibinfo {author} {\bibfnamefont {M.~A.}\ \bibnamefont {Troxel}} \emph {et~al.} (\bibinfo {collaboration} {DES}),\ }\href {\doibase 10.1103/PhysRevD.98.043528} {\bibfield  {journal} {\bibinfo  {journal} {Phys. Rev. D}\ }\textbf {\bibinfo {volume} {98}},\ \bibinfo {pages} {043528} (\bibinfo {year} {2018})},\ \Eprint {http://arxiv.org/abs/1708.01538} {arXiv:1708.01538 [astro-ph.CO]} \BibitemShut {NoStop}%
\bibitem [{\citenamefont {Abbott}\ \emph {et~al.}(2022)\citenamefont {Abbott} \emph {et~al.}}]{DES:2021wwk}%
  \BibitemOpen
  \bibfield  {author} {\bibinfo {author} {\bibfnamefont {T.~M.~C.}\ \bibnamefont {Abbott}} \emph {et~al.} (\bibinfo {collaboration} {DES}),\ }\href {\doibase 10.1103/PhysRevD.105.023520} {\bibfield  {journal} {\bibinfo  {journal} {Phys. Rev. D}\ }\textbf {\bibinfo {volume} {105}},\ \bibinfo {pages} {023520} (\bibinfo {year} {2022})},\ \Eprint {http://arxiv.org/abs/2105.13549} {arXiv:2105.13549 [astro-ph.CO]} \BibitemShut {NoStop}%
\bibitem [{\citenamefont {Abbott}\ \emph {et~al.}(2023{\natexlab{a}})\citenamefont {Abbott} \emph {et~al.}}]{DES:2022ccp}%
  \BibitemOpen
  \bibfield  {author} {\bibinfo {author} {\bibfnamefont {T.~M.~C.}\ \bibnamefont {Abbott}} \emph {et~al.} (\bibinfo {collaboration} {DES}),\ }\href {\doibase 10.1103/PhysRevD.107.083504} {\bibfield  {journal} {\bibinfo  {journal} {Phys. Rev. D}\ }\textbf {\bibinfo {volume} {107}},\ \bibinfo {pages} {083504} (\bibinfo {year} {2023}{\natexlab{a}})},\ \Eprint {http://arxiv.org/abs/2207.05766} {arXiv:2207.05766 [astro-ph.CO]} \BibitemShut {NoStop}%
\bibitem [{\citenamefont {Asgari}\ \emph {et~al.}(2021)\citenamefont {Asgari} \emph {et~al.}}]{KiDS:2020suj}%
  \BibitemOpen
  \bibfield  {author} {\bibinfo {author} {\bibfnamefont {M.}~\bibnamefont {Asgari}} \emph {et~al.} (\bibinfo {collaboration} {KiDS}),\ }\href {\doibase 10.1051/0004-6361/202039070} {\bibfield  {journal} {\bibinfo  {journal} {Astron. Astrophys.}\ }\textbf {\bibinfo {volume} {645}},\ \bibinfo {pages} {A104} (\bibinfo {year} {2021})},\ \Eprint {http://arxiv.org/abs/2007.15633} {arXiv:2007.15633 [astro-ph.CO]} \BibitemShut {NoStop}%
\bibitem [{\citenamefont {de~Jong}\ \emph {et~al.}(2013)\citenamefont {de~Jong}, \citenamefont {Verdoes~Kleijn}, \citenamefont {Kuijken},\ and\ \citenamefont {Valentijn}}]{deJong:2012zb}%
  \BibitemOpen
  \bibfield  {author} {\bibinfo {author} {\bibfnamefont {J.~T.~A.}\ \bibnamefont {de~Jong}}, \bibinfo {author} {\bibfnamefont {G.~A.}\ \bibnamefont {Verdoes~Kleijn}}, \bibinfo {author} {\bibfnamefont {K.~H.}\ \bibnamefont {Kuijken}}, \ and\ \bibinfo {author} {\bibfnamefont {E.~A.}\ \bibnamefont {Valentijn}} (\bibinfo {collaboration} {Astro-WISE, KiDS}),\ }\href {\doibase 10.1007/s10686-012-9306-1} {\bibfield  {journal} {\bibinfo  {journal} {Exper. Astron.}\ }\textbf {\bibinfo {volume} {35}},\ \bibinfo {pages} {25} (\bibinfo {year} {2013})},\ \Eprint {http://arxiv.org/abs/1206.1254} {arXiv:1206.1254 [astro-ph.CO]} \BibitemShut {NoStop}%
\bibitem [{\citenamefont {Tr\"oster}\ \emph {et~al.}(2021)\citenamefont {Tr\"oster} \emph {et~al.}}]{KiDS:2020ghu}%
  \BibitemOpen
  \bibfield  {author} {\bibinfo {author} {\bibfnamefont {T.}~\bibnamefont {Tr\"oster}} \emph {et~al.} (\bibinfo {collaboration} {KiDS}),\ }\href {\doibase 10.1051/0004-6361/202039805} {\bibfield  {journal} {\bibinfo  {journal} {Astron. Astrophys.}\ }\textbf {\bibinfo {volume} {649}},\ \bibinfo {pages} {A88} (\bibinfo {year} {2021})},\ \Eprint {http://arxiv.org/abs/2010.16416} {arXiv:2010.16416 [astro-ph.CO]} \BibitemShut {NoStop}%
\bibitem [{\citenamefont {Abbott}\ \emph {et~al.}(2023{\natexlab{b}})\citenamefont {Abbott} \emph {et~al.}}]{Kilo-DegreeSurvey:2023gfr}%
  \BibitemOpen
  \bibfield  {author} {\bibinfo {author} {\bibfnamefont {T.~M.~C.}\ \bibnamefont {Abbott}} \emph {et~al.} (\bibinfo {collaboration} {Kilo-Degree Survey, Dark Energy Survey}),\ }\href {\doibase 10.21105/astro.2305.17173} {\bibfield  {journal} {\bibinfo  {journal} {Open J. Astrophys.}\ }\textbf {\bibinfo {volume} {6}},\ \bibinfo {pages} {2305.17173} (\bibinfo {year} {2023}{\natexlab{b}})},\ \Eprint {http://arxiv.org/abs/2305.17173} {arXiv:2305.17173 [astro-ph.CO]} \BibitemShut {NoStop}%
\bibitem [{\citenamefont {Scolnic}\ \emph {et~al.}(2018)\citenamefont {Scolnic} \emph {et~al.}}]{Pan-STARRS1:2017jku}%
  \BibitemOpen
  \bibfield  {author} {\bibinfo {author} {\bibfnamefont {D.~M.}\ \bibnamefont {Scolnic}} \emph {et~al.} (\bibinfo {collaboration} {Pan-STARRS1}),\ }\href {\doibase 10.3847/1538-4357/aab9bb} {\bibfield  {journal} {\bibinfo  {journal} {Astrophys. J.}\ }\textbf {\bibinfo {volume} {859}},\ \bibinfo {pages} {101} (\bibinfo {year} {2018})},\ \Eprint {http://arxiv.org/abs/1710.00845} {arXiv:1710.00845 [astro-ph.CO]} \BibitemShut {NoStop}%
\bibitem [{\citenamefont {Brout}\ \emph {et~al.}(2022)\citenamefont {Brout} \emph {et~al.}}]{Brout:2022vxf}%
  \BibitemOpen
  \bibfield  {author} {\bibinfo {author} {\bibfnamefont {D.}~\bibnamefont {Brout}} \emph {et~al.},\ }\href {\doibase 10.3847/1538-4357/ac8e04} {\bibfield  {journal} {\bibinfo  {journal} {Astrophys. J.}\ }\textbf {\bibinfo {volume} {938}},\ \bibinfo {pages} {110} (\bibinfo {year} {2022})},\ \Eprint {http://arxiv.org/abs/2202.04077} {arXiv:2202.04077 [astro-ph.CO]} \BibitemShut {NoStop}%
\bibitem [{\citenamefont {Abbott}\ \emph {et~al.}(2017{\natexlab{a}})\citenamefont {Abbott} \emph {et~al.}}]{LIGO_SGWB-2017}%
  \BibitemOpen
  \bibfield  {author} {\bibinfo {author} {\bibfnamefont {B.~P.}\ \bibnamefont {Abbott}} \emph {et~al.} (\bibinfo {collaboration} {LIGO Scientific Collaboration and Virgo Collaboration}),\ }\href {\doibase 10.1103/PhysRevLett.118.121101} {\bibfield  {journal} {\bibinfo  {journal} {Phys. Rev. Lett.}\ }\textbf {\bibinfo {volume} {118}},\ \bibinfo {pages} {121101} (\bibinfo {year} {2017}{\natexlab{a}})}\BibitemShut {NoStop}%
\bibitem [{\citenamefont {{Abbott}}\ \emph {et~al.}(2019)\citenamefont {{Abbott}} \emph {et~al.}}]{LIGO_SGWB-2019}%
  \BibitemOpen
  \bibfield  {author} {\bibinfo {author} {\bibfnamefont {B.~P.}\ \bibnamefont {{Abbott}}} \emph {et~al.} (\bibinfo {collaboration} {LIGO Scientific Collaboration and Virgo Collaboration}),\ }\href@noop {} {\bibfield  {journal} {\bibinfo  {journal} {arXiv}\ } (\bibinfo {year} {2019})},\ \Eprint {http://arxiv.org/abs/1903.02886} {arXiv:1903.02886 [gr-qc]} \BibitemShut {NoStop}%
\bibitem [{\citenamefont {Abbott}\ \emph {et~al.}(2017{\natexlab{b}})\citenamefont {Abbott} \emph {et~al.}}]{TheLIGOScientific:2016dpb}%
  \BibitemOpen
  \bibfield  {author} {\bibinfo {author} {\bibfnamefont {B.~P.}\ \bibnamefont {Abbott}} \emph {et~al.} (\bibinfo {collaboration} {LIGO Scientific, Virgo}),\ }\href {\doibase 10.1103/PhysRevLett.118.121101, 10.1103/PhysRevLett.119.029901} {\bibfield  {journal} {\bibinfo  {journal} {Phys. Rev. Lett.}\ }\textbf {\bibinfo {volume} {118}},\ \bibinfo {pages} {121101} (\bibinfo {year} {2017}{\natexlab{b}})},\ \bibinfo {note} {[Erratum: Phys. Rev. Lett.119,no.2,029901(2017)]},\ \Eprint {http://arxiv.org/abs/1612.02029} {arXiv:1612.02029 [gr-qc]} \BibitemShut {NoStop}%
\bibitem [{\citenamefont {Leach}\ \emph {et~al.}(2002)\citenamefont {Leach}, \citenamefont {Liddle}, \citenamefont {Martin},\ and\ \citenamefont {Schwarz}}]{Leach:2002ar}%
  \BibitemOpen
  \bibfield  {author} {\bibinfo {author} {\bibfnamefont {S.~M.}\ \bibnamefont {Leach}}, \bibinfo {author} {\bibfnamefont {A.~R.}\ \bibnamefont {Liddle}}, \bibinfo {author} {\bibfnamefont {J.}~\bibnamefont {Martin}}, \ and\ \bibinfo {author} {\bibfnamefont {D.~J.}\ \bibnamefont {Schwarz}},\ }\href {\doibase 10.1103/PhysRevD.66.023515} {\bibfield  {journal} {\bibinfo  {journal} {Phys. Rev. D}\ }\textbf {\bibinfo {volume} {66}},\ \bibinfo {pages} {023515} (\bibinfo {year} {2002})},\ \Eprint {http://arxiv.org/abs/astro-ph/0202094} {arXiv:astro-ph/0202094} \BibitemShut {NoStop}%
\bibitem [{\citenamefont {Boubekeur}\ and\ \citenamefont {Lyth}(2005)}]{Boubekeur:2005zm}%
  \BibitemOpen
  \bibfield  {author} {\bibinfo {author} {\bibfnamefont {L.}~\bibnamefont {Boubekeur}}\ and\ \bibinfo {author} {\bibfnamefont {D.~H.}\ \bibnamefont {Lyth}},\ }\href {\doibase 10.1088/1475-7516/2005/07/010} {\bibfield  {journal} {\bibinfo  {journal} {JCAP}\ }\textbf {\bibinfo {volume} {07}},\ \bibinfo {pages} {010} (\bibinfo {year} {2005})},\ \Eprint {http://arxiv.org/abs/hep-ph/0502047} {arXiv:hep-ph/0502047} \BibitemShut {NoStop}%
\bibitem [{\citenamefont {Martin}\ and\ \citenamefont {Ringeval}(2006)}]{Martin:2006rs}%
  \BibitemOpen
  \bibfield  {author} {\bibinfo {author} {\bibfnamefont {J.}~\bibnamefont {Martin}}\ and\ \bibinfo {author} {\bibfnamefont {C.}~\bibnamefont {Ringeval}},\ }\href {\doibase 10.1088/1475-7516/2006/08/009} {\bibfield  {journal} {\bibinfo  {journal} {JCAP}\ }\textbf {\bibinfo {volume} {08}},\ \bibinfo {pages} {009} (\bibinfo {year} {2006})},\ \Eprint {http://arxiv.org/abs/astro-ph/0605367} {arXiv:astro-ph/0605367} \BibitemShut {NoStop}%
\bibitem [{\citenamefont {Moss}\ and\ \citenamefont {Graham}(2007)}]{Moss:2007qd}%
  \BibitemOpen
  \bibfield  {author} {\bibinfo {author} {\bibfnamefont {I.}~\bibnamefont {Moss}}\ and\ \bibinfo {author} {\bibfnamefont {C.}~\bibnamefont {Graham}},\ }\href {\doibase 10.1088/1475-7516/2007/11/004} {\bibfield  {journal} {\bibinfo  {journal} {JCAP}\ }\textbf {\bibinfo {volume} {11}},\ \bibinfo {pages} {004} (\bibinfo {year} {2007})},\ \Eprint {http://arxiv.org/abs/0707.1647} {arXiv:0707.1647 [astro-ph]} \BibitemShut {NoStop}%
\bibitem [{\citenamefont {Bezrukov}\ \emph {et~al.}(2011)\citenamefont {Bezrukov}, \citenamefont {Magnin}, \citenamefont {Shaposhnikov},\ and\ \citenamefont {Sibiryakov}}]{Bezrukov:2010jz}%
  \BibitemOpen
  \bibfield  {author} {\bibinfo {author} {\bibfnamefont {F.}~\bibnamefont {Bezrukov}}, \bibinfo {author} {\bibfnamefont {A.}~\bibnamefont {Magnin}}, \bibinfo {author} {\bibfnamefont {M.}~\bibnamefont {Shaposhnikov}}, \ and\ \bibinfo {author} {\bibfnamefont {S.}~\bibnamefont {Sibiryakov}},\ }\href {\doibase 10.1007/JHEP01(2011)016} {\bibfield  {journal} {\bibinfo  {journal} {JHEP}\ }\textbf {\bibinfo {volume} {01}},\ \bibinfo {pages} {016} (\bibinfo {year} {2011})},\ \Eprint {http://arxiv.org/abs/1008.5157} {arXiv:1008.5157 [hep-ph]} \BibitemShut {NoStop}%
\bibitem [{\citenamefont {Zhao}\ and\ \citenamefont {Huang}(2011)}]{Zhao:2011zb}%
  \BibitemOpen
  \bibfield  {author} {\bibinfo {author} {\bibfnamefont {W.}~\bibnamefont {Zhao}}\ and\ \bibinfo {author} {\bibfnamefont {Q.-G.}\ \bibnamefont {Huang}},\ }\href {\doibase 10.1088/0264-9381/28/23/235003} {\bibfield  {journal} {\bibinfo  {journal} {Class. Quant. Grav.}\ }\textbf {\bibinfo {volume} {28}},\ \bibinfo {pages} {235003} (\bibinfo {year} {2011})},\ \Eprint {http://arxiv.org/abs/1101.3163} {arXiv:1101.3163 [astro-ph.CO]} \BibitemShut {NoStop}%
\bibitem [{\citenamefont {Martin}\ \emph {et~al.}(2014{\natexlab{a}})\citenamefont {Martin}, \citenamefont {Ringeval}, \citenamefont {Trotta},\ and\ \citenamefont {Vennin}}]{Martin:2013nzq}%
  \BibitemOpen
  \bibfield  {author} {\bibinfo {author} {\bibfnamefont {J.}~\bibnamefont {Martin}}, \bibinfo {author} {\bibfnamefont {C.}~\bibnamefont {Ringeval}}, \bibinfo {author} {\bibfnamefont {R.}~\bibnamefont {Trotta}}, \ and\ \bibinfo {author} {\bibfnamefont {V.}~\bibnamefont {Vennin}},\ }\href {\doibase 10.1088/1475-7516/2014/03/039} {\bibfield  {journal} {\bibinfo  {journal} {JCAP}\ }\textbf {\bibinfo {volume} {03}},\ \bibinfo {pages} {039} (\bibinfo {year} {2014}{\natexlab{a}})},\ \Eprint {http://arxiv.org/abs/1312.3529} {arXiv:1312.3529 [astro-ph.CO]} \BibitemShut {NoStop}%
\bibitem [{\citenamefont {Martin}\ \emph {et~al.}(2014{\natexlab{b}})\citenamefont {Martin}, \citenamefont {Ringeval},\ and\ \citenamefont {Vennin}}]{Martin:2014rqa}%
  \BibitemOpen
  \bibfield  {author} {\bibinfo {author} {\bibfnamefont {J.}~\bibnamefont {Martin}}, \bibinfo {author} {\bibfnamefont {C.}~\bibnamefont {Ringeval}}, \ and\ \bibinfo {author} {\bibfnamefont {V.}~\bibnamefont {Vennin}},\ }\href {\doibase 10.1088/1475-7516/2014/10/038} {\bibfield  {journal} {\bibinfo  {journal} {JCAP}\ }\textbf {\bibinfo {volume} {10}},\ \bibinfo {pages} {038} (\bibinfo {year} {2014}{\natexlab{b}})},\ \Eprint {http://arxiv.org/abs/1407.4034} {arXiv:1407.4034 [astro-ph.CO]} \BibitemShut {NoStop}%
\bibitem [{\citenamefont {Martin}\ \emph {et~al.}(2014{\natexlab{c}})\citenamefont {Martin}, \citenamefont {Ringeval}, \citenamefont {Trotta},\ and\ \citenamefont {Vennin}}]{Martin:2014lra}%
  \BibitemOpen
  \bibfield  {author} {\bibinfo {author} {\bibfnamefont {J.}~\bibnamefont {Martin}}, \bibinfo {author} {\bibfnamefont {C.}~\bibnamefont {Ringeval}}, \bibinfo {author} {\bibfnamefont {R.}~\bibnamefont {Trotta}}, \ and\ \bibinfo {author} {\bibfnamefont {V.}~\bibnamefont {Vennin}},\ }\href {\doibase 10.1103/PhysRevD.90.063501} {\bibfield  {journal} {\bibinfo  {journal} {Phys. Rev. D}\ }\textbf {\bibinfo {volume} {90}},\ \bibinfo {pages} {063501} (\bibinfo {year} {2014}{\natexlab{c}})},\ \Eprint {http://arxiv.org/abs/1405.7272} {arXiv:1405.7272 [astro-ph.CO]} \BibitemShut {NoStop}%
\bibitem [{\citenamefont {Carrillo-Gonz\'alez}\ \emph {et~al.}(2014)\citenamefont {Carrillo-Gonz\'alez}, \citenamefont {Germ\'an-Velarde}, \citenamefont {Herrera-Aguilar}, \citenamefont {Hidalgo},\ and\ \citenamefont {Sussman}}]{Carrillo-Gonzalez:2014tia}%
  \BibitemOpen
  \bibfield  {author} {\bibinfo {author} {\bibfnamefont {M.}~\bibnamefont {Carrillo-Gonz\'alez}}, \bibinfo {author} {\bibfnamefont {G.}~\bibnamefont {Germ\'an-Velarde}}, \bibinfo {author} {\bibfnamefont {A.}~\bibnamefont {Herrera-Aguilar}}, \bibinfo {author} {\bibfnamefont {J.~C.}\ \bibnamefont {Hidalgo}}, \ and\ \bibinfo {author} {\bibfnamefont {R.}~\bibnamefont {Sussman}},\ }\href {\doibase 10.1016/j.physletb.2014.05.062} {\bibfield  {journal} {\bibinfo  {journal} {Phys. Lett. B}\ }\textbf {\bibinfo {volume} {734}},\ \bibinfo {pages} {345} (\bibinfo {year} {2014})},\ \Eprint {http://arxiv.org/abs/1404.1122} {arXiv:1404.1122 [astro-ph.CO]} \BibitemShut {NoStop}%
\bibitem [{\citenamefont {Creminelli}\ \emph {et~al.}(2014)\citenamefont {Creminelli}, \citenamefont {L\'opez~Nacir}, \citenamefont {Simonovi\'c}, \citenamefont {Trevisan},\ and\ \citenamefont {Zaldarriaga}}]{Creminelli:2014oaa}%
  \BibitemOpen
  \bibfield  {author} {\bibinfo {author} {\bibfnamefont {P.}~\bibnamefont {Creminelli}}, \bibinfo {author} {\bibfnamefont {D.}~\bibnamefont {L\'opez~Nacir}}, \bibinfo {author} {\bibfnamefont {M.}~\bibnamefont {Simonovi\'c}}, \bibinfo {author} {\bibfnamefont {G.}~\bibnamefont {Trevisan}}, \ and\ \bibinfo {author} {\bibfnamefont {M.}~\bibnamefont {Zaldarriaga}},\ }\href {\doibase 10.1103/PhysRevLett.112.241303} {\bibfield  {journal} {\bibinfo  {journal} {Phys. Rev. Lett.}\ }\textbf {\bibinfo {volume} {112}},\ \bibinfo {pages} {241303} (\bibinfo {year} {2014})},\ \Eprint {http://arxiv.org/abs/1404.1065} {arXiv:1404.1065 [astro-ph.CO]} \BibitemShut {NoStop}%
\bibitem [{\citenamefont {Di~Valentino}\ and\ \citenamefont {Mersini-Houghton}(2017{\natexlab{a}})}]{DiValentino:2016nni}%
  \BibitemOpen
  \bibfield  {author} {\bibinfo {author} {\bibfnamefont {E.}~\bibnamefont {Di~Valentino}}\ and\ \bibinfo {author} {\bibfnamefont {L.}~\bibnamefont {Mersini-Houghton}},\ }\href {\doibase 10.1088/1475-7516/2017/03/002} {\bibfield  {journal} {\bibinfo  {journal} {JCAP}\ }\textbf {\bibinfo {volume} {03}},\ \bibinfo {pages} {002} (\bibinfo {year} {2017}{\natexlab{a}})},\ \Eprint {http://arxiv.org/abs/1612.09588} {arXiv:1612.09588 [astro-ph.CO]} \BibitemShut {NoStop}%
\bibitem [{\citenamefont {Di~Valentino}\ and\ \citenamefont {Mersini-Houghton}(2017{\natexlab{b}})}]{DiValentino:2016ziq}%
  \BibitemOpen
  \bibfield  {author} {\bibinfo {author} {\bibfnamefont {E.}~\bibnamefont {Di~Valentino}}\ and\ \bibinfo {author} {\bibfnamefont {L.}~\bibnamefont {Mersini-Houghton}},\ }\href {\doibase 10.1088/1475-7516/2017/03/020} {\bibfield  {journal} {\bibinfo  {journal} {JCAP}\ }\textbf {\bibinfo {volume} {03}},\ \bibinfo {pages} {020} (\bibinfo {year} {2017}{\natexlab{b}})},\ \Eprint {http://arxiv.org/abs/1612.08334} {arXiv:1612.08334 [astro-ph.CO]} \BibitemShut {NoStop}%
\bibitem [{\citenamefont {Campista}\ \emph {et~al.}(2017)\citenamefont {Campista}, \citenamefont {Benetti},\ and\ \citenamefont {Alcaniz}}]{Campista:2017ovq}%
  \BibitemOpen
  \bibfield  {author} {\bibinfo {author} {\bibfnamefont {M.}~\bibnamefont {Campista}}, \bibinfo {author} {\bibfnamefont {M.}~\bibnamefont {Benetti}}, \ and\ \bibinfo {author} {\bibfnamefont {J.}~\bibnamefont {Alcaniz}},\ }\href {\doibase 10.1088/1475-7516/2017/09/010} {\bibfield  {journal} {\bibinfo  {journal} {JCAP}\ }\textbf {\bibinfo {volume} {09}},\ \bibinfo {pages} {010} (\bibinfo {year} {2017})},\ \Eprint {http://arxiv.org/abs/1705.08877} {arXiv:1705.08877 [astro-ph.CO]} \BibitemShut {NoStop}%
\bibitem [{\citenamefont {Giar\`e}\ \emph {et~al.}(2019)\citenamefont {Giar\`e}, \citenamefont {Di~Valentino},\ and\ \citenamefont {Melchiorri}}]{Giare:2019snj}%
  \BibitemOpen
  \bibfield  {author} {\bibinfo {author} {\bibfnamefont {W.}~\bibnamefont {Giar\`e}}, \bibinfo {author} {\bibfnamefont {E.}~\bibnamefont {Di~Valentino}}, \ and\ \bibinfo {author} {\bibfnamefont {A.}~\bibnamefont {Melchiorri}},\ }\href {\doibase 10.1103/PhysRevD.99.123522} {\bibfield  {journal} {\bibinfo  {journal} {Phys. Rev. D}\ }\textbf {\bibinfo {volume} {99}},\ \bibinfo {pages} {123522} (\bibinfo {year} {2019})}\BibitemShut {NoStop}%
\bibitem [{\citenamefont {Forconi}\ \emph {et~al.}(2021)\citenamefont {Forconi}, \citenamefont {Giar\`e}, \citenamefont {Di~Valentino},\ and\ \citenamefont {Melchiorri}}]{Forconi:2021que}%
  \BibitemOpen
  \bibfield  {author} {\bibinfo {author} {\bibfnamefont {M.}~\bibnamefont {Forconi}}, \bibinfo {author} {\bibfnamefont {W.}~\bibnamefont {Giar\`e}}, \bibinfo {author} {\bibfnamefont {E.}~\bibnamefont {Di~Valentino}}, \ and\ \bibinfo {author} {\bibfnamefont {A.}~\bibnamefont {Melchiorri}},\ }\href {\doibase 10.1103/PhysRevD.104.103528} {\bibfield  {journal} {\bibinfo  {journal} {Phys. Rev. D}\ }\textbf {\bibinfo {volume} {104}},\ \bibinfo {pages} {103528} (\bibinfo {year} {2021})},\ \Eprint {http://arxiv.org/abs/2110.01695} {arXiv:2110.01695 [astro-ph.CO]} \BibitemShut {NoStop}%
\bibitem [{\citenamefont {Dai}\ and\ \citenamefont {Zhu}(2020)}]{Dai:2019ejv}%
  \BibitemOpen
  \bibfield  {author} {\bibinfo {author} {\bibfnamefont {R.}~\bibnamefont {Dai}}\ and\ \bibinfo {author} {\bibfnamefont {Y.}~\bibnamefont {Zhu}},\ }\href {\doibase 10.1088/1475-7516/2020/05/017} {\bibfield  {journal} {\bibinfo  {journal} {JCAP}\ }\textbf {\bibinfo {volume} {05}},\ \bibinfo {pages} {017} (\bibinfo {year} {2020})},\ \Eprint {http://arxiv.org/abs/1911.05973} {arXiv:1911.05973 [astro-ph.CO]} \BibitemShut {NoStop}%
\bibitem [{\citenamefont {Baumann}\ \emph {et~al.}(2016)\citenamefont {Baumann}, \citenamefont {Lee},\ and\ \citenamefont {Pimentel}}]{Baumann:2015xxa}%
  \BibitemOpen
  \bibfield  {author} {\bibinfo {author} {\bibfnamefont {D.}~\bibnamefont {Baumann}}, \bibinfo {author} {\bibfnamefont {H.}~\bibnamefont {Lee}}, \ and\ \bibinfo {author} {\bibfnamefont {G.~L.}\ \bibnamefont {Pimentel}},\ }\href {\doibase 10.1007/JHEP01(2016)101} {\bibfield  {journal} {\bibinfo  {journal} {JHEP}\ }\textbf {\bibinfo {volume} {01}},\ \bibinfo {pages} {101} (\bibinfo {year} {2016})},\ \Eprint {http://arxiv.org/abs/1507.07250} {arXiv:1507.07250 [hep-th]} \BibitemShut {NoStop}%
\bibitem [{\citenamefont {Odintsov}\ \emph {et~al.}(2021)\citenamefont {Odintsov}, \citenamefont {Oikonomou},\ and\ \citenamefont {Fronimos}}]{Odintsov:2020ilr}%
  \BibitemOpen
  \bibfield  {author} {\bibinfo {author} {\bibfnamefont {S.~D.}\ \bibnamefont {Odintsov}}, \bibinfo {author} {\bibfnamefont {V.~K.}\ \bibnamefont {Oikonomou}}, \ and\ \bibinfo {author} {\bibfnamefont {F.~P.}\ \bibnamefont {Fronimos}},\ }\href {\doibase 10.1016/j.aop.2020.168359} {\bibfield  {journal} {\bibinfo  {journal} {Annals Phys.}\ }\textbf {\bibinfo {volume} {424}},\ \bibinfo {pages} {168359} (\bibinfo {year} {2021})},\ \Eprint {http://arxiv.org/abs/2011.08680} {arXiv:2011.08680 [gr-qc]} \BibitemShut {NoStop}%
\bibitem [{\citenamefont {Giar\`e}\ \emph {et~al.}(2021{\natexlab{a}})\citenamefont {Giar\`e}, \citenamefont {Renzi},\ and\ \citenamefont {Melchiorri}}]{Giare:2020plo}%
  \BibitemOpen
  \bibfield  {author} {\bibinfo {author} {\bibfnamefont {W.}~\bibnamefont {Giar\`e}}, \bibinfo {author} {\bibfnamefont {F.}~\bibnamefont {Renzi}}, \ and\ \bibinfo {author} {\bibfnamefont {A.}~\bibnamefont {Melchiorri}},\ }\href {\doibase 10.1103/PhysRevD.103.043515} {\bibfield  {journal} {\bibinfo  {journal} {Phys. Rev. D}\ }\textbf {\bibinfo {volume} {103}},\ \bibinfo {pages} {043515} (\bibinfo {year} {2021}{\natexlab{a}})},\ \Eprint {http://arxiv.org/abs/2012.00527} {arXiv:2012.00527 [astro-ph.CO]} \BibitemShut {NoStop}%
\bibitem [{\citenamefont {Oikonomou}(2021)}]{Oikonomou:2021kql}%
  \BibitemOpen
  \bibfield  {author} {\bibinfo {author} {\bibfnamefont {V.~K.}\ \bibnamefont {Oikonomou}},\ }\href {\doibase 10.1088/1361-6382/ac2168} {\bibfield  {journal} {\bibinfo  {journal} {Class. Quant. Grav.}\ }\textbf {\bibinfo {volume} {38}},\ \bibinfo {pages} {195025} (\bibinfo {year} {2021})},\ \Eprint {http://arxiv.org/abs/2108.10460} {arXiv:2108.10460 [gr-qc]} \BibitemShut {NoStop}%
\bibitem [{\citenamefont {Odintsov}\ \emph {et~al.}(2022)\citenamefont {Odintsov}, \citenamefont {Oikonomou},\ and\ \citenamefont {Myrzakulov}}]{Odintsov:2022cbm}%
  \BibitemOpen
  \bibfield  {author} {\bibinfo {author} {\bibfnamefont {S.~D.}\ \bibnamefont {Odintsov}}, \bibinfo {author} {\bibfnamefont {V.~K.}\ \bibnamefont {Oikonomou}}, \ and\ \bibinfo {author} {\bibfnamefont {R.}~\bibnamefont {Myrzakulov}},\ }\href {\doibase 10.3390/sym14040729} {\bibfield  {journal} {\bibinfo  {journal} {Symmetry}\ }\textbf {\bibinfo {volume} {14}},\ \bibinfo {pages} {729} (\bibinfo {year} {2022})},\ \Eprint {http://arxiv.org/abs/2204.00876} {arXiv:2204.00876 [gr-qc]} \BibitemShut {NoStop}%
\bibitem [{\citenamefont {Namba}\ \emph {et~al.}(2016)\citenamefont {Namba}, \citenamefont {Peloso}, \citenamefont {Shiraishi}, \citenamefont {Sorbo},\ and\ \citenamefont {Unal}}]{Namba:2015gja}%
  \BibitemOpen
  \bibfield  {author} {\bibinfo {author} {\bibfnamefont {R.}~\bibnamefont {Namba}}, \bibinfo {author} {\bibfnamefont {M.}~\bibnamefont {Peloso}}, \bibinfo {author} {\bibfnamefont {M.}~\bibnamefont {Shiraishi}}, \bibinfo {author} {\bibfnamefont {L.}~\bibnamefont {Sorbo}}, \ and\ \bibinfo {author} {\bibfnamefont {C.}~\bibnamefont {Unal}},\ }\href {\doibase 10.1088/1475-7516/2016/01/041} {\bibfield  {journal} {\bibinfo  {journal} {JCAP}\ }\textbf {\bibinfo {volume} {01}},\ \bibinfo {pages} {041} (\bibinfo {year} {2016})},\ \Eprint {http://arxiv.org/abs/1509.07521} {arXiv:1509.07521 [astro-ph.CO]} \BibitemShut {NoStop}%
\bibitem [{\citenamefont {Peloso}\ \emph {et~al.}(2016)\citenamefont {Peloso}, \citenamefont {Sorbo},\ and\ \citenamefont {Unal}}]{Peloso:2016gqs}%
  \BibitemOpen
  \bibfield  {author} {\bibinfo {author} {\bibfnamefont {M.}~\bibnamefont {Peloso}}, \bibinfo {author} {\bibfnamefont {L.}~\bibnamefont {Sorbo}}, \ and\ \bibinfo {author} {\bibfnamefont {C.}~\bibnamefont {Unal}},\ }\href {\doibase 10.1088/1475-7516/2016/09/001} {\bibfield  {journal} {\bibinfo  {journal} {JCAP}\ }\textbf {\bibinfo {volume} {09}},\ \bibinfo {pages} {001} (\bibinfo {year} {2016})},\ \Eprint {http://arxiv.org/abs/1606.00459} {arXiv:1606.00459 [astro-ph.CO]} \BibitemShut {NoStop}%
\bibitem [{\citenamefont {Pi}\ \emph {et~al.}(2019)\citenamefont {Pi}, \citenamefont {Sasaki},\ and\ \citenamefont {Zhang}}]{Pi:2019ihn}%
  \BibitemOpen
  \bibfield  {author} {\bibinfo {author} {\bibfnamefont {S.}~\bibnamefont {Pi}}, \bibinfo {author} {\bibfnamefont {M.}~\bibnamefont {Sasaki}}, \ and\ \bibinfo {author} {\bibfnamefont {Y.-l.}\ \bibnamefont {Zhang}},\ }\href {\doibase 10.1088/1475-7516/2019/06/049} {\bibfield  {journal} {\bibinfo  {journal} {JCAP}\ }\textbf {\bibinfo {volume} {06}},\ \bibinfo {pages} {049} (\bibinfo {year} {2019})},\ \Eprint {http://arxiv.org/abs/1904.06304} {arXiv:1904.06304 [gr-qc]} \BibitemShut {NoStop}%
\bibitem [{\citenamefont {\"Ozsoy}(2021)}]{Ozsoy:2020ccy}%
  \BibitemOpen
  \bibfield  {author} {\bibinfo {author} {\bibfnamefont {O.}~\bibnamefont {\"Ozsoy}},\ }\href {\doibase 10.1088/1475-7516/2021/04/040} {\bibfield  {journal} {\bibinfo  {journal} {JCAP}\ }\textbf {\bibinfo {volume} {04}},\ \bibinfo {pages} {040} (\bibinfo {year} {2021})},\ \Eprint {http://arxiv.org/abs/2005.10280} {arXiv:2005.10280 [astro-ph.CO]} \BibitemShut {NoStop}%
\bibitem [{\citenamefont {Stewart}\ and\ \citenamefont {Brandenberger}(2008)}]{Stewart:2007fu}%
  \BibitemOpen
  \bibfield  {author} {\bibinfo {author} {\bibfnamefont {A.}~\bibnamefont {Stewart}}\ and\ \bibinfo {author} {\bibfnamefont {R.}~\bibnamefont {Brandenberger}},\ }\href {\doibase 10.1088/1475-7516/2008/08/012} {\bibfield  {journal} {\bibinfo  {journal} {JCAP}\ }\textbf {\bibinfo {volume} {08}},\ \bibinfo {pages} {012} (\bibinfo {year} {2008})},\ \Eprint {http://arxiv.org/abs/0711.4602} {arXiv:0711.4602 [astro-ph]} \BibitemShut {NoStop}%
\bibitem [{\citenamefont {Mukohyama}\ \emph {et~al.}(2014)\citenamefont {Mukohyama}, \citenamefont {Namba}, \citenamefont {Peloso},\ and\ \citenamefont {Shiu}}]{Mukohyama:2014gba}%
  \BibitemOpen
  \bibfield  {author} {\bibinfo {author} {\bibfnamefont {S.}~\bibnamefont {Mukohyama}}, \bibinfo {author} {\bibfnamefont {R.}~\bibnamefont {Namba}}, \bibinfo {author} {\bibfnamefont {M.}~\bibnamefont {Peloso}}, \ and\ \bibinfo {author} {\bibfnamefont {G.}~\bibnamefont {Shiu}},\ }\href {\doibase 10.1088/1475-7516/2014/08/036} {\bibfield  {journal} {\bibinfo  {journal} {JCAP}\ }\textbf {\bibinfo {volume} {08}},\ \bibinfo {pages} {036} (\bibinfo {year} {2014})},\ \Eprint {http://arxiv.org/abs/1405.0346} {arXiv:1405.0346 [astro-ph.CO]} \BibitemShut {NoStop}%
\bibitem [{\citenamefont {Giovannini}(2016)}]{Giovannini:2015kfa}%
  \BibitemOpen
  \bibfield  {author} {\bibinfo {author} {\bibfnamefont {M.}~\bibnamefont {Giovannini}},\ }\href {\doibase 10.1088/0264-9381/33/12/125002} {\bibfield  {journal} {\bibinfo  {journal} {Class. Quant. Grav.}\ }\textbf {\bibinfo {volume} {33}},\ \bibinfo {pages} {125002} (\bibinfo {year} {2016})},\ \Eprint {http://arxiv.org/abs/1507.03456} {arXiv:1507.03456 [astro-ph.CO]} \BibitemShut {NoStop}%
\bibitem [{\citenamefont {Giovannini}(2018{\natexlab{a}})}]{Giovannini:2018dob}%
  \BibitemOpen
  \bibfield  {author} {\bibinfo {author} {\bibfnamefont {M.}~\bibnamefont {Giovannini}},\ }\href {\doibase 10.1103/PhysRevD.98.103509} {\bibfield  {journal} {\bibinfo  {journal} {Phys. Rev. D}\ }\textbf {\bibinfo {volume} {98}},\ \bibinfo {pages} {103509} (\bibinfo {year} {2018}{\natexlab{a}})},\ \Eprint {http://arxiv.org/abs/1806.01937} {arXiv:1806.01937 [gr-qc]} \BibitemShut {NoStop}%
\bibitem [{\citenamefont {Giovannini}(2019)}]{Giovannini:2018nkt}%
  \BibitemOpen
  \bibfield  {author} {\bibinfo {author} {\bibfnamefont {M.}~\bibnamefont {Giovannini}},\ }\href {\doibase 10.1016/j.physletb.2018.12.068} {\bibfield  {journal} {\bibinfo  {journal} {Phys. Lett. B}\ }\textbf {\bibinfo {volume} {789}},\ \bibinfo {pages} {502} (\bibinfo {year} {2019})},\ \Eprint {http://arxiv.org/abs/1805.08142} {arXiv:1805.08142 [astro-ph.CO]} \BibitemShut {NoStop}%
\bibitem [{\citenamefont {Giovannini}(2018{\natexlab{b}})}]{Giovannini:2018zbf}%
  \BibitemOpen
  \bibfield  {author} {\bibinfo {author} {\bibfnamefont {M.}~\bibnamefont {Giovannini}},\ }\href {\doibase 10.1140/epjc/s10052-018-5931-9} {\bibfield  {journal} {\bibinfo  {journal} {Eur. Phys. J. C}\ }\textbf {\bibinfo {volume} {78}},\ \bibinfo {pages} {442} (\bibinfo {year} {2018}{\natexlab{b}})},\ \Eprint {http://arxiv.org/abs/1803.05203} {arXiv:1803.05203 [gr-qc]} \BibitemShut {NoStop}%
\bibitem [{\citenamefont {Giar\`e}\ and\ \citenamefont {Melchiorri}(2021)}]{Giare:2020vhn}%
  \BibitemOpen
  \bibfield  {author} {\bibinfo {author} {\bibfnamefont {W.}~\bibnamefont {Giar\`e}}\ and\ \bibinfo {author} {\bibfnamefont {A.}~\bibnamefont {Melchiorri}},\ }\href {\doibase 10.1016/j.physletb.2021.136137} {\bibfield  {journal} {\bibinfo  {journal} {Phys. Lett. B}\ }\textbf {\bibinfo {volume} {815}},\ \bibinfo {pages} {136137} (\bibinfo {year} {2021})},\ \Eprint {http://arxiv.org/abs/2003.04783} {arXiv:2003.04783 [astro-ph.CO]} \BibitemShut {NoStop}%
\bibitem [{\citenamefont {Giar\`e}\ and\ \citenamefont {Renzi}(2020)}]{Giare:2020vss}%
  \BibitemOpen
  \bibfield  {author} {\bibinfo {author} {\bibfnamefont {W.}~\bibnamefont {Giar\`e}}\ and\ \bibinfo {author} {\bibfnamefont {F.}~\bibnamefont {Renzi}},\ }\href {\doibase 10.1103/PhysRevD.102.083530} {\bibfield  {journal} {\bibinfo  {journal} {Phys. Rev. D}\ }\textbf {\bibinfo {volume} {102}},\ \bibinfo {pages} {083530} (\bibinfo {year} {2020})},\ \Eprint {http://arxiv.org/abs/2007.04256} {arXiv:2007.04256 [astro-ph.CO]} \BibitemShut {NoStop}%
\bibitem [{\citenamefont {Giar\`e}\ \emph {et~al.}(2023{\natexlab{a}})\citenamefont {Giar\`e}, \citenamefont {Forconi}, \citenamefont {Di~Valentino},\ and\ \citenamefont {Melchiorri}}]{Giare:2022wxq}%
  \BibitemOpen
  \bibfield  {author} {\bibinfo {author} {\bibfnamefont {W.}~\bibnamefont {Giar\`e}}, \bibinfo {author} {\bibfnamefont {M.}~\bibnamefont {Forconi}}, \bibinfo {author} {\bibfnamefont {E.}~\bibnamefont {Di~Valentino}}, \ and\ \bibinfo {author} {\bibfnamefont {A.}~\bibnamefont {Melchiorri}},\ }\href {\doibase 10.1093/mnras/stad258} {\bibfield  {journal} {\bibinfo  {journal} {Mon. Not. Roy. Astron. Soc.}\ }\textbf {\bibinfo {volume} {520}},\ \bibinfo {pages} {2} (\bibinfo {year} {2023}{\natexlab{a}})},\ \Eprint {http://arxiv.org/abs/2210.14159} {arXiv:2210.14159 [astro-ph.CO]} \BibitemShut {NoStop}%
\bibitem [{\citenamefont {Baumgart}\ \emph {et~al.}(2022)\citenamefont {Baumgart}, \citenamefont {Heckman},\ and\ \citenamefont {Thomas}}]{Baumgart:2021ptt}%
  \BibitemOpen
  \bibfield  {author} {\bibinfo {author} {\bibfnamefont {M.}~\bibnamefont {Baumgart}}, \bibinfo {author} {\bibfnamefont {J.~J.}\ \bibnamefont {Heckman}}, \ and\ \bibinfo {author} {\bibfnamefont {L.}~\bibnamefont {Thomas}},\ }\href {\doibase 10.1088/1475-7516/2022/07/034} {\bibfield  {journal} {\bibinfo  {journal} {JCAP}\ }\textbf {\bibinfo {volume} {07}},\ \bibinfo {pages} {034} (\bibinfo {year} {2022})},\ \Eprint {http://arxiv.org/abs/2109.08166} {arXiv:2109.08166 [hep-ph]} \BibitemShut {NoStop}%
\bibitem [{\citenamefont {Franciolini}\ \emph {et~al.}(2019)\citenamefont {Franciolini}, \citenamefont {Giudice}, \citenamefont {Racco},\ and\ \citenamefont {Riotto}}]{Franciolini:2018ebs}%
  \BibitemOpen
  \bibfield  {author} {\bibinfo {author} {\bibfnamefont {G.}~\bibnamefont {Franciolini}}, \bibinfo {author} {\bibfnamefont {G.~F.}\ \bibnamefont {Giudice}}, \bibinfo {author} {\bibfnamefont {D.}~\bibnamefont {Racco}}, \ and\ \bibinfo {author} {\bibfnamefont {A.}~\bibnamefont {Riotto}},\ }\href {\doibase 10.1088/1475-7516/2019/05/022} {\bibfield  {journal} {\bibinfo  {journal} {JCAP}\ }\textbf {\bibinfo {volume} {05}},\ \bibinfo {pages} {022} (\bibinfo {year} {2019})},\ \Eprint {http://arxiv.org/abs/1811.08118} {arXiv:1811.08118 [hep-ph]} \BibitemShut {NoStop}%
\bibitem [{\citenamefont {D'Eramo}\ and\ \citenamefont {Schmitz}(2019)}]{DEramo:2019tit}%
  \BibitemOpen
  \bibfield  {author} {\bibinfo {author} {\bibfnamefont {F.}~\bibnamefont {D'Eramo}}\ and\ \bibinfo {author} {\bibfnamefont {K.}~\bibnamefont {Schmitz}},\ }\href {\doibase 10.1103/PhysRevResearch.1.013010} {\bibfield  {journal} {\bibinfo  {journal} {Phys. Rev. Research.}\ }\textbf {\bibinfo {volume} {1}},\ \bibinfo {pages} {013010} (\bibinfo {year} {2019})},\ \Eprint {http://arxiv.org/abs/1904.07870} {arXiv:1904.07870 [hep-ph]} \BibitemShut {NoStop}%
\bibitem [{\citenamefont {Caldwell}\ \emph {et~al.}(2019)\citenamefont {Caldwell}, \citenamefont {Smith},\ and\ \citenamefont {Walker}}]{Caldwell:2018giq}%
  \BibitemOpen
  \bibfield  {author} {\bibinfo {author} {\bibfnamefont {R.~R.}\ \bibnamefont {Caldwell}}, \bibinfo {author} {\bibfnamefont {T.~L.}\ \bibnamefont {Smith}}, \ and\ \bibinfo {author} {\bibfnamefont {D.~G.~E.}\ \bibnamefont {Walker}},\ }\href {\doibase 10.1103/PhysRevD.100.043513} {\bibfield  {journal} {\bibinfo  {journal} {Phys. Rev. D}\ }\textbf {\bibinfo {volume} {100}},\ \bibinfo {pages} {043513} (\bibinfo {year} {2019})},\ \Eprint {http://arxiv.org/abs/1812.07577} {arXiv:1812.07577 [astro-ph.CO]} \BibitemShut {NoStop}%
\bibitem [{\citenamefont {Clarke}\ \emph {et~al.}(2020)\citenamefont {Clarke}, \citenamefont {Copeland},\ and\ \citenamefont {Moss}}]{Clarke:2020bil}%
  \BibitemOpen
  \bibfield  {author} {\bibinfo {author} {\bibfnamefont {T.~J.}\ \bibnamefont {Clarke}}, \bibinfo {author} {\bibfnamefont {E.~J.}\ \bibnamefont {Copeland}}, \ and\ \bibinfo {author} {\bibfnamefont {A.}~\bibnamefont {Moss}},\ }\href {\doibase 10.1088/1475-7516/2020/10/002} {\bibfield  {journal} {\bibinfo  {journal} {JCAP}\ }\textbf {\bibinfo {volume} {10}},\ \bibinfo {pages} {002} (\bibinfo {year} {2020})},\ \Eprint {http://arxiv.org/abs/2004.11396} {arXiv:2004.11396 [astro-ph.CO]} \BibitemShut {NoStop}%
\bibitem [{\citenamefont {Caprini}\ and\ \citenamefont {Figueroa}(2018)}]{Caprini:2018mtu}%
  \BibitemOpen
  \bibfield  {author} {\bibinfo {author} {\bibfnamefont {C.}~\bibnamefont {Caprini}}\ and\ \bibinfo {author} {\bibfnamefont {D.~G.}\ \bibnamefont {Figueroa}},\ }\href {\doibase 10.1088/1361-6382/aac608} {\bibfield  {journal} {\bibinfo  {journal} {Class. Quant. Grav.}\ }\textbf {\bibinfo {volume} {35}},\ \bibinfo {pages} {163001} (\bibinfo {year} {2018})},\ \Eprint {http://arxiv.org/abs/1801.04268} {arXiv:1801.04268 [astro-ph.CO]} \BibitemShut {NoStop}%
\bibitem [{\citenamefont {Allen}\ and\ \citenamefont {Romano}(1999)}]{Allen:1997ad}%
  \BibitemOpen
  \bibfield  {author} {\bibinfo {author} {\bibfnamefont {B.}~\bibnamefont {Allen}}\ and\ \bibinfo {author} {\bibfnamefont {J.~D.}\ \bibnamefont {Romano}},\ }\href {\doibase 10.1103/PhysRevD.59.102001} {\bibfield  {journal} {\bibinfo  {journal} {Phys. Rev. D}\ }\textbf {\bibinfo {volume} {59}},\ \bibinfo {pages} {102001} (\bibinfo {year} {1999})},\ \Eprint {http://arxiv.org/abs/gr-qc/9710117} {arXiv:gr-qc/9710117} \BibitemShut {NoStop}%
\bibitem [{\citenamefont {Smith}\ \emph {et~al.}(2006)\citenamefont {Smith}, \citenamefont {Pierpaoli},\ and\ \citenamefont {Kamionkowski}}]{Smith:2006nka}%
  \BibitemOpen
  \bibfield  {author} {\bibinfo {author} {\bibfnamefont {T.~L.}\ \bibnamefont {Smith}}, \bibinfo {author} {\bibfnamefont {E.}~\bibnamefont {Pierpaoli}}, \ and\ \bibinfo {author} {\bibfnamefont {M.}~\bibnamefont {Kamionkowski}},\ }\href {\doibase 10.1103/PhysRevLett.97.021301} {\bibfield  {journal} {\bibinfo  {journal} {Phys. Rev. Lett.}\ }\textbf {\bibinfo {volume} {97}},\ \bibinfo {pages} {021301} (\bibinfo {year} {2006})},\ \Eprint {http://arxiv.org/abs/astro-ph/0603144} {arXiv:astro-ph/0603144} \BibitemShut {NoStop}%
\bibitem [{\citenamefont {Boyle}\ and\ \citenamefont {Buonanno}(2008)}]{Boyle:2007zx}%
  \BibitemOpen
  \bibfield  {author} {\bibinfo {author} {\bibfnamefont {L.~A.}\ \bibnamefont {Boyle}}\ and\ \bibinfo {author} {\bibfnamefont {A.}~\bibnamefont {Buonanno}},\ }\href {\doibase 10.1103/PhysRevD.78.043531} {\bibfield  {journal} {\bibinfo  {journal} {Phys. Rev. D}\ }\textbf {\bibinfo {volume} {78}},\ \bibinfo {pages} {043531} (\bibinfo {year} {2008})},\ \Eprint {http://arxiv.org/abs/0708.2279} {arXiv:0708.2279 [astro-ph]} \BibitemShut {NoStop}%
\bibitem [{\citenamefont {Kuroyanagi}\ \emph {et~al.}(2015)\citenamefont {Kuroyanagi}, \citenamefont {Takahashi},\ and\ \citenamefont {Yokoyama}}]{Kuroyanagi:2014nba}%
  \BibitemOpen
  \bibfield  {author} {\bibinfo {author} {\bibfnamefont {S.}~\bibnamefont {Kuroyanagi}}, \bibinfo {author} {\bibfnamefont {T.}~\bibnamefont {Takahashi}}, \ and\ \bibinfo {author} {\bibfnamefont {S.}~\bibnamefont {Yokoyama}},\ }\href {\doibase 10.1088/1475-7516/2015/02/003} {\bibfield  {journal} {\bibinfo  {journal} {JCAP}\ }\textbf {\bibinfo {volume} {02}},\ \bibinfo {pages} {003} (\bibinfo {year} {2015})},\ \Eprint {http://arxiv.org/abs/1407.4785} {arXiv:1407.4785 [astro-ph.CO]} \BibitemShut {NoStop}%
\bibitem [{\citenamefont {Ben-Dayan}\ \emph {et~al.}(2019)\citenamefont {Ben-Dayan}, \citenamefont {Keating}, \citenamefont {Leon},\ and\ \citenamefont {Wolfson}}]{Ben-Dayan:2019gll}%
  \BibitemOpen
  \bibfield  {author} {\bibinfo {author} {\bibfnamefont {I.}~\bibnamefont {Ben-Dayan}}, \bibinfo {author} {\bibfnamefont {B.}~\bibnamefont {Keating}}, \bibinfo {author} {\bibfnamefont {D.}~\bibnamefont {Leon}}, \ and\ \bibinfo {author} {\bibfnamefont {I.}~\bibnamefont {Wolfson}},\ }\href {\doibase 10.1088/1475-7516/2019/06/007} {\bibfield  {journal} {\bibinfo  {journal} {JCAP}\ }\textbf {\bibinfo {volume} {06}},\ \bibinfo {pages} {007} (\bibinfo {year} {2019})},\ \Eprint {http://arxiv.org/abs/1903.11843} {arXiv:1903.11843 [astro-ph.CO]} \BibitemShut {NoStop}%
\bibitem [{\citenamefont {Aich}\ \emph {et~al.}(2020)\citenamefont {Aich}, \citenamefont {Ma}, \citenamefont {Dai},\ and\ \citenamefont {Xia}}]{Aich:2019obd}%
  \BibitemOpen
  \bibfield  {author} {\bibinfo {author} {\bibfnamefont {M.}~\bibnamefont {Aich}}, \bibinfo {author} {\bibfnamefont {Y.-Z.}\ \bibnamefont {Ma}}, \bibinfo {author} {\bibfnamefont {W.-M.}\ \bibnamefont {Dai}}, \ and\ \bibinfo {author} {\bibfnamefont {J.-Q.}\ \bibnamefont {Xia}},\ }\href {\doibase 10.1103/PhysRevD.101.063536} {\bibfield  {journal} {\bibinfo  {journal} {Phys. Rev. D}\ }\textbf {\bibinfo {volume} {101}},\ \bibinfo {pages} {063536} (\bibinfo {year} {2020})},\ \Eprint {http://arxiv.org/abs/1912.00995} {arXiv:1912.00995 [astro-ph.CO]} \BibitemShut {NoStop}%
\bibitem [{\citenamefont {Cabass}\ \emph {et~al.}(2016)\citenamefont {Cabass}, \citenamefont {Pagano}, \citenamefont {Salvati}, \citenamefont {Gerbino}, \citenamefont {Giusarma},\ and\ \citenamefont {Melchiorri}}]{Cabass:2015jwe}%
  \BibitemOpen
  \bibfield  {author} {\bibinfo {author} {\bibfnamefont {G.}~\bibnamefont {Cabass}}, \bibinfo {author} {\bibfnamefont {L.}~\bibnamefont {Pagano}}, \bibinfo {author} {\bibfnamefont {L.}~\bibnamefont {Salvati}}, \bibinfo {author} {\bibfnamefont {M.}~\bibnamefont {Gerbino}}, \bibinfo {author} {\bibfnamefont {E.}~\bibnamefont {Giusarma}}, \ and\ \bibinfo {author} {\bibfnamefont {A.}~\bibnamefont {Melchiorri}},\ }\href {\doibase 10.1103/PhysRevD.93.063508} {\bibfield  {journal} {\bibinfo  {journal} {Phys. Rev. D}\ }\textbf {\bibinfo {volume} {93}},\ \bibinfo {pages} {063508} (\bibinfo {year} {2016})},\ \Eprint {http://arxiv.org/abs/1511.05146} {arXiv:1511.05146 [astro-ph.CO]} \BibitemShut {NoStop}%
\bibitem [{\citenamefont {Vagnozzi}(2021)}]{Vagnozzi:2020gtf}%
  \BibitemOpen
  \bibfield  {author} {\bibinfo {author} {\bibfnamefont {S.}~\bibnamefont {Vagnozzi}},\ }\href {\doibase 10.1093/mnrasl/slaa203} {\bibfield  {journal} {\bibinfo  {journal} {Mon. Not. Roy. Astron. Soc.}\ }\textbf {\bibinfo {volume} {502}},\ \bibinfo {pages} {L11} (\bibinfo {year} {2021})},\ \Eprint {http://arxiv.org/abs/2009.13432} {arXiv:2009.13432 [astro-ph.CO]} \BibitemShut {NoStop}%
\bibitem [{\citenamefont {Benetti}\ \emph {et~al.}(2022)\citenamefont {Benetti}, \citenamefont {Graef},\ and\ \citenamefont {Vagnozzi}}]{Benetti:2021uea}%
  \BibitemOpen
  \bibfield  {author} {\bibinfo {author} {\bibfnamefont {M.}~\bibnamefont {Benetti}}, \bibinfo {author} {\bibfnamefont {L.~L.}\ \bibnamefont {Graef}}, \ and\ \bibinfo {author} {\bibfnamefont {S.}~\bibnamefont {Vagnozzi}},\ }\href {\doibase 10.1103/PhysRevD.105.043520} {\bibfield  {journal} {\bibinfo  {journal} {Phys. Rev. D}\ }\textbf {\bibinfo {volume} {105}},\ \bibinfo {pages} {043520} (\bibinfo {year} {2022})},\ \Eprint {http://arxiv.org/abs/2111.04758} {arXiv:2111.04758 [astro-ph.CO]} \BibitemShut {NoStop}%
\bibitem [{\citenamefont {Calcagni}\ and\ \citenamefont {Kuroyanagi}(2021)}]{Calcagni:2020tvw}%
  \BibitemOpen
  \bibfield  {author} {\bibinfo {author} {\bibfnamefont {G.}~\bibnamefont {Calcagni}}\ and\ \bibinfo {author} {\bibfnamefont {S.}~\bibnamefont {Kuroyanagi}},\ }\href {\doibase 10.1088/1475-7516/2021/03/019} {\bibfield  {journal} {\bibinfo  {journal} {JCAP}\ }\textbf {\bibinfo {volume} {03}},\ \bibinfo {pages} {019} (\bibinfo {year} {2021})},\ \Eprint {http://arxiv.org/abs/2012.00170} {arXiv:2012.00170 [gr-qc]} \BibitemShut {NoStop}%
\bibitem [{\citenamefont {Oikonomou}(2023{\natexlab{a}})}]{Oikonomou:2022ijs}%
  \BibitemOpen
  \bibfield  {author} {\bibinfo {author} {\bibfnamefont {V.~K.}\ \bibnamefont {Oikonomou}},\ }\href {\doibase 10.1016/j.astropartphys.2022.102777} {\bibfield  {journal} {\bibinfo  {journal} {Astropart. Phys.}\ }\textbf {\bibinfo {volume} {144}},\ \bibinfo {pages} {102777} (\bibinfo {year} {2023}{\natexlab{a}})},\ \Eprint {http://arxiv.org/abs/2209.09781} {arXiv:2209.09781 [gr-qc]} \BibitemShut {NoStop}%
\bibitem [{\citenamefont {Barrow}\ \emph {et~al.}(1993)\citenamefont {Barrow}, \citenamefont {Mimoso},\ and\ \citenamefont {de~Garcia~Maia}}]{Barrow:1993ad}%
  \BibitemOpen
  \bibfield  {author} {\bibinfo {author} {\bibfnamefont {J.~D.}\ \bibnamefont {Barrow}}, \bibinfo {author} {\bibfnamefont {J.~P.}\ \bibnamefont {Mimoso}}, \ and\ \bibinfo {author} {\bibfnamefont {M.~R.}\ \bibnamefont {de~Garcia~Maia}},\ }\href {\doibase 10.1103/PhysRevD.48.3630} {\bibfield  {journal} {\bibinfo  {journal} {Phys. Rev. D}\ }\textbf {\bibinfo {volume} {48}},\ \bibinfo {pages} {3630} (\bibinfo {year} {1993})},\ \bibinfo {note} {[Erratum: Phys.Rev.D 51, 5967 (1995)]}\BibitemShut {NoStop}%
\bibitem [{\citenamefont {Peng}\ \emph {et~al.}(2021)\citenamefont {Peng}, \citenamefont {Fu}, \citenamefont {Liu}, \citenamefont {Guo},\ and\ \citenamefont {Cai}}]{Peng:2021zon}%
  \BibitemOpen
  \bibfield  {author} {\bibinfo {author} {\bibfnamefont {Z.-Z.}\ \bibnamefont {Peng}}, \bibinfo {author} {\bibfnamefont {C.}~\bibnamefont {Fu}}, \bibinfo {author} {\bibfnamefont {J.}~\bibnamefont {Liu}}, \bibinfo {author} {\bibfnamefont {Z.-K.}\ \bibnamefont {Guo}}, \ and\ \bibinfo {author} {\bibfnamefont {R.-G.}\ \bibnamefont {Cai}},\ }\href {\doibase 10.1088/1475-7516/2021/10/050} {\bibfield  {journal} {\bibinfo  {journal} {JCAP}\ }\textbf {\bibinfo {volume} {10}},\ \bibinfo {pages} {050} (\bibinfo {year} {2021})},\ \Eprint {http://arxiv.org/abs/2106.11816} {arXiv:2106.11816 [astro-ph.CO]} \BibitemShut {NoStop}%
\bibitem [{\citenamefont {Ota}\ \emph {et~al.}(2023)\citenamefont {Ota}, \citenamefont {Sasaki},\ and\ \citenamefont {Wang}}]{Ota:2022hvh}%
  \BibitemOpen
  \bibfield  {author} {\bibinfo {author} {\bibfnamefont {A.}~\bibnamefont {Ota}}, \bibinfo {author} {\bibfnamefont {M.}~\bibnamefont {Sasaki}}, \ and\ \bibinfo {author} {\bibfnamefont {Y.}~\bibnamefont {Wang}},\ }\href {\doibase 10.1142/S0217732323500633} {\bibfield  {journal} {\bibinfo  {journal} {Mod. Phys. Lett. A}\ }\textbf {\bibinfo {volume} {38}},\ \bibinfo {pages} {2350063} (\bibinfo {year} {2023})},\ \Eprint {http://arxiv.org/abs/2209.02272} {arXiv:2209.02272 [astro-ph.CO]} \BibitemShut {NoStop}%
\bibitem [{\citenamefont {Odintsov}\ and\ \citenamefont {Oikonomou}(2022{\natexlab{a}})}]{Odintsov:2022sdk}%
  \BibitemOpen
  \bibfield  {author} {\bibinfo {author} {\bibfnamefont {S.~D.}\ \bibnamefont {Odintsov}}\ and\ \bibinfo {author} {\bibfnamefont {V.~K.}\ \bibnamefont {Oikonomou}},\ }\href {\doibase 10.1002/prop.202100167} {\bibfield  {journal} {\bibinfo  {journal} {Fortsch. Phys.}\ }\textbf {\bibinfo {volume} {70}},\ \bibinfo {pages} {2100167} (\bibinfo {year} {2022}{\natexlab{a}})},\ \Eprint {http://arxiv.org/abs/2203.10599} {arXiv:2203.10599 [gr-qc]} \BibitemShut {NoStop}%
\bibitem [{\citenamefont {Capurri}\ \emph {et~al.}(2020)\citenamefont {Capurri}, \citenamefont {Bartolo}, \citenamefont {Maino},\ and\ \citenamefont {Matarrese}}]{Capurri:2020qgz}%
  \BibitemOpen
  \bibfield  {author} {\bibinfo {author} {\bibfnamefont {G.}~\bibnamefont {Capurri}}, \bibinfo {author} {\bibfnamefont {N.}~\bibnamefont {Bartolo}}, \bibinfo {author} {\bibfnamefont {D.}~\bibnamefont {Maino}}, \ and\ \bibinfo {author} {\bibfnamefont {S.}~\bibnamefont {Matarrese}},\ }\href {\doibase 10.1088/1475-7516/2020/11/037} {\bibfield  {journal} {\bibinfo  {journal} {JCAP}\ }\textbf {\bibinfo {volume} {11}},\ \bibinfo {pages} {037} (\bibinfo {year} {2020})},\ \Eprint {http://arxiv.org/abs/2006.10781} {arXiv:2006.10781 [astro-ph.CO]} \BibitemShut {NoStop}%
\bibitem [{\citenamefont {Ca\~nas Herrera}\ and\ \citenamefont {Renzi}(2021)}]{Canas-Herrera:2021sjs}%
  \BibitemOpen
  \bibfield  {author} {\bibinfo {author} {\bibfnamefont {G.}~\bibnamefont {Ca\~nas Herrera}}\ and\ \bibinfo {author} {\bibfnamefont {F.}~\bibnamefont {Renzi}},\ }\href {\doibase 10.1103/PhysRevD.104.103512} {\bibfield  {journal} {\bibinfo  {journal} {Phys. Rev. D}\ }\textbf {\bibinfo {volume} {104}},\ \bibinfo {pages} {103512} (\bibinfo {year} {2021})},\ \Eprint {http://arxiv.org/abs/2104.06398} {arXiv:2104.06398 [astro-ph.CO]} \BibitemShut {NoStop}%
\bibitem [{\citenamefont {Odintsov}\ \emph {et~al.}(2023)\citenamefont {Odintsov}, \citenamefont {Oikonomou},\ and\ \citenamefont {Fronimos}}]{Odintsov:2023aaw}%
  \BibitemOpen
  \bibfield  {author} {\bibinfo {author} {\bibfnamefont {S.~D.}\ \bibnamefont {Odintsov}}, \bibinfo {author} {\bibfnamefont {V.~K.}\ \bibnamefont {Oikonomou}}, \ and\ \bibinfo {author} {\bibfnamefont {F.~P.}\ \bibnamefont {Fronimos}},\ }\href {\doibase 10.1103/PhysRevD.107.084007} {\bibfield  {journal} {\bibinfo  {journal} {Phys. Rev. D}\ }\textbf {\bibinfo {volume} {107}},\ \bibinfo {pages} {08} (\bibinfo {year} {2023})},\ \Eprint {http://arxiv.org/abs/2303.14594} {arXiv:2303.14594 [gr-qc]} \BibitemShut {NoStop}%
\bibitem [{\citenamefont {Oikonomou}(2023{\natexlab{b}})}]{Oikonomou:2023bah}%
  \BibitemOpen
  \bibfield  {author} {\bibinfo {author} {\bibfnamefont {V.~K.}\ \bibnamefont {Oikonomou}},\ }\href {\doibase 10.1103/PhysRevD.107.064071} {\bibfield  {journal} {\bibinfo  {journal} {Phys. Rev. D}\ }\textbf {\bibinfo {volume} {107}},\ \bibinfo {pages} {064071} (\bibinfo {year} {2023}{\natexlab{b}})},\ \Eprint {http://arxiv.org/abs/2303.05889} {arXiv:2303.05889 [hep-ph]} \BibitemShut {NoStop}%
\bibitem [{\citenamefont {Fronimos}\ and\ \citenamefont {Venikoudis}(2023)}]{Fronimos:2023tim}%
  \BibitemOpen
  \bibfield  {author} {\bibinfo {author} {\bibfnamefont {F.~P.}\ \bibnamefont {Fronimos}}\ and\ \bibinfo {author} {\bibfnamefont {S.~A.}\ \bibnamefont {Venikoudis}},\ }\href {\doibase 10.1140/epjp/s13360-023-04149-0} {\bibfield  {journal} {\bibinfo  {journal} {Eur. Phys. J. Plus}\ }\textbf {\bibinfo {volume} {138}},\ \bibinfo {pages} {529} (\bibinfo {year} {2023})},\ \Eprint {http://arxiv.org/abs/2302.05173} {arXiv:2302.05173 [gr-qc]} \BibitemShut {NoStop}%
\bibitem [{\citenamefont {Cai}(2023)}]{Cai:2022lec}%
  \BibitemOpen
  \bibfield  {author} {\bibinfo {author} {\bibfnamefont {Y.}~\bibnamefont {Cai}},\ }\href {\doibase 10.1103/PhysRevD.107.063512} {\bibfield  {journal} {\bibinfo  {journal} {Phys. Rev. D}\ }\textbf {\bibinfo {volume} {107}},\ \bibinfo {pages} {063512} (\bibinfo {year} {2023})},\ \Eprint {http://arxiv.org/abs/2212.10893} {arXiv:2212.10893 [gr-qc]} \BibitemShut {NoStop}%
\bibitem [{\citenamefont {Oikonomou}(2022)}]{Oikonomou:2022irx}%
  \BibitemOpen
  \bibfield  {author} {\bibinfo {author} {\bibfnamefont {V.~K.}\ \bibnamefont {Oikonomou}},\ }\href {\doibase 10.1016/j.nuclphysb.2022.115985} {\bibfield  {journal} {\bibinfo  {journal} {Nucl. Phys. B}\ }\textbf {\bibinfo {volume} {984}},\ \bibinfo {pages} {115985} (\bibinfo {year} {2022})},\ \Eprint {http://arxiv.org/abs/2210.02861} {arXiv:2210.02861 [gr-qc]} \BibitemShut {NoStop}%
\bibitem [{\citenamefont {Gangopadhyay}\ \emph {et~al.}(2023)\citenamefont {Gangopadhyay}, \citenamefont {Khan},\ and\ \citenamefont {Yogesh}}]{Gangopadhyay:2022vgh}%
  \BibitemOpen
  \bibfield  {author} {\bibinfo {author} {\bibfnamefont {M.~R.}\ \bibnamefont {Gangopadhyay}}, \bibinfo {author} {\bibfnamefont {H.~A.}\ \bibnamefont {Khan}}, \ and\ \bibinfo {author} {\bibnamefont {Yogesh}},\ }\href {\doibase 10.1016/j.dark.2023.101177} {\bibfield  {journal} {\bibinfo  {journal} {Phys. Dark Univ.}\ }\textbf {\bibinfo {volume} {40}},\ \bibinfo {pages} {101177} (\bibinfo {year} {2023})},\ \Eprint {http://arxiv.org/abs/2205.15261} {arXiv:2205.15261 [astro-ph.CO]} \BibitemShut {NoStop}%
\bibitem [{\citenamefont {Odintsov}\ and\ \citenamefont {Oikonomou}(2022{\natexlab{b}})}]{Odintsov:2022hxu}%
  \BibitemOpen
  \bibfield  {author} {\bibinfo {author} {\bibfnamefont {S.~D.}\ \bibnamefont {Odintsov}}\ and\ \bibinfo {author} {\bibfnamefont {V.~K.}\ \bibnamefont {Oikonomou}},\ }\href {\doibase 10.1103/PhysRevD.105.104054} {\bibfield  {journal} {\bibinfo  {journal} {Phys. Rev. D}\ }\textbf {\bibinfo {volume} {105}},\ \bibinfo {pages} {104054} (\bibinfo {year} {2022}{\natexlab{b}})},\ \Eprint {http://arxiv.org/abs/2205.07304} {arXiv:2205.07304 [gr-qc]} \BibitemShut {NoStop}%
\bibitem [{\citenamefont {Odintsov}\ \emph {et~al.}(2020)\citenamefont {Odintsov}, \citenamefont {Oikonomou}, \citenamefont {Fronimos},\ and\ \citenamefont {Venikoudis}}]{Odintsov:2020mkz}%
  \BibitemOpen
  \bibfield  {author} {\bibinfo {author} {\bibfnamefont {S.~D.}\ \bibnamefont {Odintsov}}, \bibinfo {author} {\bibfnamefont {V.~K.}\ \bibnamefont {Oikonomou}}, \bibinfo {author} {\bibfnamefont {F.~P.}\ \bibnamefont {Fronimos}}, \ and\ \bibinfo {author} {\bibfnamefont {S.~A.}\ \bibnamefont {Venikoudis}},\ }\href {\doibase 10.1016/j.dark.2020.100718} {\bibfield  {journal} {\bibinfo  {journal} {Phys. Dark Univ.}\ }\textbf {\bibinfo {volume} {30}},\ \bibinfo {pages} {100718} (\bibinfo {year} {2020})},\ \Eprint {http://arxiv.org/abs/2009.06113} {arXiv:2009.06113 [gr-qc]} \BibitemShut {NoStop}%
\bibitem [{\citenamefont {Galloni}\ \emph {et~al.}(2023)\citenamefont {Galloni}, \citenamefont {Bartolo}, \citenamefont {Matarrese}, \citenamefont {Migliaccio}, \citenamefont {Ricciardone},\ and\ \citenamefont {Vittorio}}]{Galloni:2022mok}%
  \BibitemOpen
  \bibfield  {author} {\bibinfo {author} {\bibfnamefont {G.}~\bibnamefont {Galloni}}, \bibinfo {author} {\bibfnamefont {N.}~\bibnamefont {Bartolo}}, \bibinfo {author} {\bibfnamefont {S.}~\bibnamefont {Matarrese}}, \bibinfo {author} {\bibfnamefont {M.}~\bibnamefont {Migliaccio}}, \bibinfo {author} {\bibfnamefont {A.}~\bibnamefont {Ricciardone}}, \ and\ \bibinfo {author} {\bibfnamefont {N.}~\bibnamefont {Vittorio}},\ }\href {\doibase 10.1088/1475-7516/2023/04/062} {\bibfield  {journal} {\bibinfo  {journal} {JCAP}\ }\textbf {\bibinfo {volume} {04}},\ \bibinfo {pages} {062} (\bibinfo {year} {2023})},\ \Eprint {http://arxiv.org/abs/2208.00188} {arXiv:2208.00188 [astro-ph.CO]} \BibitemShut {NoStop}%
\bibitem [{\citenamefont {Braglia}\ \emph {et~al.}(2023)\citenamefont {Braglia}, \citenamefont {Linde}, \citenamefont {Kallosh},\ and\ \citenamefont {Finelli}}]{Braglia:2022phb}%
  \BibitemOpen
  \bibfield  {author} {\bibinfo {author} {\bibfnamefont {M.}~\bibnamefont {Braglia}}, \bibinfo {author} {\bibfnamefont {A.}~\bibnamefont {Linde}}, \bibinfo {author} {\bibfnamefont {R.}~\bibnamefont {Kallosh}}, \ and\ \bibinfo {author} {\bibfnamefont {F.}~\bibnamefont {Finelli}},\ }\href {\doibase 10.1088/1475-7516/2023/04/033} {\bibfield  {journal} {\bibinfo  {journal} {JCAP}\ }\textbf {\bibinfo {volume} {04}},\ \bibinfo {pages} {033} (\bibinfo {year} {2023})},\ \Eprint {http://arxiv.org/abs/2211.14262} {arXiv:2211.14262 [astro-ph.CO]} \BibitemShut {NoStop}%
\bibitem [{\citenamefont {Giar\`e}\ \emph {et~al.}(2023{\natexlab{b}})\citenamefont {Giar\`e}, \citenamefont {De~Angelis}, \citenamefont {van~de Bruck},\ and\ \citenamefont {Di~Valentino}}]{Giare:2023kiv}%
  \BibitemOpen
  \bibfield  {author} {\bibinfo {author} {\bibfnamefont {W.}~\bibnamefont {Giar\`e}}, \bibinfo {author} {\bibfnamefont {M.}~\bibnamefont {De~Angelis}}, \bibinfo {author} {\bibfnamefont {C.}~\bibnamefont {van~de Bruck}}, \ and\ \bibinfo {author} {\bibfnamefont {E.}~\bibnamefont {Di~Valentino}},\ }\href@noop {} {\  (\bibinfo {year} {2023}{\natexlab{b}})},\ \Eprint {http://arxiv.org/abs/2306.12414} {arXiv:2306.12414 [astro-ph.CO]} \BibitemShut {NoStop}%
\bibitem [{\citenamefont {Antoniadis}\ \emph {et~al.}(2023)\citenamefont {Antoniadis} \emph {et~al.}}]{Antoniadis:2023zhi}%
  \BibitemOpen
  \bibfield  {author} {\bibinfo {author} {\bibfnamefont {J.}~\bibnamefont {Antoniadis}} \emph {et~al.} (\bibinfo {collaboration} {EPTA}),\ }\href@noop {} {\  (\bibinfo {year} {2023})},\ \Eprint {http://arxiv.org/abs/2306.16227} {arXiv:2306.16227 [astro-ph.CO]} \BibitemShut {NoStop}%
\bibitem [{\citenamefont {Agazie}\ \emph {et~al.}(2023)\citenamefont {Agazie} \emph {et~al.}}]{NANOGrav:2023gor}%
  \BibitemOpen
  \bibfield  {author} {\bibinfo {author} {\bibfnamefont {G.}~\bibnamefont {Agazie}} \emph {et~al.} (\bibinfo {collaboration} {NANOGrav}),\ }\href {\doibase 10.3847/2041-8213/acdac6} {\bibfield  {journal} {\bibinfo  {journal} {Astrophys. J. Lett.}\ }\textbf {\bibinfo {volume} {951}},\ \bibinfo {pages} {L8} (\bibinfo {year} {2023})},\ \Eprint {http://arxiv.org/abs/2306.16213} {arXiv:2306.16213 [astro-ph.HE]} \BibitemShut {NoStop}%
\bibitem [{\citenamefont {Afzal}\ \emph {et~al.}(2023)\citenamefont {Afzal} \emph {et~al.}}]{NANOGrav:2023hvm}%
  \BibitemOpen
  \bibfield  {author} {\bibinfo {author} {\bibfnamefont {A.}~\bibnamefont {Afzal}} \emph {et~al.} (\bibinfo {collaboration} {NANOGrav}),\ }\href {\doibase 10.3847/2041-8213/acdc91} {\bibfield  {journal} {\bibinfo  {journal} {Astrophys. J. Lett.}\ }\textbf {\bibinfo {volume} {951}},\ \bibinfo {pages} {L11} (\bibinfo {year} {2023})},\ \Eprint {http://arxiv.org/abs/2306.16219} {arXiv:2306.16219 [astro-ph.HE]} \BibitemShut {NoStop}%
\bibitem [{\citenamefont {Vagnozzi}(2023{\natexlab{a}})}]{Vagnozzi:2023lwo}%
  \BibitemOpen
  \bibfield  {author} {\bibinfo {author} {\bibfnamefont {S.}~\bibnamefont {Vagnozzi}},\ }\href {\doibase 10.1016/j.jheap.2023.07.001} {\bibfield  {journal} {\bibinfo  {journal} {JHEAp}\ }\textbf {\bibinfo {volume} {39}},\ \bibinfo {pages} {81} (\bibinfo {year} {2023}{\natexlab{a}})},\ \Eprint {http://arxiv.org/abs/2306.16912} {arXiv:2306.16912 [astro-ph.CO]} \BibitemShut {NoStop}%
\bibitem [{\citenamefont {Oikonomou}(2023{\natexlab{c}})}]{Oikonomou:2023qfz}%
  \BibitemOpen
  \bibfield  {author} {\bibinfo {author} {\bibfnamefont {V.~K.}\ \bibnamefont {Oikonomou}},\ }\href {\doibase 10.1103/PhysRevD.108.043516} {\bibfield  {journal} {\bibinfo  {journal} {Phys. Rev. D}\ }\textbf {\bibinfo {volume} {108}},\ \bibinfo {pages} {043516} (\bibinfo {year} {2023}{\natexlab{c}})},\ \Eprint {http://arxiv.org/abs/2306.17351} {arXiv:2306.17351 [astro-ph.CO]} \BibitemShut {NoStop}%
\bibitem [{\citenamefont {Iacconi}\ \emph {et~al.}(2020)\citenamefont {Iacconi}, \citenamefont {Fasiello}, \citenamefont {Assadullahi},\ and\ \citenamefont {Wands}}]{Iacconi:2020yxn}%
  \BibitemOpen
  \bibfield  {author} {\bibinfo {author} {\bibfnamefont {L.}~\bibnamefont {Iacconi}}, \bibinfo {author} {\bibfnamefont {M.}~\bibnamefont {Fasiello}}, \bibinfo {author} {\bibfnamefont {H.}~\bibnamefont {Assadullahi}}, \ and\ \bibinfo {author} {\bibfnamefont {D.}~\bibnamefont {Wands}},\ }\href {\doibase 10.1088/1475-7516/2020/12/005} {\bibfield  {journal} {\bibinfo  {journal} {JCAP}\ }\textbf {\bibinfo {volume} {12}},\ \bibinfo {pages} {005} (\bibinfo {year} {2020})},\ \Eprint {http://arxiv.org/abs/2008.00452} {arXiv:2008.00452 [astro-ph.CO]} \BibitemShut {NoStop}%
\bibitem [{\citenamefont {Iacconi}\ \emph {et~al.}(2022)\citenamefont {Iacconi}, \citenamefont {Assadullahi}, \citenamefont {Fasiello},\ and\ \citenamefont {Wands}}]{Iacconi:2021ltm}%
  \BibitemOpen
  \bibfield  {author} {\bibinfo {author} {\bibfnamefont {L.}~\bibnamefont {Iacconi}}, \bibinfo {author} {\bibfnamefont {H.}~\bibnamefont {Assadullahi}}, \bibinfo {author} {\bibfnamefont {M.}~\bibnamefont {Fasiello}}, \ and\ \bibinfo {author} {\bibfnamefont {D.}~\bibnamefont {Wands}},\ }\href {\doibase 10.1088/1475-7516/2022/06/007} {\bibfield  {journal} {\bibinfo  {journal} {JCAP}\ }\textbf {\bibinfo {volume} {06}},\ \bibinfo {pages} {007} (\bibinfo {year} {2022})},\ \Eprint {http://arxiv.org/abs/2112.05092} {arXiv:2112.05092 [astro-ph.CO]} \BibitemShut {NoStop}%
\bibitem [{\citenamefont {Iacconi}\ \emph {et~al.}(2023)\citenamefont {Iacconi}, \citenamefont {Fasiello}, \citenamefont {V\"aliviita},\ and\ \citenamefont {Wands}}]{Iacconi:2023mnw}%
  \BibitemOpen
  \bibfield  {author} {\bibinfo {author} {\bibfnamefont {L.}~\bibnamefont {Iacconi}}, \bibinfo {author} {\bibfnamefont {M.}~\bibnamefont {Fasiello}}, \bibinfo {author} {\bibfnamefont {J.}~\bibnamefont {V\"aliviita}}, \ and\ \bibinfo {author} {\bibfnamefont {D.}~\bibnamefont {Wands}},\ }\href {\doibase 10.1088/1475-7516/2023/10/015} {\bibfield  {journal} {\bibinfo  {journal} {JCAP}\ }\textbf {\bibinfo {volume} {10}},\ \bibinfo {pages} {015} (\bibinfo {year} {2023})},\ \Eprint {http://arxiv.org/abs/2306.00918} {arXiv:2306.00918 [astro-ph.CO]} \BibitemShut {NoStop}%
\bibitem [{\citenamefont {Santos}\ \emph {et~al.}(2023)\citenamefont {Santos}, \citenamefont {Rodrigues}, \citenamefont {Rodrigues}, \citenamefont {de~Souza},\ and\ \citenamefont {Alcaniz}}]{Santos:2023bnu}%
  \BibitemOpen
  \bibfield  {author} {\bibinfo {author} {\bibfnamefont {F.~B. M.~d.}\ \bibnamefont {Santos}}, \bibinfo {author} {\bibfnamefont {G.}~\bibnamefont {Rodrigues}}, \bibinfo {author} {\bibfnamefont {J.~G.}\ \bibnamefont {Rodrigues}}, \bibinfo {author} {\bibfnamefont {R.}~\bibnamefont {de~Souza}}, \ and\ \bibinfo {author} {\bibfnamefont {J.~S.}\ \bibnamefont {Alcaniz}},\ }\href@noop {} {\  (\bibinfo {year} {2023})},\ \Eprint {http://arxiv.org/abs/2312.12286} {arXiv:2312.12286 [astro-ph.CO]} \BibitemShut {NoStop}%
\bibitem [{\citenamefont {Pozo}\ \emph {et~al.}(2023)\citenamefont {Pozo}, \citenamefont {Zambrano}, \citenamefont {Villegas}, \citenamefont {Hern\'andez-Jim\'enez},\ and\ \citenamefont {Rojas}}]{Pozo:2023dto}%
  \BibitemOpen
  \bibfield  {author} {\bibinfo {author} {\bibfnamefont {D.}~\bibnamefont {Pozo}}, \bibinfo {author} {\bibfnamefont {J.}~\bibnamefont {Zambrano}}, \bibinfo {author} {\bibfnamefont {I.}~\bibnamefont {Villegas}}, \bibinfo {author} {\bibfnamefont {R.}~\bibnamefont {Hern\'andez-Jim\'enez}}, \ and\ \bibinfo {author} {\bibfnamefont {C.}~\bibnamefont {Rojas}},\ }\href@noop {} {\  (\bibinfo {year} {2023})},\ \Eprint {http://arxiv.org/abs/2311.04683} {arXiv:2311.04683 [gr-qc]} \BibitemShut {NoStop}%
\bibitem [{\citenamefont {Giar\`e}\ \emph {et~al.}(2024{\natexlab{a}})\citenamefont {Giar\`e}, \citenamefont {Di~Valentino}, \citenamefont {Linder},\ and\ \citenamefont {Specogna}}]{Giare:2024sdl}%
  \BibitemOpen
  \bibfield  {author} {\bibinfo {author} {\bibfnamefont {W.}~\bibnamefont {Giar\`e}}, \bibinfo {author} {\bibfnamefont {E.}~\bibnamefont {Di~Valentino}}, \bibinfo {author} {\bibfnamefont {E.~V.}\ \bibnamefont {Linder}}, \ and\ \bibinfo {author} {\bibfnamefont {E.}~\bibnamefont {Specogna}},\ }\href@noop {} {\  (\bibinfo {year} {2024}{\natexlab{a}})},\ \Eprint {http://arxiv.org/abs/2402.01560} {arXiv:2402.01560 [astro-ph.CO]} \BibitemShut {NoStop}%
\bibitem [{\citenamefont {Cecchini}\ \emph {et~al.}(2024)\citenamefont {Cecchini}, \citenamefont {De~Angelis}, \citenamefont {Giar\`e}, \citenamefont {Rinaldi},\ and\ \citenamefont {Vagnozzi}}]{Cecchini:2024xoq}%
  \BibitemOpen
  \bibfield  {author} {\bibinfo {author} {\bibfnamefont {C.}~\bibnamefont {Cecchini}}, \bibinfo {author} {\bibfnamefont {M.}~\bibnamefont {De~Angelis}}, \bibinfo {author} {\bibfnamefont {W.}~\bibnamefont {Giar\`e}}, \bibinfo {author} {\bibfnamefont {M.}~\bibnamefont {Rinaldi}}, \ and\ \bibinfo {author} {\bibfnamefont {S.}~\bibnamefont {Vagnozzi}},\ }\href@noop {} {\  (\bibinfo {year} {2024})},\ \Eprint {http://arxiv.org/abs/2403.04316} {arXiv:2403.04316 [astro-ph.CO]} \BibitemShut {NoStop}%
\bibitem [{\citenamefont {Wang}\ \emph {et~al.}(2024)\citenamefont {Wang}, \citenamefont {Zhang},\ and\ \citenamefont {Sasaki}}]{Wang:2024vfv}%
  \BibitemOpen
  \bibfield  {author} {\bibinfo {author} {\bibfnamefont {X.}~\bibnamefont {Wang}}, \bibinfo {author} {\bibfnamefont {Y.-l.}\ \bibnamefont {Zhang}}, \ and\ \bibinfo {author} {\bibfnamefont {M.}~\bibnamefont {Sasaki}},\ }\href@noop {} {\  (\bibinfo {year} {2024})},\ \Eprint {http://arxiv.org/abs/2404.02492} {arXiv:2404.02492 [astro-ph.CO]} \BibitemShut {NoStop}%
\bibitem [{\citenamefont {Martin}\ \emph {et~al.}(2024)\citenamefont {Martin}, \citenamefont {Ringeval},\ and\ \citenamefont {Vennin}}]{Martin:2024qnn}%
  \BibitemOpen
  \bibfield  {author} {\bibinfo {author} {\bibfnamefont {J.}~\bibnamefont {Martin}}, \bibinfo {author} {\bibfnamefont {C.}~\bibnamefont {Ringeval}}, \ and\ \bibinfo {author} {\bibfnamefont {V.}~\bibnamefont {Vennin}},\ }\href@noop {} {\  (\bibinfo {year} {2024})},\ \Eprint {http://arxiv.org/abs/2404.10647} {arXiv:2404.10647 [astro-ph.CO]} \BibitemShut {NoStop}%
\bibitem [{\citenamefont {Lyth}\ and\ \citenamefont {Riotto}(1999)}]{Lyth:1998xn}%
  \BibitemOpen
  \bibfield  {author} {\bibinfo {author} {\bibfnamefont {D.~H.}\ \bibnamefont {Lyth}}\ and\ \bibinfo {author} {\bibfnamefont {A.}~\bibnamefont {Riotto}},\ }\href {\doibase 10.1016/S0370-1573(98)00128-8} {\bibfield  {journal} {\bibinfo  {journal} {Phys. Rept.}\ }\textbf {\bibinfo {volume} {314}},\ \bibinfo {pages} {1} (\bibinfo {year} {1999})},\ \Eprint {http://arxiv.org/abs/hep-ph/9807278} {arXiv:hep-ph/9807278} \BibitemShut {NoStop}%
\bibitem [{\citenamefont {Linde}(2008)}]{Linde:2007fr}%
  \BibitemOpen
  \bibfield  {author} {\bibinfo {author} {\bibfnamefont {A.~D.}\ \bibnamefont {Linde}},\ }\href {\doibase 10.1007/978-3-540-74353-8_1} {\bibfield  {journal} {\bibinfo  {journal} {Lect. Notes Phys.}\ }\textbf {\bibinfo {volume} {738}},\ \bibinfo {pages} {1} (\bibinfo {year} {2008})},\ \Eprint {http://arxiv.org/abs/0705.0164} {arXiv:0705.0164 [hep-th]} \BibitemShut {NoStop}%
\bibitem [{\citenamefont {Baumann}\ and\ \citenamefont {McAllister}(2015)}]{Baumann:2014nda}%
  \BibitemOpen
  \bibfield  {author} {\bibinfo {author} {\bibfnamefont {D.}~\bibnamefont {Baumann}}\ and\ \bibinfo {author} {\bibfnamefont {L.}~\bibnamefont {McAllister}},\ }\href {\doibase 10.1017/CBO9781316105733} {\emph {\bibinfo {title} {{Inflation and String Theory}}}},\ Cambridge Monographs on Mathematical Physics\ (\bibinfo  {publisher} {Cambridge University Press},\ \bibinfo {year} {2015})\ \Eprint {http://arxiv.org/abs/1404.2601} {arXiv:1404.2601 [hep-th]} \BibitemShut {NoStop}%
\bibitem [{\citenamefont {Akrami}\ \emph {et~al.}(2020{\natexlab{a}})\citenamefont {Akrami} \emph {et~al.}}]{Planck:2018jri}%
  \BibitemOpen
  \bibfield  {author} {\bibinfo {author} {\bibfnamefont {Y.}~\bibnamefont {Akrami}} \emph {et~al.} (\bibinfo {collaboration} {Planck}),\ }\href {\doibase 10.1051/0004-6361/201833887} {\bibfield  {journal} {\bibinfo  {journal} {Astron. Astrophys.}\ }\textbf {\bibinfo {volume} {641}},\ \bibinfo {pages} {A10} (\bibinfo {year} {2020}{\natexlab{a}})},\ \Eprint {http://arxiv.org/abs/1807.06211} {arXiv:1807.06211 [astro-ph.CO]} \BibitemShut {NoStop}%
\bibitem [{\citenamefont {Starobinsky}\ \emph {et~al.}(2001)\citenamefont {Starobinsky}, \citenamefont {Tsujikawa},\ and\ \citenamefont {Yokoyama}}]{Starobinsky:2001xq}%
  \BibitemOpen
  \bibfield  {author} {\bibinfo {author} {\bibfnamefont {A.~A.}\ \bibnamefont {Starobinsky}}, \bibinfo {author} {\bibfnamefont {S.}~\bibnamefont {Tsujikawa}}, \ and\ \bibinfo {author} {\bibfnamefont {J.}~\bibnamefont {Yokoyama}},\ }\href {\doibase 10.1016/S0550-3213(01)00322-4} {\bibfield  {journal} {\bibinfo  {journal} {Nucl. Phys. B}\ }\textbf {\bibinfo {volume} {610}},\ \bibinfo {pages} {383} (\bibinfo {year} {2001})},\ \Eprint {http://arxiv.org/abs/astro-ph/0107555} {arXiv:astro-ph/0107555} \BibitemShut {NoStop}%
\bibitem [{\citenamefont {Tsujikawa}\ \emph {et~al.}(2003)\citenamefont {Tsujikawa}, \citenamefont {Parkinson},\ and\ \citenamefont {Bassett}}]{Tsujikawa:2002qx}%
  \BibitemOpen
  \bibfield  {author} {\bibinfo {author} {\bibfnamefont {S.}~\bibnamefont {Tsujikawa}}, \bibinfo {author} {\bibfnamefont {D.}~\bibnamefont {Parkinson}}, \ and\ \bibinfo {author} {\bibfnamefont {B.~A.}\ \bibnamefont {Bassett}},\ }\href {\doibase 10.1103/PhysRevD.67.083516} {\bibfield  {journal} {\bibinfo  {journal} {Phys. Rev. D}\ }\textbf {\bibinfo {volume} {67}},\ \bibinfo {pages} {083516} (\bibinfo {year} {2003})},\ \Eprint {http://arxiv.org/abs/astro-ph/0210322} {arXiv:astro-ph/0210322} \BibitemShut {NoStop}%
\bibitem [{\citenamefont {Di~Marco}\ \emph {et~al.}(2003)\citenamefont {Di~Marco}, \citenamefont {Finelli},\ and\ \citenamefont {Brandenberger}}]{DiMarco:2002eb}%
  \BibitemOpen
  \bibfield  {author} {\bibinfo {author} {\bibfnamefont {F.}~\bibnamefont {Di~Marco}}, \bibinfo {author} {\bibfnamefont {F.}~\bibnamefont {Finelli}}, \ and\ \bibinfo {author} {\bibfnamefont {R.}~\bibnamefont {Brandenberger}},\ }\href {\doibase 10.1103/PhysRevD.67.063512} {\bibfield  {journal} {\bibinfo  {journal} {Phys. Rev. D}\ }\textbf {\bibinfo {volume} {67}},\ \bibinfo {pages} {063512} (\bibinfo {year} {2003})},\ \Eprint {http://arxiv.org/abs/astro-ph/0211276} {arXiv:astro-ph/0211276} \BibitemShut {NoStop}%
\bibitem [{\citenamefont {Kaiser}(2010)}]{Kaiser:2010ps}%
  \BibitemOpen
  \bibfield  {author} {\bibinfo {author} {\bibfnamefont {D.~I.}\ \bibnamefont {Kaiser}},\ }\href {\doibase 10.1103/PhysRevD.81.084044} {\bibfield  {journal} {\bibinfo  {journal} {Phys. Rev. D}\ }\textbf {\bibinfo {volume} {81}},\ \bibinfo {pages} {084044} (\bibinfo {year} {2010})},\ \Eprint {http://arxiv.org/abs/1003.1159} {arXiv:1003.1159 [gr-qc]} \BibitemShut {NoStop}%
\bibitem [{\citenamefont {Achucarro}\ \emph {et~al.}(2011)\citenamefont {Achucarro}, \citenamefont {Gong}, \citenamefont {Hardeman}, \citenamefont {Palma},\ and\ \citenamefont {Patil}}]{Achucarro:2010da}%
  \BibitemOpen
  \bibfield  {author} {\bibinfo {author} {\bibfnamefont {A.}~\bibnamefont {Achucarro}}, \bibinfo {author} {\bibfnamefont {J.-O.}\ \bibnamefont {Gong}}, \bibinfo {author} {\bibfnamefont {S.}~\bibnamefont {Hardeman}}, \bibinfo {author} {\bibfnamefont {G.~A.}\ \bibnamefont {Palma}}, \ and\ \bibinfo {author} {\bibfnamefont {S.~P.}\ \bibnamefont {Patil}},\ }\href {\doibase 10.1088/1475-7516/2011/01/030} {\bibfield  {journal} {\bibinfo  {journal} {JCAP}\ }\textbf {\bibinfo {volume} {01}},\ \bibinfo {pages} {030} (\bibinfo {year} {2011})},\ \Eprint {http://arxiv.org/abs/1010.3693} {arXiv:1010.3693 [hep-ph]} \BibitemShut {NoStop}%
\bibitem [{\citenamefont {van~de Bruck}\ \emph {et~al.}(2011)\citenamefont {van~de Bruck}, \citenamefont {Mota},\ and\ \citenamefont {Weller}}]{vandeBruck:2010yw}%
  \BibitemOpen
  \bibfield  {author} {\bibinfo {author} {\bibfnamefont {C.}~\bibnamefont {van~de Bruck}}, \bibinfo {author} {\bibfnamefont {D.~F.}\ \bibnamefont {Mota}}, \ and\ \bibinfo {author} {\bibfnamefont {J.~M.}\ \bibnamefont {Weller}},\ }\href {\doibase 10.1088/1475-7516/2011/03/034} {\bibfield  {journal} {\bibinfo  {journal} {JCAP}\ }\textbf {\bibinfo {volume} {03}},\ \bibinfo {pages} {034} (\bibinfo {year} {2011})},\ \Eprint {http://arxiv.org/abs/1012.1567} {arXiv:1012.1567 [astro-ph.CO]} \BibitemShut {NoStop}%
\bibitem [{\citenamefont {Kaiser}\ and\ \citenamefont {Sfakianakis}(2014)}]{Kaiser:2013sna}%
  \BibitemOpen
  \bibfield  {author} {\bibinfo {author} {\bibfnamefont {D.~I.}\ \bibnamefont {Kaiser}}\ and\ \bibinfo {author} {\bibfnamefont {E.~I.}\ \bibnamefont {Sfakianakis}},\ }\href {\doibase 10.1103/PhysRevLett.112.011302} {\bibfield  {journal} {\bibinfo  {journal} {Phys. Rev. Lett.}\ }\textbf {\bibinfo {volume} {112}},\ \bibinfo {pages} {011302} (\bibinfo {year} {2014})},\ \Eprint {http://arxiv.org/abs/1304.0363} {arXiv:1304.0363 [astro-ph.CO]} \BibitemShut {NoStop}%
\bibitem [{\citenamefont {van~de Bruck}\ \emph {et~al.}(2016)\citenamefont {van~de Bruck}, \citenamefont {Koivisto},\ and\ \citenamefont {Longden}}]{vandeBruck:2015tna}%
  \BibitemOpen
  \bibfield  {author} {\bibinfo {author} {\bibfnamefont {C.}~\bibnamefont {van~de Bruck}}, \bibinfo {author} {\bibfnamefont {T.}~\bibnamefont {Koivisto}}, \ and\ \bibinfo {author} {\bibfnamefont {C.}~\bibnamefont {Longden}},\ }\href {\doibase 10.1088/1475-7516/2016/03/006} {\bibfield  {journal} {\bibinfo  {journal} {JCAP}\ }\textbf {\bibinfo {volume} {03}},\ \bibinfo {pages} {006} (\bibinfo {year} {2016})},\ \Eprint {http://arxiv.org/abs/1510.01650} {arXiv:1510.01650 [astro-ph.CO]} \BibitemShut {NoStop}%
\bibitem [{\citenamefont {van~de Bruck}\ and\ \citenamefont {Paduraru}(2015)}]{vandeBruck:2015xpa}%
  \BibitemOpen
  \bibfield  {author} {\bibinfo {author} {\bibfnamefont {C.}~\bibnamefont {van~de Bruck}}\ and\ \bibinfo {author} {\bibfnamefont {L.~E.}\ \bibnamefont {Paduraru}},\ }\href {\doibase 10.1103/PhysRevD.92.083513} {\bibfield  {journal} {\bibinfo  {journal} {Phys. Rev. D}\ }\textbf {\bibinfo {volume} {92}},\ \bibinfo {pages} {083513} (\bibinfo {year} {2015})},\ \Eprint {http://arxiv.org/abs/1505.01727} {arXiv:1505.01727 [hep-th]} \BibitemShut {NoStop}%
\bibitem [{\citenamefont {van~de Bruck}\ \emph {et~al.}(2017)\citenamefont {van~de Bruck}, \citenamefont {Koivisto},\ and\ \citenamefont {Longden}}]{vandeBruck:2016vlw}%
  \BibitemOpen
  \bibfield  {author} {\bibinfo {author} {\bibfnamefont {C.}~\bibnamefont {van~de Bruck}}, \bibinfo {author} {\bibfnamefont {T.}~\bibnamefont {Koivisto}}, \ and\ \bibinfo {author} {\bibfnamefont {C.}~\bibnamefont {Longden}},\ }\href {\doibase 10.1088/1475-7516/2017/02/029} {\bibfield  {journal} {\bibinfo  {journal} {JCAP}\ }\textbf {\bibinfo {volume} {02}},\ \bibinfo {pages} {029} (\bibinfo {year} {2017})},\ \Eprint {http://arxiv.org/abs/1608.08801} {arXiv:1608.08801 [astro-ph.CO]} \BibitemShut {NoStop}%
\bibitem [{\citenamefont {Carrilho}\ \emph {et~al.}(2018)\citenamefont {Carrilho}, \citenamefont {Mulryne}, \citenamefont {Ronayne},\ and\ \citenamefont {Tenkanen}}]{Carrilho:2018ffi}%
  \BibitemOpen
  \bibfield  {author} {\bibinfo {author} {\bibfnamefont {P.}~\bibnamefont {Carrilho}}, \bibinfo {author} {\bibfnamefont {D.}~\bibnamefont {Mulryne}}, \bibinfo {author} {\bibfnamefont {J.}~\bibnamefont {Ronayne}}, \ and\ \bibinfo {author} {\bibfnamefont {T.}~\bibnamefont {Tenkanen}},\ }\href {\doibase 10.1088/1475-7516/2018/06/032} {\bibfield  {journal} {\bibinfo  {journal} {JCAP}\ }\textbf {\bibinfo {volume} {06}},\ \bibinfo {pages} {032} (\bibinfo {year} {2018})},\ \Eprint {http://arxiv.org/abs/1804.10489} {arXiv:1804.10489 [astro-ph.CO]} \BibitemShut {NoStop}%
\bibitem [{\citenamefont {Ach\'ucarro}\ \emph {et~al.}(2020)\citenamefont {Ach\'ucarro}, \citenamefont {Copeland}, \citenamefont {Iarygina}, \citenamefont {Palma}, \citenamefont {Wang},\ and\ \citenamefont {Welling}}]{Achucarro:2019pux}%
  \BibitemOpen
  \bibfield  {author} {\bibinfo {author} {\bibfnamefont {A.}~\bibnamefont {Ach\'ucarro}}, \bibinfo {author} {\bibfnamefont {E.~J.}\ \bibnamefont {Copeland}}, \bibinfo {author} {\bibfnamefont {O.}~\bibnamefont {Iarygina}}, \bibinfo {author} {\bibfnamefont {G.~A.}\ \bibnamefont {Palma}}, \bibinfo {author} {\bibfnamefont {D.-G.}\ \bibnamefont {Wang}}, \ and\ \bibinfo {author} {\bibfnamefont {Y.}~\bibnamefont {Welling}},\ }\href {\doibase 10.1103/PhysRevD.102.021302} {\bibfield  {journal} {\bibinfo  {journal} {Phys. Rev. D}\ }\textbf {\bibinfo {volume} {102}},\ \bibinfo {pages} {021302} (\bibinfo {year} {2020})},\ \Eprint {http://arxiv.org/abs/1901.03657} {arXiv:1901.03657 [astro-ph.CO]} \BibitemShut {NoStop}%
\bibitem [{\citenamefont {Pinol}(2021)}]{Pinol:2020kvw}%
  \BibitemOpen
  \bibfield  {author} {\bibinfo {author} {\bibfnamefont {L.}~\bibnamefont {Pinol}},\ }\href {\doibase 10.1088/1475-7516/2021/04/002} {\bibfield  {journal} {\bibinfo  {journal} {JCAP}\ }\textbf {\bibinfo {volume} {04}},\ \bibinfo {pages} {002} (\bibinfo {year} {2021})},\ \Eprint {http://arxiv.org/abs/2011.05930} {arXiv:2011.05930 [astro-ph.CO]} \BibitemShut {NoStop}%
\bibitem [{\citenamefont {Ach\'ucarro}\ and\ \citenamefont {Palma}(2019)}]{Achucarro:2018vey}%
  \BibitemOpen
  \bibfield  {author} {\bibinfo {author} {\bibfnamefont {A.}~\bibnamefont {Ach\'ucarro}}\ and\ \bibinfo {author} {\bibfnamefont {G.~A.}\ \bibnamefont {Palma}},\ }\href {\doibase 10.1088/1475-7516/2019/02/041} {\bibfield  {journal} {\bibinfo  {journal} {JCAP}\ }\textbf {\bibinfo {volume} {02}},\ \bibinfo {pages} {041} (\bibinfo {year} {2019})},\ \Eprint {http://arxiv.org/abs/1807.04390} {arXiv:1807.04390 [hep-th]} \BibitemShut {NoStop}%
\bibitem [{\citenamefont {van~de Bruck}\ and\ \citenamefont {Daniel}(2021)}]{vandeBruck:2021xkm}%
  \BibitemOpen
  \bibfield  {author} {\bibinfo {author} {\bibfnamefont {C.}~\bibnamefont {van~de Bruck}}\ and\ \bibinfo {author} {\bibfnamefont {R.}~\bibnamefont {Daniel}},\ }\href {\doibase 10.1103/PhysRevD.103.123506} {\bibfield  {journal} {\bibinfo  {journal} {Phys. Rev. D}\ }\textbf {\bibinfo {volume} {103}},\ \bibinfo {pages} {123506} (\bibinfo {year} {2021})},\ \Eprint {http://arxiv.org/abs/2102.11719} {arXiv:2102.11719 [gr-qc]} \BibitemShut {NoStop}%
\bibitem [{\citenamefont {De~Angelis}\ and\ \citenamefont {van~de Bruck}(2023)}]{DeAngelis:2023fdu}%
  \BibitemOpen
  \bibfield  {author} {\bibinfo {author} {\bibfnamefont {M.}~\bibnamefont {De~Angelis}}\ and\ \bibinfo {author} {\bibfnamefont {C.}~\bibnamefont {van~de Bruck}},\ }\href {\doibase 10.1088/1475-7516/2023/10/023} {\bibfield  {journal} {\bibinfo  {journal} {JCAP}\ }\textbf {\bibinfo {volume} {10}},\ \bibinfo {pages} {023} (\bibinfo {year} {2023})},\ \Eprint {http://arxiv.org/abs/2304.12364} {arXiv:2304.12364 [hep-th]} \BibitemShut {NoStop}%
\bibitem [{\citenamefont {Tsujikawa}\ and\ \citenamefont {Yajima}(2000)}]{Tsujikawa:2000wc}%
  \BibitemOpen
  \bibfield  {author} {\bibinfo {author} {\bibfnamefont {S.}~\bibnamefont {Tsujikawa}}\ and\ \bibinfo {author} {\bibfnamefont {H.}~\bibnamefont {Yajima}},\ }\href {\doibase 10.1103/PhysRevD.62.123512} {\bibfield  {journal} {\bibinfo  {journal} {Phys. Rev. D}\ }\textbf {\bibinfo {volume} {62}},\ \bibinfo {pages} {123512} (\bibinfo {year} {2000})},\ \Eprint {http://arxiv.org/abs/hep-ph/0007351} {arXiv:hep-ph/0007351} \BibitemShut {NoStop}%
\bibitem [{\citenamefont {Weinberg}(2004)}]{Weinberg:2004kf}%
  \BibitemOpen
  \bibfield  {author} {\bibinfo {author} {\bibfnamefont {S.}~\bibnamefont {Weinberg}},\ }\href {\doibase 10.1103/PhysRevD.70.083522} {\bibfield  {journal} {\bibinfo  {journal} {Phys. Rev. D}\ }\textbf {\bibinfo {volume} {70}},\ \bibinfo {pages} {083522} (\bibinfo {year} {2004})},\ \Eprint {http://arxiv.org/abs/astro-ph/0405397} {arXiv:astro-ph/0405397} \BibitemShut {NoStop}%
\bibitem [{\citenamefont {Kaiser}\ and\ \citenamefont {Todhunter}(2010)}]{Kaiser:2010yu}%
  \BibitemOpen
  \bibfield  {author} {\bibinfo {author} {\bibfnamefont {D.~I.}\ \bibnamefont {Kaiser}}\ and\ \bibinfo {author} {\bibfnamefont {A.~T.}\ \bibnamefont {Todhunter}},\ }\href {\doibase 10.1103/PhysRevD.81.124037} {\bibfield  {journal} {\bibinfo  {journal} {Phys. Rev. D}\ }\textbf {\bibinfo {volume} {81}},\ \bibinfo {pages} {124037} (\bibinfo {year} {2010})},\ \Eprint {http://arxiv.org/abs/1004.3805} {arXiv:1004.3805 [astro-ph.CO]} \BibitemShut {NoStop}%
\bibitem [{\citenamefont {Frazer}(2014)}]{Frazer:2013zoa}%
  \BibitemOpen
  \bibfield  {author} {\bibinfo {author} {\bibfnamefont {J.}~\bibnamefont {Frazer}},\ }\href {\doibase 10.1088/1475-7516/2014/01/028} {\bibfield  {journal} {\bibinfo  {journal} {JCAP}\ }\textbf {\bibinfo {volume} {01}},\ \bibinfo {pages} {028} (\bibinfo {year} {2014})},\ \Eprint {http://arxiv.org/abs/1303.3611} {arXiv:1303.3611 [astro-ph.CO]} \BibitemShut {NoStop}%
\bibitem [{\citenamefont {Ach\'ucarro}\ \emph {et~al.}(2013)\citenamefont {Ach\'ucarro}, \citenamefont {Gong}, \citenamefont {Palma},\ and\ \citenamefont {Patil}}]{Achucarro:2012fd}%
  \BibitemOpen
  \bibfield  {author} {\bibinfo {author} {\bibfnamefont {A.}~\bibnamefont {Ach\'ucarro}}, \bibinfo {author} {\bibfnamefont {J.-O.}\ \bibnamefont {Gong}}, \bibinfo {author} {\bibfnamefont {G.~A.}\ \bibnamefont {Palma}}, \ and\ \bibinfo {author} {\bibfnamefont {S.~P.}\ \bibnamefont {Patil}},\ }\href {\doibase 10.1103/PhysRevD.87.121301} {\bibfield  {journal} {\bibinfo  {journal} {Phys. Rev. D}\ }\textbf {\bibinfo {volume} {87}},\ \bibinfo {pages} {121301} (\bibinfo {year} {2013})},\ \Eprint {http://arxiv.org/abs/1211.5619} {arXiv:1211.5619 [astro-ph.CO]} \BibitemShut {NoStop}%
\bibitem [{\citenamefont {van~de Bruck}\ and\ \citenamefont {Robinson}(2014)}]{vandeBruck:2014ata}%
  \BibitemOpen
  \bibfield  {author} {\bibinfo {author} {\bibfnamefont {C.}~\bibnamefont {van~de Bruck}}\ and\ \bibinfo {author} {\bibfnamefont {M.}~\bibnamefont {Robinson}},\ }\href {\doibase 10.1088/1475-7516/2014/08/024} {\bibfield  {journal} {\bibinfo  {journal} {JCAP}\ }\textbf {\bibinfo {volume} {08}},\ \bibinfo {pages} {024} (\bibinfo {year} {2014})},\ \Eprint {http://arxiv.org/abs/1404.7806} {arXiv:1404.7806 [astro-ph.CO]} \BibitemShut {NoStop}%
\bibitem [{\citenamefont {Dias}\ \emph {et~al.}(2015)\citenamefont {Dias}, \citenamefont {Frazer},\ and\ \citenamefont {Seery}}]{Dias:2015rca}%
  \BibitemOpen
  \bibfield  {author} {\bibinfo {author} {\bibfnamefont {M.}~\bibnamefont {Dias}}, \bibinfo {author} {\bibfnamefont {J.}~\bibnamefont {Frazer}}, \ and\ \bibinfo {author} {\bibfnamefont {D.}~\bibnamefont {Seery}},\ }\href {\doibase 10.1088/1475-7516/2015/12/030} {\bibfield  {journal} {\bibinfo  {journal} {JCAP}\ }\textbf {\bibinfo {volume} {12}},\ \bibinfo {pages} {030} (\bibinfo {year} {2015})},\ \Eprint {http://arxiv.org/abs/1502.03125} {arXiv:1502.03125 [astro-ph.CO]} \BibitemShut {NoStop}%
\bibitem [{\citenamefont {Dias}\ \emph {et~al.}(2016)\citenamefont {Dias}, \citenamefont {Frazer}, \citenamefont {Mulryne},\ and\ \citenamefont {Seery}}]{Dias:2016rjq}%
  \BibitemOpen
  \bibfield  {author} {\bibinfo {author} {\bibfnamefont {M.}~\bibnamefont {Dias}}, \bibinfo {author} {\bibfnamefont {J.}~\bibnamefont {Frazer}}, \bibinfo {author} {\bibfnamefont {D.~J.}\ \bibnamefont {Mulryne}}, \ and\ \bibinfo {author} {\bibfnamefont {D.}~\bibnamefont {Seery}},\ }\href {\doibase 10.1088/1475-7516/2016/12/033} {\bibfield  {journal} {\bibinfo  {journal} {JCAP}\ }\textbf {\bibinfo {volume} {12}},\ \bibinfo {pages} {033} (\bibinfo {year} {2016})},\ \Eprint {http://arxiv.org/abs/1609.00379} {arXiv:1609.00379 [astro-ph.CO]} \BibitemShut {NoStop}%
\bibitem [{\citenamefont {Braglia}\ \emph {et~al.}(2020)\citenamefont {Braglia}, \citenamefont {Hazra}, \citenamefont {Sriramkumar},\ and\ \citenamefont {Finelli}}]{Braglia:2020fms}%
  \BibitemOpen
  \bibfield  {author} {\bibinfo {author} {\bibfnamefont {M.}~\bibnamefont {Braglia}}, \bibinfo {author} {\bibfnamefont {D.~K.}\ \bibnamefont {Hazra}}, \bibinfo {author} {\bibfnamefont {L.}~\bibnamefont {Sriramkumar}}, \ and\ \bibinfo {author} {\bibfnamefont {F.}~\bibnamefont {Finelli}},\ }\href {\doibase 10.1088/1475-7516/2020/08/025} {\bibfield  {journal} {\bibinfo  {journal} {JCAP}\ }\textbf {\bibinfo {volume} {08}},\ \bibinfo {pages} {025} (\bibinfo {year} {2020})},\ \Eprint {http://arxiv.org/abs/2004.00672} {arXiv:2004.00672 [astro-ph.CO]} \BibitemShut {NoStop}%
\bibitem [{\citenamefont {Braglia}\ \emph {et~al.}(2021)\citenamefont {Braglia}, \citenamefont {Chen},\ and\ \citenamefont {Hazra}}]{Braglia:2021ckn}%
  \BibitemOpen
  \bibfield  {author} {\bibinfo {author} {\bibfnamefont {M.}~\bibnamefont {Braglia}}, \bibinfo {author} {\bibfnamefont {X.}~\bibnamefont {Chen}}, \ and\ \bibinfo {author} {\bibfnamefont {D.~K.}\ \bibnamefont {Hazra}},\ }\href {\doibase 10.1088/1475-7516/2021/06/005} {\bibfield  {journal} {\bibinfo  {journal} {JCAP}\ }\textbf {\bibinfo {volume} {06}},\ \bibinfo {pages} {005} (\bibinfo {year} {2021})},\ \Eprint {http://arxiv.org/abs/2103.03025} {arXiv:2103.03025 [astro-ph.CO]} \BibitemShut {NoStop}%
\bibitem [{\citenamefont {Cabass}\ \emph {et~al.}(2022)\citenamefont {Cabass}, \citenamefont {Ivanov}, \citenamefont {Philcox}, \citenamefont {Simonovi\'c},\ and\ \citenamefont {Zaldarriaga}}]{Cabass:2022ymb}%
  \BibitemOpen
  \bibfield  {author} {\bibinfo {author} {\bibfnamefont {G.}~\bibnamefont {Cabass}}, \bibinfo {author} {\bibfnamefont {M.~M.}\ \bibnamefont {Ivanov}}, \bibinfo {author} {\bibfnamefont {O.~H.~E.}\ \bibnamefont {Philcox}}, \bibinfo {author} {\bibfnamefont {M.}~\bibnamefont {Simonovi\'c}}, \ and\ \bibinfo {author} {\bibfnamefont {M.}~\bibnamefont {Zaldarriaga}},\ }\href {\doibase 10.1103/PhysRevD.106.043506} {\bibfield  {journal} {\bibinfo  {journal} {Phys. Rev. D}\ }\textbf {\bibinfo {volume} {106}},\ \bibinfo {pages} {043506} (\bibinfo {year} {2022})},\ \Eprint {http://arxiv.org/abs/2204.01781} {arXiv:2204.01781 [astro-ph.CO]} \BibitemShut {NoStop}%
\bibitem [{\citenamefont {Geller}\ \emph {et~al.}(2022)\citenamefont {Geller}, \citenamefont {Qin}, \citenamefont {McDonough},\ and\ \citenamefont {Kaiser}}]{Geller:2022nkr}%
  \BibitemOpen
  \bibfield  {author} {\bibinfo {author} {\bibfnamefont {S.~R.}\ \bibnamefont {Geller}}, \bibinfo {author} {\bibfnamefont {W.}~\bibnamefont {Qin}}, \bibinfo {author} {\bibfnamefont {E.}~\bibnamefont {McDonough}}, \ and\ \bibinfo {author} {\bibfnamefont {D.~I.}\ \bibnamefont {Kaiser}},\ }\href {\doibase 10.1103/PhysRevD.106.063535} {\bibfield  {journal} {\bibinfo  {journal} {Phys. Rev. D}\ }\textbf {\bibinfo {volume} {106}},\ \bibinfo {pages} {063535} (\bibinfo {year} {2022})},\ \Eprint {http://arxiv.org/abs/2205.04471} {arXiv:2205.04471 [hep-th]} \BibitemShut {NoStop}%
\bibitem [{\citenamefont {Wang}\ \emph {et~al.}(2023)\citenamefont {Wang}, \citenamefont {Pimentel},\ and\ \citenamefont {Ach\'ucarro}}]{Wang:2022eop}%
  \BibitemOpen
  \bibfield  {author} {\bibinfo {author} {\bibfnamefont {D.-G.}\ \bibnamefont {Wang}}, \bibinfo {author} {\bibfnamefont {G.~L.}\ \bibnamefont {Pimentel}}, \ and\ \bibinfo {author} {\bibfnamefont {A.}~\bibnamefont {Ach\'ucarro}},\ }\href {\doibase 10.1088/1475-7516/2023/05/043} {\bibfield  {journal} {\bibinfo  {journal} {JCAP}\ }\textbf {\bibinfo {volume} {05}},\ \bibinfo {pages} {043} (\bibinfo {year} {2023})},\ \Eprint {http://arxiv.org/abs/2212.14035} {arXiv:2212.14035 [astro-ph.CO]} \BibitemShut {NoStop}%
\bibitem [{\citenamefont {Iacconi}\ and\ \citenamefont {Mulryne}(2023)}]{Iacconi:2023slv}%
  \BibitemOpen
  \bibfield  {author} {\bibinfo {author} {\bibfnamefont {L.}~\bibnamefont {Iacconi}}\ and\ \bibinfo {author} {\bibfnamefont {D.~J.}\ \bibnamefont {Mulryne}},\ }\href {\doibase 10.1088/1475-7516/2023/09/033} {\bibfield  {journal} {\bibinfo  {journal} {JCAP}\ }\textbf {\bibinfo {volume} {09}},\ \bibinfo {pages} {033} (\bibinfo {year} {2023})},\ \Eprint {http://arxiv.org/abs/2304.14260} {arXiv:2304.14260 [astro-ph.CO]} \BibitemShut {NoStop}%
\bibitem [{\citenamefont {Qin}\ \emph {et~al.}(2023)\citenamefont {Qin}, \citenamefont {Geller}, \citenamefont {Balaji}, \citenamefont {McDonough},\ and\ \citenamefont {Kaiser}}]{Qin:2023lgo}%
  \BibitemOpen
  \bibfield  {author} {\bibinfo {author} {\bibfnamefont {W.}~\bibnamefont {Qin}}, \bibinfo {author} {\bibfnamefont {S.~R.}\ \bibnamefont {Geller}}, \bibinfo {author} {\bibfnamefont {S.}~\bibnamefont {Balaji}}, \bibinfo {author} {\bibfnamefont {E.}~\bibnamefont {McDonough}}, \ and\ \bibinfo {author} {\bibfnamefont {D.~I.}\ \bibnamefont {Kaiser}},\ }\href {\doibase 10.1103/PhysRevD.108.043508} {\bibfield  {journal} {\bibinfo  {journal} {Phys. Rev. D}\ }\textbf {\bibinfo {volume} {108}},\ \bibinfo {pages} {043508} (\bibinfo {year} {2023})},\ \Eprint {http://arxiv.org/abs/2303.02168} {arXiv:2303.02168 [astro-ph.CO]} \BibitemShut {NoStop}%
\bibitem [{\citenamefont {Freytsis}\ \emph {et~al.}(2023)\citenamefont {Freytsis}, \citenamefont {Kumar}, \citenamefont {Remmen},\ and\ \citenamefont {Rodd}}]{Freytsis:2022aho}%
  \BibitemOpen
  \bibfield  {author} {\bibinfo {author} {\bibfnamefont {M.}~\bibnamefont {Freytsis}}, \bibinfo {author} {\bibfnamefont {S.}~\bibnamefont {Kumar}}, \bibinfo {author} {\bibfnamefont {G.~N.}\ \bibnamefont {Remmen}}, \ and\ \bibinfo {author} {\bibfnamefont {N.~L.}\ \bibnamefont {Rodd}},\ }\href {\doibase 10.1007/JHEP09(2023)041} {\bibfield  {journal} {\bibinfo  {journal} {JHEP}\ }\textbf {\bibinfo {volume} {09}},\ \bibinfo {pages} {041} (\bibinfo {year} {2023})},\ \Eprint {http://arxiv.org/abs/2210.10791} {arXiv:2210.10791 [hep-th]} \BibitemShut {NoStop}%
\bibitem [{\citenamefont {Cicoli}\ \emph {et~al.}(2023)\citenamefont {Cicoli}, \citenamefont {Guidetti}, \citenamefont {Muia}, \citenamefont {Pedro},\ and\ \citenamefont {Vacca}}]{Cicoli:2021yhb}%
  \BibitemOpen
  \bibfield  {author} {\bibinfo {author} {\bibfnamefont {M.}~\bibnamefont {Cicoli}}, \bibinfo {author} {\bibfnamefont {V.}~\bibnamefont {Guidetti}}, \bibinfo {author} {\bibfnamefont {F.}~\bibnamefont {Muia}}, \bibinfo {author} {\bibfnamefont {F.~G.}\ \bibnamefont {Pedro}}, \ and\ \bibinfo {author} {\bibfnamefont {G.~P.}\ \bibnamefont {Vacca}},\ }\href {\doibase 10.1088/1361-6382/acabf7} {\bibfield  {journal} {\bibinfo  {journal} {Class. Quant. Grav.}\ }\textbf {\bibinfo {volume} {40}},\ \bibinfo {pages} {025008} (\bibinfo {year} {2023})},\ \Eprint {http://arxiv.org/abs/2107.03391} {arXiv:2107.03391 [astro-ph.CO]} \BibitemShut {NoStop}%
\bibitem [{\citenamefont {Guerrero}\ \emph {et~al.}(2020)\citenamefont {Guerrero}, \citenamefont {Rubiera-Garcia},\ and\ \citenamefont {Saez-Chillon~Gomez}}]{Guerrero:2020lng}%
  \BibitemOpen
  \bibfield  {author} {\bibinfo {author} {\bibfnamefont {M.}~\bibnamefont {Guerrero}}, \bibinfo {author} {\bibfnamefont {D.}~\bibnamefont {Rubiera-Garcia}}, \ and\ \bibinfo {author} {\bibfnamefont {D.}~\bibnamefont {Saez-Chillon~Gomez}},\ }\href {\doibase 10.1103/PhysRevD.102.123528} {\bibfield  {journal} {\bibinfo  {journal} {Phys. Rev. D}\ }\textbf {\bibinfo {volume} {102}},\ \bibinfo {pages} {123528} (\bibinfo {year} {2020})},\ \Eprint {http://arxiv.org/abs/2008.07260} {arXiv:2008.07260 [gr-qc]} \BibitemShut {NoStop}%
\bibitem [{\citenamefont {Garcia-Saenz}\ \emph {et~al.}(2020)\citenamefont {Garcia-Saenz}, \citenamefont {Pinol},\ and\ \citenamefont {Renaux-Petel}}]{Garcia-Saenz:2019njm}%
  \BibitemOpen
  \bibfield  {author} {\bibinfo {author} {\bibfnamefont {S.}~\bibnamefont {Garcia-Saenz}}, \bibinfo {author} {\bibfnamefont {L.}~\bibnamefont {Pinol}}, \ and\ \bibinfo {author} {\bibfnamefont {S.}~\bibnamefont {Renaux-Petel}},\ }\href {\doibase 10.1007/JHEP01(2020)073} {\bibfield  {journal} {\bibinfo  {journal} {JHEP}\ }\textbf {\bibinfo {volume} {01}},\ \bibinfo {pages} {073} (\bibinfo {year} {2020})},\ \Eprint {http://arxiv.org/abs/1907.10403} {arXiv:1907.10403 [hep-th]} \BibitemShut {NoStop}%
\bibitem [{\citenamefont {Nguyen}\ \emph {et~al.}(2019)\citenamefont {Nguyen}, \citenamefont {van~de Vis}, \citenamefont {Sfakianakis}, \citenamefont {Giblin},\ and\ \citenamefont {Kaiser}}]{Nguyen:2019kbm}%
  \BibitemOpen
  \bibfield  {author} {\bibinfo {author} {\bibfnamefont {R.}~\bibnamefont {Nguyen}}, \bibinfo {author} {\bibfnamefont {J.}~\bibnamefont {van~de Vis}}, \bibinfo {author} {\bibfnamefont {E.~I.}\ \bibnamefont {Sfakianakis}}, \bibinfo {author} {\bibfnamefont {J.~T.}\ \bibnamefont {Giblin}}, \ and\ \bibinfo {author} {\bibfnamefont {D.~I.}\ \bibnamefont {Kaiser}},\ }\href {\doibase 10.1103/PhysRevLett.123.171301} {\bibfield  {journal} {\bibinfo  {journal} {Phys. Rev. Lett.}\ }\textbf {\bibinfo {volume} {123}},\ \bibinfo {pages} {171301} (\bibinfo {year} {2019})},\ \Eprint {http://arxiv.org/abs/1905.12562} {arXiv:1905.12562 [hep-ph]} \BibitemShut {NoStop}%
\bibitem [{\citenamefont {Li}\ \emph {et~al.}(2019)\citenamefont {Li}, \citenamefont {Zheng},\ and\ \citenamefont {Zhu}}]{Li:2019zbk}%
  \BibitemOpen
  \bibfield  {author} {\bibinfo {author} {\bibfnamefont {X.-B.}\ \bibnamefont {Li}}, \bibinfo {author} {\bibfnamefont {X.-G.}\ \bibnamefont {Zheng}}, \ and\ \bibinfo {author} {\bibfnamefont {J.-Y.}\ \bibnamefont {Zhu}},\ }\href {\doibase 10.1103/PhysRevD.99.043528} {\bibfield  {journal} {\bibinfo  {journal} {Phys. Rev. D}\ }\textbf {\bibinfo {volume} {99}},\ \bibinfo {pages} {043528} (\bibinfo {year} {2019})},\ \Eprint {http://arxiv.org/abs/1902.01017} {arXiv:1902.01017 [gr-qc]} \BibitemShut {NoStop}%
\bibitem [{\citenamefont {Bernardeau}\ and\ \citenamefont {Uzan}(2002)}]{Bernardeau:2002jy}%
  \BibitemOpen
  \bibfield  {author} {\bibinfo {author} {\bibfnamefont {F.}~\bibnamefont {Bernardeau}}\ and\ \bibinfo {author} {\bibfnamefont {J.-P.}\ \bibnamefont {Uzan}},\ }\href {\doibase 10.1103/PhysRevD.66.103506} {\bibfield  {journal} {\bibinfo  {journal} {Phys. Rev. D}\ }\textbf {\bibinfo {volume} {66}},\ \bibinfo {pages} {103506} (\bibinfo {year} {2002})},\ \Eprint {http://arxiv.org/abs/hep-ph/0207295} {arXiv:hep-ph/0207295} \BibitemShut {NoStop}%
\bibitem [{\citenamefont {Kaiser}\ \emph {et~al.}(2013)\citenamefont {Kaiser}, \citenamefont {Mazenc},\ and\ \citenamefont {Sfakianakis}}]{Kaiser:2012ak}%
  \BibitemOpen
  \bibfield  {author} {\bibinfo {author} {\bibfnamefont {D.~I.}\ \bibnamefont {Kaiser}}, \bibinfo {author} {\bibfnamefont {E.~A.}\ \bibnamefont {Mazenc}}, \ and\ \bibinfo {author} {\bibfnamefont {E.~I.}\ \bibnamefont {Sfakianakis}},\ }\href {\doibase 10.1103/PhysRevD.87.064004} {\bibfield  {journal} {\bibinfo  {journal} {Phys. Rev. D}\ }\textbf {\bibinfo {volume} {87}},\ \bibinfo {pages} {064004} (\bibinfo {year} {2013})},\ \Eprint {http://arxiv.org/abs/1210.7487} {arXiv:1210.7487 [astro-ph.CO]} \BibitemShut {NoStop}%
\bibitem [{\citenamefont {McAllister}\ \emph {et~al.}(2012)\citenamefont {McAllister}, \citenamefont {Renaux-Petel},\ and\ \citenamefont {Xu}}]{McAllister:2012am}%
  \BibitemOpen
  \bibfield  {author} {\bibinfo {author} {\bibfnamefont {L.}~\bibnamefont {McAllister}}, \bibinfo {author} {\bibfnamefont {S.}~\bibnamefont {Renaux-Petel}}, \ and\ \bibinfo {author} {\bibfnamefont {G.}~\bibnamefont {Xu}},\ }\href {\doibase 10.1088/1475-7516/2012/10/046} {\bibfield  {journal} {\bibinfo  {journal} {JCAP}\ }\textbf {\bibinfo {volume} {10}},\ \bibinfo {pages} {046} (\bibinfo {year} {2012})},\ \Eprint {http://arxiv.org/abs/1207.0317} {arXiv:1207.0317 [astro-ph.CO]} \BibitemShut {NoStop}%
\bibitem [{\citenamefont {Peterson}\ and\ \citenamefont {Tegmark}(2013)}]{Peterson:2011yt}%
  \BibitemOpen
  \bibfield  {author} {\bibinfo {author} {\bibfnamefont {C.~M.}\ \bibnamefont {Peterson}}\ and\ \bibinfo {author} {\bibfnamefont {M.}~\bibnamefont {Tegmark}},\ }\href {\doibase 10.1103/PhysRevD.87.103507} {\bibfield  {journal} {\bibinfo  {journal} {Phys. Rev. D}\ }\textbf {\bibinfo {volume} {87}},\ \bibinfo {pages} {103507} (\bibinfo {year} {2013})},\ \Eprint {http://arxiv.org/abs/1111.0927} {arXiv:1111.0927 [astro-ph.CO]} \BibitemShut {NoStop}%
\bibitem [{\citenamefont {Dias}\ \emph {et~al.}(2012)\citenamefont {Dias}, \citenamefont {Frazer},\ and\ \citenamefont {Liddle}}]{Dias:2012nf}%
  \BibitemOpen
  \bibfield  {author} {\bibinfo {author} {\bibfnamefont {M.}~\bibnamefont {Dias}}, \bibinfo {author} {\bibfnamefont {J.}~\bibnamefont {Frazer}}, \ and\ \bibinfo {author} {\bibfnamefont {A.~R.}\ \bibnamefont {Liddle}},\ }\href {\doibase 10.1088/1475-7516/2012/06/020} {\bibfield  {journal} {\bibinfo  {journal} {JCAP}\ }\textbf {\bibinfo {volume} {06}},\ \bibinfo {pages} {020} (\bibinfo {year} {2012})},\ \bibinfo {note} {[Erratum: JCAP 03, E01 (2013)]},\ \Eprint {http://arxiv.org/abs/1203.3792} {arXiv:1203.3792 [astro-ph.CO]} \BibitemShut {NoStop}%
\bibitem [{\citenamefont {Kehagias}\ and\ \citenamefont {Riotto}(2013)}]{Kehagias:2012td}%
  \BibitemOpen
  \bibfield  {author} {\bibinfo {author} {\bibfnamefont {A.}~\bibnamefont {Kehagias}}\ and\ \bibinfo {author} {\bibfnamefont {A.}~\bibnamefont {Riotto}},\ }\href {\doibase 10.1016/j.nuclphysb.2012.11.025} {\bibfield  {journal} {\bibinfo  {journal} {Nucl. Phys. B}\ }\textbf {\bibinfo {volume} {868}},\ \bibinfo {pages} {577} (\bibinfo {year} {2013})},\ \Eprint {http://arxiv.org/abs/1210.1918} {arXiv:1210.1918 [hep-th]} \BibitemShut {NoStop}%
\bibitem [{\citenamefont {Leung}\ \emph {et~al.}(2012)\citenamefont {Leung}, \citenamefont {Tarrant}, \citenamefont {Byrnes},\ and\ \citenamefont {Copeland}}]{Leung:2012ve}%
  \BibitemOpen
  \bibfield  {author} {\bibinfo {author} {\bibfnamefont {G.}~\bibnamefont {Leung}}, \bibinfo {author} {\bibfnamefont {E.~R.~M.}\ \bibnamefont {Tarrant}}, \bibinfo {author} {\bibfnamefont {C.~T.}\ \bibnamefont {Byrnes}}, \ and\ \bibinfo {author} {\bibfnamefont {E.~J.}\ \bibnamefont {Copeland}},\ }\href {\doibase 10.1088/1475-7516/2012/09/008} {\bibfield  {journal} {\bibinfo  {journal} {JCAP}\ }\textbf {\bibinfo {volume} {09}},\ \bibinfo {pages} {008} (\bibinfo {year} {2012})},\ \Eprint {http://arxiv.org/abs/1206.5196} {arXiv:1206.5196 [astro-ph.CO]} \BibitemShut {NoStop}%
\bibitem [{\citenamefont {Meyers}\ and\ \citenamefont {Sivanandam}(2011)}]{Meyers:2010rg}%
  \BibitemOpen
  \bibfield  {author} {\bibinfo {author} {\bibfnamefont {J.}~\bibnamefont {Meyers}}\ and\ \bibinfo {author} {\bibfnamefont {N.}~\bibnamefont {Sivanandam}},\ }\href {\doibase 10.1103/PhysRevD.83.103517} {\bibfield  {journal} {\bibinfo  {journal} {Phys. Rev. D}\ }\textbf {\bibinfo {volume} {83}},\ \bibinfo {pages} {103517} (\bibinfo {year} {2011})},\ \Eprint {http://arxiv.org/abs/1011.4934} {arXiv:1011.4934 [astro-ph.CO]} \BibitemShut {NoStop}%
\bibitem [{\citenamefont {Price}\ \emph {et~al.}(2015)\citenamefont {Price}, \citenamefont {Peiris}, \citenamefont {Frazer},\ and\ \citenamefont {Easther}}]{Price:2014ufa}%
  \BibitemOpen
  \bibfield  {author} {\bibinfo {author} {\bibfnamefont {L.~C.}\ \bibnamefont {Price}}, \bibinfo {author} {\bibfnamefont {H.~V.}\ \bibnamefont {Peiris}}, \bibinfo {author} {\bibfnamefont {J.}~\bibnamefont {Frazer}}, \ and\ \bibinfo {author} {\bibfnamefont {R.}~\bibnamefont {Easther}},\ }\href {\doibase 10.1103/PhysRevLett.114.031301} {\bibfield  {journal} {\bibinfo  {journal} {Phys. Rev. Lett.}\ }\textbf {\bibinfo {volume} {114}},\ \bibinfo {pages} {031301} (\bibinfo {year} {2015})},\ \Eprint {http://arxiv.org/abs/1409.2498} {arXiv:1409.2498 [astro-ph.CO]} \BibitemShut {NoStop}%
\bibitem [{\citenamefont {Battefeld}\ \emph {et~al.}(2011)\citenamefont {Battefeld}, \citenamefont {Battefeld}, \citenamefont {Byrnes},\ and\ \citenamefont {Langlois}}]{Battefeld:2011yj}%
  \BibitemOpen
  \bibfield  {author} {\bibinfo {author} {\bibfnamefont {D.}~\bibnamefont {Battefeld}}, \bibinfo {author} {\bibfnamefont {T.}~\bibnamefont {Battefeld}}, \bibinfo {author} {\bibfnamefont {C.}~\bibnamefont {Byrnes}}, \ and\ \bibinfo {author} {\bibfnamefont {D.}~\bibnamefont {Langlois}},\ }\href {\doibase 10.1088/1475-7516/2011/08/025} {\bibfield  {journal} {\bibinfo  {journal} {JCAP}\ }\textbf {\bibinfo {volume} {08}},\ \bibinfo {pages} {025} (\bibinfo {year} {2011})},\ \Eprint {http://arxiv.org/abs/1106.1891} {arXiv:1106.1891 [astro-ph.CO]} \BibitemShut {NoStop}%
\bibitem [{\citenamefont {Kaiser}(2016)}]{Kaiser:2015usz}%
  \BibitemOpen
  \bibfield  {author} {\bibinfo {author} {\bibfnamefont {D.~I.}\ \bibnamefont {Kaiser}},\ }\href {\doibase 10.1007/978-3-319-31299-6_2} {\bibfield  {journal} {\bibinfo  {journal} {Fundam. Theor. Phys.}\ }\textbf {\bibinfo {volume} {183}},\ \bibinfo {pages} {41} (\bibinfo {year} {2016})},\ \Eprint {http://arxiv.org/abs/1511.09148} {arXiv:1511.09148 [astro-ph.CO]} \BibitemShut {NoStop}%
\bibitem [{\citenamefont {Ashcroft}\ \emph {et~al.}(2002)\citenamefont {Ashcroft}, \citenamefont {van~de Bruck},\ and\ \citenamefont {Davis}}]{Ashcroft:2002ap}%
  \BibitemOpen
  \bibfield  {author} {\bibinfo {author} {\bibfnamefont {P.~R.}\ \bibnamefont {Ashcroft}}, \bibinfo {author} {\bibfnamefont {C.}~\bibnamefont {van~de Bruck}}, \ and\ \bibinfo {author} {\bibfnamefont {A.~C.}\ \bibnamefont {Davis}},\ }\href {\doibase 10.1103/PhysRevD.66.121302} {\bibfield  {journal} {\bibinfo  {journal} {Phys. Rev. D}\ }\textbf {\bibinfo {volume} {66}},\ \bibinfo {pages} {121302} (\bibinfo {year} {2002})},\ \Eprint {http://arxiv.org/abs/astro-ph/0208411} {arXiv:astro-ph/0208411} \BibitemShut {NoStop}%
\bibitem [{\citenamefont {Paliathanasis}\ and\ \citenamefont {Leon}(2022)}]{Paliathanasis:2021fxi}%
  \BibitemOpen
  \bibfield  {author} {\bibinfo {author} {\bibfnamefont {A.}~\bibnamefont {Paliathanasis}}\ and\ \bibinfo {author} {\bibfnamefont {G.}~\bibnamefont {Leon}},\ }\href {\doibase 10.1140/epjp/s13360-022-02383-6} {\bibfield  {journal} {\bibinfo  {journal} {Eur. Phys. J. Plus}\ }\textbf {\bibinfo {volume} {137}},\ \bibinfo {pages} {165} (\bibinfo {year} {2022})},\ \Eprint {http://arxiv.org/abs/2105.03261} {arXiv:2105.03261 [gr-qc]} \BibitemShut {NoStop}%
\bibitem [{\citenamefont {Paliathanasis}\ and\ \citenamefont {Leon}(2020)}]{Paliathanasis:2020abu}%
  \BibitemOpen
  \bibfield  {author} {\bibinfo {author} {\bibfnamefont {A.}~\bibnamefont {Paliathanasis}}\ and\ \bibinfo {author} {\bibfnamefont {G.}~\bibnamefont {Leon}},\ }\href {\doibase 10.1140/epjc/s10052-020-8423-7} {\bibfield  {journal} {\bibinfo  {journal} {Eur. Phys. J. C}\ }\textbf {\bibinfo {volume} {80}},\ \bibinfo {pages} {847} (\bibinfo {year} {2020})},\ \Eprint {http://arxiv.org/abs/2007.13223} {arXiv:2007.13223 [gr-qc]} \BibitemShut {NoStop}%
\bibitem [{\citenamefont {Paliathanasis}(2020)}]{Paliathanasis:2020wjl}%
  \BibitemOpen
  \bibfield  {author} {\bibinfo {author} {\bibfnamefont {A.}~\bibnamefont {Paliathanasis}},\ }\href {\doibase 10.1088/1361-6382/aba667} {\bibfield  {journal} {\bibinfo  {journal} {Class. Quant. Grav.}\ }\textbf {\bibinfo {volume} {37}},\ \bibinfo {pages} {195014} (\bibinfo {year} {2020})},\ \Eprint {http://arxiv.org/abs/2003.05342} {arXiv:2003.05342 [gr-qc]} \BibitemShut {NoStop}%
\bibitem [{\citenamefont {Christodoulidis}\ and\ \citenamefont {Paliathanasis}(2021)}]{Christodoulidis:2021vye}%
  \BibitemOpen
  \bibfield  {author} {\bibinfo {author} {\bibfnamefont {P.}~\bibnamefont {Christodoulidis}}\ and\ \bibinfo {author} {\bibfnamefont {A.}~\bibnamefont {Paliathanasis}},\ }\href {\doibase 10.1088/1475-7516/2021/05/038} {\bibfield  {journal} {\bibinfo  {journal} {JCAP}\ }\textbf {\bibinfo {volume} {05}},\ \bibinfo {pages} {038} (\bibinfo {year} {2021})},\ \Eprint {http://arxiv.org/abs/2101.09582} {arXiv:2101.09582 [gr-qc]} \BibitemShut {NoStop}%
\bibitem [{\citenamefont {Piao}(2006)}]{Piao:2006nm}%
  \BibitemOpen
  \bibfield  {author} {\bibinfo {author} {\bibfnamefont {Y.-S.}\ \bibnamefont {Piao}},\ }\href {\doibase 10.1103/PhysRevD.74.047302} {\bibfield  {journal} {\bibinfo  {journal} {Phys. Rev. D}\ }\textbf {\bibinfo {volume} {74}},\ \bibinfo {pages} {047302} (\bibinfo {year} {2006})},\ \Eprint {http://arxiv.org/abs/gr-qc/0606034} {arXiv:gr-qc/0606034} \BibitemShut {NoStop}%
\bibitem [{\citenamefont {Rinaldi}\ \emph {et~al.}(2023)\citenamefont {Rinaldi}, \citenamefont {Cecchini}, \citenamefont {Ghoshal},\ and\ \citenamefont {Mukherjee}}]{Rinaldi:2023mdf}%
  \BibitemOpen
  \bibfield  {author} {\bibinfo {author} {\bibfnamefont {M.}~\bibnamefont {Rinaldi}}, \bibinfo {author} {\bibfnamefont {C.}~\bibnamefont {Cecchini}}, \bibinfo {author} {\bibfnamefont {A.}~\bibnamefont {Ghoshal}}, \ and\ \bibinfo {author} {\bibfnamefont {D.}~\bibnamefont {Mukherjee}},\ }\href {\doibase 10.1088/1742-6596/2531/1/012012} {\bibfield  {journal} {\bibinfo  {journal} {J. Phys. Conf. Ser.}\ }\textbf {\bibinfo {volume} {2531}},\ \bibinfo {pages} {012012} (\bibinfo {year} {2023})},\ \Eprint {http://arxiv.org/abs/2303.16107} {arXiv:2303.16107 [gr-qc]} \BibitemShut {NoStop}%
\bibitem [{\citenamefont {Ijaz}\ \emph {et~al.}(2023)\citenamefont {Ijaz}, \citenamefont {Mehmood},\ and\ \citenamefont {Rehman}}]{Ijaz:2023cvc}%
  \BibitemOpen
  \bibfield  {author} {\bibinfo {author} {\bibfnamefont {N.}~\bibnamefont {Ijaz}}, \bibinfo {author} {\bibfnamefont {M.}~\bibnamefont {Mehmood}}, \ and\ \bibinfo {author} {\bibfnamefont {M.~U.}\ \bibnamefont {Rehman}},\ }\href@noop {} {\  (\bibinfo {year} {2023})},\ \Eprint {http://arxiv.org/abs/2308.14908} {arXiv:2308.14908 [astro-ph.CO]} \BibitemShut {NoStop}%
\bibitem [{\citenamefont {Akrami}\ \emph {et~al.}(2020{\natexlab{b}})\citenamefont {Akrami} \emph {et~al.}}]{Planck:2019kim}%
  \BibitemOpen
  \bibfield  {author} {\bibinfo {author} {\bibfnamefont {Y.}~\bibnamefont {Akrami}} \emph {et~al.} (\bibinfo {collaboration} {Planck}),\ }\href {\doibase 10.1051/0004-6361/201935891} {\bibfield  {journal} {\bibinfo  {journal} {Astron. Astrophys.}\ }\textbf {\bibinfo {volume} {641}},\ \bibinfo {pages} {A9} (\bibinfo {year} {2020}{\natexlab{b}})},\ \Eprint {http://arxiv.org/abs/1905.05697} {arXiv:1905.05697 [astro-ph.CO]} \BibitemShut {NoStop}%
\bibitem [{\citenamefont {Harrison}(1970)}]{Harrison:1969fb}%
  \BibitemOpen
  \bibfield  {author} {\bibinfo {author} {\bibfnamefont {E.~R.}\ \bibnamefont {Harrison}},\ }\href {\doibase 10.1103/PhysRevD.1.2726} {\bibfield  {journal} {\bibinfo  {journal} {Phys. Rev. D}\ }\textbf {\bibinfo {volume} {1}},\ \bibinfo {pages} {2726} (\bibinfo {year} {1970})}\BibitemShut {NoStop}%
\bibitem [{\citenamefont {Zeldovich}(1972)}]{Zeldovich:1972zz}%
  \BibitemOpen
  \bibfield  {author} {\bibinfo {author} {\bibfnamefont {Y.~B.}\ \bibnamefont {Zeldovich}},\ }\href {\doibase 10.1093/mnras/160.1.1P} {\bibfield  {journal} {\bibinfo  {journal} {Mon. Not. Roy. Astron. Soc.}\ }\textbf {\bibinfo {volume} {160}},\ \bibinfo {pages} {1P} (\bibinfo {year} {1972})}\BibitemShut {NoStop}%
\bibitem [{\citenamefont {Peebles}\ and\ \citenamefont {Yu}(1970)}]{Peebles:1970ag}%
  \BibitemOpen
  \bibfield  {author} {\bibinfo {author} {\bibfnamefont {P.~J.~E.}\ \bibnamefont {Peebles}}\ and\ \bibinfo {author} {\bibfnamefont {J.~T.}\ \bibnamefont {Yu}},\ }\href {\doibase 10.1086/150713} {\bibfield  {journal} {\bibinfo  {journal} {Astrophys. J.}\ }\textbf {\bibinfo {volume} {162}},\ \bibinfo {pages} {815} (\bibinfo {year} {1970})}\BibitemShut {NoStop}%
\bibitem [{\citenamefont {Martin}\ \emph {et~al.}(2014{\natexlab{d}})\citenamefont {Martin}, \citenamefont {Ringeval},\ and\ \citenamefont {Vennin}}]{Martin:2013tda}%
  \BibitemOpen
  \bibfield  {author} {\bibinfo {author} {\bibfnamefont {J.}~\bibnamefont {Martin}}, \bibinfo {author} {\bibfnamefont {C.}~\bibnamefont {Ringeval}}, \ and\ \bibinfo {author} {\bibfnamefont {V.}~\bibnamefont {Vennin}},\ }\href {\doibase 10.1016/j.dark.2014.01.003} {\bibfield  {journal} {\bibinfo  {journal} {Phys. Dark Univ.}\ }\textbf {\bibinfo {volume} {5-6}},\ \bibinfo {pages} {75} (\bibinfo {year} {2014}{\natexlab{d}})},\ \Eprint {http://arxiv.org/abs/1303.3787} {arXiv:1303.3787 [astro-ph.CO]} \BibitemShut {NoStop}%
\bibitem [{\citenamefont {Ade}\ \emph {et~al.}(2021)\citenamefont {Ade} \emph {et~al.}}]{BICEP:2021xfz}%
  \BibitemOpen
  \bibfield  {author} {\bibinfo {author} {\bibfnamefont {P.~A.~R.}\ \bibnamefont {Ade}} \emph {et~al.} (\bibinfo {collaboration} {BICEP, Keck}),\ }\href {\doibase 10.1103/PhysRevLett.127.151301} {\bibfield  {journal} {\bibinfo  {journal} {Phys. Rev. Lett.}\ }\textbf {\bibinfo {volume} {127}},\ \bibinfo {pages} {151301} (\bibinfo {year} {2021})},\ \Eprint {http://arxiv.org/abs/2110.00483} {arXiv:2110.00483 [astro-ph.CO]} \BibitemShut {NoStop}%
\bibitem [{\citenamefont {Grayson}\ \emph {et~al.}(2016)\citenamefont {Grayson} \emph {et~al.}}]{BICEP3}%
  \BibitemOpen
  \bibfield  {author} {\bibinfo {author} {\bibfnamefont {J.~A.}\ \bibnamefont {Grayson}} \emph {et~al.} (\bibinfo {collaboration} {BICEP3}),\ }\href {\doibase 10.1117/12.2233894} {\bibfield  {journal} {\bibinfo  {journal} {Proc. SPIE Int. Soc. Opt. Eng.}\ }\textbf {\bibinfo {volume} {9914}},\ \bibinfo {pages} {99140S} (\bibinfo {year} {2016})},\ \Eprint {http://arxiv.org/abs/1607.04668} {arXiv:1607.04668 [astro-ph.IM]} \BibitemShut {NoStop}%
\bibitem [{\citenamefont {Essinger-Hileman}\ \emph {et~al.}(2014)\citenamefont {Essinger-Hileman} \emph {et~al.}}]{CLASS}%
  \BibitemOpen
  \bibfield  {author} {\bibinfo {author} {\bibfnamefont {T.}~\bibnamefont {Essinger-Hileman}} \emph {et~al.},\ }\href {\doibase 10.1117/12.2056701} {\bibfield  {journal} {\bibinfo  {journal} {Proc. SPIE Int. Soc. Opt. Eng.}\ }\textbf {\bibinfo {volume} {9153}},\ \bibinfo {pages} {91531I} (\bibinfo {year} {2014})},\ \Eprint {http://arxiv.org/abs/1408.4788} {arXiv:1408.4788 [astro-ph.IM]} \BibitemShut {NoStop}%
\bibitem [{\citenamefont {Benson}\ \emph {et~al.}(2014)\citenamefont {Benson} \emph {et~al.}}]{SPT-3G}%
  \BibitemOpen
  \bibfield  {author} {\bibinfo {author} {\bibfnamefont {B.~A.}\ \bibnamefont {Benson}} \emph {et~al.} (\bibinfo {collaboration} {SPT-3G}),\ }\href {\doibase 10.1117/12.2057305} {\bibfield  {journal} {\bibinfo  {journal} {Proc. SPIE Int. Soc. Opt. Eng.}\ }\textbf {\bibinfo {volume} {9153}},\ \bibinfo {pages} {91531P} (\bibinfo {year} {2014})},\ \Eprint {http://arxiv.org/abs/1407.2973} {arXiv:1407.2973 [astro-ph.IM]} \BibitemShut {NoStop}%
\bibitem [{\citenamefont {Henderson}\ \emph {et~al.}(2016)\citenamefont {Henderson} \emph {et~al.}}]{ACTPol}%
  \BibitemOpen
  \bibfield  {author} {\bibinfo {author} {\bibfnamefont {S.~W.}\ \bibnamefont {Henderson}} \emph {et~al.},\ }\href {\doibase 10.1007/s10909-016-1575-z} {\bibfield  {journal} {\bibinfo  {journal} {J. Low. Temp. Phys.}\ }\textbf {\bibinfo {volume} {184}},\ \bibinfo {pages} {772} (\bibinfo {year} {2016})},\ \Eprint {http://arxiv.org/abs/1510.02809} {arXiv:1510.02809 [astro-ph.IM]} \BibitemShut {NoStop}%
\bibitem [{\citenamefont {Suzuki}\ \emph {et~al.}(2018)\citenamefont {Suzuki} \emph {et~al.}}]{LBIRD}%
  \BibitemOpen
  \bibfield  {author} {\bibinfo {author} {\bibfnamefont {A.}~\bibnamefont {Suzuki}} \emph {et~al.} (\bibinfo {collaboration} {LiteBIRD}),\ }\href@noop {} {\bibfield  {journal} {\bibinfo  {journal} {Journal of Low Temperature Physics}\ }\textbf {\bibinfo {volume} {193}},\ \bibinfo {pages} {1048} (\bibinfo {year} {2018})}\BibitemShut {NoStop}%
\bibitem [{\citenamefont {Abazajian}\ \emph {et~al.}(2016)\citenamefont {Abazajian} \emph {et~al.}}]{CMB-S4}%
  \BibitemOpen
  \bibfield  {author} {\bibinfo {author} {\bibfnamefont {K.~N.}\ \bibnamefont {Abazajian}} \emph {et~al.} (\bibinfo {collaboration} {CMB-S4}),\ }\href@noop {} {\  (\bibinfo {year} {2016})},\ \Eprint {http://arxiv.org/abs/1610.02743} {arXiv:1610.02743 [astro-ph.CO]} \BibitemShut {NoStop}%
\bibitem [{\citenamefont {Starobinsky}(1980)}]{Starobinsky:1980te}%
  \BibitemOpen
  \bibfield  {author} {\bibinfo {author} {\bibfnamefont {A.~A.}\ \bibnamefont {Starobinsky}},\ }\href {\doibase 10.1016/0370-2693(80)90670-X} {\bibfield  {journal} {\bibinfo  {journal} {Phys. Lett. B}\ }\textbf {\bibinfo {volume} {91}},\ \bibinfo {pages} {99} (\bibinfo {year} {1980})}\BibitemShut {NoStop}%
\bibitem [{\citenamefont {Kallosh}\ \emph {et~al.}(2013)\citenamefont {Kallosh}, \citenamefont {Linde},\ and\ \citenamefont {Roest}}]{Kallosh:2013yoa}%
  \BibitemOpen
  \bibfield  {author} {\bibinfo {author} {\bibfnamefont {R.}~\bibnamefont {Kallosh}}, \bibinfo {author} {\bibfnamefont {A.}~\bibnamefont {Linde}}, \ and\ \bibinfo {author} {\bibfnamefont {D.}~\bibnamefont {Roest}},\ }\href {\doibase 10.1007/JHEP11(2013)198} {\bibfield  {journal} {\bibinfo  {journal} {JHEP}\ }\textbf {\bibinfo {volume} {11}},\ \bibinfo {pages} {198} (\bibinfo {year} {2013})},\ \Eprint {http://arxiv.org/abs/1311.0472} {arXiv:1311.0472 [hep-th]} \BibitemShut {NoStop}%
\bibitem [{\citenamefont {Kehagias}\ \emph {et~al.}(2014)\citenamefont {Kehagias}, \citenamefont {Moradinezhad~Dizgah},\ and\ \citenamefont {Riotto}}]{Kehagias:2013mya}%
  \BibitemOpen
  \bibfield  {author} {\bibinfo {author} {\bibfnamefont {A.}~\bibnamefont {Kehagias}}, \bibinfo {author} {\bibfnamefont {A.}~\bibnamefont {Moradinezhad~Dizgah}}, \ and\ \bibinfo {author} {\bibfnamefont {A.}~\bibnamefont {Riotto}},\ }\href {\doibase 10.1103/PhysRevD.89.043527} {\bibfield  {journal} {\bibinfo  {journal} {Phys. Rev. D}\ }\textbf {\bibinfo {volume} {89}},\ \bibinfo {pages} {043527} (\bibinfo {year} {2014})},\ \Eprint {http://arxiv.org/abs/1312.1155} {arXiv:1312.1155 [hep-th]} \BibitemShut {NoStop}%
\bibitem [{\citenamefont {Giar\`e}\ \emph {et~al.}(2023{\natexlab{c}})\citenamefont {Giar\`e}, \citenamefont {Pan}, \citenamefont {Di~Valentino}, \citenamefont {Yang}, \citenamefont {de~Haro},\ and\ \citenamefont {Melchiorri}}]{Giare:2023wzl}%
  \BibitemOpen
  \bibfield  {author} {\bibinfo {author} {\bibfnamefont {W.}~\bibnamefont {Giar\`e}}, \bibinfo {author} {\bibfnamefont {S.}~\bibnamefont {Pan}}, \bibinfo {author} {\bibfnamefont {E.}~\bibnamefont {Di~Valentino}}, \bibinfo {author} {\bibfnamefont {W.}~\bibnamefont {Yang}}, \bibinfo {author} {\bibfnamefont {J.}~\bibnamefont {de~Haro}}, \ and\ \bibinfo {author} {\bibfnamefont {A.}~\bibnamefont {Melchiorri}},\ }\href {\doibase 10.1088/1475-7516/2023/09/019} {\bibfield  {journal} {\bibinfo  {journal} {JCAP}\ }\textbf {\bibinfo {volume} {09}},\ \bibinfo {pages} {019} (\bibinfo {year} {2023}{\natexlab{c}})},\ \Eprint {http://arxiv.org/abs/2305.15378} {arXiv:2305.15378 [astro-ph.CO]} \BibitemShut {NoStop}%
\bibitem [{\citenamefont {Perivolaropoulos}\ and\ \citenamefont {Skara}(2022)}]{Perivolaropoulos:2021jda}%
  \BibitemOpen
  \bibfield  {author} {\bibinfo {author} {\bibfnamefont {L.}~\bibnamefont {Perivolaropoulos}}\ and\ \bibinfo {author} {\bibfnamefont {F.}~\bibnamefont {Skara}},\ }\href {\doibase 10.1016/j.newar.2022.101659} {\bibfield  {journal} {\bibinfo  {journal} {New Astron. Rev.}\ }\textbf {\bibinfo {volume} {95}},\ \bibinfo {pages} {101659} (\bibinfo {year} {2022})},\ \Eprint {http://arxiv.org/abs/2105.05208} {arXiv:2105.05208 [astro-ph.CO]} \BibitemShut {NoStop}%
\bibitem [{\citenamefont {Di~Valentino}(2022)}]{DiValentino:2022fjm}%
  \BibitemOpen
  \bibfield  {author} {\bibinfo {author} {\bibfnamefont {E.}~\bibnamefont {Di~Valentino}},\ }\href {\doibase 10.3390/universe8080399} {\bibfield  {journal} {\bibinfo  {journal} {Universe}\ }\textbf {\bibinfo {volume} {8}},\ \bibinfo {pages} {399} (\bibinfo {year} {2022})}\BibitemShut {NoStop}%
\bibitem [{\citenamefont {Abdalla}\ \emph {et~al.}(2022)\citenamefont {Abdalla} \emph {et~al.}}]{Abdalla:2022yfr}%
  \BibitemOpen
  \bibfield  {author} {\bibinfo {author} {\bibfnamefont {E.}~\bibnamefont {Abdalla}} \emph {et~al.},\ }\href {\doibase 10.1016/j.jheap.2022.04.002} {\bibfield  {journal} {\bibinfo  {journal} {JHEAp}\ }\textbf {\bibinfo {volume} {34}},\ \bibinfo {pages} {49} (\bibinfo {year} {2022})},\ \Eprint {http://arxiv.org/abs/2203.06142} {arXiv:2203.06142 [astro-ph.CO]} \BibitemShut {NoStop}%
\bibitem [{\citenamefont {Lin}\ and\ \citenamefont {Ishak}(2021)}]{Lin:2019zdn}%
  \BibitemOpen
  \bibfield  {author} {\bibinfo {author} {\bibfnamefont {W.}~\bibnamefont {Lin}}\ and\ \bibinfo {author} {\bibfnamefont {M.}~\bibnamefont {Ishak}},\ }\href {\doibase 10.1088/1475-7516/2021/05/009} {\bibfield  {journal} {\bibinfo  {journal} {JCAP}\ }\textbf {\bibinfo {volume} {05}},\ \bibinfo {pages} {009} (\bibinfo {year} {2021})},\ \Eprint {http://arxiv.org/abs/1909.10991} {arXiv:1909.10991 [astro-ph.CO]} \BibitemShut {NoStop}%
\bibitem [{\citenamefont {Handley}\ and\ \citenamefont {Lemos}(2021)}]{Handley:2020hdp}%
  \BibitemOpen
  \bibfield  {author} {\bibinfo {author} {\bibfnamefont {W.}~\bibnamefont {Handley}}\ and\ \bibinfo {author} {\bibfnamefont {P.}~\bibnamefont {Lemos}},\ }\href {\doibase 10.1103/PhysRevD.103.063529} {\bibfield  {journal} {\bibinfo  {journal} {Phys. Rev. D}\ }\textbf {\bibinfo {volume} {103}},\ \bibinfo {pages} {063529} (\bibinfo {year} {2021})},\ \Eprint {http://arxiv.org/abs/2007.08496} {arXiv:2007.08496 [astro-ph.CO]} \BibitemShut {NoStop}%
\bibitem [{\citenamefont {La~Posta}\ \emph {et~al.}(2023)\citenamefont {La~Posta}, \citenamefont {Natale}, \citenamefont {Calabrese}, \citenamefont {Garrido},\ and\ \citenamefont {Louis}}]{LaPosta:2022llv}%
  \BibitemOpen
  \bibfield  {author} {\bibinfo {author} {\bibfnamefont {A.}~\bibnamefont {La~Posta}}, \bibinfo {author} {\bibfnamefont {U.}~\bibnamefont {Natale}}, \bibinfo {author} {\bibfnamefont {E.}~\bibnamefont {Calabrese}}, \bibinfo {author} {\bibfnamefont {X.}~\bibnamefont {Garrido}}, \ and\ \bibinfo {author} {\bibfnamefont {T.}~\bibnamefont {Louis}},\ }\href {\doibase 10.1103/PhysRevD.107.023510} {\bibfield  {journal} {\bibinfo  {journal} {Phys. Rev. D}\ }\textbf {\bibinfo {volume} {107}},\ \bibinfo {pages} {023510} (\bibinfo {year} {2023})},\ \Eprint {http://arxiv.org/abs/2204.01885} {arXiv:2204.01885 [astro-ph.CO]} \BibitemShut {NoStop}%
\bibitem [{\citenamefont {Di~Valentino}\ \emph {et~al.}(2023)\citenamefont {Di~Valentino}, \citenamefont {Giar\`e}, \citenamefont {Melchiorri},\ and\ \citenamefont {Silk}}]{DiValentino:2022rdg}%
  \BibitemOpen
  \bibfield  {author} {\bibinfo {author} {\bibfnamefont {E.}~\bibnamefont {Di~Valentino}}, \bibinfo {author} {\bibfnamefont {W.}~\bibnamefont {Giar\`e}}, \bibinfo {author} {\bibfnamefont {A.}~\bibnamefont {Melchiorri}}, \ and\ \bibinfo {author} {\bibfnamefont {J.}~\bibnamefont {Silk}},\ }\href {\doibase 10.1093/mnras/stad152} {\bibfield  {journal} {\bibinfo  {journal} {Mon. Not. Roy. Astron. Soc.}\ }\textbf {\bibinfo {volume} {520}},\ \bibinfo {pages} {210} (\bibinfo {year} {2023})},\ \Eprint {http://arxiv.org/abs/2209.14054} {arXiv:2209.14054 [astro-ph.CO]} \BibitemShut {NoStop}%
\bibitem [{\citenamefont {Di~Valentino}\ \emph {et~al.}(2022)\citenamefont {Di~Valentino}, \citenamefont {Giar\`e}, \citenamefont {Melchiorri},\ and\ \citenamefont {Silk}}]{DiValentino:2022oon}%
  \BibitemOpen
  \bibfield  {author} {\bibinfo {author} {\bibfnamefont {E.}~\bibnamefont {Di~Valentino}}, \bibinfo {author} {\bibfnamefont {W.}~\bibnamefont {Giar\`e}}, \bibinfo {author} {\bibfnamefont {A.}~\bibnamefont {Melchiorri}}, \ and\ \bibinfo {author} {\bibfnamefont {J.}~\bibnamefont {Silk}},\ }\href {\doibase 10.1103/PhysRevD.106.103506} {\bibfield  {journal} {\bibinfo  {journal} {Phys. Rev. D}\ }\textbf {\bibinfo {volume} {106}},\ \bibinfo {pages} {103506} (\bibinfo {year} {2022})},\ \Eprint {http://arxiv.org/abs/2209.12872} {arXiv:2209.12872 [astro-ph.CO]} \BibitemShut {NoStop}%
\bibitem [{\citenamefont {Giar\`e}\ \emph {et~al.}(2023{\natexlab{d}})\citenamefont {Giar\`e}, \citenamefont {Renzi}, \citenamefont {Mena}, \citenamefont {Di~Valentino},\ and\ \citenamefont {Melchiorri}}]{Giare:2022rvg}%
  \BibitemOpen
  \bibfield  {author} {\bibinfo {author} {\bibfnamefont {W.}~\bibnamefont {Giar\`e}}, \bibinfo {author} {\bibfnamefont {F.}~\bibnamefont {Renzi}}, \bibinfo {author} {\bibfnamefont {O.}~\bibnamefont {Mena}}, \bibinfo {author} {\bibfnamefont {E.}~\bibnamefont {Di~Valentino}}, \ and\ \bibinfo {author} {\bibfnamefont {A.}~\bibnamefont {Melchiorri}},\ }\href {\doibase 10.1093/mnras/stad724} {\bibfield  {journal} {\bibinfo  {journal} {Mon. Not. Roy. Astron. Soc.}\ }\textbf {\bibinfo {volume} {521}},\ \bibinfo {pages} {2911} (\bibinfo {year} {2023}{\natexlab{d}})},\ \Eprint {http://arxiv.org/abs/2210.09018} {arXiv:2210.09018 [astro-ph.CO]} \BibitemShut {NoStop}%
\bibitem [{\citenamefont {Calder\'on}\ \emph {et~al.}(2023)\citenamefont {Calder\'on}, \citenamefont {Shafieloo}, \citenamefont {Hazra},\ and\ \citenamefont {Sohn}}]{Calderon:2023obf}%
  \BibitemOpen
  \bibfield  {author} {\bibinfo {author} {\bibfnamefont {R.}~\bibnamefont {Calder\'on}}, \bibinfo {author} {\bibfnamefont {A.}~\bibnamefont {Shafieloo}}, \bibinfo {author} {\bibfnamefont {D.~K.}\ \bibnamefont {Hazra}}, \ and\ \bibinfo {author} {\bibfnamefont {W.}~\bibnamefont {Sohn}},\ }\href {\doibase 10.1088/1475-7516/2023/08/059} {\bibfield  {journal} {\bibinfo  {journal} {JCAP}\ }\textbf {\bibinfo {volume} {08}},\ \bibinfo {pages} {059} (\bibinfo {year} {2023})},\ \Eprint {http://arxiv.org/abs/2302.14300} {arXiv:2302.14300 [astro-ph.CO]} \BibitemShut {NoStop}%
\bibitem [{\citenamefont {Giar\`e}(2023)}]{Giare:2023xoc}%
  \BibitemOpen
  \bibfield  {author} {\bibinfo {author} {\bibfnamefont {W.}~\bibnamefont {Giar\`e}},\ }\href@noop {} {\  (\bibinfo {year} {2023})},\ \Eprint {http://arxiv.org/abs/2305.16919} {arXiv:2305.16919 [astro-ph.CO]} \BibitemShut {NoStop}%
\bibitem [{\citenamefont {Gariazzo}\ \emph {et~al.}(2024)\citenamefont {Gariazzo}, \citenamefont {Giar\`e}, \citenamefont {Mena},\ and\ \citenamefont {Di~Valentino}}]{Gariazzo:2024sil}%
  \BibitemOpen
  \bibfield  {author} {\bibinfo {author} {\bibfnamefont {S.}~\bibnamefont {Gariazzo}}, \bibinfo {author} {\bibfnamefont {W.}~\bibnamefont {Giar\`e}}, \bibinfo {author} {\bibfnamefont {O.}~\bibnamefont {Mena}}, \ and\ \bibinfo {author} {\bibfnamefont {E.}~\bibnamefont {Di~Valentino}},\ }\href@noop {} {\  (\bibinfo {year} {2024})},\ \Eprint {http://arxiv.org/abs/2404.11182} {arXiv:2404.11182 [astro-ph.CO]} \BibitemShut {NoStop}%
\bibitem [{\citenamefont {Di~Valentino}\ \emph {et~al.}(2018)\citenamefont {Di~Valentino}, \citenamefont {Melchiorri}, \citenamefont {Fantaye},\ and\ \citenamefont {Heavens}}]{DiValentino:2018zjj}%
  \BibitemOpen
  \bibfield  {author} {\bibinfo {author} {\bibfnamefont {E.}~\bibnamefont {Di~Valentino}}, \bibinfo {author} {\bibfnamefont {A.}~\bibnamefont {Melchiorri}}, \bibinfo {author} {\bibfnamefont {Y.}~\bibnamefont {Fantaye}}, \ and\ \bibinfo {author} {\bibfnamefont {A.}~\bibnamefont {Heavens}},\ }\href {\doibase 10.1103/PhysRevD.98.063508} {\bibfield  {journal} {\bibinfo  {journal} {Phys. Rev. D}\ }\textbf {\bibinfo {volume} {98}},\ \bibinfo {pages} {063508} (\bibinfo {year} {2018})},\ \Eprint {http://arxiv.org/abs/1808.09201} {arXiv:1808.09201 [astro-ph.CO]} \BibitemShut {NoStop}%
\bibitem [{\citenamefont {Ye}\ \emph {et~al.}(2022)\citenamefont {Ye}, \citenamefont {Jiang},\ and\ \citenamefont {Piao}}]{Ye:2022efx}%
  \BibitemOpen
  \bibfield  {author} {\bibinfo {author} {\bibfnamefont {G.}~\bibnamefont {Ye}}, \bibinfo {author} {\bibfnamefont {J.-Q.}\ \bibnamefont {Jiang}}, \ and\ \bibinfo {author} {\bibfnamefont {Y.-S.}\ \bibnamefont {Piao}},\ }\href {\doibase 10.1103/PhysRevD.106.103528} {\bibfield  {journal} {\bibinfo  {journal} {Phys. Rev. D}\ }\textbf {\bibinfo {volume} {106}},\ \bibinfo {pages} {103528} (\bibinfo {year} {2022})},\ \Eprint {http://arxiv.org/abs/2205.02478} {arXiv:2205.02478 [astro-ph.CO]} \BibitemShut {NoStop}%
\bibitem [{\citenamefont {Jiang}\ and\ \citenamefont {Piao}(2022)}]{Jiang:2022uyg}%
  \BibitemOpen
  \bibfield  {author} {\bibinfo {author} {\bibfnamefont {J.-Q.}\ \bibnamefont {Jiang}}\ and\ \bibinfo {author} {\bibfnamefont {Y.-S.}\ \bibnamefont {Piao}},\ }\href {\doibase 10.1103/PhysRevD.105.103514} {\bibfield  {journal} {\bibinfo  {journal} {Phys. Rev. D}\ }\textbf {\bibinfo {volume} {105}},\ \bibinfo {pages} {103514} (\bibinfo {year} {2022})},\ \Eprint {http://arxiv.org/abs/2202.13379} {arXiv:2202.13379 [astro-ph.CO]} \BibitemShut {NoStop}%
\bibitem [{\citenamefont {Jiang}\ \emph {et~al.}(2022)\citenamefont {Jiang}, \citenamefont {Ye},\ and\ \citenamefont {Piao}}]{Jiang:2022qlj}%
  \BibitemOpen
  \bibfield  {author} {\bibinfo {author} {\bibfnamefont {J.-Q.}\ \bibnamefont {Jiang}}, \bibinfo {author} {\bibfnamefont {G.}~\bibnamefont {Ye}}, \ and\ \bibinfo {author} {\bibfnamefont {Y.-S.}\ \bibnamefont {Piao}},\ }\href {\doibase 10.1093/mnrasl/slad137} {\  (\bibinfo {year} {2022}),\ 10.1093/mnrasl/slad137},\ \Eprint {http://arxiv.org/abs/2210.06125} {arXiv:2210.06125 [astro-ph.CO]} \BibitemShut {NoStop}%
\bibitem [{\citenamefont {Takahashi}\ and\ \citenamefont {Yin}(2022)}]{Takahashi:2021bti}%
  \BibitemOpen
  \bibfield  {author} {\bibinfo {author} {\bibfnamefont {F.}~\bibnamefont {Takahashi}}\ and\ \bibinfo {author} {\bibfnamefont {W.}~\bibnamefont {Yin}},\ }\href {\doibase 10.1016/j.physletb.2022.137143} {\bibfield  {journal} {\bibinfo  {journal} {Phys. Lett. B}\ }\textbf {\bibinfo {volume} {830}},\ \bibinfo {pages} {137143} (\bibinfo {year} {2022})},\ \Eprint {http://arxiv.org/abs/2112.06710} {arXiv:2112.06710 [astro-ph.CO]} \BibitemShut {NoStop}%
\bibitem [{\citenamefont {Lin}(2022)}]{Lin:2022gbl}%
  \BibitemOpen
  \bibfield  {author} {\bibinfo {author} {\bibfnamefont {C.-M.}\ \bibnamefont {Lin}},\ }\href {\doibase 10.1103/PhysRevD.106.103511} {\bibfield  {journal} {\bibinfo  {journal} {Phys. Rev. D}\ }\textbf {\bibinfo {volume} {106}},\ \bibinfo {pages} {103511} (\bibinfo {year} {2022})},\ \Eprint {http://arxiv.org/abs/2204.10475} {arXiv:2204.10475 [hep-th]} \BibitemShut {NoStop}%
\bibitem [{\citenamefont {Hazra}\ \emph {et~al.}(2022)\citenamefont {Hazra}, \citenamefont {Antony},\ and\ \citenamefont {Shafieloo}}]{Hazra:2022rdl}%
  \BibitemOpen
  \bibfield  {author} {\bibinfo {author} {\bibfnamefont {D.~K.}\ \bibnamefont {Hazra}}, \bibinfo {author} {\bibfnamefont {A.}~\bibnamefont {Antony}}, \ and\ \bibinfo {author} {\bibfnamefont {A.}~\bibnamefont {Shafieloo}},\ }\href {\doibase 10.1088/1475-7516/2022/08/063} {\bibfield  {journal} {\bibinfo  {journal} {JCAP}\ }\textbf {\bibinfo {volume} {08}},\ \bibinfo {pages} {063} (\bibinfo {year} {2022})},\ \Eprint {http://arxiv.org/abs/2201.12000} {arXiv:2201.12000 [astro-ph.CO]} \BibitemShut {NoStop}%
\bibitem [{\citenamefont {Braglia}\ \emph {et~al.}(2022)\citenamefont {Braglia}, \citenamefont {Chen},\ and\ \citenamefont {Hazra}}]{Braglia:2021sun}%
  \BibitemOpen
  \bibfield  {author} {\bibinfo {author} {\bibfnamefont {M.}~\bibnamefont {Braglia}}, \bibinfo {author} {\bibfnamefont {X.}~\bibnamefont {Chen}}, \ and\ \bibinfo {author} {\bibfnamefont {D.~K.}\ \bibnamefont {Hazra}},\ }\href {\doibase 10.1140/epjc/s10052-022-10461-3} {\bibfield  {journal} {\bibinfo  {journal} {Eur. Phys. J. C}\ }\textbf {\bibinfo {volume} {82}},\ \bibinfo {pages} {498} (\bibinfo {year} {2022})},\ \Eprint {http://arxiv.org/abs/2106.07546} {arXiv:2106.07546 [astro-ph.CO]} \BibitemShut {NoStop}%
\bibitem [{\citenamefont {Keeley}\ \emph {et~al.}(2020)\citenamefont {Keeley}, \citenamefont {Shafieloo}, \citenamefont {Hazra},\ and\ \citenamefont {Souradeep}}]{Keeley:2020rmo}%
  \BibitemOpen
  \bibfield  {author} {\bibinfo {author} {\bibfnamefont {R.~E.}\ \bibnamefont {Keeley}}, \bibinfo {author} {\bibfnamefont {A.}~\bibnamefont {Shafieloo}}, \bibinfo {author} {\bibfnamefont {D.~K.}\ \bibnamefont {Hazra}}, \ and\ \bibinfo {author} {\bibfnamefont {T.}~\bibnamefont {Souradeep}},\ }\href {\doibase 10.1088/1475-7516/2020/09/055} {\bibfield  {journal} {\bibinfo  {journal} {JCAP}\ }\textbf {\bibinfo {volume} {09}},\ \bibinfo {pages} {055} (\bibinfo {year} {2020})},\ \Eprint {http://arxiv.org/abs/2006.12710} {arXiv:2006.12710 [astro-ph.CO]} \BibitemShut {NoStop}%
\bibitem [{\citenamefont {Jiang}\ \emph {et~al.}(2023)\citenamefont {Jiang}, \citenamefont {Ye},\ and\ \citenamefont {Piao}}]{Jiang:2023bsz}%
  \BibitemOpen
  \bibfield  {author} {\bibinfo {author} {\bibfnamefont {J.-Q.}\ \bibnamefont {Jiang}}, \bibinfo {author} {\bibfnamefont {G.}~\bibnamefont {Ye}}, \ and\ \bibinfo {author} {\bibfnamefont {Y.-S.}\ \bibnamefont {Piao}},\ }\href@noop {} {\  (\bibinfo {year} {2023})},\ \Eprint {http://arxiv.org/abs/2303.12345} {arXiv:2303.12345 [astro-ph.CO]} \BibitemShut {NoStop}%
\bibitem [{\citenamefont {Verde}\ \emph {et~al.}(2019)\citenamefont {Verde}, \citenamefont {Treu},\ and\ \citenamefont {Riess}}]{Verde:2019ivm}%
  \BibitemOpen
  \bibfield  {author} {\bibinfo {author} {\bibfnamefont {L.}~\bibnamefont {Verde}}, \bibinfo {author} {\bibfnamefont {T.}~\bibnamefont {Treu}}, \ and\ \bibinfo {author} {\bibfnamefont {A.~G.}\ \bibnamefont {Riess}},\ }\href {\doibase 10.1038/s41550-019-0902-0} {\bibfield  {journal} {\bibinfo  {journal} {Nature Astron.}\ }\textbf {\bibinfo {volume} {3}},\ \bibinfo {pages} {891} (\bibinfo {year} {2019})},\ \Eprint {http://arxiv.org/abs/1907.10625} {arXiv:1907.10625 [astro-ph.CO]} \BibitemShut {NoStop}%
\bibitem [{\citenamefont {Di~Valentino}\ \emph {et~al.}(2021{\natexlab{a}})\citenamefont {Di~Valentino}, \citenamefont {Mena}, \citenamefont {Pan}, \citenamefont {Visinelli}, \citenamefont {Yang}, \citenamefont {Melchiorri}, \citenamefont {Mota}, \citenamefont {Riess},\ and\ \citenamefont {Silk}}]{DiValentino:2021izs}%
  \BibitemOpen
  \bibfield  {author} {\bibinfo {author} {\bibfnamefont {E.}~\bibnamefont {Di~Valentino}}, \bibinfo {author} {\bibfnamefont {O.}~\bibnamefont {Mena}}, \bibinfo {author} {\bibfnamefont {S.}~\bibnamefont {Pan}}, \bibinfo {author} {\bibfnamefont {L.}~\bibnamefont {Visinelli}}, \bibinfo {author} {\bibfnamefont {W.}~\bibnamefont {Yang}}, \bibinfo {author} {\bibfnamefont {A.}~\bibnamefont {Melchiorri}}, \bibinfo {author} {\bibfnamefont {D.~F.}\ \bibnamefont {Mota}}, \bibinfo {author} {\bibfnamefont {A.~G.}\ \bibnamefont {Riess}}, \ and\ \bibinfo {author} {\bibfnamefont {J.}~\bibnamefont {Silk}},\ }\href {\doibase 10.1088/1361-6382/ac086d} {\bibfield  {journal} {\bibinfo  {journal} {Class. Quant. Grav.}\ }\textbf {\bibinfo {volume} {38}},\ \bibinfo {pages} {153001} (\bibinfo {year} {2021}{\natexlab{a}})},\ \Eprint {http://arxiv.org/abs/2103.01183} {arXiv:2103.01183 [astro-ph.CO]} \BibitemShut {NoStop}%
\bibitem [{\citenamefont {Kamionkowski}\ and\ \citenamefont {Riess}(2023)}]{Kamionkowski:2022pkx}%
  \BibitemOpen
  \bibfield  {author} {\bibinfo {author} {\bibfnamefont {M.}~\bibnamefont {Kamionkowski}}\ and\ \bibinfo {author} {\bibfnamefont {A.~G.}\ \bibnamefont {Riess}},\ }\href@noop {} {\bibfield  {journal} {\bibinfo  {journal} {Ann. Rev. Nucl. Part. Sci.}\ }\textbf {\bibinfo {volume} {73}},\ \bibinfo {pages} {153} (\bibinfo {year} {2023})},\ \Eprint {http://arxiv.org/abs/2211.04492} {arXiv:2211.04492 [astro-ph.CO]} \BibitemShut {NoStop}%
\bibitem [{\citenamefont {Khalife}\ \emph {et~al.}(2023)\citenamefont {Khalife}, \citenamefont {Zanjani}, \citenamefont {Galli}, \citenamefont {G\"unther}, \citenamefont {Lesgourgues},\ and\ \citenamefont {Benabed}}]{Khalife:2023qbu}%
  \BibitemOpen
  \bibfield  {author} {\bibinfo {author} {\bibfnamefont {A.~R.}\ \bibnamefont {Khalife}}, \bibinfo {author} {\bibfnamefont {M.~B.}\ \bibnamefont {Zanjani}}, \bibinfo {author} {\bibfnamefont {S.}~\bibnamefont {Galli}}, \bibinfo {author} {\bibfnamefont {S.}~\bibnamefont {G\"unther}}, \bibinfo {author} {\bibfnamefont {J.}~\bibnamefont {Lesgourgues}}, \ and\ \bibinfo {author} {\bibfnamefont {K.}~\bibnamefont {Benabed}},\ }\href@noop {} {\  (\bibinfo {year} {2023})},\ \Eprint {http://arxiv.org/abs/2312.09814} {arXiv:2312.09814 [astro-ph.CO]} \BibitemShut {NoStop}%
\bibitem [{\citenamefont {Riess}\ \emph {et~al.}(2022)\citenamefont {Riess} \emph {et~al.}}]{Riess:2021jrx}%
  \BibitemOpen
  \bibfield  {author} {\bibinfo {author} {\bibfnamefont {A.~G.}\ \bibnamefont {Riess}} \emph {et~al.},\ }\href {\doibase 10.3847/2041-8213/ac5c5b} {\bibfield  {journal} {\bibinfo  {journal} {Astrophys. J. Lett.}\ }\textbf {\bibinfo {volume} {934}},\ \bibinfo {pages} {L7} (\bibinfo {year} {2022})},\ \Eprint {http://arxiv.org/abs/2112.04510} {arXiv:2112.04510 [astro-ph.CO]} \BibitemShut {NoStop}%
\bibitem [{\citenamefont {Wetterich}(2004)}]{Wetterich:2004pv}%
  \BibitemOpen
  \bibfield  {author} {\bibinfo {author} {\bibfnamefont {C.}~\bibnamefont {Wetterich}},\ }\href {\doibase 10.1016/j.physletb.2004.05.008} {\bibfield  {journal} {\bibinfo  {journal} {Phys. Lett. B}\ }\textbf {\bibinfo {volume} {594}},\ \bibinfo {pages} {17} (\bibinfo {year} {2004})},\ \Eprint {http://arxiv.org/abs/astro-ph/0403289} {arXiv:astro-ph/0403289} \BibitemShut {NoStop}%
\bibitem [{\citenamefont {Doran}\ and\ \citenamefont {Robbers}(2006)}]{Doran:2006kp}%
  \BibitemOpen
  \bibfield  {author} {\bibinfo {author} {\bibfnamefont {M.}~\bibnamefont {Doran}}\ and\ \bibinfo {author} {\bibfnamefont {G.}~\bibnamefont {Robbers}},\ }\href {\doibase 10.1088/1475-7516/2006/06/026} {\bibfield  {journal} {\bibinfo  {journal} {JCAP}\ }\textbf {\bibinfo {volume} {06}},\ \bibinfo {pages} {026} (\bibinfo {year} {2006})},\ \Eprint {http://arxiv.org/abs/astro-ph/0601544} {arXiv:astro-ph/0601544} \BibitemShut {NoStop}%
\bibitem [{\citenamefont {Hollenstein}\ \emph {et~al.}(2009)\citenamefont {Hollenstein}, \citenamefont {Sapone}, \citenamefont {Crittenden},\ and\ \citenamefont {Schaefer}}]{Hollenstein:2009ph}%
  \BibitemOpen
  \bibfield  {author} {\bibinfo {author} {\bibfnamefont {L.}~\bibnamefont {Hollenstein}}, \bibinfo {author} {\bibfnamefont {D.}~\bibnamefont {Sapone}}, \bibinfo {author} {\bibfnamefont {R.}~\bibnamefont {Crittenden}}, \ and\ \bibinfo {author} {\bibfnamefont {B.~M.}\ \bibnamefont {Schaefer}},\ }\href {\doibase 10.1088/1475-7516/2009/04/012} {\bibfield  {journal} {\bibinfo  {journal} {JCAP}\ }\textbf {\bibinfo {volume} {04}},\ \bibinfo {pages} {012} (\bibinfo {year} {2009})},\ \Eprint {http://arxiv.org/abs/0902.1494} {arXiv:0902.1494 [astro-ph.CO]} \BibitemShut {NoStop}%
\bibitem [{\citenamefont {Calabrese}\ \emph {et~al.}(2011{\natexlab{a}})\citenamefont {Calabrese}, \citenamefont {de~Putter}, \citenamefont {Huterer}, \citenamefont {Linder},\ and\ \citenamefont {Melchiorri}}]{Calabrese:2010uf}%
  \BibitemOpen
  \bibfield  {author} {\bibinfo {author} {\bibfnamefont {E.}~\bibnamefont {Calabrese}}, \bibinfo {author} {\bibfnamefont {R.}~\bibnamefont {de~Putter}}, \bibinfo {author} {\bibfnamefont {D.}~\bibnamefont {Huterer}}, \bibinfo {author} {\bibfnamefont {E.~V.}\ \bibnamefont {Linder}}, \ and\ \bibinfo {author} {\bibfnamefont {A.}~\bibnamefont {Melchiorri}},\ }\href {\doibase 10.1103/PhysRevD.83.023011} {\bibfield  {journal} {\bibinfo  {journal} {Phys. Rev. D}\ }\textbf {\bibinfo {volume} {83}},\ \bibinfo {pages} {023011} (\bibinfo {year} {2011}{\natexlab{a}})},\ \Eprint {http://arxiv.org/abs/1010.5612} {arXiv:1010.5612 [astro-ph.CO]} \BibitemShut {NoStop}%
\bibitem [{\citenamefont {Calabrese}\ \emph {et~al.}(2011{\natexlab{b}})\citenamefont {Calabrese}, \citenamefont {Huterer}, \citenamefont {Linder}, \citenamefont {Melchiorri},\ and\ \citenamefont {Pagano}}]{Calabrese:2011hg}%
  \BibitemOpen
  \bibfield  {author} {\bibinfo {author} {\bibfnamefont {E.}~\bibnamefont {Calabrese}}, \bibinfo {author} {\bibfnamefont {D.}~\bibnamefont {Huterer}}, \bibinfo {author} {\bibfnamefont {E.~V.}\ \bibnamefont {Linder}}, \bibinfo {author} {\bibfnamefont {A.}~\bibnamefont {Melchiorri}}, \ and\ \bibinfo {author} {\bibfnamefont {L.}~\bibnamefont {Pagano}},\ }\href {\doibase 10.1103/PhysRevD.83.123504} {\bibfield  {journal} {\bibinfo  {journal} {Phys. Rev. D}\ }\textbf {\bibinfo {volume} {83}},\ \bibinfo {pages} {123504} (\bibinfo {year} {2011}{\natexlab{b}})},\ \Eprint {http://arxiv.org/abs/1103.4132} {arXiv:1103.4132 [astro-ph.CO]} \BibitemShut {NoStop}%
\bibitem [{\citenamefont {Pettorino}\ \emph {et~al.}(2013)\citenamefont {Pettorino}, \citenamefont {Amendola},\ and\ \citenamefont {Wetterich}}]{Pettorino:2013ia}%
  \BibitemOpen
  \bibfield  {author} {\bibinfo {author} {\bibfnamefont {V.}~\bibnamefont {Pettorino}}, \bibinfo {author} {\bibfnamefont {L.}~\bibnamefont {Amendola}}, \ and\ \bibinfo {author} {\bibfnamefont {C.}~\bibnamefont {Wetterich}},\ }\href {\doibase 10.1103/PhysRevD.87.083009} {\bibfield  {journal} {\bibinfo  {journal} {Phys. Rev. D}\ }\textbf {\bibinfo {volume} {87}},\ \bibinfo {pages} {083009} (\bibinfo {year} {2013})},\ \Eprint {http://arxiv.org/abs/1301.5279} {arXiv:1301.5279 [astro-ph.CO]} \BibitemShut {NoStop}%
\bibitem [{\citenamefont {Archidiacono}\ \emph {et~al.}(2014)\citenamefont {Archidiacono}, \citenamefont {Lopez-Honorez},\ and\ \citenamefont {Mena}}]{Archidiacono:2014msa}%
  \BibitemOpen
  \bibfield  {author} {\bibinfo {author} {\bibfnamefont {M.}~\bibnamefont {Archidiacono}}, \bibinfo {author} {\bibfnamefont {L.}~\bibnamefont {Lopez-Honorez}}, \ and\ \bibinfo {author} {\bibfnamefont {O.}~\bibnamefont {Mena}},\ }\href {\doibase 10.1103/PhysRevD.90.123016} {\bibfield  {journal} {\bibinfo  {journal} {Phys. Rev. D}\ }\textbf {\bibinfo {volume} {90}},\ \bibinfo {pages} {123016} (\bibinfo {year} {2014})},\ \Eprint {http://arxiv.org/abs/1409.1802} {arXiv:1409.1802 [astro-ph.CO]} \BibitemShut {NoStop}%
\bibitem [{\citenamefont {Poulin}\ \emph {et~al.}(2023)\citenamefont {Poulin}, \citenamefont {Smith},\ and\ \citenamefont {Karwal}}]{Poulin:2023lkg}%
  \BibitemOpen
  \bibfield  {author} {\bibinfo {author} {\bibfnamefont {V.}~\bibnamefont {Poulin}}, \bibinfo {author} {\bibfnamefont {T.~L.}\ \bibnamefont {Smith}}, \ and\ \bibinfo {author} {\bibfnamefont {T.}~\bibnamefont {Karwal}},\ }\href {\doibase 10.1016/j.dark.2023.101348} {\bibfield  {journal} {\bibinfo  {journal} {Phys. Dark Univ.}\ }\textbf {\bibinfo {volume} {42}},\ \bibinfo {pages} {101348} (\bibinfo {year} {2023})},\ \Eprint {http://arxiv.org/abs/2302.09032} {arXiv:2302.09032 [astro-ph.CO]} \BibitemShut {NoStop}%
\bibitem [{\citenamefont {Poulin}\ \emph {et~al.}(2018{\natexlab{a}})\citenamefont {Poulin}, \citenamefont {Smith}, \citenamefont {Grin}, \citenamefont {Karwal},\ and\ \citenamefont {Kamionkowski}}]{Poulin:2018dzj}%
  \BibitemOpen
  \bibfield  {author} {\bibinfo {author} {\bibfnamefont {V.}~\bibnamefont {Poulin}}, \bibinfo {author} {\bibfnamefont {T.~L.}\ \bibnamefont {Smith}}, \bibinfo {author} {\bibfnamefont {D.}~\bibnamefont {Grin}}, \bibinfo {author} {\bibfnamefont {T.}~\bibnamefont {Karwal}}, \ and\ \bibinfo {author} {\bibfnamefont {M.}~\bibnamefont {Kamionkowski}},\ }\href {\doibase 10.1103/PhysRevD.98.083525} {\bibfield  {journal} {\bibinfo  {journal} {Phys. Rev. D}\ }\textbf {\bibinfo {volume} {98}},\ \bibinfo {pages} {083525} (\bibinfo {year} {2018}{\natexlab{a}})},\ \Eprint {http://arxiv.org/abs/1806.10608} {arXiv:1806.10608 [astro-ph.CO]} \BibitemShut {NoStop}%
\bibitem [{\citenamefont {Poulin}\ \emph {et~al.}(2018{\natexlab{b}})\citenamefont {Poulin}, \citenamefont {Boddy}, \citenamefont {Bird},\ and\ \citenamefont {Kamionkowski}}]{Poulin:2018zxs}%
  \BibitemOpen
  \bibfield  {author} {\bibinfo {author} {\bibfnamefont {V.}~\bibnamefont {Poulin}}, \bibinfo {author} {\bibfnamefont {K.~K.}\ \bibnamefont {Boddy}}, \bibinfo {author} {\bibfnamefont {S.}~\bibnamefont {Bird}}, \ and\ \bibinfo {author} {\bibfnamefont {M.}~\bibnamefont {Kamionkowski}},\ }\href {\doibase 10.1103/PhysRevD.97.123504} {\bibfield  {journal} {\bibinfo  {journal} {Phys. Rev. D}\ }\textbf {\bibinfo {volume} {97}},\ \bibinfo {pages} {123504} (\bibinfo {year} {2018}{\natexlab{b}})},\ \Eprint {http://arxiv.org/abs/1803.02474} {arXiv:1803.02474 [astro-ph.CO]} \BibitemShut {NoStop}%
\bibitem [{\citenamefont {Smith}\ \emph {et~al.}(2020)\citenamefont {Smith}, \citenamefont {Poulin},\ and\ \citenamefont {Amin}}]{Smith:2019ihp}%
  \BibitemOpen
  \bibfield  {author} {\bibinfo {author} {\bibfnamefont {T.~L.}\ \bibnamefont {Smith}}, \bibinfo {author} {\bibfnamefont {V.}~\bibnamefont {Poulin}}, \ and\ \bibinfo {author} {\bibfnamefont {M.~A.}\ \bibnamefont {Amin}},\ }\href {\doibase 10.1103/PhysRevD.101.063523} {\bibfield  {journal} {\bibinfo  {journal} {Phys. Rev. D}\ }\textbf {\bibinfo {volume} {101}},\ \bibinfo {pages} {063523} (\bibinfo {year} {2020})},\ \Eprint {http://arxiv.org/abs/1908.06995} {arXiv:1908.06995 [astro-ph.CO]} \BibitemShut {NoStop}%
\bibitem [{\citenamefont {Niedermann}\ and\ \citenamefont {Sloth}(2021)}]{Niedermann:2019olb}%
  \BibitemOpen
  \bibfield  {author} {\bibinfo {author} {\bibfnamefont {F.}~\bibnamefont {Niedermann}}\ and\ \bibinfo {author} {\bibfnamefont {M.~S.}\ \bibnamefont {Sloth}},\ }\href {\doibase 10.1103/PhysRevD.103.L041303} {\bibfield  {journal} {\bibinfo  {journal} {Phys. Rev. D}\ }\textbf {\bibinfo {volume} {103}},\ \bibinfo {pages} {L041303} (\bibinfo {year} {2021})},\ \Eprint {http://arxiv.org/abs/1910.10739} {arXiv:1910.10739 [astro-ph.CO]} \BibitemShut {NoStop}%
\bibitem [{\citenamefont {Niedermann}\ and\ \citenamefont {Sloth}(2020)}]{Niedermann:2020dwg}%
  \BibitemOpen
  \bibfield  {author} {\bibinfo {author} {\bibfnamefont {F.}~\bibnamefont {Niedermann}}\ and\ \bibinfo {author} {\bibfnamefont {M.~S.}\ \bibnamefont {Sloth}},\ }\href {\doibase 10.1103/PhysRevD.102.063527} {\bibfield  {journal} {\bibinfo  {journal} {Phys. Rev. D}\ }\textbf {\bibinfo {volume} {102}},\ \bibinfo {pages} {063527} (\bibinfo {year} {2020})},\ \Eprint {http://arxiv.org/abs/2006.06686} {arXiv:2006.06686 [astro-ph.CO]} \BibitemShut {NoStop}%
\bibitem [{\citenamefont {Murgia}\ \emph {et~al.}(2021)\citenamefont {Murgia}, \citenamefont {Abell\'an},\ and\ \citenamefont {Poulin}}]{Murgia:2020ryi}%
  \BibitemOpen
  \bibfield  {author} {\bibinfo {author} {\bibfnamefont {R.}~\bibnamefont {Murgia}}, \bibinfo {author} {\bibfnamefont {G.~F.}\ \bibnamefont {Abell\'an}}, \ and\ \bibinfo {author} {\bibfnamefont {V.}~\bibnamefont {Poulin}},\ }\href {\doibase 10.1103/PhysRevD.103.063502} {\bibfield  {journal} {\bibinfo  {journal} {Phys. Rev. D}\ }\textbf {\bibinfo {volume} {103}},\ \bibinfo {pages} {063502} (\bibinfo {year} {2021})},\ \Eprint {http://arxiv.org/abs/2009.10733} {arXiv:2009.10733 [astro-ph.CO]} \BibitemShut {NoStop}%
\bibitem [{\citenamefont {Ye}\ and\ \citenamefont {Piao}(2020)}]{Ye:2020btb}%
  \BibitemOpen
  \bibfield  {author} {\bibinfo {author} {\bibfnamefont {G.}~\bibnamefont {Ye}}\ and\ \bibinfo {author} {\bibfnamefont {Y.-S.}\ \bibnamefont {Piao}},\ }\href {\doibase 10.1103/PhysRevD.101.083507} {\bibfield  {journal} {\bibinfo  {journal} {Phys. Rev. D}\ }\textbf {\bibinfo {volume} {101}},\ \bibinfo {pages} {083507} (\bibinfo {year} {2020})},\ \Eprint {http://arxiv.org/abs/2001.02451} {arXiv:2001.02451 [astro-ph.CO]} \BibitemShut {NoStop}%
\bibitem [{\citenamefont {Klypin}\ \emph {et~al.}(2021)\citenamefont {Klypin}, \citenamefont {Poulin}, \citenamefont {Prada}, \citenamefont {Primack}, \citenamefont {Kamionkowski}, \citenamefont {Avila-Reese}, \citenamefont {Rodriguez-Puebla}, \citenamefont {Behroozi}, \citenamefont {Hellinger},\ and\ \citenamefont {Smith}}]{Klypin:2020tud}%
  \BibitemOpen
  \bibfield  {author} {\bibinfo {author} {\bibfnamefont {A.}~\bibnamefont {Klypin}}, \bibinfo {author} {\bibfnamefont {V.}~\bibnamefont {Poulin}}, \bibinfo {author} {\bibfnamefont {F.}~\bibnamefont {Prada}}, \bibinfo {author} {\bibfnamefont {J.}~\bibnamefont {Primack}}, \bibinfo {author} {\bibfnamefont {M.}~\bibnamefont {Kamionkowski}}, \bibinfo {author} {\bibfnamefont {V.}~\bibnamefont {Avila-Reese}}, \bibinfo {author} {\bibfnamefont {A.}~\bibnamefont {Rodriguez-Puebla}}, \bibinfo {author} {\bibfnamefont {P.}~\bibnamefont {Behroozi}}, \bibinfo {author} {\bibfnamefont {D.}~\bibnamefont {Hellinger}}, \ and\ \bibinfo {author} {\bibfnamefont {T.~L.}\ \bibnamefont {Smith}},\ }\href {\doibase 10.1093/mnras/stab769} {\bibfield  {journal} {\bibinfo  {journal} {Mon. Not. Roy. Astron. Soc.}\ }\textbf {\bibinfo {volume} {504}},\ \bibinfo {pages} {769} (\bibinfo {year} {2021})},\ \Eprint {http://arxiv.org/abs/2006.14910} {arXiv:2006.14910 [astro-ph.CO]} \BibitemShut {NoStop}%
\bibitem [{\citenamefont {Hill}\ \emph {et~al.}(2020)\citenamefont {Hill}, \citenamefont {McDonough}, \citenamefont {Toomey},\ and\ \citenamefont {Alexander}}]{Hill:2020osr}%
  \BibitemOpen
  \bibfield  {author} {\bibinfo {author} {\bibfnamefont {J.~C.}\ \bibnamefont {Hill}}, \bibinfo {author} {\bibfnamefont {E.}~\bibnamefont {McDonough}}, \bibinfo {author} {\bibfnamefont {M.~W.}\ \bibnamefont {Toomey}}, \ and\ \bibinfo {author} {\bibfnamefont {S.}~\bibnamefont {Alexander}},\ }\href {\doibase 10.1103/PhysRevD.102.043507} {\bibfield  {journal} {\bibinfo  {journal} {Phys. Rev. D}\ }\textbf {\bibinfo {volume} {102}},\ \bibinfo {pages} {043507} (\bibinfo {year} {2020})},\ \Eprint {http://arxiv.org/abs/2003.07355} {arXiv:2003.07355 [astro-ph.CO]} \BibitemShut {NoStop}%
\bibitem [{\citenamefont {Herold}\ \emph {et~al.}(2022)\citenamefont {Herold}, \citenamefont {Ferreira},\ and\ \citenamefont {Komatsu}}]{Herold:2021ksg}%
  \BibitemOpen
  \bibfield  {author} {\bibinfo {author} {\bibfnamefont {L.}~\bibnamefont {Herold}}, \bibinfo {author} {\bibfnamefont {E.~G.~M.}\ \bibnamefont {Ferreira}}, \ and\ \bibinfo {author} {\bibfnamefont {E.}~\bibnamefont {Komatsu}},\ }\href {\doibase 10.3847/2041-8213/ac63a3} {\bibfield  {journal} {\bibinfo  {journal} {Astrophys. J. Lett.}\ }\textbf {\bibinfo {volume} {929}},\ \bibinfo {pages} {L16} (\bibinfo {year} {2022})},\ \Eprint {http://arxiv.org/abs/2112.12140} {arXiv:2112.12140 [astro-ph.CO]} \BibitemShut {NoStop}%
\bibitem [{\citenamefont {Herold}\ and\ \citenamefont {Ferreira}(2023)}]{Herold:2022iib}%
  \BibitemOpen
  \bibfield  {author} {\bibinfo {author} {\bibfnamefont {L.}~\bibnamefont {Herold}}\ and\ \bibinfo {author} {\bibfnamefont {E.~G.~M.}\ \bibnamefont {Ferreira}},\ }\href {\doibase 10.1103/PhysRevD.108.043513} {\bibfield  {journal} {\bibinfo  {journal} {Phys. Rev. D}\ }\textbf {\bibinfo {volume} {108}},\ \bibinfo {pages} {043513} (\bibinfo {year} {2023})},\ \Eprint {http://arxiv.org/abs/2210.16296} {arXiv:2210.16296 [astro-ph.CO]} \BibitemShut {NoStop}%
\bibitem [{\citenamefont {Reeves}\ \emph {et~al.}(2023)\citenamefont {Reeves}, \citenamefont {Herold}, \citenamefont {Vagnozzi}, \citenamefont {Sherwin},\ and\ \citenamefont {Ferreira}}]{Reeves:2022aoi}%
  \BibitemOpen
  \bibfield  {author} {\bibinfo {author} {\bibfnamefont {A.}~\bibnamefont {Reeves}}, \bibinfo {author} {\bibfnamefont {L.}~\bibnamefont {Herold}}, \bibinfo {author} {\bibfnamefont {S.}~\bibnamefont {Vagnozzi}}, \bibinfo {author} {\bibfnamefont {B.~D.}\ \bibnamefont {Sherwin}}, \ and\ \bibinfo {author} {\bibfnamefont {E.~G.~M.}\ \bibnamefont {Ferreira}},\ }\href {\doibase 10.1093/mnras/stad317} {\bibfield  {journal} {\bibinfo  {journal} {Mon. Not. Roy. Astron. Soc.}\ }\textbf {\bibinfo {volume} {520}},\ \bibinfo {pages} {3688} (\bibinfo {year} {2023})},\ \Eprint {http://arxiv.org/abs/2207.01501} {arXiv:2207.01501 [astro-ph.CO]} \BibitemShut {NoStop}%
\bibitem [{\citenamefont {Simon}\ \emph {et~al.}(2023)\citenamefont {Simon}, \citenamefont {Zhang}, \citenamefont {Poulin},\ and\ \citenamefont {Smith}}]{Simon:2022adh}%
  \BibitemOpen
  \bibfield  {author} {\bibinfo {author} {\bibfnamefont {T.}~\bibnamefont {Simon}}, \bibinfo {author} {\bibfnamefont {P.}~\bibnamefont {Zhang}}, \bibinfo {author} {\bibfnamefont {V.}~\bibnamefont {Poulin}}, \ and\ \bibinfo {author} {\bibfnamefont {T.~L.}\ \bibnamefont {Smith}},\ }\href {\doibase 10.1103/PhysRevD.107.063505} {\bibfield  {journal} {\bibinfo  {journal} {Phys. Rev. D}\ }\textbf {\bibinfo {volume} {107}},\ \bibinfo {pages} {063505} (\bibinfo {year} {2023})},\ \Eprint {http://arxiv.org/abs/2208.05930} {arXiv:2208.05930 [astro-ph.CO]} \BibitemShut {NoStop}%
\bibitem [{\citenamefont {Smith}\ \emph {et~al.}(2022)\citenamefont {Smith}, \citenamefont {Lucca}, \citenamefont {Poulin}, \citenamefont {Abellan}, \citenamefont {Balkenhol}, \citenamefont {Benabed}, \citenamefont {Galli},\ and\ \citenamefont {Murgia}}]{Smith:2022hwi}%
  \BibitemOpen
  \bibfield  {author} {\bibinfo {author} {\bibfnamefont {T.~L.}\ \bibnamefont {Smith}}, \bibinfo {author} {\bibfnamefont {M.}~\bibnamefont {Lucca}}, \bibinfo {author} {\bibfnamefont {V.}~\bibnamefont {Poulin}}, \bibinfo {author} {\bibfnamefont {G.~F.}\ \bibnamefont {Abellan}}, \bibinfo {author} {\bibfnamefont {L.}~\bibnamefont {Balkenhol}}, \bibinfo {author} {\bibfnamefont {K.}~\bibnamefont {Benabed}}, \bibinfo {author} {\bibfnamefont {S.}~\bibnamefont {Galli}}, \ and\ \bibinfo {author} {\bibfnamefont {R.}~\bibnamefont {Murgia}},\ }\href {\doibase 10.1103/PhysRevD.106.043526} {\bibfield  {journal} {\bibinfo  {journal} {Phys. Rev. D}\ }\textbf {\bibinfo {volume} {106}},\ \bibinfo {pages} {043526} (\bibinfo {year} {2022})},\ \Eprint {http://arxiv.org/abs/2202.09379} {arXiv:2202.09379 [astro-ph.CO]} \BibitemShut {NoStop}%
\bibitem [{\citenamefont {Nakagawa}\ \emph {et~al.}(2023)\citenamefont {Nakagawa}, \citenamefont {Takahashi},\ and\ \citenamefont {Yin}}]{Nakagawa:2022knn}%
  \BibitemOpen
  \bibfield  {author} {\bibinfo {author} {\bibfnamefont {S.}~\bibnamefont {Nakagawa}}, \bibinfo {author} {\bibfnamefont {F.}~\bibnamefont {Takahashi}}, \ and\ \bibinfo {author} {\bibfnamefont {W.}~\bibnamefont {Yin}},\ }\href {\doibase 10.1103/PhysRevD.107.063016} {\bibfield  {journal} {\bibinfo  {journal} {Phys. Rev. D}\ }\textbf {\bibinfo {volume} {107}},\ \bibinfo {pages} {063016} (\bibinfo {year} {2023})},\ \Eprint {http://arxiv.org/abs/2209.01107} {arXiv:2209.01107 [astro-ph.CO]} \BibitemShut {NoStop}%
\bibitem [{\citenamefont {Niedermann}\ and\ \citenamefont {Sloth}(2023)}]{Niedermann:2023ssr}%
  \BibitemOpen
  \bibfield  {author} {\bibinfo {author} {\bibfnamefont {F.}~\bibnamefont {Niedermann}}\ and\ \bibinfo {author} {\bibfnamefont {M.~S.}\ \bibnamefont {Sloth}},\ }\href@noop {} {\  (\bibinfo {year} {2023})},\ \Eprint {http://arxiv.org/abs/2307.03481} {arXiv:2307.03481 [hep-ph]} \BibitemShut {NoStop}%
\bibitem [{\citenamefont {Cruz}\ \emph {et~al.}(2023)\citenamefont {Cruz}, \citenamefont {Niedermann},\ and\ \citenamefont {Sloth}}]{Cruz:2023lmn}%
  \BibitemOpen
  \bibfield  {author} {\bibinfo {author} {\bibfnamefont {J.~S.}\ \bibnamefont {Cruz}}, \bibinfo {author} {\bibfnamefont {F.}~\bibnamefont {Niedermann}}, \ and\ \bibinfo {author} {\bibfnamefont {M.~S.}\ \bibnamefont {Sloth}},\ }\href {\doibase 10.1088/1475-7516/2023/11/033} {\bibfield  {journal} {\bibinfo  {journal} {JCAP}\ }\textbf {\bibinfo {volume} {11}},\ \bibinfo {pages} {033} (\bibinfo {year} {2023})},\ \Eprint {http://arxiv.org/abs/2305.08895} {arXiv:2305.08895 [astro-ph.CO]} \BibitemShut {NoStop}%
\bibitem [{\citenamefont {Eskilt}\ \emph {et~al.}(2023)\citenamefont {Eskilt}, \citenamefont {Herold}, \citenamefont {Komatsu}, \citenamefont {Murai}, \citenamefont {Namikawa},\ and\ \citenamefont {Naokawa}}]{Eskilt:2023nxm}%
  \BibitemOpen
  \bibfield  {author} {\bibinfo {author} {\bibfnamefont {J.~R.}\ \bibnamefont {Eskilt}}, \bibinfo {author} {\bibfnamefont {L.}~\bibnamefont {Herold}}, \bibinfo {author} {\bibfnamefont {E.}~\bibnamefont {Komatsu}}, \bibinfo {author} {\bibfnamefont {K.}~\bibnamefont {Murai}}, \bibinfo {author} {\bibfnamefont {T.}~\bibnamefont {Namikawa}}, \ and\ \bibinfo {author} {\bibfnamefont {F.}~\bibnamefont {Naokawa}},\ }\href {\doibase 10.1103/PhysRevLett.131.121001} {\bibfield  {journal} {\bibinfo  {journal} {Phys. Rev. Lett.}\ }\textbf {\bibinfo {volume} {131}},\ \bibinfo {pages} {121001} (\bibinfo {year} {2023})},\ \Eprint {http://arxiv.org/abs/2303.15369} {arXiv:2303.15369 [astro-ph.CO]} \BibitemShut {NoStop}%
\bibitem [{\citenamefont {Smith}\ and\ \citenamefont {Poulin}(2023)}]{Smith:2023oop}%
  \BibitemOpen
  \bibfield  {author} {\bibinfo {author} {\bibfnamefont {T.~L.}\ \bibnamefont {Smith}}\ and\ \bibinfo {author} {\bibfnamefont {V.}~\bibnamefont {Poulin}},\ }\href@noop {} {\  (\bibinfo {year} {2023})},\ \Eprint {http://arxiv.org/abs/2309.03265} {arXiv:2309.03265 [astro-ph.CO]} \BibitemShut {NoStop}%
\bibitem [{\citenamefont {Sharma}\ \emph {et~al.}(2023)\citenamefont {Sharma}, \citenamefont {Das},\ and\ \citenamefont {Poulin}}]{Sharma:2023kzr}%
  \BibitemOpen
  \bibfield  {author} {\bibinfo {author} {\bibfnamefont {R.~K.}\ \bibnamefont {Sharma}}, \bibinfo {author} {\bibfnamefont {S.}~\bibnamefont {Das}}, \ and\ \bibinfo {author} {\bibfnamefont {V.}~\bibnamefont {Poulin}},\ }\href@noop {} {\  (\bibinfo {year} {2023})},\ \Eprint {http://arxiv.org/abs/2309.00401} {arXiv:2309.00401 [astro-ph.CO]} \BibitemShut {NoStop}%
\bibitem [{\citenamefont {Efstathiou}\ \emph {et~al.}(2023)\citenamefont {Efstathiou}, \citenamefont {Rosenberg},\ and\ \citenamefont {Poulin}}]{Efstathiou:2023fbn}%
  \BibitemOpen
  \bibfield  {author} {\bibinfo {author} {\bibfnamefont {G.}~\bibnamefont {Efstathiou}}, \bibinfo {author} {\bibfnamefont {E.}~\bibnamefont {Rosenberg}}, \ and\ \bibinfo {author} {\bibfnamefont {V.}~\bibnamefont {Poulin}},\ }\href@noop {} {\  (\bibinfo {year} {2023})},\ \Eprint {http://arxiv.org/abs/2311.00524} {arXiv:2311.00524 [astro-ph.CO]} \BibitemShut {NoStop}%
\bibitem [{\citenamefont {Gsponer}\ \emph {et~al.}(2023)\citenamefont {Gsponer}, \citenamefont {Zhao}, \citenamefont {Donald-McCann}, \citenamefont {Bacon}, \citenamefont {Koyama}, \citenamefont {Crittenden}, \citenamefont {Simon},\ and\ \citenamefont {Mueller}}]{Gsponer:2023wpm}%
  \BibitemOpen
  \bibfield  {author} {\bibinfo {author} {\bibfnamefont {R.}~\bibnamefont {Gsponer}}, \bibinfo {author} {\bibfnamefont {R.}~\bibnamefont {Zhao}}, \bibinfo {author} {\bibfnamefont {J.}~\bibnamefont {Donald-McCann}}, \bibinfo {author} {\bibfnamefont {D.}~\bibnamefont {Bacon}}, \bibinfo {author} {\bibfnamefont {K.}~\bibnamefont {Koyama}}, \bibinfo {author} {\bibfnamefont {R.}~\bibnamefont {Crittenden}}, \bibinfo {author} {\bibfnamefont {T.}~\bibnamefont {Simon}}, \ and\ \bibinfo {author} {\bibfnamefont {E.-M.}\ \bibnamefont {Mueller}},\ }\href@noop {} {\  (\bibinfo {year} {2023})},\ \Eprint {http://arxiv.org/abs/2312.01977} {arXiv:2312.01977 [astro-ph.CO]} \BibitemShut {NoStop}%
\bibitem [{\citenamefont {Goldstein}\ \emph {et~al.}(2023)\citenamefont {Goldstein}, \citenamefont {Hill}, \citenamefont {Ir\v{s}i\v{c}},\ and\ \citenamefont {Sherwin}}]{Goldstein:2023gnw}%
  \BibitemOpen
  \bibfield  {author} {\bibinfo {author} {\bibfnamefont {S.}~\bibnamefont {Goldstein}}, \bibinfo {author} {\bibfnamefont {J.~C.}\ \bibnamefont {Hill}}, \bibinfo {author} {\bibfnamefont {V.}~\bibnamefont {Ir\v{s}i\v{c}}}, \ and\ \bibinfo {author} {\bibfnamefont {B.~D.}\ \bibnamefont {Sherwin}},\ }\href {\doibase 10.1103/PhysRevLett.131.201001} {\bibfield  {journal} {\bibinfo  {journal} {Phys. Rev. Lett.}\ }\textbf {\bibinfo {volume} {131}},\ \bibinfo {pages} {201001} (\bibinfo {year} {2023})},\ \Eprint {http://arxiv.org/abs/2303.00746} {arXiv:2303.00746 [astro-ph.CO]} \BibitemShut {NoStop}%
\bibitem [{\citenamefont {Forconi}\ \emph {et~al.}(2023)\citenamefont {Forconi}, \citenamefont {Giar\`e}, \citenamefont {Mena}, \citenamefont {Ruchika}, \citenamefont {Di~Valentino}, \citenamefont {Melchiorri},\ and\ \citenamefont {Nunes}}]{Forconi:2023hsj}%
  \BibitemOpen
  \bibfield  {author} {\bibinfo {author} {\bibfnamefont {M.}~\bibnamefont {Forconi}}, \bibinfo {author} {\bibfnamefont {W.}~\bibnamefont {Giar\`e}}, \bibinfo {author} {\bibfnamefont {O.}~\bibnamefont {Mena}}, \bibinfo {author} {\bibnamefont {Ruchika}}, \bibinfo {author} {\bibfnamefont {E.}~\bibnamefont {Di~Valentino}}, \bibinfo {author} {\bibfnamefont {A.}~\bibnamefont {Melchiorri}}, \ and\ \bibinfo {author} {\bibfnamefont {R.~C.}\ \bibnamefont {Nunes}},\ }\href@noop {} {\  (\bibinfo {year} {2023})},\ \Eprint {http://arxiv.org/abs/2312.11074} {arXiv:2312.11074 [astro-ph.CO]} \BibitemShut {NoStop}%
\bibitem [{\citenamefont {Fu}\ and\ \citenamefont {Wang}(2024)}]{Fu:2023tfo}%
  \BibitemOpen
  \bibfield  {author} {\bibinfo {author} {\bibfnamefont {C.}~\bibnamefont {Fu}}\ and\ \bibinfo {author} {\bibfnamefont {S.-J.}\ \bibnamefont {Wang}},\ }\href {\doibase 10.1103/PhysRevD.109.L041304} {\bibfield  {journal} {\bibinfo  {journal} {Phys. Rev. D}\ }\textbf {\bibinfo {volume} {109}},\ \bibinfo {pages} {L041304} (\bibinfo {year} {2024})},\ \Eprint {http://arxiv.org/abs/2310.12932} {arXiv:2310.12932 [astro-ph.CO]} \BibitemShut {NoStop}%
\bibitem [{\citenamefont {Ye}\ \emph {et~al.}(2021)\citenamefont {Ye}, \citenamefont {Hu},\ and\ \citenamefont {Piao}}]{Ye:2021nej}%
  \BibitemOpen
  \bibfield  {author} {\bibinfo {author} {\bibfnamefont {G.}~\bibnamefont {Ye}}, \bibinfo {author} {\bibfnamefont {B.}~\bibnamefont {Hu}}, \ and\ \bibinfo {author} {\bibfnamefont {Y.-S.}\ \bibnamefont {Piao}},\ }\href {\doibase 10.1103/PhysRevD.104.063510} {\bibfield  {journal} {\bibinfo  {journal} {Phys. Rev. D}\ }\textbf {\bibinfo {volume} {104}},\ \bibinfo {pages} {063510} (\bibinfo {year} {2021})},\ \Eprint {http://arxiv.org/abs/2103.09729} {arXiv:2103.09729 [astro-ph.CO]} \BibitemShut {NoStop}%
\bibitem [{\citenamefont {Peng}\ and\ \citenamefont {Piao}(2023)}]{Peng:2023bik}%
  \BibitemOpen
  \bibfield  {author} {\bibinfo {author} {\bibfnamefont {Z.-Y.}\ \bibnamefont {Peng}}\ and\ \bibinfo {author} {\bibfnamefont {Y.-S.}\ \bibnamefont {Piao}},\ }\href@noop {} {\  (\bibinfo {year} {2023})},\ \Eprint {http://arxiv.org/abs/2308.01012} {arXiv:2308.01012 [astro-ph.CO]} \BibitemShut {NoStop}%
\bibitem [{\citenamefont {Vagnozzi}(2020)}]{Vagnozzi:2019ezj}%
  \BibitemOpen
  \bibfield  {author} {\bibinfo {author} {\bibfnamefont {S.}~\bibnamefont {Vagnozzi}},\ }\href {\doibase 10.1103/PhysRevD.102.023518} {\bibfield  {journal} {\bibinfo  {journal} {Phys. Rev. D}\ }\textbf {\bibinfo {volume} {102}},\ \bibinfo {pages} {023518} (\bibinfo {year} {2020})},\ \Eprint {http://arxiv.org/abs/1907.07569} {arXiv:1907.07569 [astro-ph.CO]} \BibitemShut {NoStop}%
\bibitem [{\citenamefont {Pedreira}\ \emph {et~al.}(2023)\citenamefont {Pedreira}, \citenamefont {Benetti}, \citenamefont {Ferreira}, \citenamefont {Graef},\ and\ \citenamefont {Herold}}]{Pedreira:2023qqt}%
  \BibitemOpen
  \bibfield  {author} {\bibinfo {author} {\bibfnamefont {I.~d. O.~C.}\ \bibnamefont {Pedreira}}, \bibinfo {author} {\bibfnamefont {M.}~\bibnamefont {Benetti}}, \bibinfo {author} {\bibfnamefont {E.~G.~M.}\ \bibnamefont {Ferreira}}, \bibinfo {author} {\bibfnamefont {L.~L.}\ \bibnamefont {Graef}}, \ and\ \bibinfo {author} {\bibfnamefont {L.}~\bibnamefont {Herold}},\ }\href@noop {} {\  (\bibinfo {year} {2023})},\ \Eprint {http://arxiv.org/abs/2311.04977} {arXiv:2311.04977 [astro-ph.CO]} \BibitemShut {NoStop}%
\bibitem [{\citenamefont {Di~Valentino}\ \emph {et~al.}(2021{\natexlab{b}})\citenamefont {Di~Valentino} \emph {et~al.}}]{DiValentino:2020vvd}%
  \BibitemOpen
  \bibfield  {author} {\bibinfo {author} {\bibfnamefont {E.}~\bibnamefont {Di~Valentino}} \emph {et~al.},\ }\href {\doibase 10.1016/j.astropartphys.2021.102604} {\bibfield  {journal} {\bibinfo  {journal} {Astropart. Phys.}\ }\textbf {\bibinfo {volume} {131}},\ \bibinfo {pages} {102604} (\bibinfo {year} {2021}{\natexlab{b}})},\ \Eprint {http://arxiv.org/abs/2008.11285} {arXiv:2008.11285 [astro-ph.CO]} \BibitemShut {NoStop}%
\bibitem [{\citenamefont {Knox}\ and\ \citenamefont {Millea}(2020)}]{Knox:2019rjx}%
  \BibitemOpen
  \bibfield  {author} {\bibinfo {author} {\bibfnamefont {L.}~\bibnamefont {Knox}}\ and\ \bibinfo {author} {\bibfnamefont {M.}~\bibnamefont {Millea}},\ }\href {\doibase 10.1103/PhysRevD.101.043533} {\bibfield  {journal} {\bibinfo  {journal} {Phys. Rev. D}\ }\textbf {\bibinfo {volume} {101}},\ \bibinfo {pages} {043533} (\bibinfo {year} {2020})},\ \Eprint {http://arxiv.org/abs/1908.03663} {arXiv:1908.03663 [astro-ph.CO]} \BibitemShut {NoStop}%
\bibitem [{\citenamefont {Krishnan}\ \emph {et~al.}(2021)\citenamefont {Krishnan}, \citenamefont {Mohayaee}, \citenamefont {Colg\'ain}, \citenamefont {Sheikh-Jabbari},\ and\ \citenamefont {Yin}}]{Krishnan:2021dyb}%
  \BibitemOpen
  \bibfield  {author} {\bibinfo {author} {\bibfnamefont {C.}~\bibnamefont {Krishnan}}, \bibinfo {author} {\bibfnamefont {R.}~\bibnamefont {Mohayaee}}, \bibinfo {author} {\bibfnamefont {E.~O.}\ \bibnamefont {Colg\'ain}}, \bibinfo {author} {\bibfnamefont {M.~M.}\ \bibnamefont {Sheikh-Jabbari}}, \ and\ \bibinfo {author} {\bibfnamefont {L.}~\bibnamefont {Yin}},\ }\href {\doibase 10.1088/1361-6382/ac1a81} {\bibfield  {journal} {\bibinfo  {journal} {Class. Quant. Grav.}\ }\textbf {\bibinfo {volume} {38}},\ \bibinfo {pages} {184001} (\bibinfo {year} {2021})},\ \Eprint {http://arxiv.org/abs/2105.09790} {arXiv:2105.09790 [astro-ph.CO]} \BibitemShut {NoStop}%
\bibitem [{\citenamefont {Keeley}\ and\ \citenamefont {Shafieloo}(2023)}]{Keeley:2022ojz}%
  \BibitemOpen
  \bibfield  {author} {\bibinfo {author} {\bibfnamefont {R.~E.}\ \bibnamefont {Keeley}}\ and\ \bibinfo {author} {\bibfnamefont {A.}~\bibnamefont {Shafieloo}},\ }\href {\doibase 10.1103/PhysRevLett.131.111002} {\bibfield  {journal} {\bibinfo  {journal} {Phys. Rev. Lett.}\ }\textbf {\bibinfo {volume} {131}},\ \bibinfo {pages} {111002} (\bibinfo {year} {2023})},\ \Eprint {http://arxiv.org/abs/2206.08440} {arXiv:2206.08440 [astro-ph.CO]} \BibitemShut {NoStop}%
\bibitem [{\citenamefont {Vagnozzi}(2023{\natexlab{b}})}]{Vagnozzi:2023nrq}%
  \BibitemOpen
  \bibfield  {author} {\bibinfo {author} {\bibfnamefont {S.}~\bibnamefont {Vagnozzi}},\ }\href {\doibase 10.3390/universe9090393} {\bibfield  {journal} {\bibinfo  {journal} {Universe}\ }\textbf {\bibinfo {volume} {9}},\ \bibinfo {pages} {393} (\bibinfo {year} {2023}{\natexlab{b}})},\ \Eprint {http://arxiv.org/abs/2308.16628} {arXiv:2308.16628 [astro-ph.CO]} \BibitemShut {NoStop}%
\bibitem [{\citenamefont {Mangano}\ \emph {et~al.}(2005)\citenamefont {Mangano}, \citenamefont {Miele}, \citenamefont {Pastor}, \citenamefont {Pinto}, \citenamefont {Pisanti},\ and\ \citenamefont {Serpico}}]{Mangano:2005cc}%
  \BibitemOpen
  \bibfield  {author} {\bibinfo {author} {\bibfnamefont {G.}~\bibnamefont {Mangano}}, \bibinfo {author} {\bibfnamefont {G.}~\bibnamefont {Miele}}, \bibinfo {author} {\bibfnamefont {S.}~\bibnamefont {Pastor}}, \bibinfo {author} {\bibfnamefont {T.}~\bibnamefont {Pinto}}, \bibinfo {author} {\bibfnamefont {O.}~\bibnamefont {Pisanti}}, \ and\ \bibinfo {author} {\bibfnamefont {P.~D.}\ \bibnamefont {Serpico}},\ }\href {\doibase 10.1016/j.nuclphysb.2005.09.041} {\bibfield  {journal} {\bibinfo  {journal} {Nucl. Phys. B}\ }\textbf {\bibinfo {volume} {729}},\ \bibinfo {pages} {221} (\bibinfo {year} {2005})},\ \Eprint {http://arxiv.org/abs/hep-ph/0506164} {arXiv:hep-ph/0506164} \BibitemShut {NoStop}%
\bibitem [{\citenamefont {Akita}\ and\ \citenamefont {Yamaguchi}(2020)}]{Akita:2020szl}%
  \BibitemOpen
  \bibfield  {author} {\bibinfo {author} {\bibfnamefont {K.}~\bibnamefont {Akita}}\ and\ \bibinfo {author} {\bibfnamefont {M.}~\bibnamefont {Yamaguchi}},\ }\href {\doibase 10.1088/1475-7516/2020/08/012} {\bibfield  {journal} {\bibinfo  {journal} {JCAP}\ }\textbf {\bibinfo {volume} {08}},\ \bibinfo {pages} {012} (\bibinfo {year} {2020})},\ \Eprint {http://arxiv.org/abs/2005.07047} {arXiv:2005.07047 [hep-ph]} \BibitemShut {NoStop}%
\bibitem [{\citenamefont {Froustey}\ \emph {et~al.}(2020)\citenamefont {Froustey}, \citenamefont {Pitrou},\ and\ \citenamefont {Volpe}}]{Froustey:2020mcq}%
  \BibitemOpen
  \bibfield  {author} {\bibinfo {author} {\bibfnamefont {J.}~\bibnamefont {Froustey}}, \bibinfo {author} {\bibfnamefont {C.}~\bibnamefont {Pitrou}}, \ and\ \bibinfo {author} {\bibfnamefont {M.~C.}\ \bibnamefont {Volpe}},\ }\href {\doibase 10.1088/1475-7516/2020/12/015} {\bibfield  {journal} {\bibinfo  {journal} {JCAP}\ }\textbf {\bibinfo {volume} {12}},\ \bibinfo {pages} {015} (\bibinfo {year} {2020})},\ \Eprint {http://arxiv.org/abs/2008.01074} {arXiv:2008.01074 [hep-ph]} \BibitemShut {NoStop}%
\bibitem [{\citenamefont {Bennett}\ \emph {et~al.}(2021)\citenamefont {Bennett}, \citenamefont {Buldgen}, \citenamefont {De~Salas}, \citenamefont {Drewes}, \citenamefont {Gariazzo}, \citenamefont {Pastor},\ and\ \citenamefont {Wong}}]{Bennett:2020zkv}%
  \BibitemOpen
  \bibfield  {author} {\bibinfo {author} {\bibfnamefont {J.~J.}\ \bibnamefont {Bennett}}, \bibinfo {author} {\bibfnamefont {G.}~\bibnamefont {Buldgen}}, \bibinfo {author} {\bibfnamefont {P.~F.}\ \bibnamefont {De~Salas}}, \bibinfo {author} {\bibfnamefont {M.}~\bibnamefont {Drewes}}, \bibinfo {author} {\bibfnamefont {S.}~\bibnamefont {Gariazzo}}, \bibinfo {author} {\bibfnamefont {S.}~\bibnamefont {Pastor}}, \ and\ \bibinfo {author} {\bibfnamefont {Y.~Y.~Y.}\ \bibnamefont {Wong}},\ }\href {\doibase 10.1088/1475-7516/2021/04/073} {\bibfield  {journal} {\bibinfo  {journal} {JCAP}\ }\textbf {\bibinfo {volume} {04}},\ \bibinfo {pages} {073} (\bibinfo {year} {2021})},\ \Eprint {http://arxiv.org/abs/2012.02726} {arXiv:2012.02726 [hep-ph]} \BibitemShut {NoStop}%
\bibitem [{\citenamefont {Giar\`e}\ \emph {et~al.}(2021{\natexlab{b}})\citenamefont {Giar\`e}, \citenamefont {Di~Valentino}, \citenamefont {Melchiorri},\ and\ \citenamefont {Mena}}]{Giare:2020vzo}%
  \BibitemOpen
  \bibfield  {author} {\bibinfo {author} {\bibfnamefont {W.}~\bibnamefont {Giar\`e}}, \bibinfo {author} {\bibfnamefont {E.}~\bibnamefont {Di~Valentino}}, \bibinfo {author} {\bibfnamefont {A.}~\bibnamefont {Melchiorri}}, \ and\ \bibinfo {author} {\bibfnamefont {O.}~\bibnamefont {Mena}},\ }\href {\doibase 10.1093/mnras/stab1442} {\bibfield  {journal} {\bibinfo  {journal} {Mon. Not. Roy. Astron. Soc.}\ }\textbf {\bibinfo {volume} {505}},\ \bibinfo {pages} {2703} (\bibinfo {year} {2021}{\natexlab{b}})},\ \Eprint {http://arxiv.org/abs/2011.14704} {arXiv:2011.14704 [astro-ph.CO]} \BibitemShut {NoStop}%
\bibitem [{\citenamefont {Giar\`e}\ \emph {et~al.}(2022)\citenamefont {Giar\`e}, \citenamefont {Renzi}, \citenamefont {Melchiorri}, \citenamefont {Mena},\ and\ \citenamefont {Di~Valentino}}]{Giare:2021cqr}%
  \BibitemOpen
  \bibfield  {author} {\bibinfo {author} {\bibfnamefont {W.}~\bibnamefont {Giar\`e}}, \bibinfo {author} {\bibfnamefont {F.}~\bibnamefont {Renzi}}, \bibinfo {author} {\bibfnamefont {A.}~\bibnamefont {Melchiorri}}, \bibinfo {author} {\bibfnamefont {O.}~\bibnamefont {Mena}}, \ and\ \bibinfo {author} {\bibfnamefont {E.}~\bibnamefont {Di~Valentino}},\ }\href {\doibase 10.1093/mnras/stac126} {\bibfield  {journal} {\bibinfo  {journal} {Mon. Not. Roy. Astron. Soc.}\ }\textbf {\bibinfo {volume} {511}},\ \bibinfo {pages} {1373} (\bibinfo {year} {2022})},\ \Eprint {http://arxiv.org/abs/2110.00340} {arXiv:2110.00340 [astro-ph.CO]} \BibitemShut {NoStop}%
\bibitem [{\citenamefont {D'Eramo}\ \emph {et~al.}(2022)\citenamefont {D'Eramo}, \citenamefont {Di~Valentino}, \citenamefont {Giar\`e}, \citenamefont {Hajkarim}, \citenamefont {Melchiorri}, \citenamefont {Mena}, \citenamefont {Renzi},\ and\ \citenamefont {Yun}}]{DEramo:2022nvb}%
  \BibitemOpen
  \bibfield  {author} {\bibinfo {author} {\bibfnamefont {F.}~\bibnamefont {D'Eramo}}, \bibinfo {author} {\bibfnamefont {E.}~\bibnamefont {Di~Valentino}}, \bibinfo {author} {\bibfnamefont {W.}~\bibnamefont {Giar\`e}}, \bibinfo {author} {\bibfnamefont {F.}~\bibnamefont {Hajkarim}}, \bibinfo {author} {\bibfnamefont {A.}~\bibnamefont {Melchiorri}}, \bibinfo {author} {\bibfnamefont {O.}~\bibnamefont {Mena}}, \bibinfo {author} {\bibfnamefont {F.}~\bibnamefont {Renzi}}, \ and\ \bibinfo {author} {\bibfnamefont {S.}~\bibnamefont {Yun}},\ }\href {\doibase 10.1088/1475-7516/2022/09/022} {\bibfield  {journal} {\bibinfo  {journal} {JCAP}\ }\textbf {\bibinfo {volume} {09}},\ \bibinfo {pages} {022} (\bibinfo {year} {2022})},\ \Eprint {http://arxiv.org/abs/2205.07849} {arXiv:2205.07849 [astro-ph.CO]} \BibitemShut {NoStop}%
\bibitem [{\citenamefont {Papanikolaou}(2023)}]{Papanikolaou:2023oxq}%
  \BibitemOpen
  \bibfield  {author} {\bibinfo {author} {\bibfnamefont {T.}~\bibnamefont {Papanikolaou}},\ }\href {\doibase 10.22323/1.436.0265} {\bibfield  {journal} {\bibinfo  {journal} {PoS}\ }\textbf {\bibinfo {volume} {CORFU2022}},\ \bibinfo {pages} {265} (\bibinfo {year} {2023})},\ \Eprint {http://arxiv.org/abs/2303.00600} {arXiv:2303.00600 [astro-ph.CO]} \BibitemShut {NoStop}%
\bibitem [{\citenamefont {Garny}\ \emph {et~al.}(2024)\citenamefont {Garny}, \citenamefont {Niedermann}, \citenamefont {Rubira},\ and\ \citenamefont {Sloth}}]{Garny:2024ums}%
  \BibitemOpen
  \bibfield  {author} {\bibinfo {author} {\bibfnamefont {M.}~\bibnamefont {Garny}}, \bibinfo {author} {\bibfnamefont {F.}~\bibnamefont {Niedermann}}, \bibinfo {author} {\bibfnamefont {H.}~\bibnamefont {Rubira}}, \ and\ \bibinfo {author} {\bibfnamefont {M.~S.}\ \bibnamefont {Sloth}},\ }\href@noop {} {\  (\bibinfo {year} {2024})},\ \Eprint {http://arxiv.org/abs/2404.07256} {arXiv:2404.07256 [astro-ph.CO]} \BibitemShut {NoStop}%
\bibitem [{\citenamefont {Gavela}\ \emph {et~al.}(2009)\citenamefont {Gavela}, \citenamefont {Hernandez}, \citenamefont {Lopez~Honorez}, \citenamefont {Mena},\ and\ \citenamefont {Rigolin}}]{Gavela:2009cy}%
  \BibitemOpen
  \bibfield  {author} {\bibinfo {author} {\bibfnamefont {M.~B.}\ \bibnamefont {Gavela}}, \bibinfo {author} {\bibfnamefont {D.}~\bibnamefont {Hernandez}}, \bibinfo {author} {\bibfnamefont {L.}~\bibnamefont {Lopez~Honorez}}, \bibinfo {author} {\bibfnamefont {O.}~\bibnamefont {Mena}}, \ and\ \bibinfo {author} {\bibfnamefont {S.}~\bibnamefont {Rigolin}},\ }\href {\doibase 10.1088/1475-7516/2009/07/034} {\bibfield  {journal} {\bibinfo  {journal} {JCAP}\ }\textbf {\bibinfo {volume} {07}},\ \bibinfo {pages} {034} (\bibinfo {year} {2009})},\ \bibinfo {note} {[Erratum: JCAP 05, E01 (2010)]},\ \Eprint {http://arxiv.org/abs/0901.1611} {arXiv:0901.1611 [astro-ph.CO]} \BibitemShut {NoStop}%
\bibitem [{\citenamefont {Di~Valentino}\ \emph {et~al.}(2017)\citenamefont {Di~Valentino}, \citenamefont {Melchiorri},\ and\ \citenamefont {Mena}}]{DiValentino:2017iww}%
  \BibitemOpen
  \bibfield  {author} {\bibinfo {author} {\bibfnamefont {E.}~\bibnamefont {Di~Valentino}}, \bibinfo {author} {\bibfnamefont {A.}~\bibnamefont {Melchiorri}}, \ and\ \bibinfo {author} {\bibfnamefont {O.}~\bibnamefont {Mena}},\ }\href {\doibase 10.1103/PhysRevD.96.043503} {\bibfield  {journal} {\bibinfo  {journal} {Phys. Rev. D}\ }\textbf {\bibinfo {volume} {96}},\ \bibinfo {pages} {043503} (\bibinfo {year} {2017})},\ \Eprint {http://arxiv.org/abs/1704.08342} {arXiv:1704.08342 [astro-ph.CO]} \BibitemShut {NoStop}%
\bibitem [{\citenamefont {Kumar}\ and\ \citenamefont {Nunes}(2017)}]{Kumar:2017dnp}%
  \BibitemOpen
  \bibfield  {author} {\bibinfo {author} {\bibfnamefont {S.}~\bibnamefont {Kumar}}\ and\ \bibinfo {author} {\bibfnamefont {R.~C.}\ \bibnamefont {Nunes}},\ }\href {\doibase 10.1103/PhysRevD.96.103511} {\bibfield  {journal} {\bibinfo  {journal} {Phys. Rev. D}\ }\textbf {\bibinfo {volume} {96}},\ \bibinfo {pages} {103511} (\bibinfo {year} {2017})},\ \Eprint {http://arxiv.org/abs/1702.02143} {arXiv:1702.02143 [astro-ph.CO]} \BibitemShut {NoStop}%
\bibitem [{\citenamefont {Wang}\ \emph {et~al.}(2016)\citenamefont {Wang}, \citenamefont {Abdalla}, \citenamefont {Atrio-Barandela},\ and\ \citenamefont {Pavon}}]{Wang:2016lxa}%
  \BibitemOpen
  \bibfield  {author} {\bibinfo {author} {\bibfnamefont {B.}~\bibnamefont {Wang}}, \bibinfo {author} {\bibfnamefont {E.}~\bibnamefont {Abdalla}}, \bibinfo {author} {\bibfnamefont {F.}~\bibnamefont {Atrio-Barandela}}, \ and\ \bibinfo {author} {\bibfnamefont {D.}~\bibnamefont {Pavon}},\ }\href {\doibase 10.1088/0034-4885/79/9/096901} {\bibfield  {journal} {\bibinfo  {journal} {Rept. Prog. Phys.}\ }\textbf {\bibinfo {volume} {79}},\ \bibinfo {pages} {096901} (\bibinfo {year} {2016})},\ \Eprint {http://arxiv.org/abs/1603.08299} {arXiv:1603.08299 [astro-ph.CO]} \BibitemShut {NoStop}%
\bibitem [{\citenamefont {Di~Valentino}\ \emph {et~al.}(2020{\natexlab{a}})\citenamefont {Di~Valentino}, \citenamefont {Melchiorri}, \citenamefont {Mena},\ and\ \citenamefont {Vagnozzi}}]{DiValentino:2019ffd}%
  \BibitemOpen
  \bibfield  {author} {\bibinfo {author} {\bibfnamefont {E.}~\bibnamefont {Di~Valentino}}, \bibinfo {author} {\bibfnamefont {A.}~\bibnamefont {Melchiorri}}, \bibinfo {author} {\bibfnamefont {O.}~\bibnamefont {Mena}}, \ and\ \bibinfo {author} {\bibfnamefont {S.}~\bibnamefont {Vagnozzi}},\ }\href {\doibase 10.1016/j.dark.2020.100666} {\bibfield  {journal} {\bibinfo  {journal} {Phys. Dark Univ.}\ }\textbf {\bibinfo {volume} {30}},\ \bibinfo {pages} {100666} (\bibinfo {year} {2020}{\natexlab{a}})},\ \Eprint {http://arxiv.org/abs/1908.04281} {arXiv:1908.04281 [astro-ph.CO]} \BibitemShut {NoStop}%
\bibitem [{\citenamefont {Pan}\ \emph {et~al.}(2019{\natexlab{a}})\citenamefont {Pan}, \citenamefont {Yang}, \citenamefont {Singha},\ and\ \citenamefont {Saridakis}}]{Pan:2019jqh}%
  \BibitemOpen
  \bibfield  {author} {\bibinfo {author} {\bibfnamefont {S.}~\bibnamefont {Pan}}, \bibinfo {author} {\bibfnamefont {W.}~\bibnamefont {Yang}}, \bibinfo {author} {\bibfnamefont {C.}~\bibnamefont {Singha}}, \ and\ \bibinfo {author} {\bibfnamefont {E.~N.}\ \bibnamefont {Saridakis}},\ }\href {\doibase 10.1103/PhysRevD.100.083539} {\bibfield  {journal} {\bibinfo  {journal} {Phys. Rev. D}\ }\textbf {\bibinfo {volume} {100}},\ \bibinfo {pages} {083539} (\bibinfo {year} {2019}{\natexlab{a}})},\ \Eprint {http://arxiv.org/abs/1903.10969} {arXiv:1903.10969 [astro-ph.CO]} \BibitemShut {NoStop}%
\bibitem [{\citenamefont {Yang}\ \emph {et~al.}(2018)\citenamefont {Yang}, \citenamefont {Pan}, \citenamefont {Di~Valentino}, \citenamefont {Nunes}, \citenamefont {Vagnozzi},\ and\ \citenamefont {Mota}}]{Yang:2018euj}%
  \BibitemOpen
  \bibfield  {author} {\bibinfo {author} {\bibfnamefont {W.}~\bibnamefont {Yang}}, \bibinfo {author} {\bibfnamefont {S.}~\bibnamefont {Pan}}, \bibinfo {author} {\bibfnamefont {E.}~\bibnamefont {Di~Valentino}}, \bibinfo {author} {\bibfnamefont {R.~C.}\ \bibnamefont {Nunes}}, \bibinfo {author} {\bibfnamefont {S.}~\bibnamefont {Vagnozzi}}, \ and\ \bibinfo {author} {\bibfnamefont {D.~F.}\ \bibnamefont {Mota}},\ }\href {\doibase 10.1088/1475-7516/2018/09/019} {\bibfield  {journal} {\bibinfo  {journal} {JCAP}\ }\textbf {\bibinfo {volume} {09}},\ \bibinfo {pages} {019} (\bibinfo {year} {2018})},\ \Eprint {http://arxiv.org/abs/1805.08252} {arXiv:1805.08252 [astro-ph.CO]} \BibitemShut {NoStop}%
\bibitem [{\citenamefont {Murgia}\ \emph {et~al.}(2016)\citenamefont {Murgia}, \citenamefont {Gariazzo},\ and\ \citenamefont {Fornengo}}]{Murgia:2016ccp}%
  \BibitemOpen
  \bibfield  {author} {\bibinfo {author} {\bibfnamefont {R.}~\bibnamefont {Murgia}}, \bibinfo {author} {\bibfnamefont {S.}~\bibnamefont {Gariazzo}}, \ and\ \bibinfo {author} {\bibfnamefont {N.}~\bibnamefont {Fornengo}},\ }\href {\doibase 10.1088/1475-7516/2016/04/014} {\bibfield  {journal} {\bibinfo  {journal} {JCAP}\ }\textbf {\bibinfo {volume} {04}},\ \bibinfo {pages} {014} (\bibinfo {year} {2016})},\ \Eprint {http://arxiv.org/abs/1602.01765} {arXiv:1602.01765 [astro-ph.CO]} \BibitemShut {NoStop}%
\bibitem [{\citenamefont {Yang}\ \emph {et~al.}(2019)\citenamefont {Yang}, \citenamefont {Mena}, \citenamefont {Pan},\ and\ \citenamefont {Di~Valentino}}]{Yang:2019uzo}%
  \BibitemOpen
  \bibfield  {author} {\bibinfo {author} {\bibfnamefont {W.}~\bibnamefont {Yang}}, \bibinfo {author} {\bibfnamefont {O.}~\bibnamefont {Mena}}, \bibinfo {author} {\bibfnamefont {S.}~\bibnamefont {Pan}}, \ and\ \bibinfo {author} {\bibfnamefont {E.}~\bibnamefont {Di~Valentino}},\ }\href {\doibase 10.1103/PhysRevD.100.083509} {\bibfield  {journal} {\bibinfo  {journal} {Phys. Rev. D}\ }\textbf {\bibinfo {volume} {100}},\ \bibinfo {pages} {083509} (\bibinfo {year} {2019})},\ \Eprint {http://arxiv.org/abs/1906.11697} {arXiv:1906.11697 [astro-ph.CO]} \BibitemShut {NoStop}%
\bibitem [{\citenamefont {Pan}\ \emph {et~al.}(2019{\natexlab{b}})\citenamefont {Pan}, \citenamefont {Yang}, \citenamefont {Di~Valentino}, \citenamefont {Saridakis},\ and\ \citenamefont {Chakraborty}}]{Pan:2019gop}%
  \BibitemOpen
  \bibfield  {author} {\bibinfo {author} {\bibfnamefont {S.}~\bibnamefont {Pan}}, \bibinfo {author} {\bibfnamefont {W.}~\bibnamefont {Yang}}, \bibinfo {author} {\bibfnamefont {E.}~\bibnamefont {Di~Valentino}}, \bibinfo {author} {\bibfnamefont {E.~N.}\ \bibnamefont {Saridakis}}, \ and\ \bibinfo {author} {\bibfnamefont {S.}~\bibnamefont {Chakraborty}},\ }\href {\doibase 10.1103/PhysRevD.100.103520} {\bibfield  {journal} {\bibinfo  {journal} {Phys. Rev. D}\ }\textbf {\bibinfo {volume} {100}},\ \bibinfo {pages} {103520} (\bibinfo {year} {2019}{\natexlab{b}})},\ \Eprint {http://arxiv.org/abs/1907.07540} {arXiv:1907.07540 [astro-ph.CO]} \BibitemShut {NoStop}%
\bibitem [{\citenamefont {Di~Valentino}\ \emph {et~al.}(2020{\natexlab{b}})\citenamefont {Di~Valentino}, \citenamefont {Melchiorri}, \citenamefont {Mena},\ and\ \citenamefont {Vagnozzi}}]{DiValentino:2019jae}%
  \BibitemOpen
  \bibfield  {author} {\bibinfo {author} {\bibfnamefont {E.}~\bibnamefont {Di~Valentino}}, \bibinfo {author} {\bibfnamefont {A.}~\bibnamefont {Melchiorri}}, \bibinfo {author} {\bibfnamefont {O.}~\bibnamefont {Mena}}, \ and\ \bibinfo {author} {\bibfnamefont {S.}~\bibnamefont {Vagnozzi}},\ }\href {\doibase 10.1103/PhysRevD.101.063502} {\bibfield  {journal} {\bibinfo  {journal} {Phys. Rev. D}\ }\textbf {\bibinfo {volume} {101}},\ \bibinfo {pages} {063502} (\bibinfo {year} {2020}{\natexlab{b}})},\ \Eprint {http://arxiv.org/abs/1910.09853} {arXiv:1910.09853 [astro-ph.CO]} \BibitemShut {NoStop}%
\bibitem [{\citenamefont {Lucca}\ and\ \citenamefont {Hooper}(2020)}]{Lucca:2020zjb}%
  \BibitemOpen
  \bibfield  {author} {\bibinfo {author} {\bibfnamefont {M.}~\bibnamefont {Lucca}}\ and\ \bibinfo {author} {\bibfnamefont {D.~C.}\ \bibnamefont {Hooper}},\ }\href {\doibase 10.1103/PhysRevD.102.123502} {\bibfield  {journal} {\bibinfo  {journal} {Phys. Rev. D}\ }\textbf {\bibinfo {volume} {102}},\ \bibinfo {pages} {123502} (\bibinfo {year} {2020})},\ \Eprint {http://arxiv.org/abs/2002.06127} {arXiv:2002.06127 [astro-ph.CO]} \BibitemShut {NoStop}%
\bibitem [{\citenamefont {G\'omez-Valent}\ \emph {et~al.}(2020)\citenamefont {G\'omez-Valent}, \citenamefont {Pettorino},\ and\ \citenamefont {Amendola}}]{Gomez-Valent:2020mqn}%
  \BibitemOpen
  \bibfield  {author} {\bibinfo {author} {\bibfnamefont {A.}~\bibnamefont {G\'omez-Valent}}, \bibinfo {author} {\bibfnamefont {V.}~\bibnamefont {Pettorino}}, \ and\ \bibinfo {author} {\bibfnamefont {L.}~\bibnamefont {Amendola}},\ }\href {\doibase 10.1103/PhysRevD.101.123513} {\bibfield  {journal} {\bibinfo  {journal} {Phys. Rev. D}\ }\textbf {\bibinfo {volume} {101}},\ \bibinfo {pages} {123513} (\bibinfo {year} {2020})},\ \Eprint {http://arxiv.org/abs/2004.00610} {arXiv:2004.00610 [astro-ph.CO]} \BibitemShut {NoStop}%
\bibitem [{\citenamefont {Di~Valentino}(2021)}]{DiValentino:2020vnx}%
  \BibitemOpen
  \bibfield  {author} {\bibinfo {author} {\bibfnamefont {E.}~\bibnamefont {Di~Valentino}},\ }\href {\doibase 10.1093/mnras/stab187} {\bibfield  {journal} {\bibinfo  {journal} {Mon. Not. Roy. Astron. Soc.}\ }\textbf {\bibinfo {volume} {502}},\ \bibinfo {pages} {2065} (\bibinfo {year} {2021})},\ \Eprint {http://arxiv.org/abs/2011.00246} {arXiv:2011.00246 [astro-ph.CO]} \BibitemShut {NoStop}%
\bibitem [{\citenamefont {Yang}\ \emph {et~al.}(2021{\natexlab{a}})\citenamefont {Yang}, \citenamefont {Pan}, \citenamefont {Di~Valentino}, \citenamefont {Mena},\ and\ \citenamefont {Melchiorri}}]{Yang:2021hxg}%
  \BibitemOpen
  \bibfield  {author} {\bibinfo {author} {\bibfnamefont {W.}~\bibnamefont {Yang}}, \bibinfo {author} {\bibfnamefont {S.}~\bibnamefont {Pan}}, \bibinfo {author} {\bibfnamefont {E.}~\bibnamefont {Di~Valentino}}, \bibinfo {author} {\bibfnamefont {O.}~\bibnamefont {Mena}}, \ and\ \bibinfo {author} {\bibfnamefont {A.}~\bibnamefont {Melchiorri}},\ }\href {\doibase 10.1088/1475-7516/2021/10/008} {\bibfield  {journal} {\bibinfo  {journal} {JCAP}\ }\textbf {\bibinfo {volume} {10}},\ \bibinfo {pages} {008} (\bibinfo {year} {2021}{\natexlab{a}})},\ \Eprint {http://arxiv.org/abs/2101.03129} {arXiv:2101.03129 [astro-ph.CO]} \BibitemShut {NoStop}%
\bibitem [{\citenamefont {Gariazzo}\ \emph {et~al.}(2022)\citenamefont {Gariazzo}, \citenamefont {Di~Valentino}, \citenamefont {Mena},\ and\ \citenamefont {Nunes}}]{Gariazzo:2021qtg}%
  \BibitemOpen
  \bibfield  {author} {\bibinfo {author} {\bibfnamefont {S.}~\bibnamefont {Gariazzo}}, \bibinfo {author} {\bibfnamefont {E.}~\bibnamefont {Di~Valentino}}, \bibinfo {author} {\bibfnamefont {O.}~\bibnamefont {Mena}}, \ and\ \bibinfo {author} {\bibfnamefont {R.~C.}\ \bibnamefont {Nunes}},\ }\href {\doibase 10.1103/PhysRevD.106.023530} {\bibfield  {journal} {\bibinfo  {journal} {Phys. Rev. D}\ }\textbf {\bibinfo {volume} {106}},\ \bibinfo {pages} {023530} (\bibinfo {year} {2022})},\ \Eprint {http://arxiv.org/abs/2111.03152} {arXiv:2111.03152 [astro-ph.CO]} \BibitemShut {NoStop}%
\bibitem [{\citenamefont {Bernui}\ \emph {et~al.}(2023)\citenamefont {Bernui}, \citenamefont {Di~Valentino}, \citenamefont {Giar\`e}, \citenamefont {Kumar},\ and\ \citenamefont {Nunes}}]{Bernui:2023byc}%
  \BibitemOpen
  \bibfield  {author} {\bibinfo {author} {\bibfnamefont {A.}~\bibnamefont {Bernui}}, \bibinfo {author} {\bibfnamefont {E.}~\bibnamefont {Di~Valentino}}, \bibinfo {author} {\bibfnamefont {W.}~\bibnamefont {Giar\`e}}, \bibinfo {author} {\bibfnamefont {S.}~\bibnamefont {Kumar}}, \ and\ \bibinfo {author} {\bibfnamefont {R.~C.}\ \bibnamefont {Nunes}},\ }\href {\doibase 10.1103/PhysRevD.107.103531} {\bibfield  {journal} {\bibinfo  {journal} {Phys. Rev. D}\ }\textbf {\bibinfo {volume} {107}},\ \bibinfo {pages} {103531} (\bibinfo {year} {2023})},\ \Eprint {http://arxiv.org/abs/2301.06097} {arXiv:2301.06097 [astro-ph.CO]} \BibitemShut {NoStop}%
\bibitem [{\citenamefont {Zhai}\ \emph {et~al.}(2023)\citenamefont {Zhai}, \citenamefont {Giar\`e}, \citenamefont {van~de Bruck}, \citenamefont {Di~Valentino}, \citenamefont {Mena},\ and\ \citenamefont {Nunes}}]{Zhai:2023yny}%
  \BibitemOpen
  \bibfield  {author} {\bibinfo {author} {\bibfnamefont {Y.}~\bibnamefont {Zhai}}, \bibinfo {author} {\bibfnamefont {W.}~\bibnamefont {Giar\`e}}, \bibinfo {author} {\bibfnamefont {C.}~\bibnamefont {van~de Bruck}}, \bibinfo {author} {\bibfnamefont {E.}~\bibnamefont {Di~Valentino}}, \bibinfo {author} {\bibfnamefont {O.}~\bibnamefont {Mena}}, \ and\ \bibinfo {author} {\bibfnamefont {R.~C.}\ \bibnamefont {Nunes}},\ }\href {\doibase 10.1088/1475-7516/2023/07/032} {\bibfield  {journal} {\bibinfo  {journal} {JCAP}\ }\textbf {\bibinfo {volume} {07}},\ \bibinfo {pages} {032} (\bibinfo {year} {2023})},\ \Eprint {http://arxiv.org/abs/2303.08201} {arXiv:2303.08201 [astro-ph.CO]} \BibitemShut {NoStop}%
\bibitem [{\citenamefont {Montani}\ \emph {et~al.}(2023{\natexlab{a}})\citenamefont {Montani}, \citenamefont {De~Angelis}, \citenamefont {Bombacigno},\ and\ \citenamefont {Carlevaro}}]{Montani:2023xpd}%
  \BibitemOpen
  \bibfield  {author} {\bibinfo {author} {\bibfnamefont {G.}~\bibnamefont {Montani}}, \bibinfo {author} {\bibfnamefont {M.}~\bibnamefont {De~Angelis}}, \bibinfo {author} {\bibfnamefont {F.}~\bibnamefont {Bombacigno}}, \ and\ \bibinfo {author} {\bibfnamefont {N.}~\bibnamefont {Carlevaro}},\ }\href {\doibase 10.1093/mnrasl/slad159} {\bibfield  {journal} {\bibinfo  {journal} {Mon. Not. Roy. Astron. Soc.}\ }\textbf {\bibinfo {volume} {527}},\ \bibinfo {pages} {L156} (\bibinfo {year} {2023}{\natexlab{a}})},\ \Eprint {http://arxiv.org/abs/2306.11101} {arXiv:2306.11101 [gr-qc]} \BibitemShut {NoStop}%
\bibitem [{\citenamefont {Schiavone}\ \emph {et~al.}(2023)\citenamefont {Schiavone}, \citenamefont {Montani},\ and\ \citenamefont {Bombacigno}}]{Schiavone:2022wvq}%
  \BibitemOpen
  \bibfield  {author} {\bibinfo {author} {\bibfnamefont {T.}~\bibnamefont {Schiavone}}, \bibinfo {author} {\bibfnamefont {G.}~\bibnamefont {Montani}}, \ and\ \bibinfo {author} {\bibfnamefont {F.}~\bibnamefont {Bombacigno}},\ }\href {\doibase 10.1093/mnrasl/slad041} {\bibfield  {journal} {\bibinfo  {journal} {Mon. Not. Roy. Astron. Soc.}\ }\textbf {\bibinfo {volume} {522}},\ \bibinfo {pages} {L72} (\bibinfo {year} {2023})},\ \Eprint {http://arxiv.org/abs/2211.16737} {arXiv:2211.16737 [gr-qc]} \BibitemShut {NoStop}%
\bibitem [{\citenamefont {Montani}\ \emph {et~al.}(2023{\natexlab{b}})\citenamefont {Montani}, \citenamefont {Carlevaro},\ and\ \citenamefont {Dainotti}}]{Montani:2023ywn}%
  \BibitemOpen
  \bibfield  {author} {\bibinfo {author} {\bibfnamefont {G.}~\bibnamefont {Montani}}, \bibinfo {author} {\bibfnamefont {N.}~\bibnamefont {Carlevaro}}, \ and\ \bibinfo {author} {\bibfnamefont {M.~G.}\ \bibnamefont {Dainotti}},\ }\href@noop {} {\  (\bibinfo {year} {2023}{\natexlab{b}})},\ \Eprint {http://arxiv.org/abs/2311.04822} {arXiv:2311.04822 [gr-qc]} \BibitemShut {NoStop}%
\bibitem [{\citenamefont {Akarsu}\ \emph {et~al.}(2021)\citenamefont {Akarsu}, \citenamefont {Kumar}, \citenamefont {\"Oz\"ulker},\ and\ \citenamefont {Vazquez}}]{Akarsu:2021fol}%
  \BibitemOpen
  \bibfield  {author} {\bibinfo {author} {\bibfnamefont {O.}~\bibnamefont {Akarsu}}, \bibinfo {author} {\bibfnamefont {S.}~\bibnamefont {Kumar}}, \bibinfo {author} {\bibfnamefont {E.}~\bibnamefont {\"Oz\"ulker}}, \ and\ \bibinfo {author} {\bibfnamefont {J.~A.}\ \bibnamefont {Vazquez}},\ }\href {\doibase 10.1103/PhysRevD.104.123512} {\bibfield  {journal} {\bibinfo  {journal} {Phys. Rev. D}\ }\textbf {\bibinfo {volume} {104}},\ \bibinfo {pages} {123512} (\bibinfo {year} {2021})},\ \Eprint {http://arxiv.org/abs/2108.09239} {arXiv:2108.09239 [astro-ph.CO]} \BibitemShut {NoStop}%
\bibitem [{\citenamefont {Akarsu}\ \emph {et~al.}(2023{\natexlab{a}})\citenamefont {Akarsu}, \citenamefont {Kumar}, \citenamefont {\"Oz\"ulker}, \citenamefont {Vazquez},\ and\ \citenamefont {Yadav}}]{Akarsu:2022typ}%
  \BibitemOpen
  \bibfield  {author} {\bibinfo {author} {\bibfnamefont {O.}~\bibnamefont {Akarsu}}, \bibinfo {author} {\bibfnamefont {S.}~\bibnamefont {Kumar}}, \bibinfo {author} {\bibfnamefont {E.}~\bibnamefont {\"Oz\"ulker}}, \bibinfo {author} {\bibfnamefont {J.~A.}\ \bibnamefont {Vazquez}}, \ and\ \bibinfo {author} {\bibfnamefont {A.}~\bibnamefont {Yadav}},\ }\href {\doibase 10.1103/PhysRevD.108.023513} {\bibfield  {journal} {\bibinfo  {journal} {Phys. Rev. D}\ }\textbf {\bibinfo {volume} {108}},\ \bibinfo {pages} {023513} (\bibinfo {year} {2023}{\natexlab{a}})},\ \Eprint {http://arxiv.org/abs/2211.05742} {arXiv:2211.05742 [astro-ph.CO]} \BibitemShut {NoStop}%
\bibitem [{\citenamefont {Akarsu}\ \emph {et~al.}(2023{\natexlab{b}})\citenamefont {Akarsu}, \citenamefont {Di~Valentino}, \citenamefont {Kumar}, \citenamefont {Nunes}, \citenamefont {Vazquez},\ and\ \citenamefont {Yadav}}]{Akarsu:2023mfb}%
  \BibitemOpen
  \bibfield  {author} {\bibinfo {author} {\bibfnamefont {O.}~\bibnamefont {Akarsu}}, \bibinfo {author} {\bibfnamefont {E.}~\bibnamefont {Di~Valentino}}, \bibinfo {author} {\bibfnamefont {S.}~\bibnamefont {Kumar}}, \bibinfo {author} {\bibfnamefont {R.~C.}\ \bibnamefont {Nunes}}, \bibinfo {author} {\bibfnamefont {J.~A.}\ \bibnamefont {Vazquez}}, \ and\ \bibinfo {author} {\bibfnamefont {A.}~\bibnamefont {Yadav}},\ }\href@noop {} {\  (\bibinfo {year} {2023}{\natexlab{b}})},\ \Eprint {http://arxiv.org/abs/2307.10899} {arXiv:2307.10899 [astro-ph.CO]} \BibitemShut {NoStop}%
\bibitem [{\citenamefont {G\'omez-Valent}\ \emph {et~al.}(2024)\citenamefont {G\'omez-Valent}, \citenamefont {Favale}, \citenamefont {Migliaccio},\ and\ \citenamefont {Sen}}]{Gomez-Valent:2023uof}%
  \BibitemOpen
  \bibfield  {author} {\bibinfo {author} {\bibfnamefont {A.}~\bibnamefont {G\'omez-Valent}}, \bibinfo {author} {\bibfnamefont {A.}~\bibnamefont {Favale}}, \bibinfo {author} {\bibfnamefont {M.}~\bibnamefont {Migliaccio}}, \ and\ \bibinfo {author} {\bibfnamefont {A.~A.}\ \bibnamefont {Sen}},\ }\href {\doibase 10.1103/PhysRevD.109.023525} {\bibfield  {journal} {\bibinfo  {journal} {Phys. Rev. D}\ }\textbf {\bibinfo {volume} {109}},\ \bibinfo {pages} {023525} (\bibinfo {year} {2024})},\ \Eprint {http://arxiv.org/abs/2309.07795} {arXiv:2309.07795 [astro-ph.CO]} \BibitemShut {NoStop}%
\bibitem [{\citenamefont {Giar\`e}\ \emph {et~al.}(2024{\natexlab{b}})\citenamefont {Giar\`e}, \citenamefont {Zhai}, \citenamefont {Pan}, \citenamefont {Di~Valentino}, \citenamefont {Nunes},\ and\ \citenamefont {van~de Bruck}}]{Giare:2024ytc}%
  \BibitemOpen
  \bibfield  {author} {\bibinfo {author} {\bibfnamefont {W.}~\bibnamefont {Giar\`e}}, \bibinfo {author} {\bibfnamefont {Y.}~\bibnamefont {Zhai}}, \bibinfo {author} {\bibfnamefont {S.}~\bibnamefont {Pan}}, \bibinfo {author} {\bibfnamefont {E.}~\bibnamefont {Di~Valentino}}, \bibinfo {author} {\bibfnamefont {R.~C.}\ \bibnamefont {Nunes}}, \ and\ \bibinfo {author} {\bibfnamefont {C.}~\bibnamefont {van~de Bruck}},\ }\href@noop {} {\  (\bibinfo {year} {2024}{\natexlab{b}})},\ \Eprint {http://arxiv.org/abs/2404.02110} {arXiv:2404.02110 [astro-ph.CO]} \BibitemShut {NoStop}%
\bibitem [{\citenamefont {Giar\`e}\ \emph {et~al.}(2024{\natexlab{c}})\citenamefont {Giar\`e}, \citenamefont {Sabogal}, \citenamefont {Nunes},\ and\ \citenamefont {Di~Valentino}}]{Giare:2024smz}%
  \BibitemOpen
  \bibfield  {author} {\bibinfo {author} {\bibfnamefont {W.}~\bibnamefont {Giar\`e}}, \bibinfo {author} {\bibfnamefont {M.~A.}\ \bibnamefont {Sabogal}}, \bibinfo {author} {\bibfnamefont {R.~C.}\ \bibnamefont {Nunes}}, \ and\ \bibinfo {author} {\bibfnamefont {E.}~\bibnamefont {Di~Valentino}},\ }\href@noop {} {\  (\bibinfo {year} {2024}{\natexlab{c}})},\ \Eprint {http://arxiv.org/abs/2404.15232} {arXiv:2404.15232 [astro-ph.CO]} \BibitemShut {NoStop}%
\bibitem [{\citenamefont {Park}\ and\ \citenamefont {Ratra}(2019)}]{Park:2017xbl}%
  \BibitemOpen
  \bibfield  {author} {\bibinfo {author} {\bibfnamefont {C.-G.}\ \bibnamefont {Park}}\ and\ \bibinfo {author} {\bibfnamefont {B.}~\bibnamefont {Ratra}},\ }\href {\doibase 10.3847/1538-4357/ab3641} {\bibfield  {journal} {\bibinfo  {journal} {Astrophys. J.}\ }\textbf {\bibinfo {volume} {882}},\ \bibinfo {pages} {158} (\bibinfo {year} {2019})},\ \Eprint {http://arxiv.org/abs/1801.00213} {arXiv:1801.00213 [astro-ph.CO]} \BibitemShut {NoStop}%
\bibitem [{\citenamefont {Handley}(2021)}]{Handley:2019tkm}%
  \BibitemOpen
  \bibfield  {author} {\bibinfo {author} {\bibfnamefont {W.}~\bibnamefont {Handley}},\ }\href {\doibase 10.1103/PhysRevD.103.L041301} {\bibfield  {journal} {\bibinfo  {journal} {Phys. Rev. D}\ }\textbf {\bibinfo {volume} {103}},\ \bibinfo {pages} {L041301} (\bibinfo {year} {2021})},\ \Eprint {http://arxiv.org/abs/1908.09139} {arXiv:1908.09139 [astro-ph.CO]} \BibitemShut {NoStop}%
\bibitem [{\citenamefont {Di~Valentino}\ \emph {et~al.}(2019)\citenamefont {Di~Valentino}, \citenamefont {Melchiorri},\ and\ \citenamefont {Silk}}]{DiValentino:2019qzk}%
  \BibitemOpen
  \bibfield  {author} {\bibinfo {author} {\bibfnamefont {E.}~\bibnamefont {Di~Valentino}}, \bibinfo {author} {\bibfnamefont {A.}~\bibnamefont {Melchiorri}}, \ and\ \bibinfo {author} {\bibfnamefont {J.}~\bibnamefont {Silk}},\ }\href {\doibase 10.1038/s41550-019-0906-9} {\bibfield  {journal} {\bibinfo  {journal} {Nature Astron.}\ }\textbf {\bibinfo {volume} {4}},\ \bibinfo {pages} {196} (\bibinfo {year} {2019})},\ \Eprint {http://arxiv.org/abs/1911.02087} {arXiv:1911.02087 [astro-ph.CO]} \BibitemShut {NoStop}%
\bibitem [{\citenamefont {Efstathiou}\ and\ \citenamefont {Gratton}(2020)}]{Efstathiou:2020wem}%
  \BibitemOpen
  \bibfield  {author} {\bibinfo {author} {\bibfnamefont {G.}~\bibnamefont {Efstathiou}}\ and\ \bibinfo {author} {\bibfnamefont {S.}~\bibnamefont {Gratton}},\ }\href {\doibase 10.1093/mnrasl/slaa093} {\bibfield  {journal} {\bibinfo  {journal} {Mon. Not. Roy. Astron. Soc.}\ }\textbf {\bibinfo {volume} {496}},\ \bibinfo {pages} {L91} (\bibinfo {year} {2020})},\ \Eprint {http://arxiv.org/abs/2002.06892} {arXiv:2002.06892 [astro-ph.CO]} \BibitemShut {NoStop}%
\bibitem [{\citenamefont {Di~Valentino}\ \emph {et~al.}(2021{\natexlab{c}})\citenamefont {Di~Valentino}, \citenamefont {Melchiorri},\ and\ \citenamefont {Silk}}]{DiValentino:2020hov}%
  \BibitemOpen
  \bibfield  {author} {\bibinfo {author} {\bibfnamefont {E.}~\bibnamefont {Di~Valentino}}, \bibinfo {author} {\bibfnamefont {A.}~\bibnamefont {Melchiorri}}, \ and\ \bibinfo {author} {\bibfnamefont {J.}~\bibnamefont {Silk}},\ }\href {\doibase 10.3847/2041-8213/abe1c4} {\bibfield  {journal} {\bibinfo  {journal} {Astrophys. J. Lett.}\ }\textbf {\bibinfo {volume} {908}},\ \bibinfo {pages} {L9} (\bibinfo {year} {2021}{\natexlab{c}})},\ \Eprint {http://arxiv.org/abs/2003.04935} {arXiv:2003.04935 [astro-ph.CO]} \BibitemShut {NoStop}%
\bibitem [{\citenamefont {Benisty}\ and\ \citenamefont {Staicova}(2021)}]{Benisty:2020otr}%
  \BibitemOpen
  \bibfield  {author} {\bibinfo {author} {\bibfnamefont {D.}~\bibnamefont {Benisty}}\ and\ \bibinfo {author} {\bibfnamefont {D.}~\bibnamefont {Staicova}},\ }\href {\doibase 10.1051/0004-6361/202039502} {\bibfield  {journal} {\bibinfo  {journal} {Astron. Astrophys.}\ }\textbf {\bibinfo {volume} {647}},\ \bibinfo {pages} {A38} (\bibinfo {year} {2021})},\ \Eprint {http://arxiv.org/abs/2009.10701} {arXiv:2009.10701 [astro-ph.CO]} \BibitemShut {NoStop}%
\bibitem [{\citenamefont {Vagnozzi}\ \emph {et~al.}(2021{\natexlab{a}})\citenamefont {Vagnozzi}, \citenamefont {Di~Valentino}, \citenamefont {Gariazzo}, \citenamefont {Melchiorri}, \citenamefont {Mena},\ and\ \citenamefont {Silk}}]{Vagnozzi:2020rcz}%
  \BibitemOpen
  \bibfield  {author} {\bibinfo {author} {\bibfnamefont {S.}~\bibnamefont {Vagnozzi}}, \bibinfo {author} {\bibfnamefont {E.}~\bibnamefont {Di~Valentino}}, \bibinfo {author} {\bibfnamefont {S.}~\bibnamefont {Gariazzo}}, \bibinfo {author} {\bibfnamefont {A.}~\bibnamefont {Melchiorri}}, \bibinfo {author} {\bibfnamefont {O.}~\bibnamefont {Mena}}, \ and\ \bibinfo {author} {\bibfnamefont {J.}~\bibnamefont {Silk}},\ }\href {\doibase 10.1016/j.dark.2021.100851} {\bibfield  {journal} {\bibinfo  {journal} {Phys. Dark Univ.}\ }\textbf {\bibinfo {volume} {33}},\ \bibinfo {pages} {100851} (\bibinfo {year} {2021}{\natexlab{a}})},\ \Eprint {http://arxiv.org/abs/2010.02230} {arXiv:2010.02230 [astro-ph.CO]} \BibitemShut {NoStop}%
\bibitem [{\citenamefont {Vagnozzi}\ \emph {et~al.}(2021{\natexlab{b}})\citenamefont {Vagnozzi}, \citenamefont {Loeb},\ and\ \citenamefont {Moresco}}]{Vagnozzi:2020dfn}%
  \BibitemOpen
  \bibfield  {author} {\bibinfo {author} {\bibfnamefont {S.}~\bibnamefont {Vagnozzi}}, \bibinfo {author} {\bibfnamefont {A.}~\bibnamefont {Loeb}}, \ and\ \bibinfo {author} {\bibfnamefont {M.}~\bibnamefont {Moresco}},\ }\href {\doibase 10.3847/1538-4357/abd4df} {\bibfield  {journal} {\bibinfo  {journal} {Astrophys. J.}\ }\textbf {\bibinfo {volume} {908}},\ \bibinfo {pages} {84} (\bibinfo {year} {2021}{\natexlab{b}})},\ \Eprint {http://arxiv.org/abs/2011.11645} {arXiv:2011.11645 [astro-ph.CO]} \BibitemShut {NoStop}%
\bibitem [{\citenamefont {Di~Valentino}\ \emph {et~al.}(2021{\natexlab{d}})\citenamefont {Di~Valentino}, \citenamefont {Melchiorri}, \citenamefont {Mena}, \citenamefont {Pan},\ and\ \citenamefont {Yang}}]{DiValentino:2020kpf}%
  \BibitemOpen
  \bibfield  {author} {\bibinfo {author} {\bibfnamefont {E.}~\bibnamefont {Di~Valentino}}, \bibinfo {author} {\bibfnamefont {A.}~\bibnamefont {Melchiorri}}, \bibinfo {author} {\bibfnamefont {O.}~\bibnamefont {Mena}}, \bibinfo {author} {\bibfnamefont {S.}~\bibnamefont {Pan}}, \ and\ \bibinfo {author} {\bibfnamefont {W.}~\bibnamefont {Yang}},\ }\href {\doibase 10.1093/mnrasl/slaa207} {\bibfield  {journal} {\bibinfo  {journal} {Mon. Not. Roy. Astron. Soc.}\ }\textbf {\bibinfo {volume} {502}},\ \bibinfo {pages} {L23} (\bibinfo {year} {2021}{\natexlab{d}})},\ \Eprint {http://arxiv.org/abs/2011.00283} {arXiv:2011.00283 [astro-ph.CO]} \BibitemShut {NoStop}%
\bibitem [{\citenamefont {Cao}\ \emph {et~al.}(2021)\citenamefont {Cao}, \citenamefont {Ryan},\ and\ \citenamefont {Ratra}}]{Cao:2021ldv}%
  \BibitemOpen
  \bibfield  {author} {\bibinfo {author} {\bibfnamefont {S.}~\bibnamefont {Cao}}, \bibinfo {author} {\bibfnamefont {J.}~\bibnamefont {Ryan}}, \ and\ \bibinfo {author} {\bibfnamefont {B.}~\bibnamefont {Ratra}},\ }\href {\doibase 10.1093/mnras/stab942} {\bibfield  {journal} {\bibinfo  {journal} {Mon. Not. Roy. Astron. Soc.}\ }\textbf {\bibinfo {volume} {504}},\ \bibinfo {pages} {300} (\bibinfo {year} {2021})},\ \Eprint {http://arxiv.org/abs/2101.08817} {arXiv:2101.08817 [astro-ph.CO]} \BibitemShut {NoStop}%
\bibitem [{\citenamefont {Dhawan}\ \emph {et~al.}(2021)\citenamefont {Dhawan}, \citenamefont {Alsing},\ and\ \citenamefont {Vagnozzi}}]{Dhawan:2021mel}%
  \BibitemOpen
  \bibfield  {author} {\bibinfo {author} {\bibfnamefont {S.}~\bibnamefont {Dhawan}}, \bibinfo {author} {\bibfnamefont {J.}~\bibnamefont {Alsing}}, \ and\ \bibinfo {author} {\bibfnamefont {S.}~\bibnamefont {Vagnozzi}},\ }\href {\doibase 10.1093/mnrasl/slab058} {\bibfield  {journal} {\bibinfo  {journal} {Mon. Not. Roy. Astron. Soc.}\ }\textbf {\bibinfo {volume} {506}},\ \bibinfo {pages} {L1} (\bibinfo {year} {2021})},\ \Eprint {http://arxiv.org/abs/2104.02485} {arXiv:2104.02485 [astro-ph.CO]} \BibitemShut {NoStop}%
\bibitem [{\citenamefont {Dinda}(2022)}]{Dinda:2021ffa}%
  \BibitemOpen
  \bibfield  {author} {\bibinfo {author} {\bibfnamefont {B.~R.}\ \bibnamefont {Dinda}},\ }\href {\doibase 10.1103/PhysRevD.105.063524} {\bibfield  {journal} {\bibinfo  {journal} {Phys. Rev. D}\ }\textbf {\bibinfo {volume} {105}},\ \bibinfo {pages} {063524} (\bibinfo {year} {2022})},\ \Eprint {http://arxiv.org/abs/2106.02963} {arXiv:2106.02963 [astro-ph.CO]} \BibitemShut {NoStop}%
\bibitem [{\citenamefont {Gonzalez}\ \emph {et~al.}(2021)\citenamefont {Gonzalez}, \citenamefont {Benetti}, \citenamefont {von Marttens},\ and\ \citenamefont {Alcaniz}}]{Gonzalez:2021ojp}%
  \BibitemOpen
  \bibfield  {author} {\bibinfo {author} {\bibfnamefont {J.~E.}\ \bibnamefont {Gonzalez}}, \bibinfo {author} {\bibfnamefont {M.}~\bibnamefont {Benetti}}, \bibinfo {author} {\bibfnamefont {R.}~\bibnamefont {von Marttens}}, \ and\ \bibinfo {author} {\bibfnamefont {J.}~\bibnamefont {Alcaniz}},\ }\href {\doibase 10.1088/1475-7516/2021/11/060} {\bibfield  {journal} {\bibinfo  {journal} {JCAP}\ }\textbf {\bibinfo {volume} {11}},\ \bibinfo {pages} {060} (\bibinfo {year} {2021})},\ \Eprint {http://arxiv.org/abs/2104.13455} {arXiv:2104.13455 [astro-ph.CO]} \BibitemShut {NoStop}%
\bibitem [{\citenamefont {Akarsu}\ \emph {et~al.}(2023{\natexlab{c}})\citenamefont {Akarsu}, \citenamefont {Di~Valentino}, \citenamefont {Kumar}, \citenamefont {Ozyigit},\ and\ \citenamefont {Sharma}}]{Akarsu:2021max}%
  \BibitemOpen
  \bibfield  {author} {\bibinfo {author} {\bibfnamefont {O.}~\bibnamefont {Akarsu}}, \bibinfo {author} {\bibfnamefont {E.}~\bibnamefont {Di~Valentino}}, \bibinfo {author} {\bibfnamefont {S.}~\bibnamefont {Kumar}}, \bibinfo {author} {\bibfnamefont {M.}~\bibnamefont {Ozyigit}}, \ and\ \bibinfo {author} {\bibfnamefont {S.}~\bibnamefont {Sharma}},\ }\href {\doibase 10.1016/j.dark.2022.101162} {\bibfield  {journal} {\bibinfo  {journal} {Phys. Dark Univ.}\ }\textbf {\bibinfo {volume} {39}},\ \bibinfo {pages} {101162} (\bibinfo {year} {2023}{\natexlab{c}})},\ \Eprint {http://arxiv.org/abs/2112.07807} {arXiv:2112.07807 [astro-ph.CO]} \BibitemShut {NoStop}%
\bibitem [{\citenamefont {Cao}\ and\ \citenamefont {Ratra}(2022)}]{Cao:2022ugh}%
  \BibitemOpen
  \bibfield  {author} {\bibinfo {author} {\bibfnamefont {S.}~\bibnamefont {Cao}}\ and\ \bibinfo {author} {\bibfnamefont {B.}~\bibnamefont {Ratra}},\ }\href {\doibase 10.1093/mnras/stac1184} {\bibfield  {journal} {\bibinfo  {journal} {Mon. Not. Roy. Astron. Soc.}\ }\textbf {\bibinfo {volume} {513}},\ \bibinfo {pages} {5686} (\bibinfo {year} {2022})},\ \Eprint {http://arxiv.org/abs/2203.10825} {arXiv:2203.10825 [astro-ph.CO]} \BibitemShut {NoStop}%
\bibitem [{\citenamefont {Glanville}\ \emph {et~al.}(2022)\citenamefont {Glanville}, \citenamefont {Howlett},\ and\ \citenamefont {Davis}}]{Glanville:2022xes}%
  \BibitemOpen
  \bibfield  {author} {\bibinfo {author} {\bibfnamefont {A.}~\bibnamefont {Glanville}}, \bibinfo {author} {\bibfnamefont {C.}~\bibnamefont {Howlett}}, \ and\ \bibinfo {author} {\bibfnamefont {T.~M.}\ \bibnamefont {Davis}},\ }\href {\doibase 10.1093/mnras/stac2891} {\bibfield  {journal} {\bibinfo  {journal} {Mon. Not. Roy. Astron. Soc.}\ }\textbf {\bibinfo {volume} {517}},\ \bibinfo {pages} {3087} (\bibinfo {year} {2022})},\ \Eprint {http://arxiv.org/abs/2205.05892} {arXiv:2205.05892 [astro-ph.CO]} \BibitemShut {NoStop}%
\bibitem [{\citenamefont {Bel}\ \emph {et~al.}(2022)\citenamefont {Bel}, \citenamefont {Larena}, \citenamefont {Maartens}, \citenamefont {Marinoni},\ and\ \citenamefont {Perenon}}]{Bel:2022iuf}%
  \BibitemOpen
  \bibfield  {author} {\bibinfo {author} {\bibfnamefont {J.}~\bibnamefont {Bel}}, \bibinfo {author} {\bibfnamefont {J.}~\bibnamefont {Larena}}, \bibinfo {author} {\bibfnamefont {R.}~\bibnamefont {Maartens}}, \bibinfo {author} {\bibfnamefont {C.}~\bibnamefont {Marinoni}}, \ and\ \bibinfo {author} {\bibfnamefont {L.}~\bibnamefont {Perenon}},\ }\href {\doibase 10.1088/1475-7516/2022/09/076} {\bibfield  {journal} {\bibinfo  {journal} {JCAP}\ }\textbf {\bibinfo {volume} {09}},\ \bibinfo {pages} {076} (\bibinfo {year} {2022})},\ \Eprint {http://arxiv.org/abs/2206.03059} {arXiv:2206.03059 [astro-ph.CO]} \BibitemShut {NoStop}%
\bibitem [{\citenamefont {Yang}\ \emph {et~al.}(2023)\citenamefont {Yang}, \citenamefont {Giar\`e}, \citenamefont {Pan}, \citenamefont {Di~Valentino}, \citenamefont {Melchiorri},\ and\ \citenamefont {Silk}}]{Yang:2022kho}%
  \BibitemOpen
  \bibfield  {author} {\bibinfo {author} {\bibfnamefont {W.}~\bibnamefont {Yang}}, \bibinfo {author} {\bibfnamefont {W.}~\bibnamefont {Giar\`e}}, \bibinfo {author} {\bibfnamefont {S.}~\bibnamefont {Pan}}, \bibinfo {author} {\bibfnamefont {E.}~\bibnamefont {Di~Valentino}}, \bibinfo {author} {\bibfnamefont {A.}~\bibnamefont {Melchiorri}}, \ and\ \bibinfo {author} {\bibfnamefont {J.}~\bibnamefont {Silk}},\ }\href {\doibase 10.1103/PhysRevD.107.063509} {\bibfield  {journal} {\bibinfo  {journal} {Phys. Rev. D}\ }\textbf {\bibinfo {volume} {107}},\ \bibinfo {pages} {063509} (\bibinfo {year} {2023})},\ \Eprint {http://arxiv.org/abs/2210.09865} {arXiv:2210.09865 [astro-ph.CO]} \BibitemShut {NoStop}%
\bibitem [{\citenamefont {Stevens}\ \emph {et~al.}(2023)\citenamefont {Stevens}, \citenamefont {Khoraminezhad},\ and\ \citenamefont {Saito}}]{Stevens:2022evv}%
  \BibitemOpen
  \bibfield  {author} {\bibinfo {author} {\bibfnamefont {J.}~\bibnamefont {Stevens}}, \bibinfo {author} {\bibfnamefont {H.}~\bibnamefont {Khoraminezhad}}, \ and\ \bibinfo {author} {\bibfnamefont {S.}~\bibnamefont {Saito}},\ }\href {\doibase 10.1088/1475-7516/2023/07/046} {\bibfield  {journal} {\bibinfo  {journal} {JCAP}\ }\textbf {\bibinfo {volume} {07}},\ \bibinfo {pages} {046} (\bibinfo {year} {2023})},\ \Eprint {http://arxiv.org/abs/2212.09804} {arXiv:2212.09804 [astro-ph.CO]} \BibitemShut {NoStop}%
\bibitem [{\citenamefont {Favale}\ \emph {et~al.}(2023)\citenamefont {Favale}, \citenamefont {G\'omez-Valent},\ and\ \citenamefont {Migliaccio}}]{Favale:2023lnp}%
  \BibitemOpen
  \bibfield  {author} {\bibinfo {author} {\bibfnamefont {A.}~\bibnamefont {Favale}}, \bibinfo {author} {\bibfnamefont {A.}~\bibnamefont {G\'omez-Valent}}, \ and\ \bibinfo {author} {\bibfnamefont {M.}~\bibnamefont {Migliaccio}},\ }\href {\doibase 10.1093/mnras/stad1621} {\bibfield  {journal} {\bibinfo  {journal} {Mon. Not. Roy. Astron. Soc.}\ }\textbf {\bibinfo {volume} {523}},\ \bibinfo {pages} {3406} (\bibinfo {year} {2023})},\ \Eprint {http://arxiv.org/abs/2301.09591} {arXiv:2301.09591 [astro-ph.CO]} \BibitemShut {NoStop}%
\bibitem [{\citenamefont {Escamilla}\ \emph {et~al.}(2023)\citenamefont {Escamilla}, \citenamefont {Giar\`e}, \citenamefont {Di~Valentino}, \citenamefont {Nunes},\ and\ \citenamefont {Vagnozzi}}]{Escamilla:2023oce}%
  \BibitemOpen
  \bibfield  {author} {\bibinfo {author} {\bibfnamefont {L.~A.}\ \bibnamefont {Escamilla}}, \bibinfo {author} {\bibfnamefont {W.}~\bibnamefont {Giar\`e}}, \bibinfo {author} {\bibfnamefont {E.}~\bibnamefont {Di~Valentino}}, \bibinfo {author} {\bibfnamefont {R.~C.}\ \bibnamefont {Nunes}}, \ and\ \bibinfo {author} {\bibfnamefont {S.}~\bibnamefont {Vagnozzi}},\ }\href@noop {} {\  (\bibinfo {year} {2023})},\ \Eprint {http://arxiv.org/abs/2307.14802} {arXiv:2307.14802 [astro-ph.CO]} \BibitemShut {NoStop}%
\bibitem [{\citenamefont {Semenaite}\ \emph {et~al.}(2023)\citenamefont {Semenaite}, \citenamefont {S\'anchez}, \citenamefont {Pezzotta}, \citenamefont {Hou}, \citenamefont {Eggemeier}, \citenamefont {Crocce}, \citenamefont {Zhao}, \citenamefont {Brownstein}, \citenamefont {Rossi},\ and\ \citenamefont {Schneider}}]{Semenaite:2022unt}%
  \BibitemOpen
  \bibfield  {author} {\bibinfo {author} {\bibfnamefont {A.}~\bibnamefont {Semenaite}}, \bibinfo {author} {\bibfnamefont {A.~G.}\ \bibnamefont {S\'anchez}}, \bibinfo {author} {\bibfnamefont {A.}~\bibnamefont {Pezzotta}}, \bibinfo {author} {\bibfnamefont {J.}~\bibnamefont {Hou}}, \bibinfo {author} {\bibfnamefont {A.}~\bibnamefont {Eggemeier}}, \bibinfo {author} {\bibfnamefont {M.}~\bibnamefont {Crocce}}, \bibinfo {author} {\bibfnamefont {C.}~\bibnamefont {Zhao}}, \bibinfo {author} {\bibfnamefont {J.~R.}\ \bibnamefont {Brownstein}}, \bibinfo {author} {\bibfnamefont {G.}~\bibnamefont {Rossi}}, \ and\ \bibinfo {author} {\bibfnamefont {D.~P.}\ \bibnamefont {Schneider}},\ }\href {\doibase 10.1093/mnras/stad849} {\bibfield  {journal} {\bibinfo  {journal} {Mon. Not. Roy. Astron. Soc.}\ }\textbf {\bibinfo {volume} {521}},\ \bibinfo {pages} {5013} (\bibinfo {year} {2023})},\ \Eprint {http://arxiv.org/abs/2210.07304} {arXiv:2210.07304 [astro-ph.CO]} \BibitemShut {NoStop}%
\bibitem [{\citenamefont {Carrilho}\ \emph {et~al.}(2023)\citenamefont {Carrilho}, \citenamefont {Moretti},\ and\ \citenamefont {Pourtsidou}}]{Carrilho:2022mon}%
  \BibitemOpen
  \bibfield  {author} {\bibinfo {author} {\bibfnamefont {P.}~\bibnamefont {Carrilho}}, \bibinfo {author} {\bibfnamefont {C.}~\bibnamefont {Moretti}}, \ and\ \bibinfo {author} {\bibfnamefont {A.}~\bibnamefont {Pourtsidou}},\ }\href {\doibase 10.1088/1475-7516/2023/01/028} {\bibfield  {journal} {\bibinfo  {journal} {JCAP}\ }\textbf {\bibinfo {volume} {01}},\ \bibinfo {pages} {028} (\bibinfo {year} {2023})},\ \Eprint {http://arxiv.org/abs/2207.14784} {arXiv:2207.14784 [astro-ph.CO]} \BibitemShut {NoStop}%
\bibitem [{\citenamefont {Chudaykin}\ \emph {et~al.}(2021)\citenamefont {Chudaykin}, \citenamefont {Dolgikh},\ and\ \citenamefont {Ivanov}}]{Chudaykin:2020ghx}%
  \BibitemOpen
  \bibfield  {author} {\bibinfo {author} {\bibfnamefont {A.}~\bibnamefont {Chudaykin}}, \bibinfo {author} {\bibfnamefont {K.}~\bibnamefont {Dolgikh}}, \ and\ \bibinfo {author} {\bibfnamefont {M.~M.}\ \bibnamefont {Ivanov}},\ }\href {\doibase 10.1103/PhysRevD.103.023507} {\bibfield  {journal} {\bibinfo  {journal} {Phys. Rev. D}\ }\textbf {\bibinfo {volume} {103}},\ \bibinfo {pages} {023507} (\bibinfo {year} {2021})},\ \Eprint {http://arxiv.org/abs/2009.10106} {arXiv:2009.10106 [astro-ph.CO]} \BibitemShut {NoStop}%
\bibitem [{\citenamefont {D'Amico}\ \emph {et~al.}(2021)\citenamefont {D'Amico}, \citenamefont {Senatore},\ and\ \citenamefont {Zhang}}]{DAmico:2020kxu}%
  \BibitemOpen
  \bibfield  {author} {\bibinfo {author} {\bibfnamefont {G.}~\bibnamefont {D'Amico}}, \bibinfo {author} {\bibfnamefont {L.}~\bibnamefont {Senatore}}, \ and\ \bibinfo {author} {\bibfnamefont {P.}~\bibnamefont {Zhang}},\ }\href {\doibase 10.1088/1475-7516/2021/01/006} {\bibfield  {journal} {\bibinfo  {journal} {JCAP}\ }\textbf {\bibinfo {volume} {01}},\ \bibinfo {pages} {006} (\bibinfo {year} {2021})},\ \Eprint {http://arxiv.org/abs/2003.07956} {arXiv:2003.07956 [astro-ph.CO]} \BibitemShut {NoStop}%
\bibitem [{\citenamefont {Brieden}\ \emph {et~al.}(2022)\citenamefont {Brieden}, \citenamefont {Gil-Mar\'\i{}n},\ and\ \citenamefont {Verde}}]{Brieden:2022lsd}%
  \BibitemOpen
  \bibfield  {author} {\bibinfo {author} {\bibfnamefont {S.}~\bibnamefont {Brieden}}, \bibinfo {author} {\bibfnamefont {H.}~\bibnamefont {Gil-Mar\'\i{}n}}, \ and\ \bibinfo {author} {\bibfnamefont {L.}~\bibnamefont {Verde}},\ }\href {\doibase 10.1088/1475-7516/2022/08/024} {\bibfield  {journal} {\bibinfo  {journal} {JCAP}\ }\textbf {\bibinfo {volume} {08}},\ \bibinfo {pages} {024} (\bibinfo {year} {2022})},\ \Eprint {http://arxiv.org/abs/2204.11868} {arXiv:2204.11868 [astro-ph.CO]} \BibitemShut {NoStop}%
\bibitem [{\citenamefont {Abbott}\ \emph {et~al.}(2019)\citenamefont {Abbott} \emph {et~al.}}]{DES:2018ufa}%
  \BibitemOpen
  \bibfield  {author} {\bibinfo {author} {\bibfnamefont {T.~M.~C.}\ \bibnamefont {Abbott}} \emph {et~al.} (\bibinfo {collaboration} {DES}),\ }\href {\doibase 10.1103/PhysRevD.99.123505} {\bibfield  {journal} {\bibinfo  {journal} {Phys. Rev. D}\ }\textbf {\bibinfo {volume} {99}},\ \bibinfo {pages} {123505} (\bibinfo {year} {2019})},\ \Eprint {http://arxiv.org/abs/1810.02499} {arXiv:1810.02499 [astro-ph.CO]} \BibitemShut {NoStop}%
\bibitem [{\citenamefont {Moresco}\ \emph {et~al.}(2016)\citenamefont {Moresco}, \citenamefont {Jimenez}, \citenamefont {Verde}, \citenamefont {Cimatti}, \citenamefont {Pozzetti}, \citenamefont {Maraston},\ and\ \citenamefont {Thomas}}]{Moresco:2016nqq}%
  \BibitemOpen
  \bibfield  {author} {\bibinfo {author} {\bibfnamefont {M.}~\bibnamefont {Moresco}}, \bibinfo {author} {\bibfnamefont {R.}~\bibnamefont {Jimenez}}, \bibinfo {author} {\bibfnamefont {L.}~\bibnamefont {Verde}}, \bibinfo {author} {\bibfnamefont {A.}~\bibnamefont {Cimatti}}, \bibinfo {author} {\bibfnamefont {L.}~\bibnamefont {Pozzetti}}, \bibinfo {author} {\bibfnamefont {C.}~\bibnamefont {Maraston}}, \ and\ \bibinfo {author} {\bibfnamefont {D.}~\bibnamefont {Thomas}},\ }\href {\doibase 10.1088/1475-7516/2016/12/039} {\bibfield  {journal} {\bibinfo  {journal} {JCAP}\ }\textbf {\bibinfo {volume} {12}},\ \bibinfo {pages} {039} (\bibinfo {year} {2016})},\ \Eprint {http://arxiv.org/abs/1604.00183} {arXiv:1604.00183 [astro-ph.CO]} \BibitemShut {NoStop}%
\bibitem [{\citenamefont {Yang}\ \emph {et~al.}(2021{\natexlab{b}})\citenamefont {Yang}, \citenamefont {Di~Valentino}, \citenamefont {Pan}, \citenamefont {Wu},\ and\ \citenamefont {Lu}}]{Yang:2021flj}%
  \BibitemOpen
  \bibfield  {author} {\bibinfo {author} {\bibfnamefont {W.}~\bibnamefont {Yang}}, \bibinfo {author} {\bibfnamefont {E.}~\bibnamefont {Di~Valentino}}, \bibinfo {author} {\bibfnamefont {S.}~\bibnamefont {Pan}}, \bibinfo {author} {\bibfnamefont {Y.}~\bibnamefont {Wu}}, \ and\ \bibinfo {author} {\bibfnamefont {J.}~\bibnamefont {Lu}},\ }\href {\doibase 10.1093/mnras/staa3914} {\bibfield  {journal} {\bibinfo  {journal} {Mon. Not. Roy. Astron. Soc.}\ }\textbf {\bibinfo {volume} {501}},\ \bibinfo {pages} {5845} (\bibinfo {year} {2021}{\natexlab{b}})},\ \Eprint {http://arxiv.org/abs/2101.02168} {arXiv:2101.02168 [astro-ph.CO]} \BibitemShut {NoStop}%
\bibitem [{\citenamefont {Moresco}\ \emph {et~al.}(2022)\citenamefont {Moresco} \emph {et~al.}}]{Moresco:2022phi}%
  \BibitemOpen
  \bibfield  {author} {\bibinfo {author} {\bibfnamefont {M.}~\bibnamefont {Moresco}} \emph {et~al.},\ }\href {\doibase 10.1007/s41114-022-00040-z} {\bibfield  {journal} {\bibinfo  {journal} {Living Rev. Rel.}\ }\textbf {\bibinfo {volume} {25}},\ \bibinfo {pages} {6} (\bibinfo {year} {2022})},\ \Eprint {http://arxiv.org/abs/2201.07241} {arXiv:2201.07241 [astro-ph.CO]} \BibitemShut {NoStop}%
\bibitem [{\citenamefont {Vagnozzi}\ \emph {et~al.}(2022)\citenamefont {Vagnozzi}, \citenamefont {Pacucci},\ and\ \citenamefont {Loeb}}]{Vagnozzi:2021tjv}%
  \BibitemOpen
  \bibfield  {author} {\bibinfo {author} {\bibfnamefont {S.}~\bibnamefont {Vagnozzi}}, \bibinfo {author} {\bibfnamefont {F.}~\bibnamefont {Pacucci}}, \ and\ \bibinfo {author} {\bibfnamefont {A.}~\bibnamefont {Loeb}},\ }\href {\doibase 10.1016/j.jheap.2022.07.004} {\bibfield  {journal} {\bibinfo  {journal} {JHEAp}\ }\textbf {\bibinfo {volume} {36}},\ \bibinfo {pages} {27} (\bibinfo {year} {2022})},\ \Eprint {http://arxiv.org/abs/2105.10421} {arXiv:2105.10421 [astro-ph.CO]} \BibitemShut {NoStop}%
\bibitem [{\citenamefont {Bargiacchi}\ \emph {et~al.}(2022)\citenamefont {Bargiacchi}, \citenamefont {Benetti}, \citenamefont {Capozziello}, \citenamefont {Lusso}, \citenamefont {Risaliti},\ and\ \citenamefont {Signorini}}]{Bargiacchi:2021hdp}%
  \BibitemOpen
  \bibfield  {author} {\bibinfo {author} {\bibfnamefont {G.}~\bibnamefont {Bargiacchi}}, \bibinfo {author} {\bibfnamefont {M.}~\bibnamefont {Benetti}}, \bibinfo {author} {\bibfnamefont {S.}~\bibnamefont {Capozziello}}, \bibinfo {author} {\bibfnamefont {E.}~\bibnamefont {Lusso}}, \bibinfo {author} {\bibfnamefont {G.}~\bibnamefont {Risaliti}}, \ and\ \bibinfo {author} {\bibfnamefont {M.}~\bibnamefont {Signorini}},\ }\href {\doibase 10.1093/mnras/stac1941} {\bibfield  {journal} {\bibinfo  {journal} {Mon. Not. Roy. Astron. Soc.}\ }\textbf {\bibinfo {volume} {515}},\ \bibinfo {pages} {1795} (\bibinfo {year} {2022})},\ \Eprint {http://arxiv.org/abs/2111.02420} {arXiv:2111.02420 [astro-ph.CO]} \BibitemShut {NoStop}%
\bibitem [{\citenamefont {Grillo}\ \emph {et~al.}(2020)\citenamefont {Grillo}, \citenamefont {Rosati}, \citenamefont {Suyu}, \citenamefont {Caminha}, \citenamefont {Mercurio},\ and\ \citenamefont {Halkola}}]{Grillo:2020yvj}%
  \BibitemOpen
  \bibfield  {author} {\bibinfo {author} {\bibfnamefont {C.}~\bibnamefont {Grillo}}, \bibinfo {author} {\bibfnamefont {P.}~\bibnamefont {Rosati}}, \bibinfo {author} {\bibfnamefont {S.~H.}\ \bibnamefont {Suyu}}, \bibinfo {author} {\bibfnamefont {G.~B.}\ \bibnamefont {Caminha}}, \bibinfo {author} {\bibfnamefont {A.}~\bibnamefont {Mercurio}}, \ and\ \bibinfo {author} {\bibfnamefont {A.}~\bibnamefont {Halkola}},\ }\href {\doibase 10.3847/1538-4357/ab9a4c} {\bibfield  {journal} {\bibinfo  {journal} {Astrophys. J.}\ }\textbf {\bibinfo {volume} {898}},\ \bibinfo {pages} {87} (\bibinfo {year} {2020})},\ \Eprint {http://arxiv.org/abs/2001.02232} {arXiv:2001.02232 [astro-ph.CO]} \BibitemShut {NoStop}%
\bibitem [{\citenamefont {Cao}\ \emph {et~al.}(2022)\citenamefont {Cao}, \citenamefont {Ryan},\ and\ \citenamefont {Ratra}}]{Cao:2021cix}%
  \BibitemOpen
  \bibfield  {author} {\bibinfo {author} {\bibfnamefont {S.}~\bibnamefont {Cao}}, \bibinfo {author} {\bibfnamefont {J.}~\bibnamefont {Ryan}}, \ and\ \bibinfo {author} {\bibfnamefont {B.}~\bibnamefont {Ratra}},\ }\href {\doibase 10.1093/mnras/stab3304} {\bibfield  {journal} {\bibinfo  {journal} {Mon. Not. Roy. Astron. Soc.}\ }\textbf {\bibinfo {volume} {509}},\ \bibinfo {pages} {4745} (\bibinfo {year} {2022})},\ \Eprint {http://arxiv.org/abs/2109.01987} {arXiv:2109.01987 [astro-ph.CO]} \BibitemShut {NoStop}%
\bibitem [{\citenamefont {Hogg}(2023)}]{Hogg:2023khs}%
  \BibitemOpen
  \bibfield  {author} {\bibinfo {author} {\bibfnamefont {N.~B.}\ \bibnamefont {Hogg}},\ }\href@noop {} {\  (\bibinfo {year} {2023})},\ \Eprint {http://arxiv.org/abs/2310.11977} {arXiv:2310.11977 [astro-ph.CO]} \BibitemShut {NoStop}%
\bibitem [{\citenamefont {Adame}\ \emph {et~al.}(2024)\citenamefont {Adame} \emph {et~al.}}]{DESI:2024mwx}%
  \BibitemOpen
  \bibfield  {author} {\bibinfo {author} {\bibfnamefont {A.~G.}\ \bibnamefont {Adame}} \emph {et~al.} (\bibinfo {collaboration} {DESI}),\ }\href@noop {} {\  (\bibinfo {year} {2024})},\ \Eprint {http://arxiv.org/abs/2404.03002} {arXiv:2404.03002 [astro-ph.CO]} \BibitemShut {NoStop}%
\bibitem [{\citenamefont {Torrado}\ and\ \citenamefont {Lewis}(2021)}]{Torrado:2020dgo}%
  \BibitemOpen
  \bibfield  {author} {\bibinfo {author} {\bibfnamefont {J.}~\bibnamefont {Torrado}}\ and\ \bibinfo {author} {\bibfnamefont {A.}~\bibnamefont {Lewis}},\ }\href {\doibase 10.1088/1475-7516/2021/05/057} {\bibfield  {journal} {\bibinfo  {journal} {JCAP}\ }\textbf {\bibinfo {volume} {05}},\ \bibinfo {pages} {057} (\bibinfo {year} {2021})},\ \Eprint {http://arxiv.org/abs/2005.05290} {arXiv:2005.05290 [astro-ph.IM]} \BibitemShut {NoStop}%
\bibitem [{\citenamefont {Lewis}\ \emph {et~al.}(2000)\citenamefont {Lewis}, \citenamefont {Challinor},\ and\ \citenamefont {Lasenby}}]{Lewis:1999bs}%
  \BibitemOpen
  \bibfield  {author} {\bibinfo {author} {\bibfnamefont {A.}~\bibnamefont {Lewis}}, \bibinfo {author} {\bibfnamefont {A.}~\bibnamefont {Challinor}}, \ and\ \bibinfo {author} {\bibfnamefont {A.}~\bibnamefont {Lasenby}},\ }\href {\doibase 10.1086/309179} {\bibfield  {journal} {\bibinfo  {journal} {Astrophys. J.}\ }\textbf {\bibinfo {volume} {538}},\ \bibinfo {pages} {473} (\bibinfo {year} {2000})},\ \Eprint {http://arxiv.org/abs/astro-ph/9911177} {arXiv:astro-ph/9911177} \BibitemShut {NoStop}%
\bibitem [{\citenamefont {Howlett}\ \emph {et~al.}(2012)\citenamefont {Howlett}, \citenamefont {Lewis}, \citenamefont {Hall},\ and\ \citenamefont {Challinor}}]{Howlett:2012mh}%
  \BibitemOpen
  \bibfield  {author} {\bibinfo {author} {\bibfnamefont {C.}~\bibnamefont {Howlett}}, \bibinfo {author} {\bibfnamefont {A.}~\bibnamefont {Lewis}}, \bibinfo {author} {\bibfnamefont {A.}~\bibnamefont {Hall}}, \ and\ \bibinfo {author} {\bibfnamefont {A.}~\bibnamefont {Challinor}},\ }\href {\doibase 10.1088/1475-7516/2012/04/027} {\bibfield  {journal} {\bibinfo  {journal} {JCAP}\ }\textbf {\bibinfo {volume} {04}},\ \bibinfo {pages} {027} (\bibinfo {year} {2012})},\ \Eprint {http://arxiv.org/abs/1201.3654} {arXiv:1201.3654 [astro-ph.CO]} \BibitemShut {NoStop}%
\bibitem [{\citenamefont {Aghanim}\ \emph {et~al.}(2020{\natexlab{c}})\citenamefont {Aghanim} \emph {et~al.}}]{Planck:2019nip}%
  \BibitemOpen
  \bibfield  {author} {\bibinfo {author} {\bibfnamefont {N.}~\bibnamefont {Aghanim}} \emph {et~al.} (\bibinfo {collaboration} {Planck}),\ }\href {\doibase 10.1051/0004-6361/201936386} {\bibfield  {journal} {\bibinfo  {journal} {Astron. Astrophys.}\ }\textbf {\bibinfo {volume} {641}},\ \bibinfo {pages} {A5} (\bibinfo {year} {2020}{\natexlab{c}})},\ \Eprint {http://arxiv.org/abs/1907.12875} {arXiv:1907.12875 [astro-ph.CO]} \BibitemShut {NoStop}%
\bibitem [{\citenamefont {Aghanim}\ \emph {et~al.}(2020{\natexlab{d}})\citenamefont {Aghanim} \emph {et~al.}}]{Planck:2018lbu}%
  \BibitemOpen
  \bibfield  {author} {\bibinfo {author} {\bibfnamefont {N.}~\bibnamefont {Aghanim}} \emph {et~al.} (\bibinfo {collaboration} {Planck}),\ }\href {\doibase 10.1051/0004-6361/201833886} {\bibfield  {journal} {\bibinfo  {journal} {Astron. Astrophys.}\ }\textbf {\bibinfo {volume} {641}},\ \bibinfo {pages} {A8} (\bibinfo {year} {2020}{\natexlab{d}})},\ \Eprint {http://arxiv.org/abs/1807.06210} {arXiv:1807.06210 [astro-ph.CO]} \BibitemShut {NoStop}%
\bibitem [{\citenamefont {Aubourg}\ \emph {et~al.}(2015)\citenamefont {Aubourg} \emph {et~al.}}]{Aubourg:2014yra}%
  \BibitemOpen
  \bibfield  {author} {\bibinfo {author} {\bibfnamefont {E.}~\bibnamefont {Aubourg}} \emph {et~al.} (\bibinfo {collaboration} {BOSS}),\ }\href {\doibase 10.1103/PhysRevD.92.123516} {\bibfield  {journal} {\bibinfo  {journal} {Phys. Rev. D}\ }\textbf {\bibinfo {volume} {92}},\ \bibinfo {pages} {123516} (\bibinfo {year} {2015})},\ \Eprint {http://arxiv.org/abs/1411.1074} {arXiv:1411.1074 [astro-ph.CO]} \BibitemShut {NoStop}%
\bibitem [{\citenamefont {Hu}\ and\ \citenamefont {Dodelson}(2002)}]{Hu:2001bc}%
  \BibitemOpen
  \bibfield  {author} {\bibinfo {author} {\bibfnamefont {W.}~\bibnamefont {Hu}}\ and\ \bibinfo {author} {\bibfnamefont {S.}~\bibnamefont {Dodelson}},\ }\href {\doibase 10.1146/annurev.astro.40.060401.093926} {\bibfield  {journal} {\bibinfo  {journal} {Ann. Rev. Astron. Astrophys.}\ }\textbf {\bibinfo {volume} {40}},\ \bibinfo {pages} {171} (\bibinfo {year} {2002})},\ \Eprint {http://arxiv.org/abs/astro-ph/0110414} {arXiv:astro-ph/0110414} \BibitemShut {NoStop}%
\bibitem [{\citenamefont {Liddle}\ and\ \citenamefont {Leach}(2003)}]{Liddle:2003as}%
  \BibitemOpen
  \bibfield  {author} {\bibinfo {author} {\bibfnamefont {A.~R.}\ \bibnamefont {Liddle}}\ and\ \bibinfo {author} {\bibfnamefont {S.~M.}\ \bibnamefont {Leach}},\ }\href {\doibase 10.1103/PhysRevD.68.103503} {\bibfield  {journal} {\bibinfo  {journal} {Phys. Rev. D}\ }\textbf {\bibinfo {volume} {68}},\ \bibinfo {pages} {103503} (\bibinfo {year} {2003})},\ \Eprint {http://arxiv.org/abs/astro-ph/0305263} {arXiv:astro-ph/0305263} \BibitemShut {NoStop}%
\bibitem [{\citenamefont {Heavens}\ \emph {et~al.}(2017{\natexlab{a}})\citenamefont {Heavens}, \citenamefont {Fantaye}, \citenamefont {Sellentin}, \citenamefont {Eggers}, \citenamefont {Hosenie}, \citenamefont {Kroon},\ and\ \citenamefont {Mootoovaloo}}]{Heavens:2017hkr}%
  \BibitemOpen
  \bibfield  {author} {\bibinfo {author} {\bibfnamefont {A.}~\bibnamefont {Heavens}}, \bibinfo {author} {\bibfnamefont {Y.}~\bibnamefont {Fantaye}}, \bibinfo {author} {\bibfnamefont {E.}~\bibnamefont {Sellentin}}, \bibinfo {author} {\bibfnamefont {H.}~\bibnamefont {Eggers}}, \bibinfo {author} {\bibfnamefont {Z.}~\bibnamefont {Hosenie}}, \bibinfo {author} {\bibfnamefont {S.}~\bibnamefont {Kroon}}, \ and\ \bibinfo {author} {\bibfnamefont {A.}~\bibnamefont {Mootoovaloo}},\ }\href {\doibase 10.1103/PhysRevLett.119.101301} {\bibfield  {journal} {\bibinfo  {journal} {Phys. Rev. Lett.}\ }\textbf {\bibinfo {volume} {119}},\ \bibinfo {pages} {101301} (\bibinfo {year} {2017}{\natexlab{a}})},\ \Eprint {http://arxiv.org/abs/1704.03467} {arXiv:1704.03467 [astro-ph.CO]} \BibitemShut {NoStop}%
\bibitem [{\citenamefont {Heavens}\ \emph {et~al.}(2017{\natexlab{b}})\citenamefont {Heavens}, \citenamefont {Fantaye}, \citenamefont {Mootoovaloo}, \citenamefont {Eggers}, \citenamefont {Hosenie}, \citenamefont {Kroon},\ and\ \citenamefont {Sellentin}}]{Heavens:2017afc}%
  \BibitemOpen
  \bibfield  {author} {\bibinfo {author} {\bibfnamefont {A.}~\bibnamefont {Heavens}}, \bibinfo {author} {\bibfnamefont {Y.}~\bibnamefont {Fantaye}}, \bibinfo {author} {\bibfnamefont {A.}~\bibnamefont {Mootoovaloo}}, \bibinfo {author} {\bibfnamefont {H.}~\bibnamefont {Eggers}}, \bibinfo {author} {\bibfnamefont {Z.}~\bibnamefont {Hosenie}}, \bibinfo {author} {\bibfnamefont {S.}~\bibnamefont {Kroon}}, \ and\ \bibinfo {author} {\bibfnamefont {E.}~\bibnamefont {Sellentin}},\ }\href@noop {} {\  (\bibinfo {year} {2017}{\natexlab{b}})},\ \Eprint {http://arxiv.org/abs/1704.03472} {arXiv:1704.03472 [stat.CO]} \BibitemShut {NoStop}%
\bibitem [{\citenamefont {Kass}\ and\ \citenamefont {Raftery}(1995)}]{Kass:1995loi}%
  \BibitemOpen
  \bibfield  {author} {\bibinfo {author} {\bibfnamefont {R.~E.}\ \bibnamefont {Kass}}\ and\ \bibinfo {author} {\bibfnamefont {A.~E.}\ \bibnamefont {Raftery}},\ }\href {\doibase 10.1080/01621459.1995.10476572} {\bibfield  {journal} {\bibinfo  {journal} {J. Am. Statist. Assoc.}\ }\textbf {\bibinfo {volume} {90}},\ \bibinfo {pages} {773} (\bibinfo {year} {1995})}\BibitemShut {NoStop}%
\bibitem [{\citenamefont {Trotta}(2008)}]{Trotta:2008qt}%
  \BibitemOpen
  \bibfield  {author} {\bibinfo {author} {\bibfnamefont {R.}~\bibnamefont {Trotta}},\ }\href {\doibase 10.1080/00107510802066753} {\bibfield  {journal} {\bibinfo  {journal} {Contemp. Phys.}\ }\textbf {\bibinfo {volume} {49}},\ \bibinfo {pages} {71} (\bibinfo {year} {2008})},\ \Eprint {http://arxiv.org/abs/0803.4089} {arXiv:0803.4089 [astro-ph]} \BibitemShut {NoStop}%
\bibitem [{\citenamefont {Ye}\ and\ \citenamefont {Piao}(2022)}]{Ye:2022afu}%
  \BibitemOpen
  \bibfield  {author} {\bibinfo {author} {\bibfnamefont {G.}~\bibnamefont {Ye}}\ and\ \bibinfo {author} {\bibfnamefont {Y.-S.}\ \bibnamefont {Piao}},\ }\href {\doibase 10.1103/PhysRevD.106.043536} {\bibfield  {journal} {\bibinfo  {journal} {Phys. Rev. D}\ }\textbf {\bibinfo {volume} {106}},\ \bibinfo {pages} {043536} (\bibinfo {year} {2022})},\ \Eprint {http://arxiv.org/abs/2202.10055} {arXiv:2202.10055 [astro-ph.CO]} \BibitemShut {NoStop}%
\bibitem [{\citenamefont {Lidsey}\ \emph {et~al.}(1997)\citenamefont {Lidsey}, \citenamefont {Liddle}, \citenamefont {Kolb}, \citenamefont {Copeland}, \citenamefont {Barreiro},\ and\ \citenamefont {Abney}}]{Lidsey:1995np}%
  \BibitemOpen
  \bibfield  {author} {\bibinfo {author} {\bibfnamefont {J.~E.}\ \bibnamefont {Lidsey}}, \bibinfo {author} {\bibfnamefont {A.~R.}\ \bibnamefont {Liddle}}, \bibinfo {author} {\bibfnamefont {E.~W.}\ \bibnamefont {Kolb}}, \bibinfo {author} {\bibfnamefont {E.~J.}\ \bibnamefont {Copeland}}, \bibinfo {author} {\bibfnamefont {T.}~\bibnamefont {Barreiro}}, \ and\ \bibinfo {author} {\bibfnamefont {M.}~\bibnamefont {Abney}},\ }\href {\doibase 10.1103/RevModPhys.69.373} {\bibfield  {journal} {\bibinfo  {journal} {Rev. Mod. Phys.}\ }\textbf {\bibinfo {volume} {69}},\ \bibinfo {pages} {373} (\bibinfo {year} {1997})},\ \Eprint {http://arxiv.org/abs/astro-ph/9508078} {arXiv:astro-ph/9508078} \BibitemShut {NoStop}%
\bibitem [{\citenamefont {Gorbunov}\ and\ \citenamefont {Panin}(2011)}]{Gorbunov:2010bn}%
  \BibitemOpen
  \bibfield  {author} {\bibinfo {author} {\bibfnamefont {D.~S.}\ \bibnamefont {Gorbunov}}\ and\ \bibinfo {author} {\bibfnamefont {A.~G.}\ \bibnamefont {Panin}},\ }\href {\doibase 10.1016/j.physletb.2011.04.067} {\bibfield  {journal} {\bibinfo  {journal} {Phys. Lett. B}\ }\textbf {\bibinfo {volume} {700}},\ \bibinfo {pages} {157} (\bibinfo {year} {2011})},\ \Eprint {http://arxiv.org/abs/1009.2448} {arXiv:1009.2448 [hep-ph]} \BibitemShut {NoStop}%
\bibitem [{\citenamefont {Bezrukov}\ and\ \citenamefont {Gorbunov}(2012)}]{Bezrukov:2011gp}%
  \BibitemOpen
  \bibfield  {author} {\bibinfo {author} {\bibfnamefont {F.~L.}\ \bibnamefont {Bezrukov}}\ and\ \bibinfo {author} {\bibfnamefont {D.~S.}\ \bibnamefont {Gorbunov}},\ }\href {\doibase 10.1016/j.physletb.2012.06.040} {\bibfield  {journal} {\bibinfo  {journal} {Phys. Lett. B}\ }\textbf {\bibinfo {volume} {713}},\ \bibinfo {pages} {365} (\bibinfo {year} {2012})},\ \Eprint {http://arxiv.org/abs/1111.4397} {arXiv:1111.4397 [hep-ph]} \BibitemShut {NoStop}%
\bibitem [{\citenamefont {Vilenkin}(1985)}]{Vilenkin:1985md}%
  \BibitemOpen
  \bibfield  {author} {\bibinfo {author} {\bibfnamefont {A.}~\bibnamefont {Vilenkin}},\ }\href {\doibase 10.1103/PhysRevD.32.2511} {\bibfield  {journal} {\bibinfo  {journal} {Phys. Rev. D}\ }\textbf {\bibinfo {volume} {32}},\ \bibinfo {pages} {2511} (\bibinfo {year} {1985})}\BibitemShut {NoStop}%
\end{thebibliography}%

\end{document}